\documentclass[]{aa}  
\usepackage{graphicx}
\usepackage{natbib}
\usepackage[varg]{txfonts}
\usepackage{amssymb}

\usepackage{multirow}
\usepackage{txfonts}
\usepackage{longtable}
\usepackage{rotating}
\usepackage{lscape}
\usepackage{subfig}
\usepackage{siunitx}
\usepackage{xcolor}
\def\apjl{ApJL}%
\def\aap{A\&A}%
\def\apjs{ApJS}%
\begin{document}

   \title{The Lyman Alpha Reference Sample XII}
   \subtitle{Morphology of extended Lyman alpha emission in star-forming galaxies}

   \author{
           Armin Rasekh \inst{1} \thanks{ \email{armin.rasekh@astro.su.se}}
           \and
           Jens Melinder \inst{1}
           \and 
           G\"oran \"Ostlin \inst{1} 
           \and
           Matthew Hayes \inst{1}
           \and
           Edmund. C. Herenz \inst{2}
           \and
           Axel Runnholm \inst{1}
           \and
           Daniel Kunth  \inst{3}
           \and
           J. Miguel Mas Hesse  \inst{4}
           \and 
           Anne Verhamme \inst{5}
           \and
           John M. Cannon \inst{6}
          }

   \institute{Department of Astronomy, Oscar Klein Centre, Stockholm University, AlbaNova, SE-106 91 Stockholm, Sweden \\
        \and
        European Southern Observatory, Av. Alonso de Córdova 3107, 763 0355 Vitacura, Santiago, Chile \\
        \and
        Institut d’Astrophysique de Paris (IAP), 98bis Boulevard Arago, 75014 Paris, France \\
        \and
        Centro de Astrobiolog\'{\i}a (CSIC--INTA), Departamento de Astrof\'{\i}sica, 28692 Villanueva de la Ca\~nada, Spain \\ 
        \and
        Observatoire de Gen\'eve, Université de Genève, 51 Ch. des Maillettes, 1290 Versoix, Switzerland \\
        \and
        Department of Physics \& Astronomy, Macalester College, 1600 Grand Avenue, Saint Paul, MN 55105, USA \\
        \\
             }

   \date{accepted October 10, 2021}

 
  \abstract
   {}
   {
   We use  Hubble space telescope data of 45 nearby star-forming galaxies to investigate properties of Lyman-alpha (Ly$\alpha$) halos, Ly$\alpha$ morphology, and the star-forming characteristics of galaxies. We study how the morphology of Ly$\alpha$ emission is related to other Ly$\alpha$ observables. Furthermore, we study the interdependencies of Ly$\alpha$ morphological quantities.
   }
   {
   We studied the spatial extent of Ly$\alpha$ using surface brightness profiles in the following two ways i) using circular apertures and ii) within faint Ly$\alpha$ isophotes. We also measured the average intensity and the size of the regions with a high star formation rate density. The morphology of the galaxies was quantified by computing centroid position, axis ratio, and position angle in the Ly$\alpha$, ultraviolet continuum, and I band maps.
   }
   {
   We found that galaxies with more extended star-forming regions possess larger Ly$\alpha$ halos. Furthermore, galaxies with more elongated Ly$\alpha$ morphology are also more extended in Ly$\alpha$.
  Our data suggest that Ly$\alpha$ bright galaxies appear rounder in their Ly$\alpha$ morphology, and there is less of a contribution from their Ly$\alpha$ halo to their overall luminosity. We compared our results with studies at high redshift and found that whilst the Ly$\alpha$ extent in the inner regions of the galaxies in our sample are similar to the high$-z$ Ly$\alpha$ emitters (LAEs), Ly$\alpha$ halos are more extended in high$-z$ LAEs.
   }
   {
  Our analysis suggests that the Ly$\alpha$ morphology affects the measurement of other observable quantities concerning Ly$\alpha$ emission, and some of the conclusions drawn from high redshift LAEs might be biased towards galaxies with specific Ly$\alpha$ shapes. In particular, faint Ly$\alpha$ emitters have larger Ly$\alpha$ scale lengths and halo fractions. This implies that faint Ly$\alpha$ emitters are harder to detect at high redshift than previously believed.
   }

     \maketitle
%
\section{Introduction} \label{intro_sec}
The Lyman alpha (Ly$\alpha$) recombination line in hydrogen (n = 2 $\rightarrow$ n = 1, with an energy difference of 10.2 eV, equivalent to the rest frame wavelength of $\lambda_0$ = 121.567 nm) is the strongest recombination line in the intrinsic spectra of galaxies. Due to the potential strength of this line, \cite{Partridge1967ApJ} proposed to use this line for detecting and studying primaeval galaxies more than fifty years ago. Since the Ly$\alpha$ emission line from galaxies is only observable with ground-based telescopes at z > 2.5, the early searches for high$-z$ galaxies showed 
little observational progress \citep{Pritchet1994PASP}. It was only in the mid-1990s that surveys started to deliver significant numbers of Ly$\alpha$ emitting star-forming galaxies (SFGs), while previously mainly quasi-stellar objects had been found \citep{Hu1996Natur, Djorgovski1996Natur}.
\par
The Ly$\alpha$ emission line is a good tracer of ionising photons since there is a 68$\%$ chance that the ionised hydrogen atoms which capture an electron eventually emit Ly$\alpha$ photons \citep{Dijkstra2014PASA}. Indeed, studies show that Ly$\alpha$ photons are either emitted from the ionised gas around the star-forming region or active galactic nuclei (AGNs) \citep{Miley2008AARv}. In other words, Ly$\alpha$ is a valuable tool for studying the SFGs in general, and ionising photons and their origins in particular. 
\par
The Ly$\alpha$ emission line is a resonant line and is subject to scattering whenever it encounters neutral hydrogen (HI). Scattering either occurs within the host galaxy due to the presence of HI in the interstellar medium (ISM), or between the source and the observer due to the presence of HI in the circumgalactic medium (CGM) and intergalactic medium (IGM). Moreover, dust grains absorb Ly$\alpha$ photons and radiate in the far-infrared (FIR). Due to the scattering in the HI media and absorption by dust at the local scale, factors such as the geometry and the distribution of the ISM in the galaxy \citep{Giavalisco1996ApJ, Hansen2006MNRAS, Hayes2019SAAS, Marchi2019AA, Jaskot2019ApJ, Charlot1993ApJ, Atek2008AA, Verhamme2008AA, Scarlata2009ApJ, Kornei2010ApJ, Hayes2011ApJ, Matthee2016MNRAS, An2017ApJ}, the HI kinematics \citep{Kunth1998AA, Mas2003ApJ, Wofford2013ApJ, Martin2015ApJ, Herenz2016AA}, and the ISM kinematics \citep{Wofford2013ApJ, Herenz2016AA} must be considered on how to interpret Ly$\alpha$ spectra and observations. Due to the complicated radiative transfer (RT), there is no clear one-to-one correlation between the Ly$\alpha$ escape fraction and other observed quantities such as Ly$\alpha$ luminosity or nebular extinction. Nevertheless, \citet{Sobral2019} found a clear correlation between the Ly$\alpha$ equivalent width and its escape fraction in Ly$\alpha$ emitters (LAEs) at redshifts $z= [0.1\sim 2.6]$, yielding an empirical relation. They find that for galaxies to follow the observed trend, high ionisation efficiencies and low dust extinction are required, consistent with very young galaxies and intense star formation episodes.
\par
Due to the processes stated above, the Ly$\alpha$ emission is expected to be more extended than the star-forming regions where most of these photons are produced. Far-ultraviolet (FUV) instrumentations capable of spatially resolving galaxies are required to observationally confirm the existence of extended Ly$\alpha$ emission. 
In the nearby universe, it was only after the installation of the Advanced Camera for Surveys (ACS) and its Solar Blind Channel (SBC) on the Hubble Space Telescope (HST) that Ly$\alpha$ imaging of galaxies began to be carried out and \cite{Kunth2003ApJ} reported observations of ESO 350-IG038 (Haro 11) and SBS 0335--052. The analysis of ESO 338-IG04 and Haro11, presented in \cite{Hayes2005AA, Hayes2007MNRAS}, clearly showed asymmetric Ly$\alpha$ halos around both galaxies. \cite{Ostlin2009AJ} summarised the findings for the first six galaxies observed (including the three above).
\par
At high redshift, \cite{Moller1998MNRAS}, and \cite{Fynbo1999MNRAS} were the first to report more extended Ly$\alpha$ emission than ultraviolet (UV) emission for galaxies at $z > 1.9 $. Later on, \cite{Steidel2011ApJ} reported that in a stack of 92 galaxies at $z = $ 2.6 extracted from deep narrow band images, Ly$\alpha$ halos are also more extended than the UV continuum. Interestingly, extended Ly$\alpha$ emission was also seen from the stack of the subset of galaxies that showed Ly$\alpha$ central absorption in spectroscopic slits. The asymmetrical shape of the Ly$\alpha$ halos is not unexpected due to Ly$\alpha$ RT as well.
\cite{Matthee2016MNRAS} show extended and asymmetric Ly$\alpha$ emission in individual galaxies. In recent years, the Multi-Unit Spectroscopic Explorer (MUSE) instrument at the Very Large Telescope (VLT) with an increased sensitivity, resolution, and Field of View (FoV), transformed our understanding of the Ly$\alpha$ halos by enabling the extragalactic community to observe so many Ly$\alpha$ halos out to low surface brightnesses. It was shown that Ly$\alpha$ halos are ubiquitous at high redshift. It was also further demonstrated that in the high$-z$ universe, Ly$\alpha$ halos are more extended than UV continuum, and the Ly$\alpha$ halos of individual galaxies are asymmetric \citep{Wisotzki2016AA, Leclercq2017AA, Wisotzki2018Natur}. 
\par
Studying the Ly$\alpha$ morphology and the geometry of the galaxies contributes to a better understanding of Ly$\alpha$ physics and the large scale distribution of the HI scattering media. To study the morphology of the galaxies at $\lessapprox$ kpc scales., high spatial resolution data from nearby galaxies is required. Thus, to study the morphology of Ly$\alpha$ in the nearby galaxies, we used data from Lyman Alpha Reference Sample (LARS) \citep{Ostlin2014ApJ, Hayes2014ApJ}, and its extension (Melinder et al. in prep). In this paper, we make use of the high-resolution HST imaging data (FUV, UV, and optical) of this dataset. In general, LARS provides the opportunity to shed more light on the complex RT of Ly$\alpha$.
Indeed, a series of papers have been already published discussing many properties of these galaxies and how they relate to Ly$\alpha$ physics, such as studying the general properties of the sample and studying the correlation between these parameters \citep{Hayes2014ApJ}, studying the properties of the neutral ISM \citep{Pardy2014ApJ}, studying the impact of neutral ISM kinematics and geometry on Ly$\alpha$ \citep{Rivera2015ApJ}, studying the kinematic of the gas and its significance in the observed Ly$\alpha$ maps \citep{Herenz2016AA}, testing and modelling the dust content in the galaxies \citep{Bridge2018ApJ} and predicting the Ly$\alpha$ properties of the SFGs \citep{2020ApJRunnholm}.
\par
An interesting question is how the morphology and orientation of galaxies affect the Ly$\alpha$ emission in star-forming galaxies. It has already been reported that in edge-on galaxies, such as M82, H$\alpha$, IR and X-ray emission is seen to extend along the minor axis \citep{Lehnert1999ApJ}. In other words, there are geometrical effects that affect radiation at different wavelengths differently. In the context of Ly$\alpha$, orientation effect has been also predicted in different studies (e.g. \cite{Laursen2007ApJ, Verhamme2012AA, Behrens2014AA}). 
\par
This study is mainly motivated by the following questions: (i) how does the distribution of the star-forming regions and stellar populations impact the Ly$\alpha$ morphology and (ii) what is the impact of Ly$\alpha$ morphology on the global Ly$\alpha$ properties such as Ly$\alpha$ luminosity, equivalent width, escape fraction.
To investigate these questions, we studied the Ly$\alpha$ and FUV morphology of galaxies in a large sample of nearby star-forming galaxies (SFGs) by looking at the light distribution of Ly$\alpha$ and how it compares to the FUV light distribution (tracing the star-forming regions). We perform measurements of the FUV and Ly$\alpha$ halos to characterise their spatial extent. We also investigate the FUV and Ly$\alpha$ morphology of these galaxies and how they relate to observables of Ly$\alpha$ radiative transfer (e.g. $\mathrm{f_{esc}}$). We devise methods to describe the surface brightness (SB) profiles and study the morphology of these galaxies in a non-parametric approach. We also check for correlations between different parameters devised and used in this study and measurements characterising the general properties of the galaxies.
\par 
In sec. \ref{data_and_obs}, we describe the observation and briefly touch upon the data reduction and photometry, in Sec. \ref{analysis_sec}, methods used in this study are described, in Sec. \ref{resuluts_sec}, we discuss the outcomes of our analysis, in Sec. \ref{dis_sec}, we discuss our results and how they compare to the previous studies, and in Sec. \ref{conclusion_sec}, we summarise our findings and results. Finally, we assume a cosmology of ($H_0$, $\Omega_M$, $\Omega_{\Lambda}$) = (70 km/s/Mpc, 0.3, 0.7).
\section{Data} \label{data_and_obs}
\begin{table}

\caption{
        Characteristics of the galaxies chosen for this sample.
        }
\label{sample_charac_tab}        
\centering
\begin{tabular}{c c c c}
\hline
\hline
ID    & 	Redshift 	& $W_{H\alpha}$	& log($\mathrm{L_{FUV}}$)	\\
 	  & 	z 			& (Å) 			& ($\mathrm{L_\odot}$)		\\ 
\hline
\hline
LARS01  &   0.028       &   560         & 9.92              \\
LARS02  &   0.030       &   312         & 9.48              \\
LARS03  &   0.031       &   238         & 9.52              \\
LARS04  &   0.033       &   234         & 9.93              \\
LARS05  &   0.034       &   333         & 10.01             \\
LARS06  &   0.034       &   455         & 9.20              \\
LARS07  &   0.038       &   423         & 9.75              \\
LARS08  &   0.038       &   167         & 10.15             \\
LARS09  &   0.047       &   505         & 10.46             \\
LARS10  &   0.057       &   99          & 9.74              \\
LARS11  &   0.084       &   105         & 10.70             \\
LARS12  &   0.102       &   408         & 10.53             \\
LARS13  &   0.147       &   201         & 10.60             \\
LARS14  &   0.181       &   578         & 10.69             \\
ELARS01 &   0.029       &   215         & 10.08             \\
ELARS02 &   0.043       &   92          & 10.03             \\
ELARS03 &   0.035       &   65          & 10.17             \\
ELARS04 &   0.029       &   103         & 10.08             \\
ELARS05 &   0.034       &   41          & 9.99             \\
ELARS06 &   0.034       &   47          & 9.68              \\
ELARS07 &   0.035       &   268         & 9.60             \\
ELARS08 &   0.031       &   48          & 9.57             \\
ELARS09 &   0.030       &   94          & 9.56             \\
ELARS10 &   0.033       &   75          & 9.56             \\
ELARS11 &   0.030       &   58          & 9.51             \\
ELARS12 &   0.032       &   46          & 9.49             \\
ELARS13 &   0.032       &   125         & 9.47             \\
ELARS14 &   0.033       &   82          & 9.37             \\
ELARS15 &   0.035       &   54          & 9.39             \\
ELARS16 &   0.035       &   64          & 9.33             \\
ELARS17 &   0.031       &   44          & 9.24             \\
ELARS18 &   0.029       &   59          & 9.03             \\
ELARS19 &   0.031       &   133         & 9.03             \\
ELARS20 &   0.031       &   73          & 9.02             \\
ELARS21 &   0.033       &   61          & 9.01             \\
ELARS22 &   0.047       &   153         & 10.09             \\
ELARS23 &   0.051       &   49          & 10.08             \\
ELARS24 &   0.048       &   83          & 10.00             \\
ELARS25 &   0.045       &   56          & 9.83             \\
ELARS26 &   0.046       &   43          & 9.69             \\
ELARS27 &   0.045       &   47          & 9.73             \\
ELARS28 &   0.046       &   131         & 9.71              \\
T1214   &   0.026       &   1644        & 9.01             \\
T1247   &   0.049       &   530         & 10.36             \\
J1156   &   0.236       &   323         & 11.05             \\ 
\hline
\hline
\end{tabular}

\tablefoot{The first, second, third, and fourth columns represent the LARS ID in this study, the SDSS spectroscopic redshift DR8), the SDSS H$\alpha$ equivalent width (DR8), the FUV luminosity ($\mathrm{L_{FUV}} = \lambda \times L_{\lambda}/L_\odot$ at $\lambda$ = 1524/(1 + $z$) \AA) measured from GALEX observations, respectively.}
\end{table}
Our sample consists of 45 galaxies, all in the nearby universe (z < 0.24). The data used to conduct this study are all obtained from the Hubble Space Telescope (HST). The first 14 galaxies- LARS01-14 -(initial LARS sample) were observed under the program GO-12310 (PI: G. \"Ostlin), the next 28 galaxies (ELARS01-28) were observed under the programs GO-11110 (PI: S. McCandliss), and GO-13483 (PI: G. \"Ostlin), Tol1214 (Tololo 1214-277) was observed under the program GO-14923 (PI: G. \"Ostlin), Tol1247 (Tololo 1247-232) was observed under the program GO-13027 (PI: G. \"Ostlin), and finally, J1156 was observed under the program GO-13656 (PI: M. Hayes) \citep{Hayes2016ApJ}.
\par
All the galaxies in this sample are nearby SFGs that were selected based on their H$\alpha$ equivalent width, and their FUV brightness (for more information see \cite{Ostlin2014ApJ} and Melinder et al. in prep), Table \ref{sample_charac_tab} lists the redshift and H$\alpha$ equivalent width based on SDSS (DR8) spectroscopic measurements, and the FUV luminosity determined from GALEX. The first 14 galaxies (LARS01-LARS14) were selected to have H$\alpha$ equivalent width equal or higher than 100 \si{\angstrom} and UV luminosity range between log ($\mathrm{L_{FUV}/L_\odot}$) = 9.2 to ($\mathrm{L_{FUV}/L_\odot}$) = 10.7, and redshift interval of $z$ = [0.028 - 0.18].
ELARS01-28 galaxies were selected to have H$\alpha$ equivalent width higher than 40 \si{\angstrom} and the luminosity function was populated homogeneously below and above $\mathrm{L_{FUV}/L_\odot}$ = 9.6, and redshift range $z$ = [0.028 - 0.051]. In addition to these galaxies, the same type of HST observations also exist for the Tololo 1214-277, Tololo 1247-232, and J1156 (which has the highest redshift in the sample), and we added them to the sample as well.
\par
In this study, we use the Ly$\alpha$ and FUV images of the galaxies to study the observed Ly$\alpha$ emission distribution and morphology of the galaxies (using Ly$\alpha$ images) compared to the sites where these photons were produced (star-forming regions) (using FUV maps). In addition, we use the I band images (using the reddest available HST filter for each galaxy) to study the morphology and distribution of the sum of both young and old stellar populations. Global measurements, such as stellar mass, Ly$\alpha$ escape fraction ($\mathrm{f_{esc}}$), nebular reddening, Ly$\alpha$ equivalent width ($\mathrm{EW_{Ly\alpha}}$), FUV and Ly$\alpha$ luminosities are taken from Melinder et al. (in prep).
\par
The method used to obtain Ly$\alpha$ maps of the first fourteen LARS galaxies (LARS01-14), which has been extended to the entire sample, was described in \cite{Ostlin2014ApJ}. The general idea behind this method is to use HST/SBC FUV long-pass filters, one containing Ly$\alpha$ emission and at least one without, to emulate a narrow band filter centred on the line. These filters also allow estimating the stellar continuum at Ly$\alpha$. This continuum subtraction method builds on the experience from the first studies on six nearby SFGs \citep{Hayes2005AA, Hayes2007MNRAS, Ostlin2009AJ} and which led to the improved method developed and suggested by \cite{Hayes2009AJ}. This method became the strategy of LARS and is implemented in the \texttt{Lyman alpha eXtraction software} (LaXs) code. This software performs accurate and reliable pixel SED fitting in order to construct Ly$\alpha$ and FUV continuum maps (among other things) of the galaxies. In this study, all measurements on the FUV continuum are made in the LaXs-produced stellar continuum maps at Ly$\alpha$ wavelength.
\par
This work is based on the latest recalibration of the HST/ACS/SBC data \citep{Avila2019acs}. The previously published  Ly$\alpha$ maps of the LARS galaxies (LARS01-14) have been re-analysed by Melinder et al. (in prep), taking advantage of the latest HST/ACS/SBC calibration.
\par
As previously mentioned in Sec. \ref{intro_sec}, due to Ly$\alpha$ RT effects, Ly$\alpha$ emission from star-forming galaxies is usually more extended than the star-forming regions (where the majority of the Ly$\alpha$ photons are produced). In this study, we use the FUV maps of the galaxies to study the morphology of the star-forming regions. We note that as discussed in \cite{Oti2012AA, Oti2014AA}, the spatial distribution of the ionised gas (where the ionising photons are reprocessed into Ly$\alpha$) is not necessarily similar to the distribution of the FUV continuum. This is only true if the star formation is still ongoing or relatively recent ($<4$ Myr) since otherwise, the ionising photon output will have dropped significantly, and massive star winds and supernovae have had enough time to repel and push out the natal gas \citep{Whitmore2011ApJ, Hollyhead2015MNRAS}. While the H$\alpha$ data is available for the galaxies in our sample, there is almost no difference between the H$\alpha$ and the FUV emission maps due to the scales we are probing in this study. However, since the FUV data are deeper, we use them to investigate the sites where Ly$\alpha$ photons originate. 
\section{Analysis and methods} \label{analysis_sec}
The cornerstone of this study is the HST imaging data which has the advantage of a larger FoV compared to available spectroscopic observations at similar redshifts. The high-resolution HST imaging data of the 45 nearby SFGs galaxies enables us to study the Ly$\alpha$ light distribution and the morphology of the galaxies out to typical distances of $\sim$ 10 kpc (40 kpc for high redshift galaxies in the sample, e.g. LARS14, J1156).
\par 
We start by investigating the Ly$\alpha$ and FUV SB profiles (Sec. \ref{SB_pr_invest_sec}). In Sec. \ref{lya_morph_sec}, we discuss the Ly$\alpha$ morphological parameters used in this study. In Sec. \ref{rsc_iso_method_sec}, we discuss the method used for studying the Ly$\alpha$ emission of the galaxies in the faint isophotes. Finally, in Sec. \ref{sf_properties_analysis_sec}, we talk about the methods used to study the star-forming properties of the galaxies in our sample.
\subsection{Fitting to the surface brightness profiles} \label{SB_pr_invest_sec}
In this section, we study the Ly$\alpha$ SB profiles of the galaxies and how they compare to their FUV SB profiles. In other words, we determine how the observed spatial distribution of Ly$\alpha$ emission differs from that of the FUV continuum, which traces the sites where most of the Ly$\alpha$ photons are produced. There are (at least) three common ways of forming and studying SB profiles, and each one has its own advantages and disadvantages. Below, we look into these options and highlight their pros and cons.
\par
\textbf{Circular annuli}
\par
The simplest approach is to integrate the light in circular annuli. The benefit is that it requires no assumption on the actual shape of a galaxy. The downside is that for galaxies that are not circular in shape, the light distribution will be artificially flattened, and the influence of background noise will be increased.
\par
\textbf{Elliptical annuli}
\par
A slightly more sophisticated method that overcomes the con above is to use elliptical annuli for integration. This works well for galaxies that are more elongated in shape. One may allow for a change in the position angle of the ellipses with radius to better capture certain features (e.g. bars and spiral arms), but for very irregular galaxies which are not well described by ellipses, it has the same con as the spherical model, without its pro.
\par
\textbf{Isophotal integration}
\par
A more general approach is isophotal integration. In this method, instead of stepping in radius, one steps in SB and calculates a characteristic radius as $r = \sqrt{A/\pi}$.  It has the advantage of not relying on any assumption on the shape of the galaxy. Its cons, however, is that it does not go as deep and will not work at levels where the isophotal level is comparable to the background noise (then the isophotes will break up, and the area becomes ill-defined). It also requires extra care to compare results from different passbands as they generally do not probe the same area. For Ly$\alpha$, there is an additional complication with isophotal integration: the Ly$\alpha$ absorption against the FUV continuum, which frequently occurs in the centre. In any case, isophotal integration is the best suited method to study the extended Ly$\alpha$ emission. 
\par
In this paper, we use the most straightforward (circular) (Sec. \ref{Sers_sec_fit}, Sec. \ref{pr_fi_sb}) and more general (isophotal) (Sec. \ref{rsc_iso_method_sec}) approach, while omitting the elliptical one, as our sample is in general quite irregular and the elliptical approach does not offer any significant advantage over the other two.
\par
The steps taken to obtain the FUV and Ly$\alpha$ SB profiles using circular annuli are as follows: we masked the noisy edge of the images, started the SB profiles from the brightest point in the FUV maps, and determined the maximum radii from the largest circular aperture that could be fitted inside the masked region (usually set by the usable area in the SBC images), we define the radius of this aperture as r$_{max}$. Next, we performed photometry using 30 circular annulus bins (with 0.1 kpc as the smallest size) by using the python package \texttt{photutils} \footnote{\url{https://photutils.readthedocs.io/en/stable/index.html}} \citep{Bradley2019zndo}. The number of bins was chosen to simultaneously provide a good signal-to-noise and radial resolution in the SB profiles. To estimate the error on the measured SB in each bin, we performed the same procedures on 100 Monte Carlo simulated science frames obtained from LaXs. It should be noted that all the aforementioned measurements were done using binned weighted Voronoi tessellated maps, utilising the Weighted Voronoi Tessellation (WVT) method developed by \cite{Diehl2006MNRAS} (For more information see \cite{Hayes2014ApJ}). 
\subsubsection {S{\'e}rsic profile fitting} \label{Sers_sec_fit}
\begin{figure*}[t!]
 \centering 
 	\includegraphics[width=\textwidth] {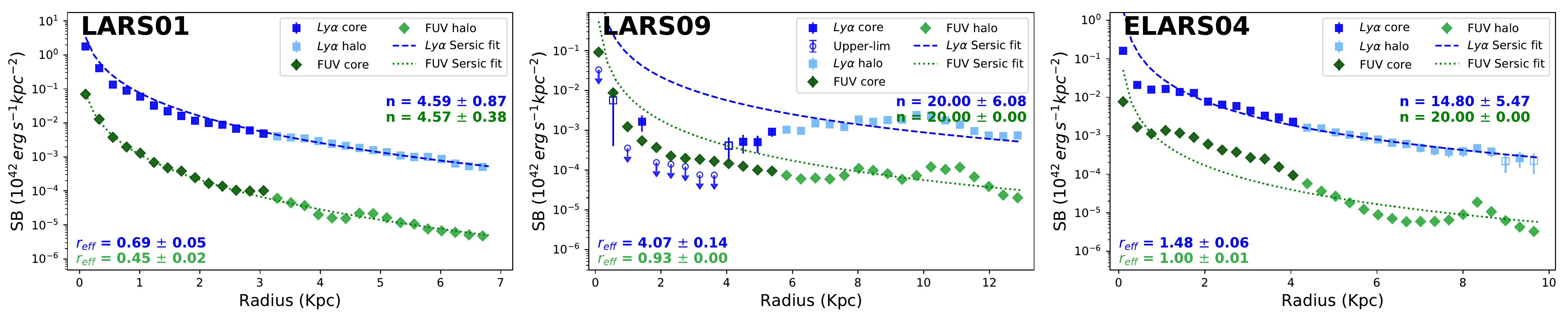}

\caption{
        S{\'e}rsic profile fitted to the Ly$\alpha$ and FUV SB profiles of three galaxies, LARS01, LARS09, and ELARS04. The resulted fits indicate S{\'e}rsic profile does not fully capture the behaviours of the SB profiles. Moreover, the limit set for the S{\'e}rsic index (e.g. 0.001 < n < 20 ) is reached in the fitted S{\'e}rsic profiles to the Ly$\alpha$ SB profile in LARS09, and FUV SB profile in both LARS09, and ELARS04.
        }
\label{sersic_fig}
\end{figure*} 
One of the most common models used to describe the SB profiles of the galaxies is S{\'e}rsic model \citep{Sersic1963BAAA, Graham2005PASA}:

\begin{equation}\label{sersic_eq}
    I(r) = I_e \, . \, exp \left( - b_n \, . \,  \left[(\frac{r}{r_{eff}})^{(\frac{1}{n})} -1 \right] \right)
\end{equation}

Where $b_n$ satisfies the relation $\gamma(2n, b_n) = \frac{1}{2} \Gamma(2n)$, where $\Gamma$, and $\gamma$ are the Gamma function and the lower incomplete function.
\par
We fit the S{\'e}rsic function to FUV and Ly$\alpha$ SB profiles of the galaxies in our sample. We used the \texttt{Sersic1D} model in the  \texttt{python astropy} package \citep{Astropy2013AA, Astropy2018AJ}. In Eq. \ref{sersic_eq}, there are three free parameters: the amplitude or the luminosity at the effective radius ($\mathrm{I_e}$), effective radius ($\mathrm{r_{eff}}$), and S{\'e}rsic index (n). We limited some of these parameters to have a more physically meaningful interpretation of the results. For instance, the effective radii were limited to vary between 0.001 to 1000 kpc, or the S{\'e}rsic index to vary between 0.001 to 20. Moreover, since we clearly observe absorption in the Ly$\alpha$ maps of the galaxies, we set the criterion that out of 30 data points in the SB profiles, there must be at least six data points with positive net emission and signal-to-noise ratio higher than 2 (S/N > 2) for fitting the S{\'e}rsic profile to each FUV and Ly$\alpha$ profile in every galaxy.
\par
Figure \ref{sersic_fig} shows examples of the fits for three galaxies. Our results clearly show that a single S{\'e}rsic profile does not describe the observed SB profiles very well. In addition, for a good fraction of the galaxies, the results indicated that the fitted parameters had reached the limits initially set as the requirements (e.g. 0.001 < n < 20 ). 
\subsubsection{Double exponential fitting function} \label{pr_fi_sb}
Around 60 years ago, \cite{de1958ApJ} used an exponential function to describe the disk region of the profile in M31. Since then, many studies have used an exponential function to successfully model the SB profiles of disk galaxies \citep{Freeman1970ApJ, Okamura1988PASP}.
\par
We found that a double exponential function describes the full SB profiles of both FUV and Ly$\alpha$ quite well in the majority of the galaxies. This double exponential function has the form described in Eq. \ref{double_exp_eq}, in this equation, $b$ is a free parameter fitted to the profiles, $\mathrm{f_1(r)}$, and $\mathrm{f_2(r)}$ describe the innermost and outermost parts of the profiles. $f_1(r)$ might not be well-constrained for some of the Ly$\alpha$ profiles due to the absorption in the inner regions of the galaxies. However, $f_2(r)$ is the main term used in our analysis which describes the outermost part of the profiles. This term is used to extrapolate the SB profiles outside of the available instrument since the angular size of these galaxies could be quite large.
\begin{equation} \label{double_exp_eq}
f(r) = 
    \begin{cases}
  f_1(r) = A_1 \, exp(\dfrac{-r}{r_{sc1}})\qquad for \quad r \leq b  \\
  \\
  f_2(r) = A_2 \, exp(\dfrac{-r}{r_{sc2}}) \qquad for \quad r \geq b \\  
    \end{cases}
\end{equation}
\par
Similar to the S{\'e}rsic profile fitting (see Sec. \ref{Sers_sec_fit}), we took several measures to enhance the fits and characterise the SB profiles better. The weighted fitting method in the \texttt{lmfit} package was used, so the data with higher S/N contribute more to the overall fits.
Moreover, the double exponential function was fitted only for those profiles where there were at least six data points with positive net emission in the full SB profiles, on top of requiring at least two data points in the halos (for the definition of the halo, see Sec. \ref{rsc_result_sec}), these points are also required to have S/N higher than two.
 In Tab. \ref{measured_paparm_this_study}, we provide for each galaxy the Ly$\alpha$ SB of the faintest (furthest) annulus bin (or upper limit if S/N < 2) measured in the Voronoi tessellated maps. 
 \par
We found that the fits fail to describe the outermost regions for several galaxies, especially in the FUV SB profiles. This is due to the relative low S/N in the outermost regions compared to the innermost and intermediate regions. In order to fit the outermost regions better, we manually down-weighted the data points in the intermediate regions, so the fits describe the outermost regions in these galaxies better. LARS01, 02, 05, 07, 12, 14, ELARS07, 09, 13 and J1156 are the galaxies with down-weighted intermediate data points in their FUV SB profiles. In LARS14 and J1156 cases, the intermediate regions were also downweighted in their Ly$\alpha$ SB profiles. Since for these two galaxies, the fit was not well-capturing the Ly$\alpha$ SB profile behaviour in the outermost regions, either. 
\par
\begin{figure*}[t!]
 \centering 
 	\includegraphics[width=\textwidth] {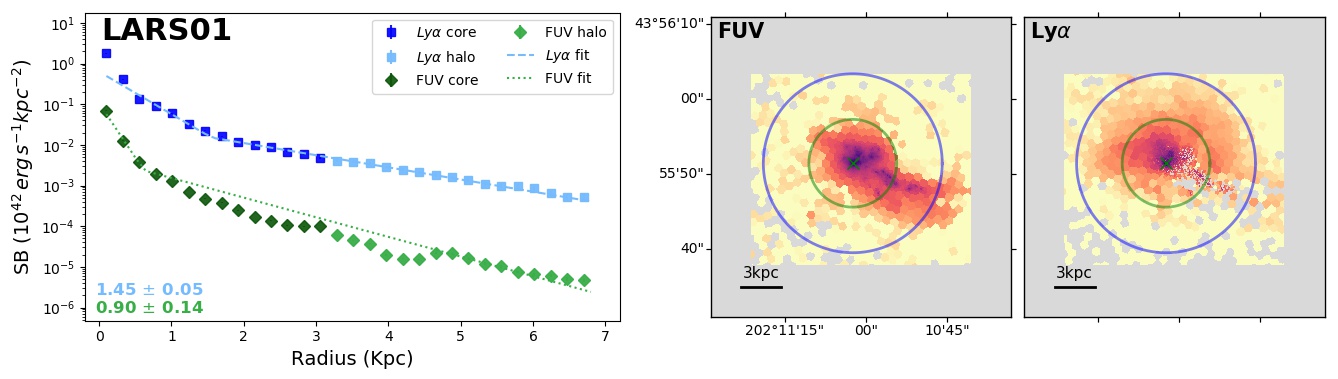}
 	\includegraphics[width=\textwidth] {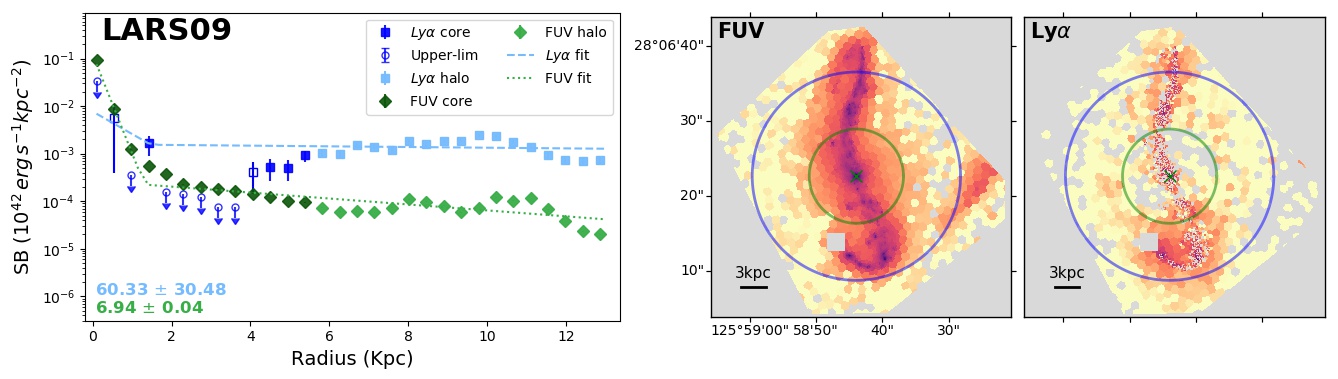}
 	\includegraphics[width=\textwidth] {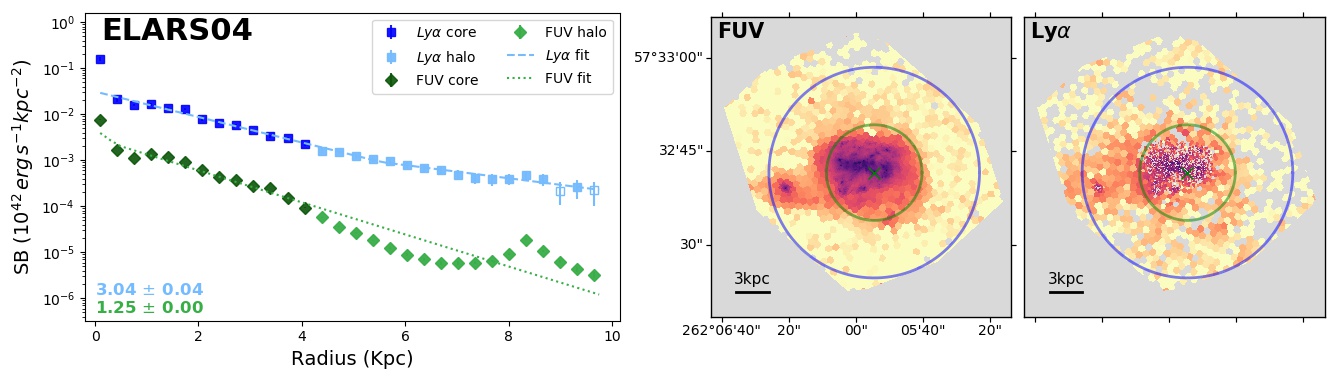}
	
\caption{
        Ly$\alpha$, FUV SB profiles, and the fitted double exponential of different galaxies (same galaxies shown in Fig. \ref{sersic_fig}) along with their FUV, and Ly$\alpha$ maps. The left panels show the Ly$\alpha$ and FUV profiles and the fitted model to them. The dark and light colours distinguish the core from the halo, respectively. For bins with S/N < 1, the 2$\sigma$ upper limit are displayed, and the bins with 1 $\leq$ S/N < 2 (data points not used in the fits) are displayed with the empty square symbol. The middle and right panels show the FUV and Ly$\alpha$ maps of each galaxy, respectively. The blue rings represent the largest radii where a circle centred at the FUV brightest pixel fits within each map, used as the last bin where the photometry was performed on. The green ring shows the radius where the SFRD drops below 0.01 $\mathrm{M_\odot \, yr^{-1} \,  kpc^{-2}}$.
        }
\label{paper_example_lya_fuv_radial}
\end{figure*}

To estimate the errors on the measured parameters, we ran 1000 Monte Carlo simulations. In these simulations, the measured SB were drawn randomly from a normal distribution based on the measured error in each bin. Figure \ref{paper_example_lya_fuv_radial} shows FUV and Ly$\alpha$ SB profiles with the fitted double exponential functions for the same galaxies shown in Fig. \ref{sersic_fig}. Appendix \ref{radia_analysis_appdix} contains the same figures for the full sample. Table. \ref{measured_paparm_this_study} lists the measured Ly$\alpha$ core and halo scale lengths and their corresponding error bars.
\par
At faint Ly$\alpha$ SB levels, there is a systematic uncertainty resulting from background subtraction in the images during the data reduction. For the galaxies at $z < 0.14$ (all but LARS13, LARS14, and J1156), Ly$\alpha$ is in the F125LP filter. This filter also contains a relatively bright geocoronal background from  O\rm{I}$\lambda 1302+1306$. The background is estimated from regions close to the edge of the chip in the images \citep{Ostlin2014ApJ}. While the statistical error on the estimated background itself is negligible, it is possible that -- if faint Ly$\alpha$ extends to the border of the chip -- this practice could lead to subtraction of actual Ly$\alpha$ emission. The subtracted background is consistent with, or lower than, the target geocoronal background for the observational setup (HST SHADOW operations) for all galaxies but five (LARS04, ELARS04, ELARS08, ELARS12, and ELARS14). However, it is still possible that Ly$\alpha$ emission from the galaxy contributes to the subtracted background.
\par 
Given that the background cannot be independently estimated, we investigate the systematic effect of subtracting an unknown flat background on the exponential profile fit in the halo. 
For this analysis, we assume that the Ly$\alpha$ emission in the halo is bright enough that an exponential can be fit to the data, or equivalently that the subtracted background emission is low enough to not wash out the real Ly$\alpha$ emission. Furthermore, we assume that either 100\% (the absolute worst case) or 10\% of the subtracted emission is Ly$\alpha$. 
In appendix \ref{bgest_appdix}, we calculate the effect of over-subtracting Ly$\alpha$ on the halo exponential that fits under these assumptions. The over-subtraction has a larger effect on the exponential fit if a) the background level is estimated close to the fitting region and b) the fitting region for the halo exponential spans a short radial range. We find that the systematic (positive) uncertainty from this effect is negligible ($<5$\%) for all of the galaxies for the 10\% assumption. With the worst case assumption (which is very unlikely to be correct), four galaxies (ELARS05, ELARS08, ELARS09, and ELARS25) show uncertainties larger than 10\%. We thus conclude that the background subtraction uncertainty does not affect the findings in this paper significantly.
\subsection{Ly$\alpha$ morphology of the galaxies} \label{lya_morph_sec}
In this work, we focus on  Ly$\alpha$ and FUV (we have also investigated the I band) maps of the galaxies. This enables us to study and investigate the differences between the sites where the ionising photons were produced and how we observe them after the recombination and the ongoing Ly$\alpha$ RT. As discussed before, there is no one-to-one mapping of the FUV to Ly$\alpha$ morphologies. In some galaxies, the morphology of the galaxies are similar, and in some cases, we can clearly spot the differences between the morphology of the galaxies in these two maps. For example, there is one small knot to the south of LARS02, which is bright in both FUV and Ly$\alpha$, while in other galaxies, FUV-bright regions tend to be associated with Ly$\alpha$ absorption, for example, ELARS08. Another example would be the spiral arms of the galaxy in ELARS05 that seem to be also evident in both FUV and Ly$\alpha$, while in ELARS06 and other galaxies, for example, ELARS12, ELARS23, ELARS25, ELARS27, and ELARS28, the correlation is either not obvious or just non-existent. 
\par
Given the high-resolution HST data available for the LARS sample, one may ask whether it is possible to relate the Ly$\alpha$ properties of the galaxies to the ionised gas media. In general, connecting Ly$\alpha$ emission to the ionised gas properties is an important subject, especially since such comparisons are not generally available at high$-z$. We note that even though the physical resolution of many galaxies in our sample is sufficient for probing the ionised gas structures, as stated in Sec. \ref{data_and_obs} the H$\alpha$ (and H$\beta$) observations of galaxies only covers star-forming regions and are not that sensitive to low surface brightness diffuse emission that tends to be present in the diffuse Ly$\alpha$ regions. A direct Ly$\alpha$-H$\alpha$ comparison could benefit from complementing the HST H$\alpha$ imaging with that greater depth but of lower spatial resolution \citep[e.g. using IFUs, see][]{Bik2015AA} and will be the scope of a future paper.   
\par
As mentioned above, LARS provides an opportunity to study the Ly$\alpha$ morphology of the nearby star-forming galaxies in $\lessapprox$ kpc scales. 
We investigate the asymmetry of the Ly$\alpha$ emission, and how it compares to the star-forming regions using three non-parametric approaches: (i) centroid shift between Ly$\alpha$, and FUV, (ii) difference in the position angles of the major axis between FUV and Ly$\alpha$, (iii)  Ly$\alpha$ axis ratio $\mathrm{(b/a)_{Ly\alpha}}$. In addition to these Ly$\alpha$ morphological parameters, we also measure the FUV and I band axis ratios.
\par
We start by introducing the regions used for measuring the aforementioned morphological parameters and continue with how these parameters are assessed within these regions.
\subsubsection{Regions used for assessing morphological parameters} \label{morph_regions_analysis_sec}
\textbf{Ly$\alpha$:}
\par
Regions where Ly$\alpha$ morphological parameters could be assessed were a significant challenge due to i) intrinsically shallow Ly$\alpha$ halos and ii ) Ly$\alpha$ absorption in the centre of the galaxies. We note that regions with high continuum flux and high column density of neutral hydrogen tend to show up as Ly$\alpha$ absorbed regions unless a high spectroscopic resolution is used, even if there is intrinsic Ly$\alpha$ emission. The narrow-band technique used for making the Ly$\alpha$ emission images for our sample is thus highly susceptible to this issue, and many galaxies show Ly$\alpha$ absorption in the central parts. 
\par
To overcome these difficulties, we explored the Ly$\alpha$ maps of the galaxies to determine the faintest encompassing isophotes (FEI). To do so, unbinned Ly$\alpha$ images were smoothed with a kernel size defined in physical units (i.e. kpc) (the kernel sizes differ based on redshift of the galaxies) provided in Table \ref{kernel_size_morph} to address the low S/N, especially in the Ly$\alpha$ halos. The Ly$\alpha$ surface brightness limit for the FEI was then chosen by-eye as the faintest isophote, which contained as much diffuse emission as possible without breaking up into individual noise peaks. $\mathrm{SB_{Ly\alpha}} = 1.5 \times 10^{39} \, \mathrm{erg/s/kpc^2}$ was found as a reasonable limit for Ly$\alpha$ FEI in our sample (except J1156, where $\mathrm{SB_{Ly\alpha}} = 5.0 \times 10^{39}$ was selected, see Table. \ref{kernel_size_morph}).
\par
Another challenge was to find a way to include the regions affected by Ly$\alpha$ absorption so that they do not bias the morphological measurements. To do so, we first identify regions with relatively high FUV surface brightness, using a threshold corresponding to a star formation rate density (SFRD) of 0.1 $\mathrm{M_\odot \, yr^{-1} \,  kpc^{-2}}$. We created a binary image for each galaxy, where the value is set to 1 if the Ly$\alpha$ brightness is higher than the FEI value, or if the SFRD is above 0.1; and 0 elsewhere. From now on, we refer to the region where the value is equal to one for measuring Ly$\alpha$ morphological parameters as Ly$\alpha$ Morphological Regions (LMR).
\par
\textbf{FUV and I band:}
\par
Following the same strategy as for Ly$\alpha$, unbinned FUV and I band images were smoothed with the same kernel size used for smoothing Ly$\alpha$ maps. Next, FUV and I band intensities in the FUV, and I band images were explored for finding the FUV and I band FEIs. $\mathrm{SB_{FUV}} = 2.5 \times 10^{37} \, \mathrm{erg/s/kpc^2erg/s/kpc^2/\si{\angstrom}}$, and $\mathrm{SB_{I \, band}} = 1.5 \times 10^{37} \, \mathrm{erg/s/kpc^2/\si{\angstrom}}$ were found as a good limit for all the galaxies in our sample.
\par 
Figure. \ref{morph_paper_example} shows examples of the regions used to determine the Ly$\alpha$ (LMR), FUV ($\mathrm{FEI_{FUV}}$), and I band ($\mathrm{FEI_{I \, band}}$) morphological parameters displayed on the Ly$\alpha$ images. 
\subsubsection{Centroid shift} \label{cf_method_sec}
\begin{figure*}[t!]
 \centering 
 	\includegraphics[width=\textwidth] {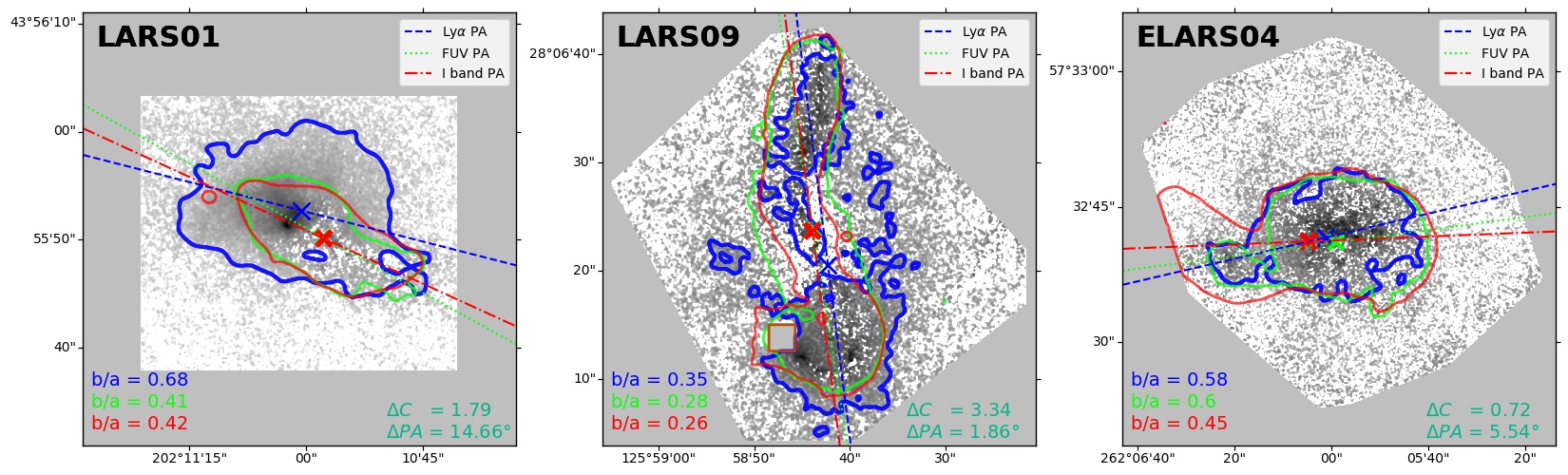}
 	
\caption{
        Regions used to determine all the morphological parameters used in this study ($\Delta$C, $\mathrm{(b/a)_{Ly\alpha}}$, $\mathrm{(b/a)_{FUV}}$, $\mathrm{(b/a)_I}$, and $\Delta$PA) displayed on the Ly$\alpha$ images for the three galaxies. The blue, green, and red contours are corresponding to the regions where the regions are brighter than $\mathrm{SB_{Ly\alpha}} = 1.5 \times 10^{39} \, \mathrm{erg/s/kpc^2}$, $\mathrm{SB_{FUV}} = 2.5 \times 10^{37} \, erg/s/kpc^2/\si{\angstrom}$, $\mathrm{SB_{I\, band}} = 1.5 \times 10^{37} \, \mathrm{erg/s/kpc^2/\si{\angstrom}}$ all displayed on the Ly$\alpha$ images (for more information on the limits, see Table. \ref{kernel_size_morph}). The blue, green, and red crosses represent the measured centroids (non-weighted first image moment) within LMR for Ly$\alpha$ and FUV, and I band FEI Ly$\alpha$, FUV, and I band, respectively. The determined PAs in the regions are also displayed with blue dashed, green dotted, and red dash-dotted lines for Ly$\alpha$, FUV, and I band, respectively. Moreover, the axis ratio $\mathrm{(b/a)_{Ly\alpha}}$ for each region is printed in the lower left part of the panels with blue, green, and red for Ly$\alpha$, FUV, and I band, respectively. Finally, the centroid shift ($\Delta$C, in kpc) between Ly$\alpha$ and FUV, and the difference between the measured PA of Ly$\alpha$, and FUV are printed on the lower right side of the panels.
        }
        
\label{morph_paper_example}    
\end{figure*}   
The difference in the light distribution and morphology of Ly$\alpha$ and FUV emissions was initially approached by measuring the centroid in each map (within LMR for Ly$\alpha$, and FUV FEI for FUV) and determining their difference. Determining the centroid position of a light distribution in an image is similar to determining the centre of mass of a distribution of mass of a system. Without any photon scattering or attenuation, the centroids of the Ly$\alpha$ ($C_{Ly\alpha}$) and FUV ($C_{FUV}$) emissions would coincide (assuming a symmetric distribution of ionised gas around the massive star clusters). But with the complex radiative transfer affecting Ly$\alpha$ morphology, the centroids do not necessarily coincide.
\par
Due to the uncertainties on Ly$\alpha$ emission level in the central parts of the galaxies (where continuum absorption is significant (see Sec. \ref{morph_regions_analysis_sec})), we determined the Ly$\alpha$ centroid as the non-weighted first image moment \citep{Stobie1980SPIE} within the region described above. Moreover, to be consistent with the Ly$\alpha$ centroid measurement, we used the non-weighted first image moment within the FUV FEI region.
The first image moment is generally defined as:
\begin{equation}
  \label{cerntroid_eq}
  (\overline{x},\overline{y}) = \left ( \frac{\sum_{i} I_{i} x_i}{\sum_{i}
      I_{i}} \, , \,  \frac{\sum_{i} I_{i} y_i}{\sum_{i}
      I_{i}} \right ) 
\end{equation}
\par
Where $x_i$, $y_i$, $\overline{x}$, and $\overline{y}$ are the x, and y coordinates in the image coordinate system, and the weighted mean in the x and y axis, respectively. Finally, $I_i$ is the flux measured in each pixel. As mentioned before, in our calculations, we used the non-weighted method where $I_i=1$. 
\par
We determined the difference between the measured Ly$\alpha$ and FUV centroid as the centroid shift ($\Delta$C): 
\begin{equation} 
    \Delta C = |C_{Ly\alpha} - C_{FUV}|
\end{equation}
This quantity indicates how the general distribution of Ly$\alpha$ photons has shifted from their main site of production. Table. \ref{measured_paparm_this_study} contains the measured centroid shift ($\Delta$C) for each galaxy. 
Finally, to have a better understanding of the distribution of the sum of both young and old stellar populations, we determined the centroid for the whole stellar populations in the galaxies. To be consistent with our measurements in both Ly$\alpha$ and FUV, the centroid was computed as the non-weighted first image moment within the region selected by the I band FEI.
\begin{table}
\caption{
        List of the galaxies, the kernel sizes used to smooth their maps, and the Ly$\alpha$ SB used for determining the faintest encompassing isophote in our morphological studies.
        }
\label{kernel_size_morph}      
\centering          
\begin{tabular}{c c c}     
\hline\hline       
ID      &  Kernel \,  size  & Ly$\alpha$ SB\\
		& (Physical scale)  & ($10^{39} \mathrm{erg/s/kpc^2}$)\\
\hline
LARS01-10 	& 	0.3 kpc     & 1.5 \\ 
LARS11-14 	& 	0.8 kpc     & 1.5 \\ 
ELARS01-28  & 	0.3 kpc     & 1.5 \\ 
Tol1214 	& 	0.3 kpc     & 1.5 \\ 
Tol1247 	& 	0.3 kpc     & 1.5 \\ 
J1156  	    &  	0.8 kpc     & 5.0 \\ 
\hline                  
\end{tabular}
\tablefoot{The kernel sizes are chosen differently based on the redshift of the galaxies.}
\end{table}
\par
In Fig. \ref{morph_paper_example}, the measured centroid for each band is shown with a cross with the same colour used to display the contour for each band. Appendix \ref{lya_morphology_appdix} contains these results for the full sample.
\subsubsection{Position angle and axis ratio} \label{ps_angle_axis_ratio_sec}
The first image moment collapses all the information on how the light is distributed in an image into a single value, the centroid position. However, by using the second image moment, the light distribution can be studied in greater detail \citep{Stobie1980SPIE}. Hence, we expanded our non-parametric morphological studies utilising the second image moments and determined parameters that reveal more information on how the light distribution varies in different wavebands within the regions discussed in Sec. \ref{morph_regions_analysis_sec} in Ly$\alpha$, FUV, and I band maps. The quantities studied through second image moments are axis ratio $\mathrm{(b/a)_{Ly\alpha}}$, and the position angle (PA) (also used in other Ly$\alpha$ studies, e.g. \citep{Herenz2020AA})(difference of the measured position angle between FUV and Ly$\alpha$), in addition to the FUV ($\mathrm{(b/a)_{FUV}}$) and I band ($\mathrm{(b/a)_{I}}$) axis ratios. Roughly, the axis ratio indicates the elongation, and the PA reveals the alignments of the light distribution in each bandpass. For instance, studying the difference in the measured PA between Ly$\alpha$ and FUV may reveal preferred directions of Ly$\alpha$ photon escape from the galaxies.
The second-order image moments are defined as in Eq. \ref{2nd_img_mom_eq}:

\begin{equation} \label{2nd_img_mom_eq}
   \begin{array}{l}

      X2 = \overline{x^2} = \dfrac{\Sigma \,  I_i x_i^2}{\Sigma \,  I_i} -   \overline{x}^2 \\
      Y2 = \overline{y^2} = \dfrac{\Sigma \,  I_i y_i^2}{\Sigma \,  I_i} -   \overline{y}^2 \\
      XY = \overline{xy}  = \dfrac{\Sigma \,  I_i y_i x_i}{\Sigma \,  I_i} - \overline{x} \overline{y} 

   \end{array}
\end{equation}
As for the first moment, we use non-weighted image moments, hence $I_i = 1$.
From the second image moments parameters in Eq. \ref{2nd_img_mom_eq}, parameters such as the minor and major axis, and the position angle of an ellipse can be analytically computed through Eq. \ref{maj_ax_eq}, to Eq. \ref{theta_eq}.

\begin{equation} \label{maj_ax_eq}
    (a/2)^2 = \frac{X2 + Y2}{2} + \sqrt{\left(\dfrac{X2 - Y2}{2}\right)^2 + XY^2}
\end{equation}

\begin{equation} \label{min_ax_eq}
    (b/2)^2 = \frac{X2 + Y2}{2} - \sqrt{\left(\dfrac{X2 - Y2}{2}\right)^2 + XY^2}
\end{equation}

\begin{equation} \label{theta_eq}
    tan(2\theta_0) = 2 \dfrac{XY}{X2 - Y2}
\end{equation}

In equation \ref{maj_ax_eq}, and \ref{min_ax_eq}, $a$, and $b$ are the semi-major, and semi-minor axis, respectively. By measuring the semi-major (a) and semi-minor (b) axis, we assess the axis ratio. In general, axis ratios reveal different elongations or symmetricities of the light distribution in the regions of interest in the galaxies. The axis ratios and PAs in different wavebands are measured in the same regions described in Sec. \ref{morph_regions_analysis_sec} in each galaxy; the results are provided in Table. \ref{measured_paparm_this_study}. From here on, we refer to the Ly$\alpha$ axis ratio ($\mathrm{(b/a)_{Ly\alpha}}$) simply as the axis ratio unless we explicitly specify the axis ratios determined in other wavebands.
\par
In Fig. \ref{morph_paper_example}, regions used for determining the axis ratios and PAs are shown with different contours, all displayed on the Ly$\alpha$ maps. The measured PAs within the region determined by Ly$\alpha$, FUV, and I band FEIs are displayed with the blue dashed, green dotted, and dash-dotted lines, respectively. Appendix \ref{lya_morphology_appdix} contains these results for the full sample.
\subsection{Isophotal analysis of the extent of the Ly$\alpha$ halos} \label{rsc_iso_method_sec}
\begin{figure*}[t!]
 \centering 
 	\includegraphics[width=\textwidth]{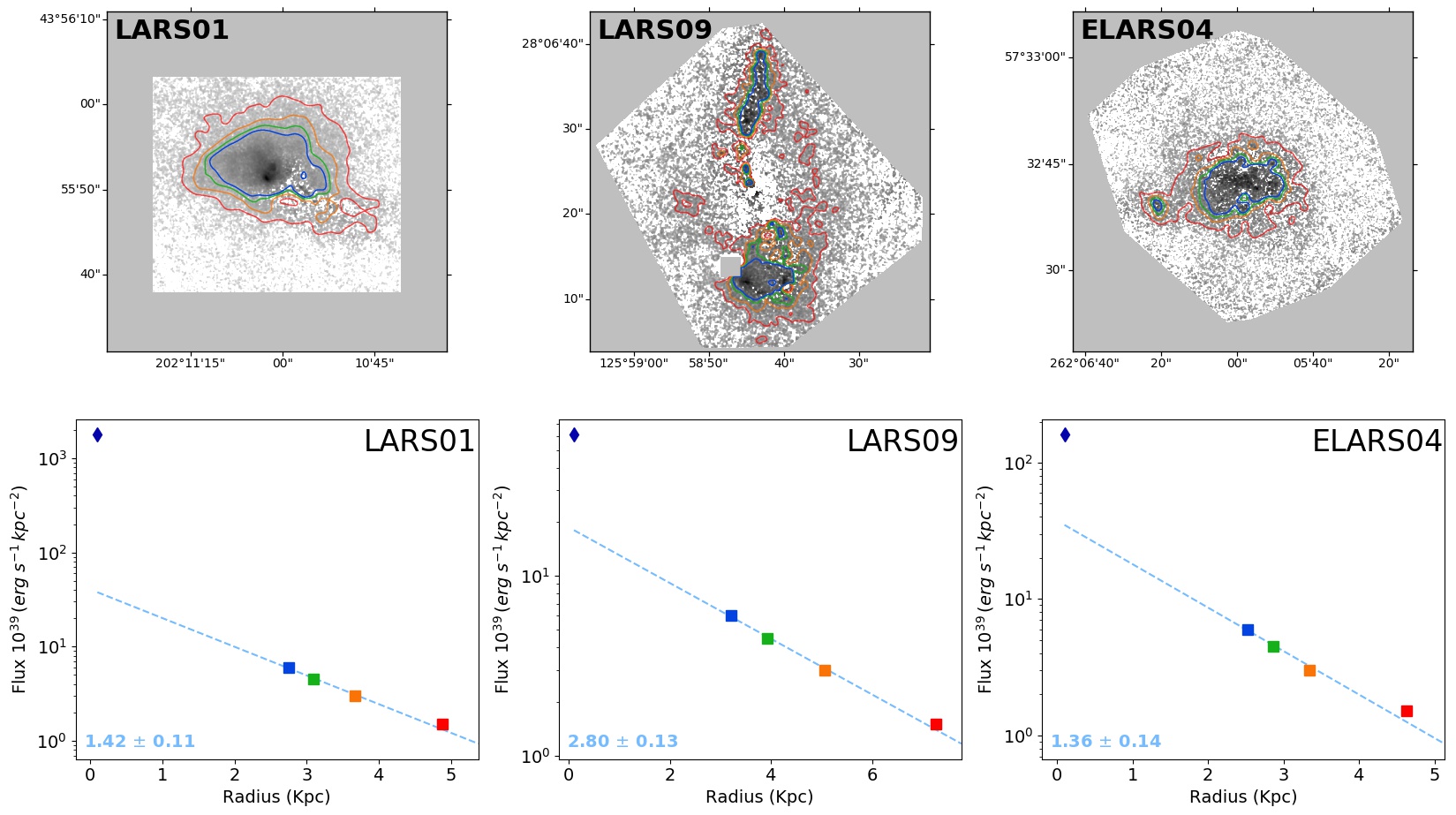}

\caption{ 
        Faint Ly$\alpha$ isophotes used to study the extent of the Ly$\alpha$ halo through the isophotal approach for each galaxy. In each figure, in the top panels, the region within the isophotal levels of even multipliers (1, 2, 3, 4) of the faintest limit ($1.5 \times 10^{39} \, \mathrm{erg/s/kpc^2}$, except J1156 $5.0 \times 10^{39} \, \mathrm{erg/s/kpc^2}$) denoted by red, orange, green and blue contours, all displayed on the Ly$\alpha$ maps of the galaxies. The bottom panels show the data points corresponding to these isophotes displayed with the same colour, and a single exponential fit (Eq. \ref{single_exp_iso}) to the points. The fitted scale lengths (and the measured error bar determined from the fit) are given in the lower left corner. The blue diamond represents the measured Ly$\alpha$ SB in the binned images at the innermost region (usually within r = 0.1 kpc from the brightest FUV point).
        }
\label{r_iso_morph_fig_main_txt}    
\end{figure*}
As discussed in Sec. \ref{intro_sec}, it has already been established that Ly$\alpha$ halos are asymmetric in both local and high redshift universe. Furthermore, the stellar morphology in most of the galaxies in our sample is also quite irregular. Hence, circular apertures may not fully characterise the complicated Ly$\alpha$ distribution resulting from the radiative transfer. To avoid this issue, we also measure the extent of the Ly$\alpha$ halo, by studying the Ly$\alpha$ emission within the faint Ly$\alpha$ isophotes. This approach has the advantage that it explores the light distribution in the halo only. It also avoids the areas where the absorption becomes more prominent.
\par
We used four isophotes, which provide us with an isophotal profile of the Ly$\alpha$ halo. The faintest encompassing Ly$\alpha$ isophote used in this analysis is $\mathrm{SB_{Ly\alpha}} = 1.5 \times 10^{39} \, \mathrm{erg/s/kpc^2}$ for all galaxies but  J1156, where Ly$\alpha$ $\mathrm{SB_{Ly\alpha}} = 5.0 \times 10^{39} \, \mathrm{erg/s/kpc^2/\si{\angstrom}}$ was selected (Sec. \ref{morph_regions_analysis_sec}). The isophotes chosen for this analysis were selected as even multipliers (1, 2, 3, 4) of the faintest limit.
Next, the areas covered within these isophotes were measured, and the equivalent radius (the radius of a circle with the same area encompassed by a given isophote) was determined.
\par
Finally, we fit a simple exponential function to the Ly$\alpha$ isophotal SB profile of the halo to find the Ly$\alpha$ isophotal halo scale length ($\mathrm{r_{sc}^{iso}}$) (Eq. \ref{single_exp_iso}).
\begin{equation} \label{single_exp_iso}
f(r) = A \, \exp(\frac{-r}{r_{sc}^{iso}})
\end{equation}
Figure \ref{r_iso_morph_fig_main_txt}, shows the Ly$\alpha$ maps and the regions used for determining the extent of the halo and the exponential fit results. 
It is clear from this approach that more than one exponential function is needed to describe the full Ly$\alpha$ SB profiles. Appendix \ref{r_iso_appx} contains these panels for the full sample.
\subsection{Size and intensity of the star forming regions} \label{sf_properties_analysis_sec}
\begin{figure*}[t!]
 \centering 
 	\includegraphics[width=\textwidth] {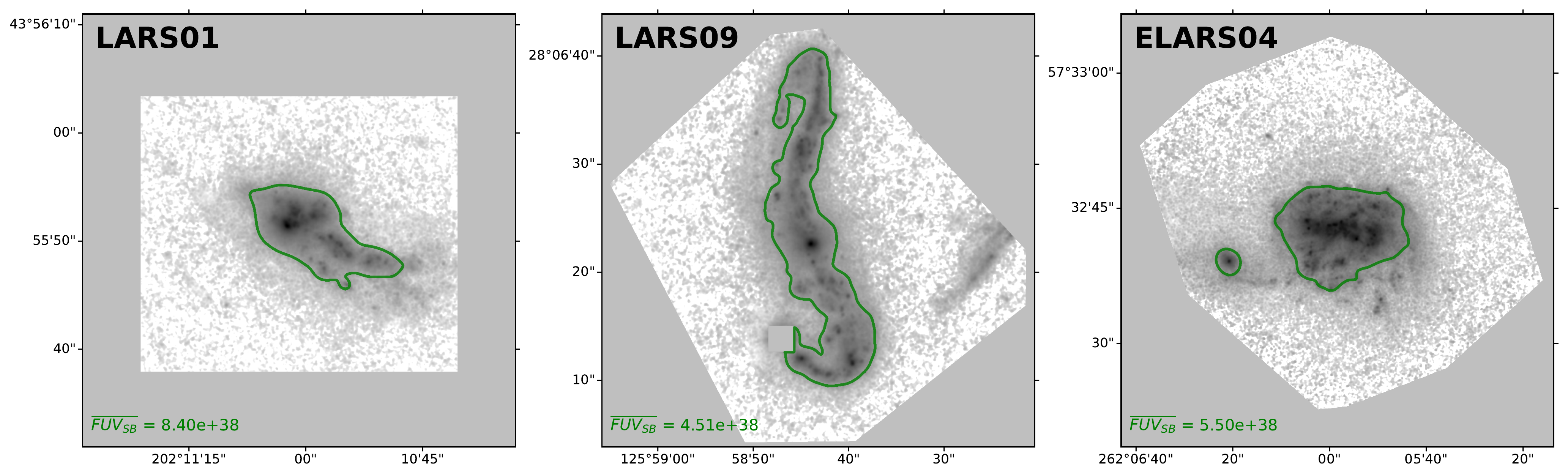}

\caption{
        Regions used to determine the SF properties of the galaxies for the three galaxies. The green contours shows the regions with SFRD > 0.01 $\mathrm{M_\odot \, yr^{-1} \,  kpc^{-2}}$ displayed on the unbinned FUV images. The measured average FUV SB ($\mathrm{\overline{FUV_{SB}}}$) are displayed on the lower left side of each panel. 
        }
\label{fuv_morph}
\end{figure*} 
So far, we have focused on measuring different quantities that quantify the Ly$\alpha$ morphology of the galaxies. With the high-resolution HST data available for our sample of galaxies, we also investigated how star-forming (SF) properties of the galaxies relate to the Ly$\alpha$ output.
\par
We use two quantities for studying the SF properties, both based on the FUV, of the galaxies: i) the size of the area with  SFRD > 0.01 $\mathrm{M_\odot \, yr^{-1} \,  kpc^{-2}}$ represented by the equivalent radius (the radius of a circle with the same area), and ii) the average FUV SB ($\mathrm{\overline{FUV_{SB}}}$) within these regions. The size of the area represents how large the SF regions are in each galaxy, while the ($\mathrm{\overline{FUV_{SB}}}$) determines how intense the SF is in each galaxy.
\par
For this analysis, we used the unbinned FUV images smoothed with the kernel sizes listed in Table. \ref{kernel_size_morph}. Figure \ref{fuv_morph} shows these regions, and the measured $\mathrm{\overline{FUV_{SB}}}$ for three galaxies. Appendix \ref{fuv_morphology_appdix} contains these results for the full sample.
\section{Results} \label{resuluts_sec}
In this section, we discuss the outcomes of the analysis using the methods presented in Sec. \ref{analysis_sec}. 
\subsection{Extent of the Ly$\alpha$ halos} \label{rsc_result_sec}
\begin{figure}
\centering
    \includegraphics[width=\linewidth]{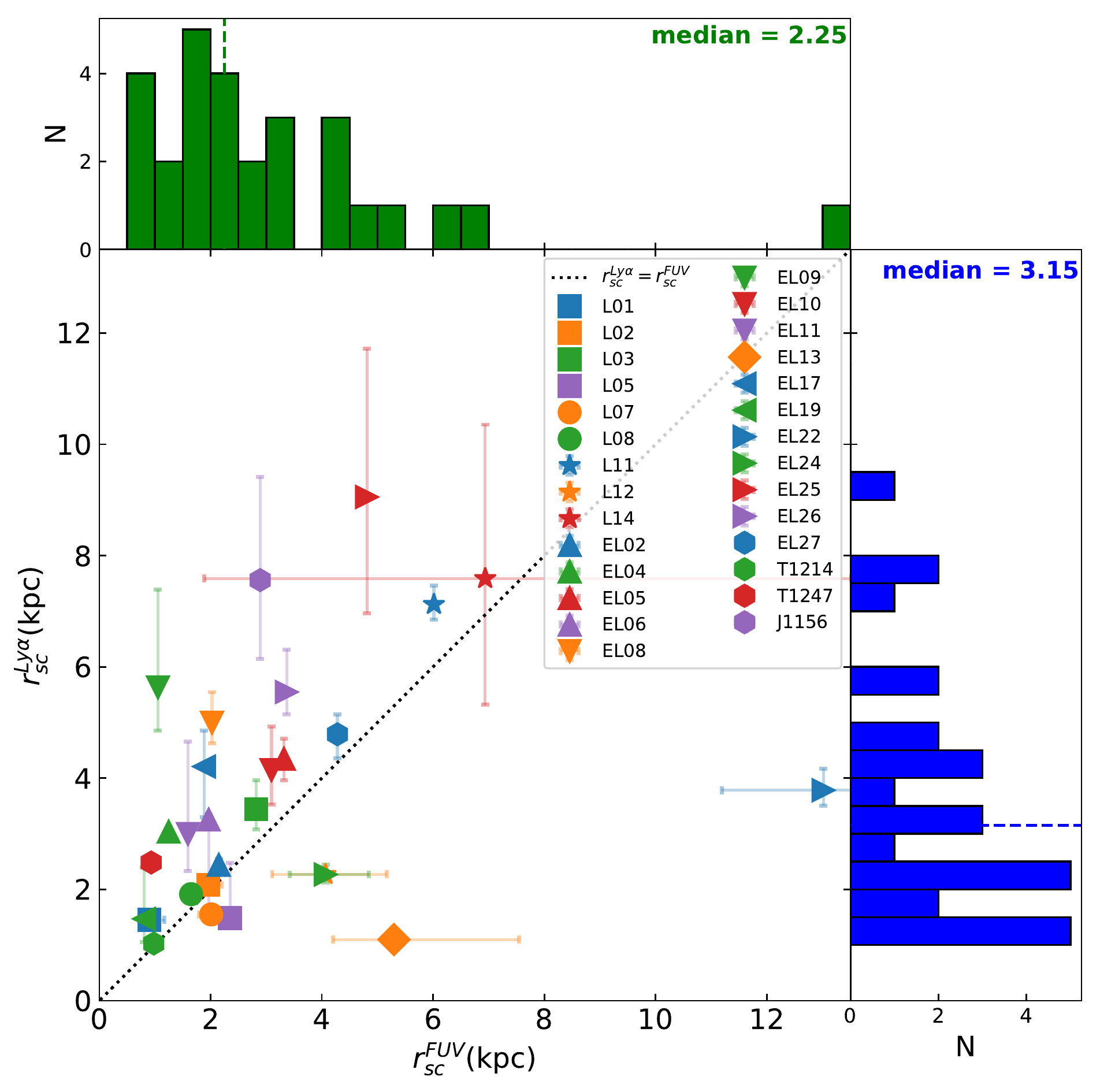}
  \caption{
          Ly$\alpha$ halo scale length ($\mathrm{r_{sc}^{Ly\alpha}}$) versus the FUV halo scale length ($\mathrm{r_{sc}^{FUV}}$). The histograms on the top and left show the distributions and the measured median for the FUV, and Ly$\alpha$ halo scale lengths, respectively. 
          }
\label{rsc_result_fig}
\end{figure}
To study the extended Ly$\alpha$ halos, we decomposed the profiles into core (inner) and halo (outer) regions.
To do so, we studied the SFRD profiles of the galaxies obtained from the FUV SB profiles using the SFR calibration from \cite{Kennicutt2012ARAA} and used a SFRD threshold of 0.01 $\mathrm{M_\odot \, yr^{-1} \,  kpc^{-2}}$ to distinguish between core and halo. This threshold is roughly the SFRD in the Kennicutt-Schmidt relation where the slope changes from the efficient SFR to the inefficient SFR \citep{Bigiel2008AJ, Micheva2018AA}. In Fig. \ref{paper_example_lya_fuv_radial}, the radius in which the SFRD falls below the 0.01 $\mathrm{M_\odot \, yr^{-1} \,  kpc^{-2}}$ is shown by a green circle on both FUV and Ly$\alpha$ images (middle, and right panels). This is also shown on the FUV and Ly$\alpha$ profiles (left panels), where the data points with the blue and sky blue in the Ly$\alpha$ profiles, and data points with the green and the light green represents core and halo, respectively. 
\par
As discussed in Sec.\ref{pr_fi_sb}, we found that a double exponential function (Eq. \ref{double_exp_eq}) describes both the Ly$\alpha$ and FUV SB profiles quite well. The second term ($f_2(r)$) in this fitting function describes the behaviour of the halo. Based on the results from the profile fitting analysis, for the rest of the analysis, we exclude 
LARS04, LARS06, LARS13, ELARS01, ELARS03, ELARS07, ELARS12, ELARS14, ELARS15, ELARS16, ELARS18, ELARS20, ELARS21, ELARS23, ELARS28.
These galaxies are faint Ly$\alpha$ emitters (the median of their Ly$\alpha$ luminosity is $2.30 \times \mathrm{10^{40} \, erg/s}$ which is fainter by almost one order of magnitude compared to the median of the full sample $1.70 \times \mathrm{10^{41} \, erg/s}$), and lack sufficient data points (less than 2) in their Ly$\alpha$ halo part of their Ly$\alpha$ SB profile with S/N > 2 for fitting the double exponential function (see Eq. \ref{double_exp_eq}). This criterion makes us more confident that our extrapolation is based on the observed Ly$\alpha$ emission in the Ly$\alpha$ halos. We also exclude LARS10: this galaxy is also faint Ly$\alpha$ emitter ($\mathrm{ L_{Ly\alpha}} = 1.10 \times \mathrm{10^{40} \, erg/s}$) and in spite of fulfilling the criterion of having more than two data points in its Ly$\alpha$ SB profiles, the measured fitted Ly$\alpha$ halo scale length has a large error bar where $ \dfrac{\mathrm{r_{sc}^{Ly\alpha}}}{\mathrm{\Delta r_{sc}^{Ly\alpha}}} $ < 1 ($\mathrm{r_{sc}^{Ly\alpha}}$ is the fitted scale length in the second term of the double exponential function defined in Eq. \ref{double_exp_eq}, to the Ly$\alpha$ SB profiles). 
Finally, LARS09 is excluded since its fitted $\mathrm{r_{sc}^{Ly\alpha}}$ is huge (due to its Ly$\alpha$ morphology, with bright emission in the north and south of the centre and little sign of decreasing SB at large radii, see Figs. \ref{paper_example_lya_fuv_radial} and \ref{morph_paper_example}). The bin size of the circular annuli used for the SB profiles will certainly affect the S/N of individual data points, but using larger bins does not change this selection.
Removing all these galaxies from the initial 45 galaxies in our sample, brings us to 28 galaxies that are explored in this study.
\par
The range covered by Ly$\alpha$ halo scale length is between 1.03 to 9.05, with a median of 3.15 kiloparsec. As previously stated, because of the physical processes involved in the Ly$\alpha$ RT, the Ly$\alpha$ SB profile usually drops more slowly than the FUV. Thus, we looked at the measured Ly$\alpha$ halo scale length $\mathrm{r_{sc}^{Ly\alpha}}$ and how they compare to the FUV halo scale length $\mathrm{r_{sc}^{FUV}}$. Figure. \ref{rsc_result_fig} shows $\mathrm{r_{sc}^{Ly\alpha}}$ versus $\mathrm{r_{sc}^{FUV}}$, and the distribution of these two quantities. We see that indeed, the distribution of Ly$\alpha$ emission is flatter than the FUV flux ($\mathrm{r_{sc}^{Ly\alpha}}$ > $\mathrm{r_{sc}^{FUV}}$). However, we note that when we compare the Ly$\alpha$ and FUV halo scale lengths, despite the difference in median value, the distributions are still consistent with each other (see the histograms). The KS test on two samples returns KS score of 0.21 corresponding to a p-value of 0.49, indicating that the distributions are similar, although this is driven by a few outliers. We also see that Ly$\alpha$ halo scale length correlates with the FUV halo scale length, indicating that more extended star-forming regions leads to more extended Ly$\alpha$ emission. Moreover, \cite{Bridge2018ApJ} finds that there is a correlation between the size of the Ly$\alpha$ halos and the scattering distance found in their analysis for the original LARS galaxies (LARS01-LARS14). We thus interpret the Ly$\alpha$ halo scale lengths to give a measure of the scattering distances of  Ly$\alpha$ emission, and, consequently, that large Ly$\alpha$ halos (large $\mathrm{r_{sc}^{Ly\alpha}}$) indicate large scattering distances. 
\subsection{Ly$\alpha$ halo fraction}  \label{hf_result_sec}
\begin{figure}
\centering
    \includegraphics[width=\linewidth]{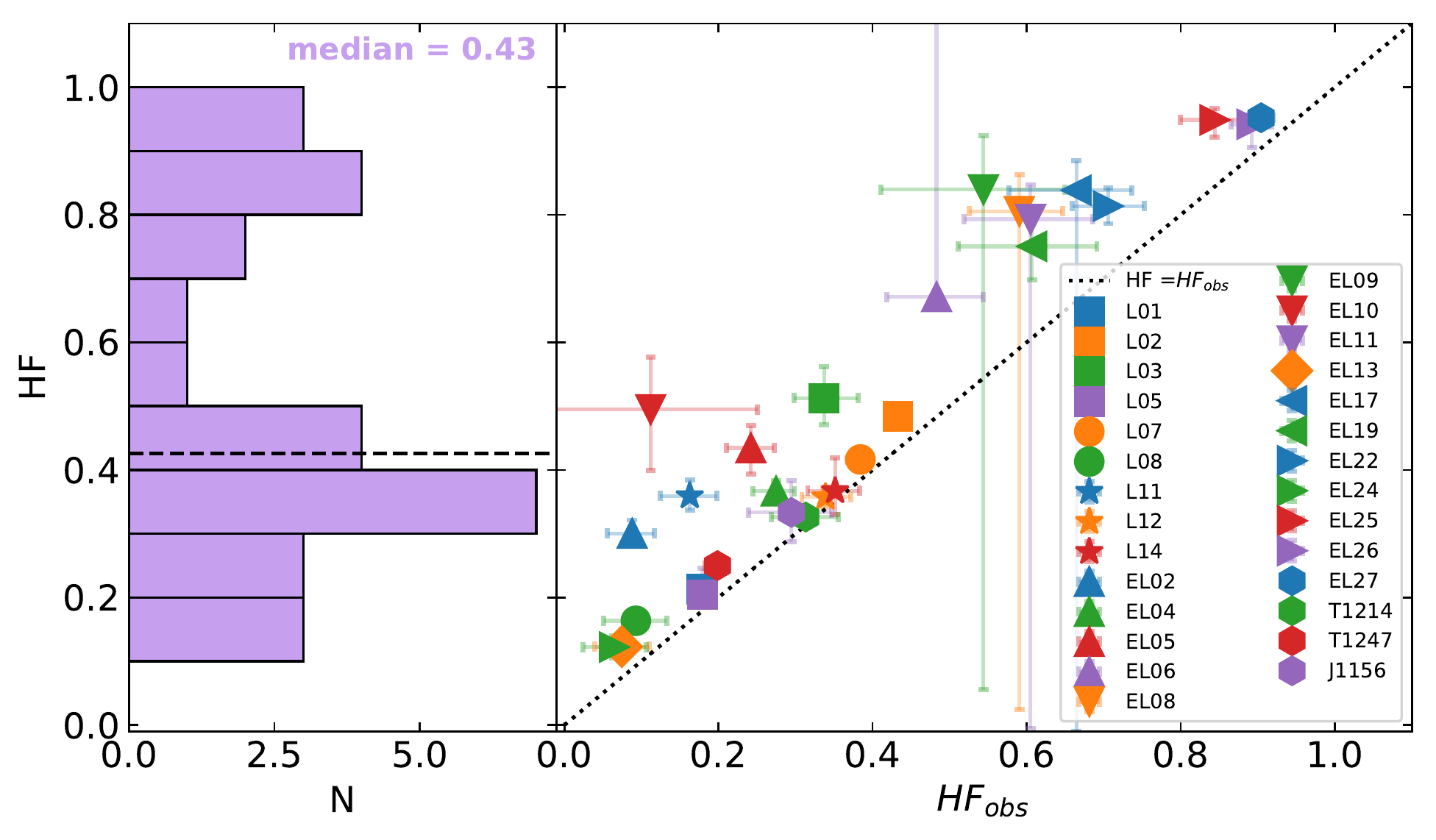}
    
  \caption{
          Left panel: histogram showing the distribution of the measured HF for our sample of galaxies. Right panel: Ly$\alpha$ halo fraction HF computed from the fit (HF) vs. the observed halo fraction ($\mathrm{HF_{obs}}$). 
          }
\label{hf_fit_vs_hf_obs}
\end{figure}
Studying Ly$\alpha$ halos is interesting on its own; however, perhaps the most important implication of the Ly$\alpha$ halo studies is to pin down the contribution of the halo to the total Ly$\alpha$ luminosity, that is the Ly$\alpha$ halo fraction. This quantity indicates what portion of the Ly$\alpha$ photons travels far away from where they were produced before escaping, and thus provide a clue on how far the ionising photons travel before ionising an H atom and how much radiative scattering goes on in each galaxy. 
\par
We define the Ly$\alpha$ halo fraction as the Ly$\alpha$ flux emitted in the halo divided by the total Ly$\alpha$ flux. The limited detector size of the SBC detector presents a problem for determining halo fractions. Many of our galaxies are close enough that their Ly$\alpha$ emission fills the detector chip. It is therefore likely that measurements in the images will miss part of the halo flux. Hence, we determine the halo fraction (HF) as: the measured Ly$\alpha$ luminosity through integrating the second term in our fitting function (which describes the halo, see Eq. \ref{double_exp_eq}) from where we define halo (SFRD drops below 0.01 $\mathrm{M_\odot \, yr^{-1} \,  kpc^{-2}}$) to infinity divided by the sum of this quantity and the observed Ly$\alpha$ luminosity in the core (Eq. \ref{hf_eq}).
\begin{equation} \label{hf_eq}
    HF = \dfrac{\int_{r_{SFRD \geq 0.01}}^{\infty}  2 \pi r f_2 (r) dr}{L_{core}^{Ly\alpha} + \int_{r_{SFRD \geq 0.01}}^{\infty}  2 \pi r f_2 (r) dr}
\end{equation}
\par
The left panel in Fig. \ref{hf_fit_vs_hf_obs} shows the distribution of HF for the galaxies in our sample. The measured HF ranges between 0.12 to 0.95 with a median size of 0.43. We looked at how HF compares with the observed halo fraction ($\mathrm{HF_{obs}}$). The observed halo fraction is determined by dividing the observed Ly$\alpha$ luminosity in the halo, to the observed Ly$\alpha$ luminosity out to the radius where $\mathrm{(S/N)_{Ly\alpha} < 1}$. The right panel in Fig. \ref{hf_fit_vs_hf_obs} shows how the measured Ly$\alpha$ HF computed using the results from the fits compares with $\mathrm{HF_{obs}}$. Unsurprisingly, we see that the observed halo fractions are smaller than the profile fitted fractions. Assuming that the exponential nature of the halo holds out to at least a few scale radii, the fitted halo fractions provide a much more secure estimate of the actual fraction. Hence, in our study, we used HF determined from the fits and from here on, when we speak of HF, we refer to the Ly$\alpha$ HF determined from the fit.
\subsection{Ly$\alpha$ morphology of the galaxies} \label{res_lya_morph_sec}
\subsubsection{Centroid shift}\label{cf_res_sec}
The centroid shift ($\Delta$C) is the offset between the measured centroid in Ly$\alpha$, and FUV maps (See Sec. \ref{cf_method_sec}). The top panel in Fig. \ref{delta_pa_result_sec_fig} shows the distribution of $\Delta$C. Overall, we see that the Ly$\alpha$ emitting region is displaced from the FUV, and the measured $\Delta$C ranges between 0.80, to 2.25, with the median size of 1.13 kiloparsec. We discuss the relation between $\Delta$C and the measured quantities in this study in addition to some Ly$\alpha$ observables in Sec. \ref{dis_sec}.
\subsubsection{Position angle difference} \label{delta_pa_res_sec}
\begin{figure}
\centering
    \includegraphics[width=\linewidth]{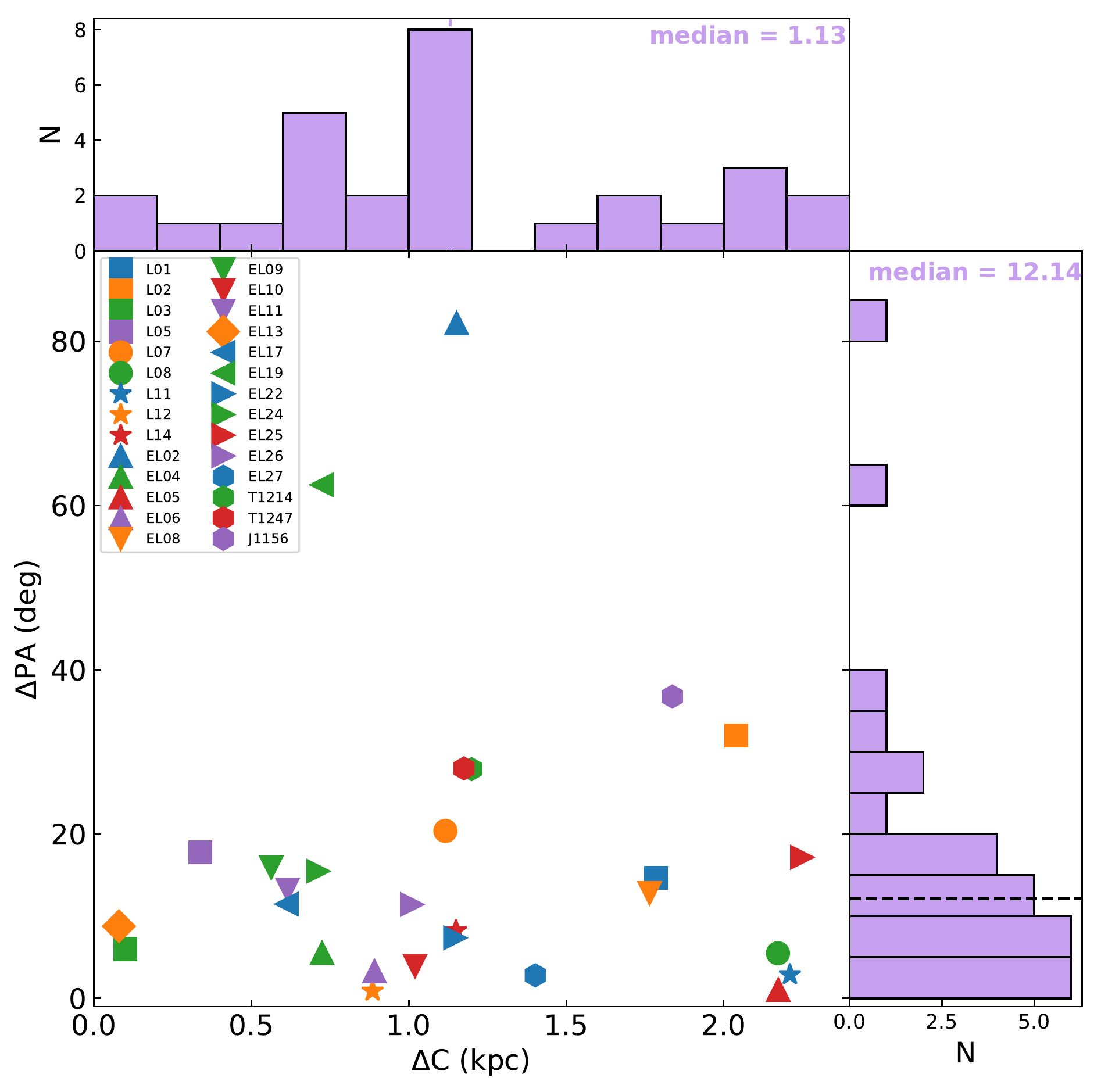}
    
  \caption{
           The difference in the position angles versus the centroid shift. The top and left histograms show the distributions and the measured median for $\Delta$C, and $\Delta$PA, respectively. 
          }
\label{delta_pa_result_sec_fig}
\end{figure}
The difference between the measured Ly$\alpha$ and FUV position angle is another Ly$\alpha$ morphological parameter, assessed from the second image moment (see Sec. \ref{ps_angle_axis_ratio_sec}). 
From here on, we refer to the difference between the measured Ly$\alpha$ and FUV position angle as the absolute value of the difference in the position angles ($\Delta$PA). This parameter indicates the difference between Ly$\alpha$ and FUV alignments. The right panel in Fig. \ref{delta_pa_result_sec_fig} shows the distribution of $\Delta$PA, $\Delta$PA ranges between one, to 82 degrees, with the median size of 12 degrees. Most of the galaxies have similar major axis in FUV and Ly$\alpha$ (small $\Delta$PA), but we note that this parameter becomes very uncertain when the axis ratio is close to one. As an example, in the bottom left panel in Fig. \ref{delta_pa_result_sec_fig} we show a scatter plot of $\Delta$PA and $\Delta$C. We discuss the relation between $\Delta$C and the other quantities in Sec. \ref{dis_sec}.
\subsubsection{Axis ratio} \label{lya_b_a_res_sec}
\begin{figure*}
\centering
    \includegraphics[width=\textwidth]{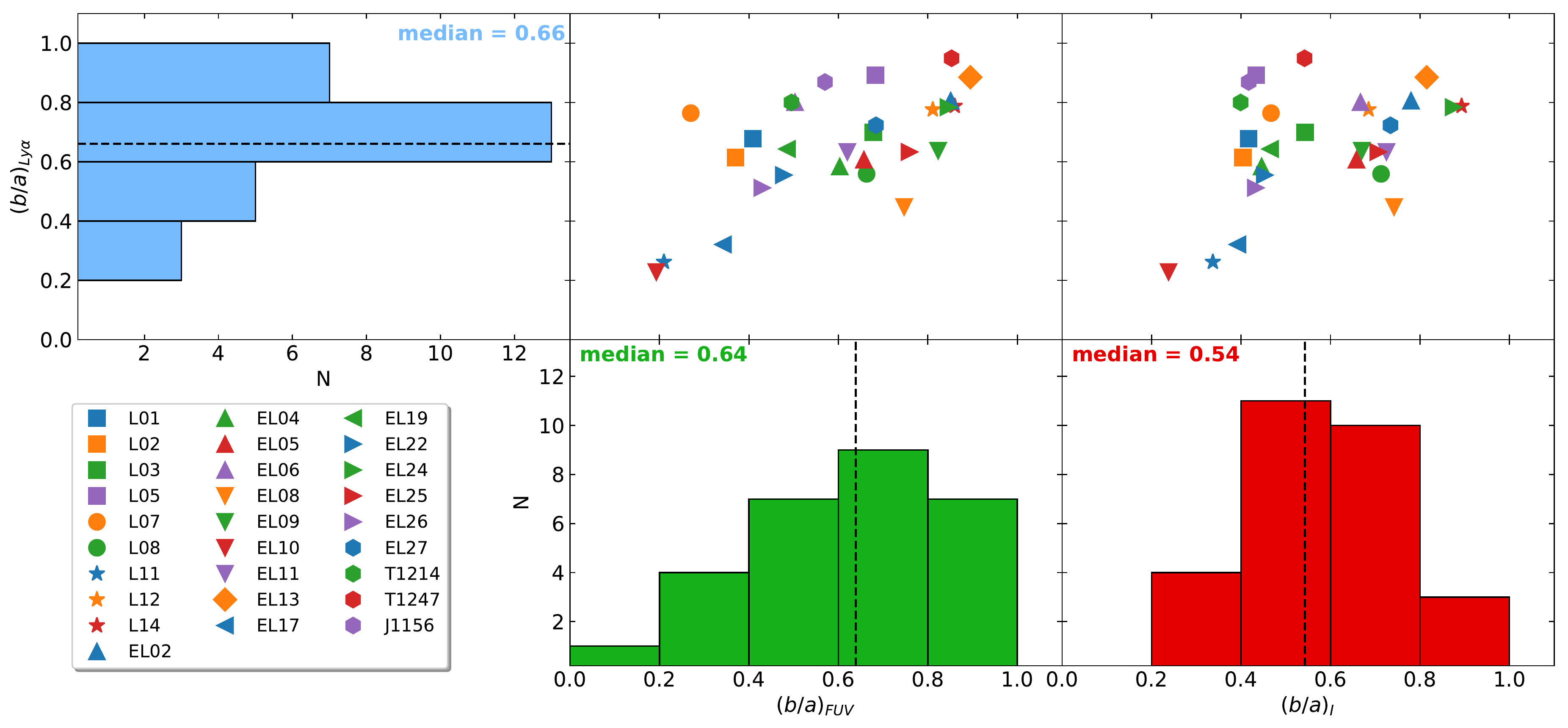}
    
  \caption{
          Top left panel: histogram showing the distribution of Ly$\alpha$ axis ratio ($\mathrm{(b/a)_{Ly\alpha}}$). Top middle panel: $\mathrm{(b/a)_{Ly\alpha}}$ versus the FUV axis ratio ($\mathrm{(b/a)_{FUV}}$). Top right panel: $\mathrm{(b/a)_{Ly\alpha}}$ versus the I band axis ratio ($\mathrm{(b/a)_I}$). Bottom middle panel: histogram showing $\mathrm{(b/a)_{FUV}}$ distribution. Bottom right panel: histogram showing $\mathrm{(b/a)_I}$ distribution.
          }
\label{b_a_result_sec_fig}
\end{figure*}
Another morphological parameter used in this study is the axis ratio. This parameter encapsulates more information about the light distribution than the centroid shift. Because, instead of collapsing all the information into a single value, it indicates how the spatial distribution (light distribution) differs in different directions. Axis ratio (b/a) is the ratio of the semi-minor axis to the semi-major axis; this parameter varies between zero and one. 
\par
The top left panel in Fig. \ref{b_a_result_sec_fig} shows the distribution of Ly$\alpha$ axis ratio ($\mathrm{(b/a)_{Ly\alpha}}$), the measured Ly$\alpha$ axis ratio ranges between 0.23, to 0.95, with the median size of 0.66. The top middle and top right panel show the measured ($\mathrm{(b/a)_{Ly\alpha}}$) versus the FUV ($\mathrm{(b/a)_{FUV}}$), and I band ($\mathrm{(b/a)_I}$) axis ratios, respectively. The FUV axis ratio shows the spatial distribution of the young stellar population, while the I band axis ratio indicate the spatial distribution of both young and old stellar populations. The bottom panels show the histograms of FUV (bottom middle panel) and I band axis ratio (bottom right). We see that the distribution of the Ly$\alpha$ axis ratio is more similar to the FUV axis ratio distribution compared to the I band axis ratio distribution. The FUV axis ratio ranges from 0.19 to 0.90 with a median size of 0.64, while the I band axis ratio ranges between 0.24 to 0.89 with a median size of 0.54. The Ly$\alpha$ emission seems to follow the general FUV morphology of the galaxy (as measured by an axis ratio), despite radiative transfer effects.
\subsection{Isophotal analysis of the Ly$\alpha$ halo} \label{rsc_iso_res_sec}
\begin{figure}
 \centering
    \includegraphics[width=\linewidth]{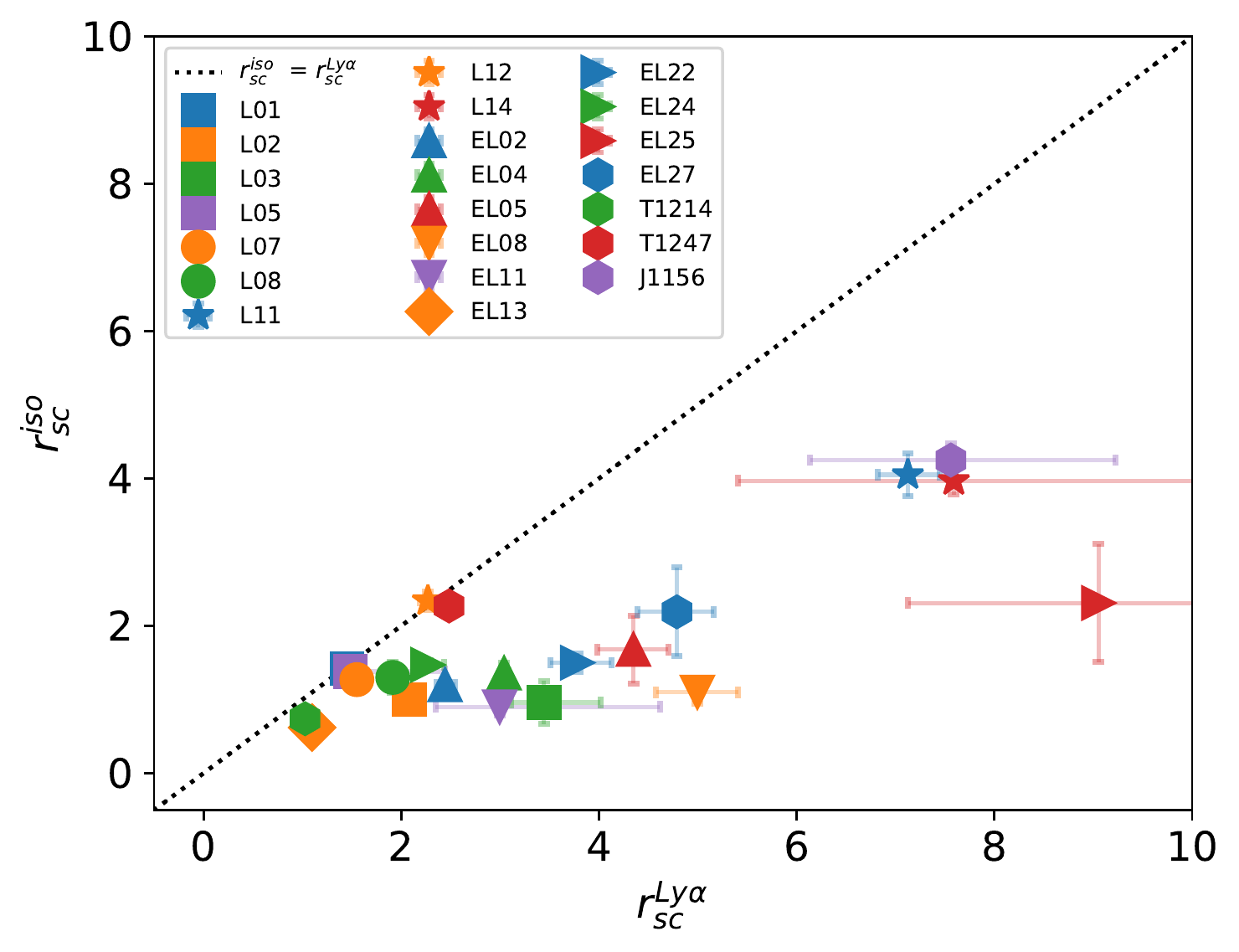}
  \caption{
           Ly$\alpha$ halo scale lengths assessed through the isophotal approach versus the Ly$\alpha$ halo scale lengths derived from fitting the double exponential function to the Ly$\alpha$ SB profiles using the circular aperture.  
          }
     \label{rsc_iso_vs_rsc_circ_fig}
\end{figure}
Figure \ref{rsc_iso_vs_rsc_circ_fig} shows the isophotal Ly$\alpha$ halo scale length ($\mathrm{r_{sc}^{iso}}$) versus the extent derived from circular aperture analysis ($\mathrm{r_{sc}^{Ly\alpha}}$). It should be noted that ELARS06, ELARS09, ELARS10, ELARS17, ELARS19, and ELARS26, are not included in this analysis (only this section and Sec. \ref{rsc_iso_discuss_sec}). These galaxies are among the faintest galaxies in our sample, and their Ly$\alpha$ emission level was not high enough to see them in all the four isophotes discussed in Sec. \ref{rsc_iso_method_sec}.
\par
The majority of the galaxies appear more extended using circular apertures. This is expected because an isophotal profile is always narrower than a circular profile (the profiles are equal for a perfectly circular symmetric source). For example, in the extreme case of the edge-on galaxy LARS09, where the measured $\mathrm{r_{sc}^{Ly\alpha}}$ is too large ($\sim$ 60 kpc, due to the morphology of the galaxy), it has a more reasonable measured $\mathrm{r_{sc}^{iso}}$ (2.8 kpc). The measured $\mathrm{r_{sc}^{iso}}$ ranges between 0.62, to 4.25, with the median size of 1.4 kiloparsec.
\subsection{Characteristics of the star forming regions} \label{res_sf_sec}
\subsubsection{Size and FUV intensity of the star forming regions}\label{rsfrd_fuvsb_res_sec}
\begin{figure}
 \centering
     \includegraphics[width=\linewidth]{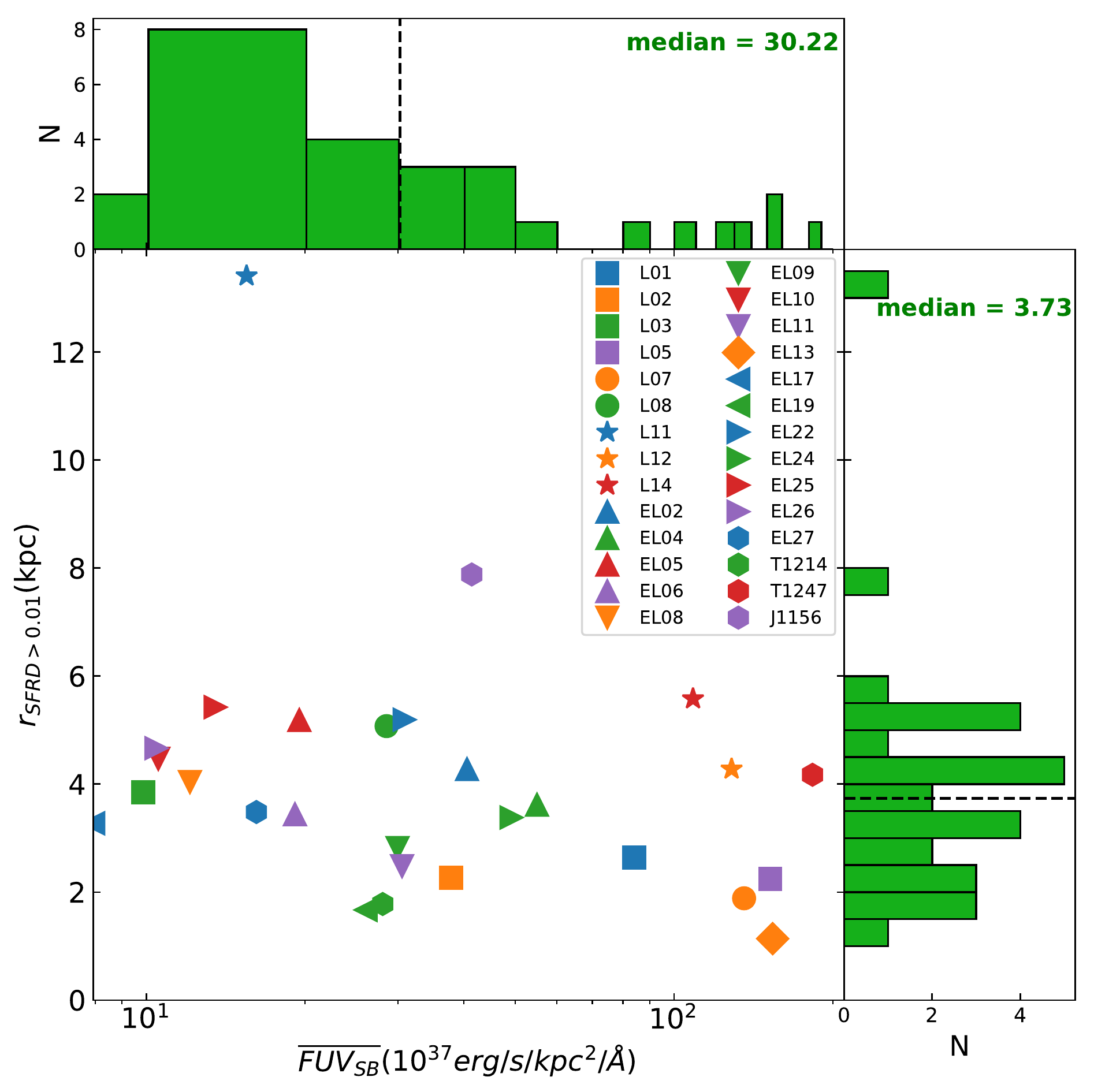}
    
  \caption{
          Size of the star forming region ($\mathrm{r_{SFRD > 0.01}}$), versus the average FUV SB ($\mathrm{\overline{FUV_{SB}}}$) within the region. The histograms on the top, and left show the distributions and the measured median for $\mathrm{\overline{FUV_{SB}}}$, and $\mathrm{r_{SFRD > 0.01}}$, respectively. 
          }
     \label{r_sfrd_fuv_sb_result_fig}
\end{figure}
The majority of the Ly$\alpha$ photons are produced in the star-forming (SF) regions. Thus, we also study the star-forming regions and their properties. First, we investigate the size of the SF regions where most of the stars are formed, by measuring the size of the SF regions where the SFRD is higher than 0.01 $\mathrm{M_\odot \, yr^{-1} \,  kpc^{-2}}$ (see Sec. \ref{sf_properties_analysis_sec}). We represent this size by the equivalent radius (the radius of a circle with the same area, $\mathrm{r_{SFRD > 0.01}}$). Second, we study the FUV intensity within this region, by measuring the average FUV SB ($\mathrm{\overline{FUV_{SB}}}$) within the region where SFRD > 0.01 $\mathrm{M_\odot \, yr^{-1} \,  kpc^{-2}}$. We use this average FUV surface brightness as a proxy for SFR density.
\par
Figure. \ref{r_sfrd_fuv_sb_result_fig} shows how $\mathrm{r_{SFRD > 0.01}}$ varies with $\mathrm{\overline{FUV_{SB}}}$. The range covered by $\mathrm{r_{SFRD > 0.01}}$ is between 1.14 to 13.41, with a median of 3.73 kiloparsec, the panel in histogram in right side shows the distribution of $\mathrm{r_{SFRD > 0.01}}$. The average FUV SB within the star forming region $\mathrm{\overline{FUV_{SB}}}$ ranges between $7.98 \times 10^{37}$ - $1.83 \times 10^{39}$ with the median size of $3.02 \times 10^{38}$ $erg/s/kpc^2/\si{\angstrom}$, the top panel in Fig. \ref{r_sfrd_fuv_sb_result_fig} shows the distribution of this quantity. We discuss the relation between $\mathrm{r_{SFRD > 0.01}}$, $\mathrm{\overline{FUV_{SB}}}$ and the other quantities in Sec. \ref{dis_sec}.
\section{Discussion} \label{dis_sec}
In this section, we explore the galaxies in our sample to find possible relation(s) between Ly$\alpha$ observables and the quantities used for studying the Ly$\alpha$ morphology. We also study the relation(s) between the Ly$\alpha$ morphological properties and host galaxy properties, such as stellar mass and I band axis ratio. In particular, we investigate how the Ly$\alpha$ morphology varies with the star-formation properties of the galaxies. Finally, we compare our results to similar measurements on the high redshift galaxies.
\par
We use the following global observables (from Melinder et al., in prep.): \textbf{stellar mass}:  estimated with the pixel SED fitting (LaXs), \textbf{Ly$\alpha$ escape fraction ($\mathrm{f_{esc}}$)}: measured using the Ly$\alpha$ and H$\alpha$ continuum maps, \textbf{nebular reddening ($\mathrm{E(B - V)_n}$}: the average dust extinction in H\rm{II} regions in the host galaxies (computed from the Balmer decrement), \textbf{Ly$\alpha$ total luminosity ($\mathrm{L_{Ly\alpha}}$}, using the Ly$\alpha$ emission map), \textbf{Ly$\alpha$ equivalent width ($\mathrm{EW_{Ly\alpha}}$}: measured using the Ly$\alpha$ and best-fit Ly$\alpha$ continuum maps), and \textbf{FUV total luminosity ($\mathrm{L_{FUV}}$}: using the best-fit Ly$\alpha$ continuum maps). The aperture used for the global observables is a circle centred on the brightest FUV pixel, and with a radius determined by growth curve analysis on the Ly$\alpha$ profiles (using the radius where the S/N of the Ly$\alpha$ SB drops below one). 
\par
In our investigation, we use the following quantities which carry information on the Ly$\alpha$ halo properties of the galaxies: Ly$\alpha$ halo scale length, Ly$\alpha$ halo fraction; and morphological parameters, such as: $\Delta$C (see Sec. \ref{cf_method_sec}), axis ratio, $\Delta$PA (see Sec. \ref{ps_angle_axis_ratio_sec}), and the isophotal Ly$\alpha$ halo scale length (see Sec. \ref{rsc_iso_method_sec}). Moreover, we use the following quantities that give insights on the star-forming properties of the galaxies: size of the SF regions, the average FUV SB within these regions, and the FUV axis ratio. We also look at the I band axis ratio as a quantity that characterises the stellar distribution of the host galaxies.
\par
To quantify possible (anti-)correlations between each two parameters, we determine the Spearman's rank correlation coefficient ($\mathrm{\rho_s}$). This parameter varies between -1 to +1 (+1 showing correlation and -1 showing anti-correlation while zero indicates no correlation at all) and p-value ($\mathrm{p_0}$) indicating the probability of an uncorrelated system with the same measured Spearman's correlation coefficient having arisen by chance \citep{Ivezi2014book}. We use \texttt{stats.Spearmanr} task from the \texttt{scipy} package in \texttt{python} for assessing the Spearman's correlation coefficients and p-values. We require a threshold of $\mathrm{p_0}$ below 0.05 ($\mathrm{p_0}$ < 0.05) for discussing the relation between any two given parameters (see Fig. \ref{lya_obs_lya_morph_pvalue_fig}). 
\par
It should be noted that we investigate the relation between many quantities, and one might be concerned that looking for correlations between too many parameters would finally result in spurious correlations exceeding a given significance. We addressed this issue by i) noting that some of the quantities investigated in our study are not independent of each other. For example, $\mathrm{L_{Ly\alpha}}$ and $\mathrm{L_{FUV}}$, or $\mathrm{f_{esc}}$ and $\mathrm{EW_{Ly\alpha}}$ correlate \citep[][Melinder et al. (in prep)]{Sobral2019}, and are not independent of each other. ii) choosing a threshold for the assessed Spearman's p-values ($\mathrm{p_0}$ < 0.05). We note that this threshold is only used as a tool to discuss the relation between different quantities. We are not claiming any physical relation between any given quantities based on the assessed $\mathrm{p_0}$.
Furthermore, we would like to stress that our study is an exploratory one. We use the $\mathrm{p_0}$ criterion to select interesting findings in the data set, which we then discuss further.
\par
It should be noted that the high spatial resolution data available for the LARS sample offers the opportunity to study the morphology of the galaxies at $\lessapprox$ kpc scales. This is a piece of the puzzle that was missing in previous Ly$\alpha$ studies. Consequently, there is not much prior information, and, knowing the complexities of Ly$\alpha$ RT, it is hard to formulate strong hypotheses that can be tested.
\subsection{Ly$\alpha$ observables and Ly$\alpha$ morphology} \label{lya_obs_discuss}
\begin{figure}
 \centering
     \includegraphics[width=\linewidth]{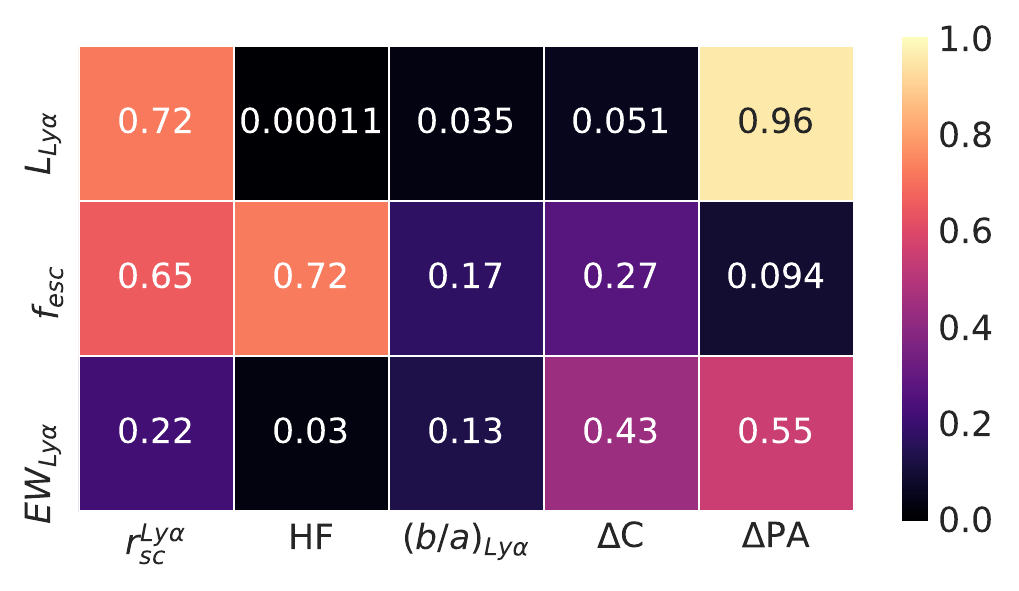}
    
  \caption{
          Spearman $\mathrm{p_0}$ value for the Ly$\alpha$ observables and the quantities used for studying Ly$\alpha$ morphology.
          }
     \label{lya_obs_lya_morph_pvalue_fig}
\end{figure}
Here, we study the relations between global Ly$\alpha$ observables: Ly$\alpha$ total luminosity, Ly$\alpha$ equivalent width and Ly$\alpha$ escape fraction, and Ly$\alpha$ morphological properties: Ly$\alpha$ halo scale length, Ly$\alpha$ halo fraction, axis ratio, $\Delta$C, and $\Delta$PA. In other words, we investigate the relation between global Ly$\alpha$ observables and the quantities that are connected to how the Ly$\alpha$ photons redistribute within the host galaxies. Figure \ref{lya_obs_lya_morph_pvalue_fig} shows the measured Spearman's p-value between the Ly$\alpha$ observables and the quantities used for studying the Ly$\alpha$ morphology. As noted above, we only discuss cases with a p-value less than 0.05. For the interested readers, we also show how Ly$\alpha$ observables vary with all the morphological measurements in appendix \ref{lya_obs_appx}.
\subsubsection{Ly$\alpha$ luminosity} \label{lya_lum_discuss}
\begin{figure*}
 \centering
     \includegraphics[width=\textwidth]{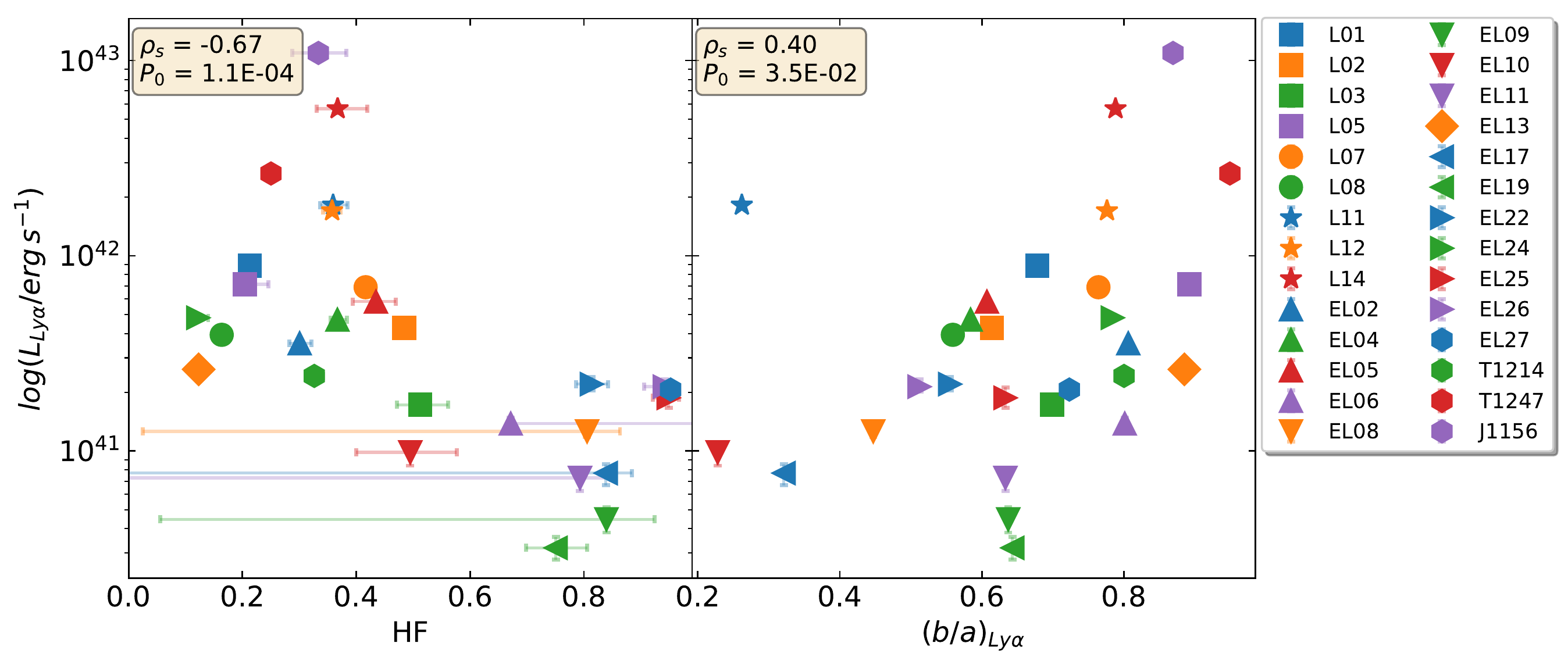}

  \caption{
          How Ly$\alpha$ luminosity ($\mathrm{L_{Ly\alpha}}$) varies with the HF, and axis ratio $\mathrm{(b/a)_{Ly\alpha}}$. The measured Spearman's coefficient and p-value between $\mathrm{L_{Ly\alpha}}$ and each quantity are displayed in the corresponding panels.
          }
     \label{lya_lum_corr_fig}
\end{figure*}
Our results show that Ly$\alpha$ luminosity ($\mathrm{L_{Ly\alpha}}$) anti-correlates with the HF (left panel in Fig. \ref{lya_lum_corr_fig}), meaning that Ly$\alpha$ halo contributes less to the overall luminosity in galaxies that are bright in Ly$\alpha$ while the majority of Ly$\alpha$ luminosity is coming from the Ly$\alpha$ halos in galaxies that are faint in Ly$\alpha$.
Thus, faint LAEs have more Ly$\alpha$ emission coming out from their halos, which causes them to exhibit overall low SB in Ly$\alpha$. Consequently, detecting faint Ly$\alpha$ emitters at high redshift is even more challenging than what their global Ly$\alpha$ fluxes would indicate, and with limited spectroscopic apertures, this emission could even be missed completely. However, we note that the trend observed between HF and $\mathrm{L_{Ly\alpha}}$ in our sample has not been observed in high$-z$ surveys \citep{Leclercq2017AA, Wisotzki2016AA}, possibly due to the lack (or absence) of faint LAEs in these surveys. This result may also suggest that conclusions drawn from studying bright LAEs only, may be biased. Keep in mind that, unlike the MUSE sample that is selected on the Ly$\alpha$ emission of galaxies, our sample selection is agnostic to the Ly$\alpha$ properties. 
\par
We also observe a correlation between the Ly$\alpha$ luminosity and the axis ratio (right panel in Fig. \ref{lya_lum_corr_fig}), suggesting that bright galaxies appear rounder than the faint galaxies in Ly$\alpha$.
\subsubsection{Ly$\alpha$ escape fraction} \label{fesc_discuss}
We do not see any relation between any of the Ly$\alpha$ morphological quantities measured in this study and the Ly$\alpha$ escape fraction.
\subsubsection{Ly$\alpha$ equivalent width} \label{lya_ew_discuss}
\begin{figure}
 \centering
     \includegraphics[width=\linewidth]{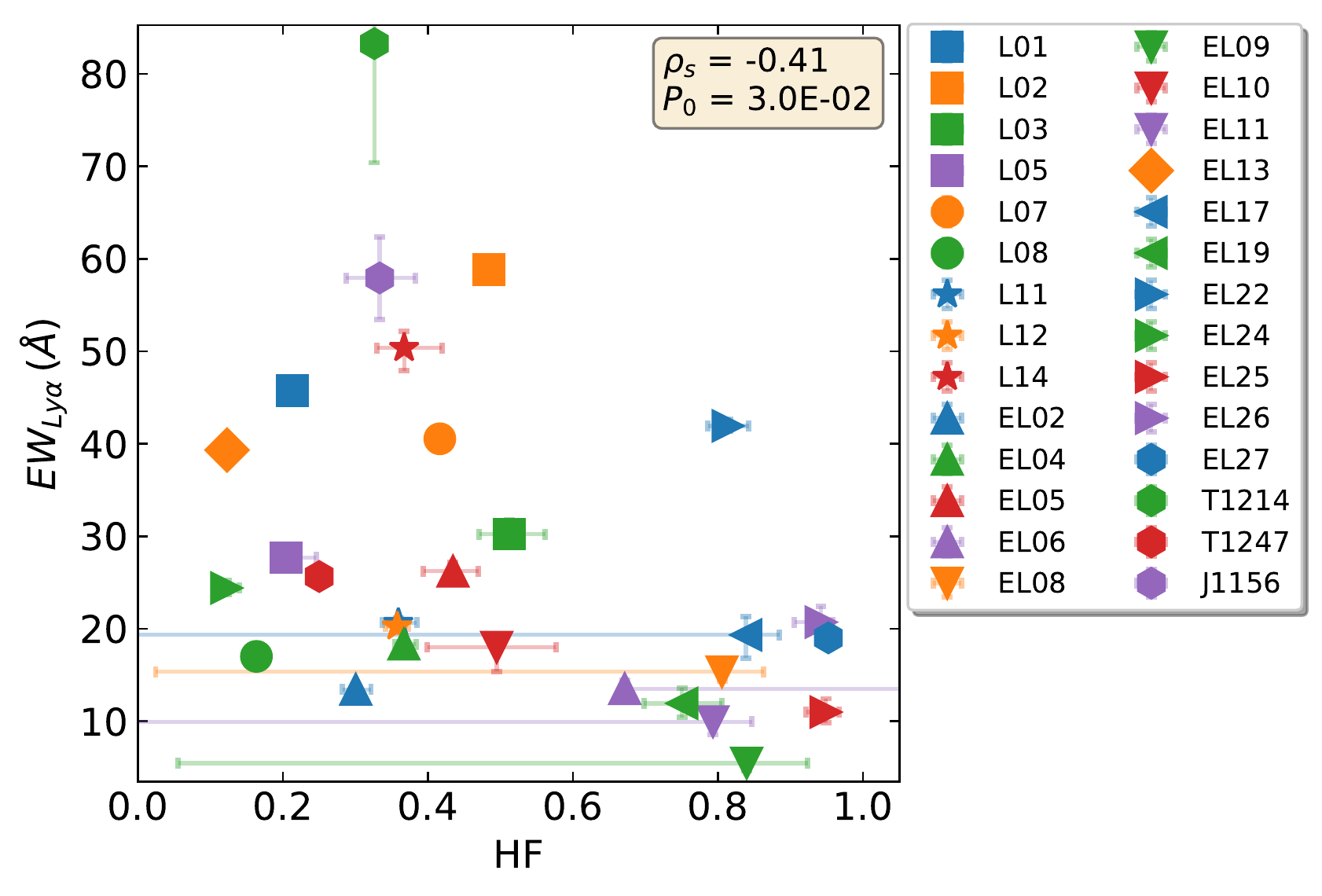}

  \caption{
          How the measured Ly$\alpha$ equivalent width luminosity ($\mathrm{EW_{Ly\alpha}}$) varies with the HF. 
          The measured Spearman's coefficient and p-value between $\mathrm{EW_{Ly\alpha}}$ and the HF are displayed in the top right.
          }
     \label{ew_corr_fig}
\end{figure}
\par
We see a weak anti-correlation between Ly$\alpha$ equivalent width ($\mathrm{EW_{Ly\alpha}}$) and the HF (see Fig. \ref{ew_corr_fig}). In other words, $\mathrm{EW_{Ly\alpha}}$ is lower in halo dominated galaxies (HF$\sim$1), indicating either lower escape fractions or an intrinsically older or higher metallicity stellar population (with a lower production efficiency of ionising photons, $\xi_{ion}$, e.g. \cite{Matthee2017MNRAS}). 
This result is in line with the high-$z$ findings reported by \cite{Steidel2011ApJ}, where they find that galaxies with weak central Ly$\alpha$ or central absorption (e.g. LBGs) still have large halos (larger halo fractions).
\subsection{Ly$\alpha$ morphology and the stellar properties of the host galaxies} \label{morph_lya_discuss}
\begin{figure*}
 \centering
     \includegraphics[width=\textwidth]{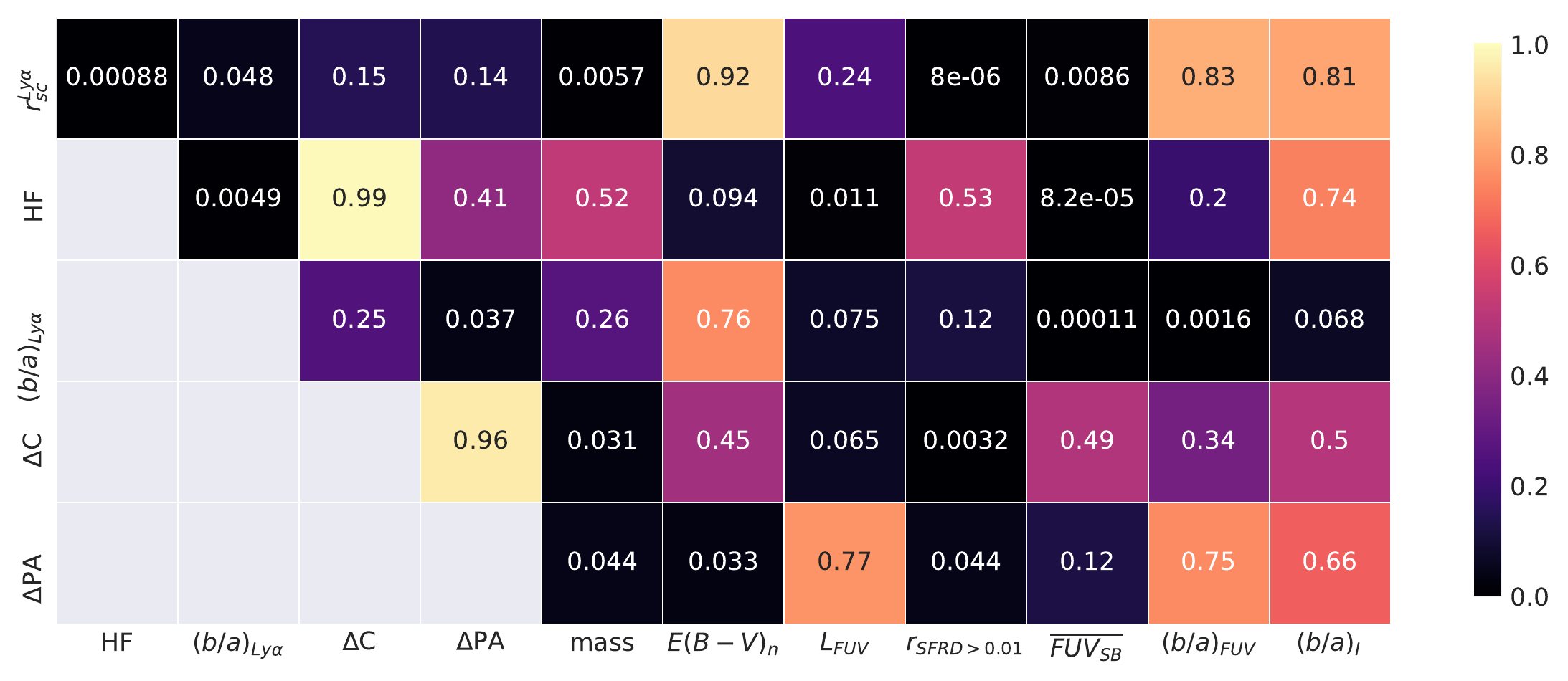}
    
  \caption{
          Spearman $\mathrm{p_0}$ value for the Ly$\alpha$ morphological quantities and the properties of the host galaxies.
          }
     \label{lya_morph_lya_morph_stell_prop_pvalue_fig}
\end{figure*}
In this subsection, we study the interdependencies of the Ly$\alpha$ morphological quantities (Ly$\alpha$ halo scale length, Ly$\alpha$ halo fraction, axis ratio, $\Delta$C, and $\Delta$PA). Moreover, we also investigate how global parameters such as the stellar mass, the nebular reddening, and star-forming characteristics (size of the SF regions and the average intensity of the FUV SB) affect the Ly$\alpha$ morphology. Figure \ref{lya_morph_lya_morph_stell_prop_pvalue_fig} shows the Spearman's p values between these quantities. 
Once again, we only discuss cases with a p-value less than 0.05. For the interested reader, all cases are presented in appendix \ref{lya_morph_appx}.
\subsubsection{Extent of the Ly$\alpha$ halos} \label{rsc_discuss}
\begin{figure*}
 \centering
     \includegraphics[width=\textwidth]{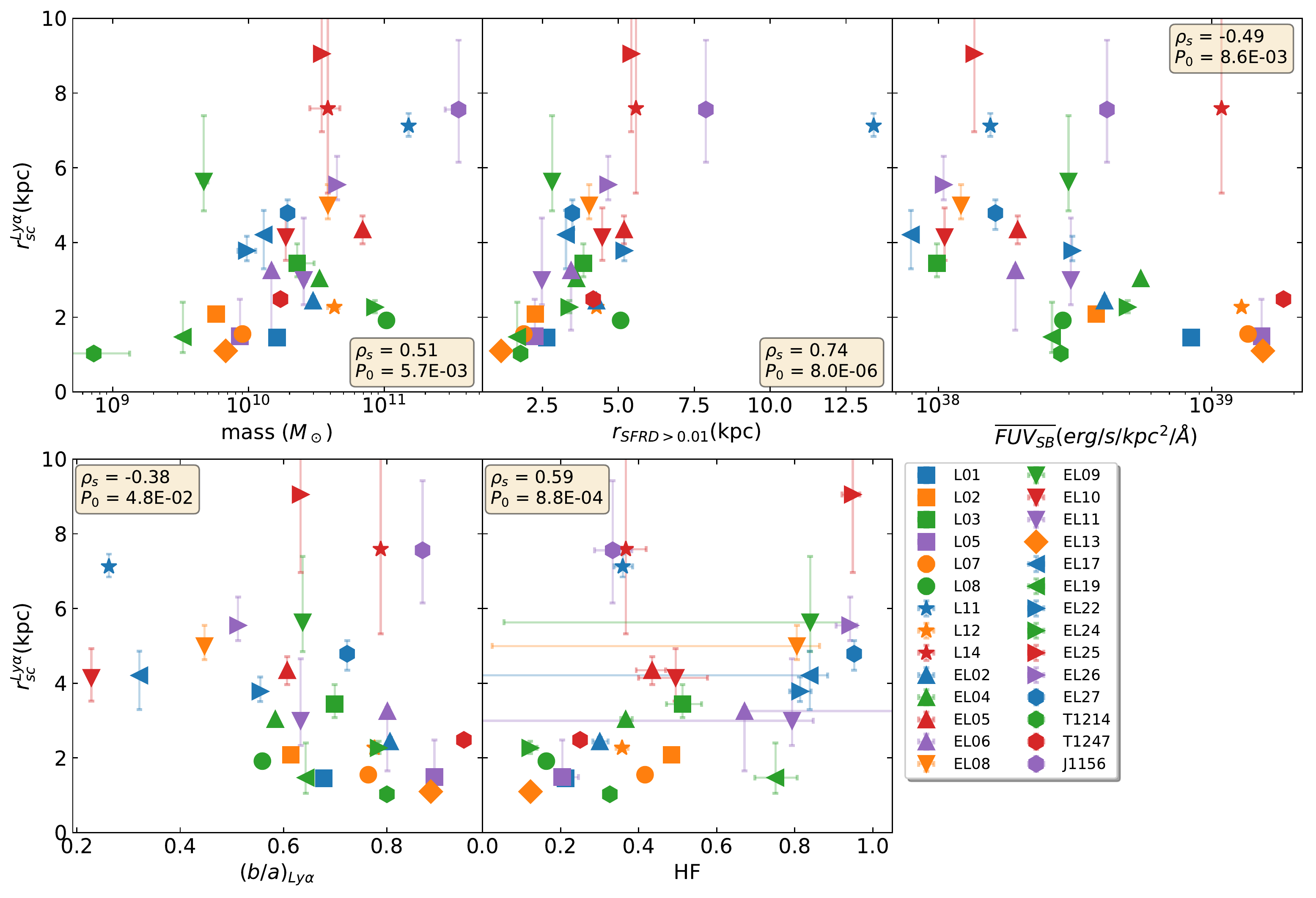}

  \caption{
          How the Ly$\alpha$ halo scale length ($\mathrm{r_{sc}^{Ly\alpha}}$) varies with the stellar mass, size of the SF regions, the average FUV SB within the SF regions, the axis ratio, and the HF. The measured Spearman's coefficient and p-value between $\mathrm{r_{sc}^{Ly\alpha}}$ and each quantity are displayed in the corresponding panels.
          }
     \label{rsc_ly_corr}
\end{figure*}
Our results show that Ly$\alpha$ halo extent ($\mathrm{r_{sc}^{Ly\alpha}}$) correlates with the stellar mass, size of the SF regions, and the HF, and anti-correlates with the average FUV SB within the SF regions, and the axis ratio (see Fig. \ref{rsc_ly_corr}). The correlation between $\mathrm{r_{sc}^{Ly\alpha}}$ and the stellar mass suggests that massive galaxies possess more extended Ly$\alpha$ halos. Perhaps this also shows that the mass-size relation observed for normal galaxies at optical wavelengths (e.g. \cite{Fathi2010MNRAS, Trujillo2020MNRAS}) also holds for Ly$\alpha$ emission. 
\par
One may expect that the Ly$\alpha$ halo extent should grow with the size of the SF regions. Since if the ionising photons cover a larger space in the galaxies, Ly$\alpha$ halos can get even more extended due to scattering. Indeed, we see a strong correlation between $\mathrm{r_{SFRD > 0.01}}$ and $\mathrm{r_{sc}^{Ly\alpha}}$. In other words, we see that galaxies with more extended star-forming regions also have larger Ly$\alpha$ halos. We see an anti-correlation between the Ly$\alpha$ halo extent and $\mathrm{\overline{FUV_{SB}}}$ suggesting that galaxies with high SFR density do not have very extended Ly$\alpha$ emission. We see an anti-correlation between $\mathrm{r_{sc}^{Ly\alpha}}$ and the axis ratio suggesting that galaxies with more elongated Ly$\alpha$ morphology are also more extended in Ly$\alpha$. This is likely driven by the use of circular annuli for the photometry (We do not see any relation between $\mathrm{r_{sc}^{iso}}$ and $\mathrm{(b/a)_{Ly\alpha}}$ see Fig. \ref{r_iso_appx_vs_all}).
\par
Finally, our data show a strong correlation between HF and $\mathrm{r_{sc}^{Ly\alpha}}$. In other words, we find galaxies that have more extended Ly$\alpha$ halos also have higher HF. This is in contrast with findings in previous studies where they see no trend between the measured Ly$\alpha$ halo scale length and the HF in their sample: \cite{Wisotzki2016AA}, and \cite{Leclercq2017AA} (see Fig. \ref{LARS_comp_MUSE_fig}). However, it is important to note that $\mathrm{r_{sc}^{Ly\alpha}}$ is used in the way we define HF (see Eq. \ref{hf_eq}), and this correlation might reflect that these two quantities are implicitly related to each other. 
\subsubsection{Ly$\alpha$ halo fraction} \label{hf_discuss}
\begin{figure*}
 \centering
    \includegraphics[width=\textwidth]{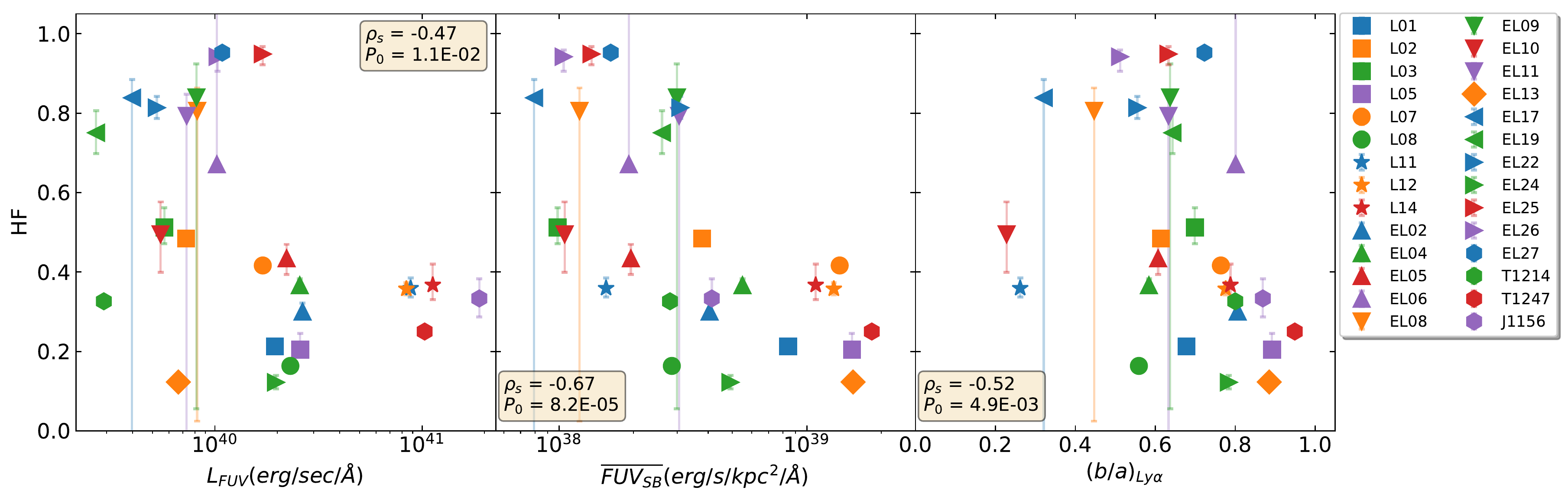}
    
  \caption{
          How the Ly$\alpha$ halo fraction (HF) varies with the total FUV luminosity ($\mathrm{L_{FUV}}$), the average FUV SB within the SF regions ($\mathrm{\overline{FUV_{SB}}}$), and the axis ratio ($\mathrm{(b/a)_{Ly\alpha}}$). The measured Spearman's coefficient and p-value between HF and each quantity are displayed in the corresponding panels.  
          }
     \label{hf_ly_corr}
\end{figure*}
Our results indicate that the Ly$\alpha$ halo fraction (HF) anti-correlates with the total FUV luminosity, the average FUV SB within the SF regions ($\mathrm{\overline{FUV_{SB}}}$) (see  Fig. \ref{hf_ly_corr}), and the axis ratio; and correlates with the Ly$\alpha$ halo scale length (see Sec. \ref{rsc_discuss}). 
\par
The anti-correlation between the HF and $\mathrm{L_{FUV}}$ implies that Ly$\alpha$ halos contribute more to the total Ly$\alpha$ luminosity in galaxies that have lower star formation rates (lower FUV luminosity) and therefore also lower intrinsic Ly$\alpha$ luminosity. Given the intrinsic relation between the Ly$\alpha$ and FUV luminosities, this relation re-portrays the anti-correlation between the $\mathrm{L_{Ly\alpha}}$ and the HF (see Sec. \ref{lya_lum_discuss}). We also see an anti-correlation between HF and $\mathrm{\overline{FUV_{SB}}}$ suggesting that the Ly$\alpha$ halos contribute more to the overall Ly$\alpha$ luminosity in SFGs galaxies that have lower SFR density. Finally, we find an anti-correlation between the HF and the axis ratio ($\mathrm{(b/a)_{Ly\alpha}}$). This indicates that the halos in galaxies with high HF are on average, more elongated. 
\subsubsection{Centroid shift} \label{cf_discus_sec}
\begin{figure*}
 \centering
     \includegraphics[width=\textwidth]{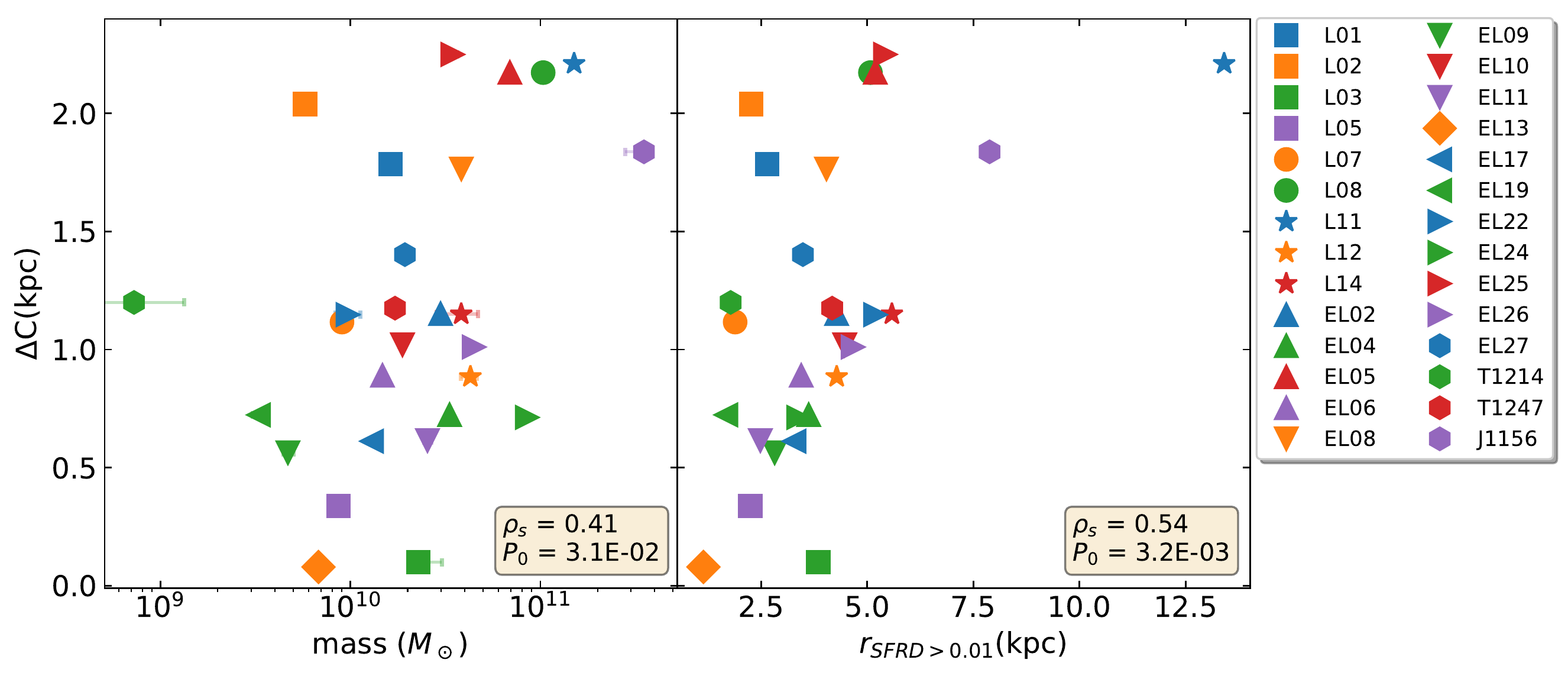}

  \caption{
         How the centroid shift ($\Delta$C) varies with the stellar mass and the size of the SF regions. The measured Spearman's coefficient and p-value between $\Delta$C and each quantity are displayed in the corresponding panels.       
          }
     \label{delta_cf_discuss_fig}
\end{figure*}
We see that the centroid shift ($\Delta$C see Eq. \ref{cerntroid_eq}) correlates with the stellar mass, and the size of the SF regions (see Fig. \ref{delta_cf_discuss_fig}). As discussed in Sec. \ref{rsc_discuss}, both stellar mass and the size of the SF regions correlate with the Ly$\alpha$ halo scale length. In other words, we see that centroid shift correlates with the quantities that are connected to more extended Ly$\alpha$ halos. This suggests that the centroid shift grows in galaxies with larger $\mathrm{r_{sc}^{Ly\alpha}}$, since Ly$\alpha$ halos do not grow symmetrically, and a larger Ly$\alpha$ halo means a larger difference between the Ly$\alpha$ and FUV measured centroid. However, it should be noted that even though we see that some galaxies with large $\mathrm{r_{sc}^{Ly\alpha}}$ possess large centroid shifts (e.g. LARS11, ELARS25, and J1156), we do not see a direct correlation between $\Delta$C and $\mathrm{r_{sc}^{Ly\alpha}}$.
\subsubsection{Position angle difference}
\begin{figure*}
 \centering
     \includegraphics[width=\textwidth]{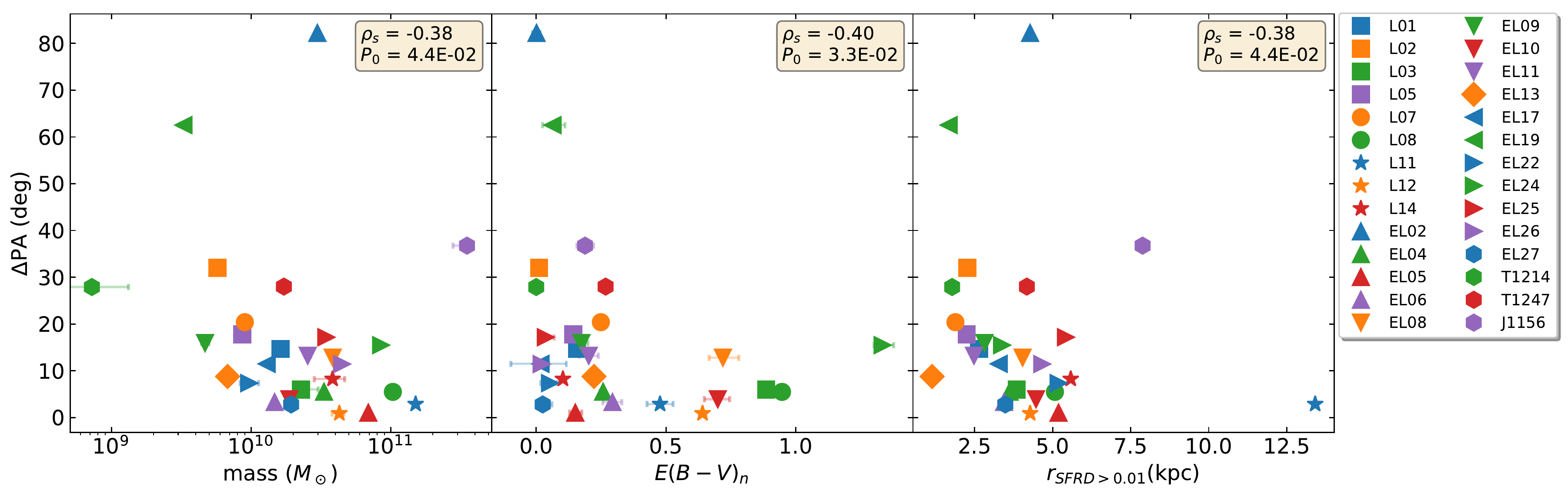}
  \caption{
          How the difference in the position angles ($\Delta$PA) varies with the stellar mass, the nebular reddening $\mathrm{E(B-V)_n}$, and the size of the SF regions. The measured Spearman's coefficient and p-value between $\Delta$PA and each quantity are displayed in the corresponding panels. 
          }
     \label{delta_pa_corr_fig}
\end{figure*}

We see that the difference in position angles ($\Delta$PA) anti-correlates with stellar mass, nebular reddening ($\mathrm{E(B-V)_n}$), and size of the SF regions (see Fig. \ref{delta_pa_corr_fig}). The anti-correlation between $\Delta$PA and the stellar mass suggests that Ly$\alpha$ and FUV misalignment is more significant in low-mass galaxies. 
We see an anti-correlation between $\Delta$PA and the nebular reddening, mainly due to a lack of dusty galaxies with high $\Delta$PA. Possibly, this is related to the paucity of dusty galaxies with high $\mathrm{r_{sc}^{Ly\alpha}}$ or HF, and that less Ly$\alpha$ scattering takes place in such galaxies. In galaxies that are dustier, the Ly$\alpha$ alignment is closer to the alignment of the star-forming regions; perhaps this is due to absorption of the Ly$\alpha$ photons by the dust particles, which results in less scattering of the Ly$\alpha$ photons, indicating that the Ly$\alpha$ photons follow the same path that the FUV photons take to escape the galaxies. Consequently, the Ly$\alpha$ and FUV alignments in these galaxies are more similar to each other. 
\par
The anti-correlation between $\Delta$PA and the size of the SF regions could be explained by there being more direct channels available for Ly$\alpha$ escape (and thus less scattering) in galaxies with larger SF regions. The Ly$\alpha$ photons are then escaping closer to where they were produced, and the overall emission is more similar to the FUV distribution. The anti-correlation between $\Delta$PA and the size of the SF regions is augmented by a small number of galaxies with high $\Delta$PA or high $r_{SFRD>0.01}$. A large $\Delta$PA in a large SF region requires a more coherent change of preferred direction than in a small one, where the effect could more likely occur stochastically.
\subsubsection{Ly$\alpha$ axis ratio} \label{b_a_discuss_sec}
\begin{figure*}
 \centering
     \includegraphics[width=\textwidth]{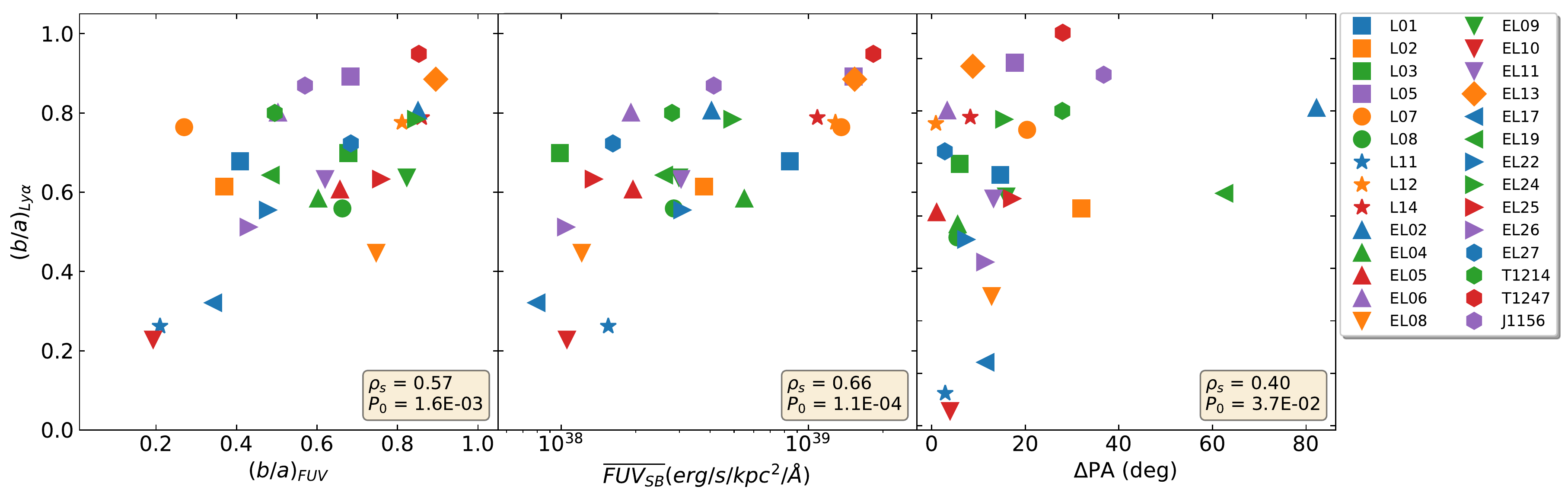}
  \caption{
          How the axis ratio ($\mathrm{(b/a)_{Ly\alpha}}$) varies with the FUV axis ratio ($\mathrm{(b/a)_{FUV}}$), the average FUV SB within the SF regions ($\mathrm{\overline{FUV_{SB}}}$), and the difference in the position angles ($\Delta$PA). The measured Spearman's coefficient and p-value between $\mathrm{(b/a)_{Ly\alpha}}$ and each quantity are displayed in the corresponding panels.
          }
     \label{axis_ratio_fig}
\end{figure*}
We see that axis ratio ($\mathrm{(b/a)_{Ly\alpha}}$) correlates with the FUV axis ratio ($\mathrm{(b/a)_{FUV}}$), the average FUV SB within the SF regions ($\mathrm{\overline{FUV_{SB}}}$), and the difference in the position angles ($\Delta$PA) (see Fig. \ref{axis_ratio_fig}).
The correlation between $\mathrm{(b/a)_{Ly\alpha}}$ and $\mathrm{(b/a)_{FUV}}$ suggests that Ly$\alpha$ distribution in the galaxies depends on the FUV distribution. Given that the majority of the ionising photons are reprocessed into Ly$\alpha$ photons, it is hardly a surprise that the Ly$\alpha$ morphology depends on the distribution of the young stellar population (if there would have not been any scattering or dust absorption, this correlation would likely be even stronger). We see that galaxies with high SFR density (high $\mathrm{\overline{FUV_{SB}}}$) appear rounder in their Ly$\alpha$ morphology (high $\mathrm{(b/a)_{Ly\alpha}}$). This could be the anti-correlation between the $\mathrm{r_{sc}^{Ly\alpha}}$ and $\mathrm{\overline{FUV_{SB}}}$ (see Sec. \ref{rsc_discuss}) from another perspective. 
Then, assuming that less scattering occurs in galaxies with high SFR density, these galaxies would also have less elongated Ly$\alpha$ morphology as well as less extended halos.
\par
Finally, the correlation between $\mathrm{(b/a)_{Ly\alpha}}$ and $\Delta$PA seem to indicate that in galaxies where Ly$\alpha$ photons escape more in off-axis direction compared to the FUV morphology, the Ly$\alpha$ morphology is rounder and more symmetric. However, we note that $\Delta$PA is unconstrained for high values of $\mathrm{(b/a)_{Ly\alpha}}$.
\subsection{Isophotal analysis of the Ly$\alpha$ halo} \label{rsc_iso_discuss_sec}
\begin{figure*}
 \centering
     \includegraphics[width=\textwidth]{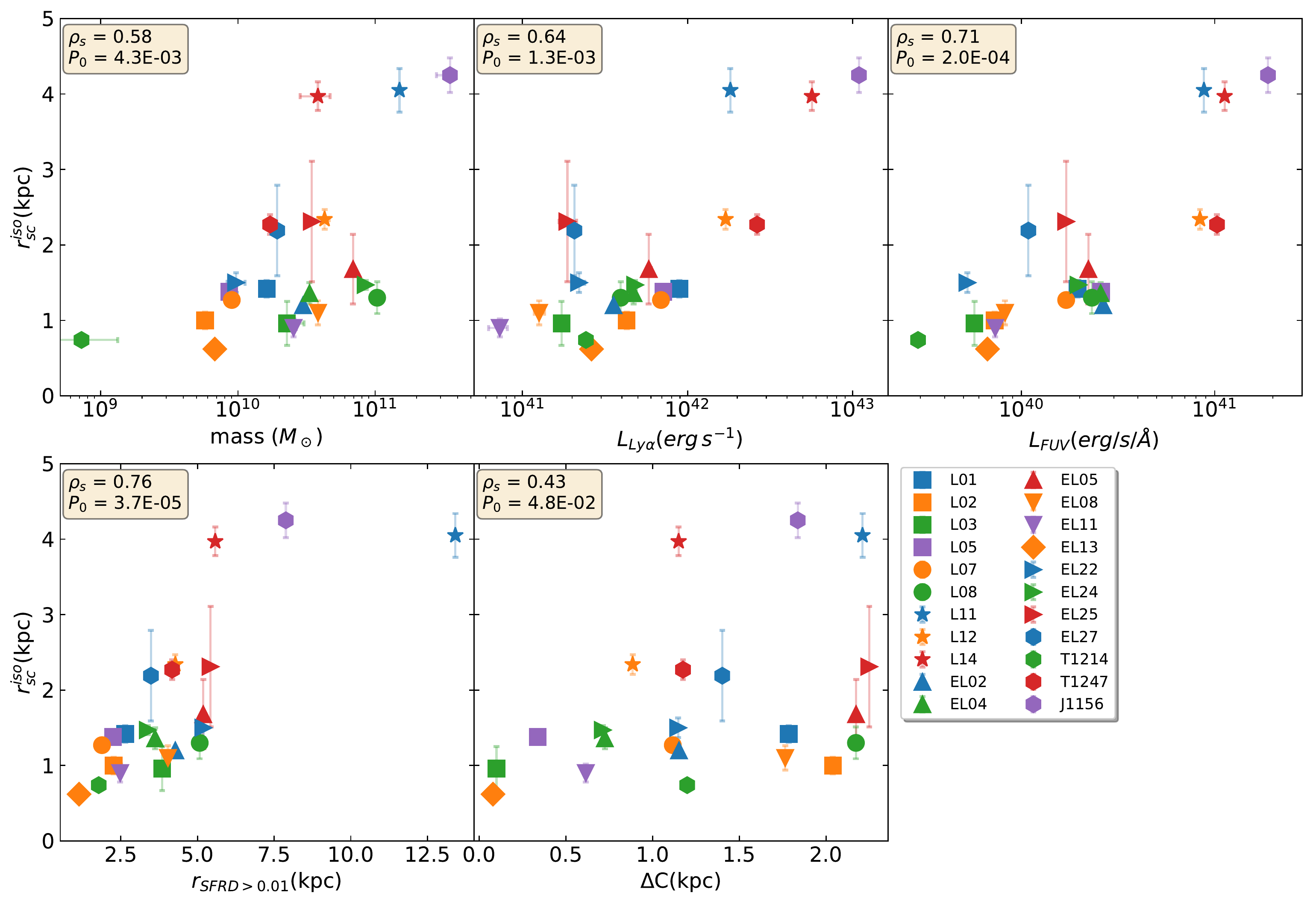}
    
  \caption{
            How the measured isophotal Ly$\alpha$ halo scale length ($\mathrm{r_{sc}^{iso}}$) varies with the stellar mass, total Ly$\alpha$ luminosity ($\mathrm{L_{Ly\alpha}}$), total FUV luminosity ($\mathrm{L_{FUV}}$), the size of the star forming regions ($\mathrm{r_{SFRD > 0.01}}$), and the centroid shift ($\Delta$C). The measured Spearman's coefficient and p-value between $\mathrm{r_{sc}^{iso}}$ and each quantity are displayed in the corresponding panels.
          }
     \label{r_iso_fig}
\end{figure*}
In this section, we look at the relation between the measured isophotal Ly$\alpha$ halo scale length ($\mathrm{r_{sc}^{iso}}$) and Ly$\alpha$ observables, quantities used for studying the Ly$\alpha$ morphology, and also the some of the global observables (stellar mass, and nebular reddening). While we only discuss the cases with a p-value less than 0.05, we provide how $\mathrm{r_{sc}^{iso}}$ varies which each quantity in appendix \ref{r_iso_appx_vs_all} for the interested readers.
\par
Our analysis indicate that the isophotal Ly$\alpha$ scale lengths ($\mathrm{r_{sc}^{iso}}$) correlate with stellar mass, total Ly$\alpha$ luminosity ($\mathrm{L_{Ly\alpha}}$), total FUV luminosity ($\mathrm{L_{FUV}}$), the size of the star forming regions, and the centroid shift ($\Delta$C) (see Fig. \ref{r_iso_fig}). 
\par
In other words, our data indicate that the size-mass relation in Ly$\alpha$ emission line holds using the isophotal Ly$\alpha$ scale lengths, too. We see that $\mathrm{r_{sc}^{iso}}$ strongly correlate with $\mathrm{L_{Ly\alpha}}$, and $\mathrm{L_{FUV}}$. We note that if a single exponential describes the isophotal Ly$\alpha$ halo SB profile well, the halo Ly$\alpha$ luminosity and the total Ly$\alpha$ luminosity will be directly proportional to the square of the scale length ($\propto (\mathrm{r_{sc}^{iso}})^2$). Hence, the strong correlation between $\mathrm{r_{sc}^{iso}}$ and $\mathrm{L_{Ly\alpha}}$ might simply reflect that a single exponential is a good choice for describing the isophotal Ly$\alpha$ halo SB profile. Due to the stellar morphology of the galaxies in our sample and the asymmetric Ly$\alpha$ halos, this was probably not captured in the circular aperture analysis. Moreover, given the intrinsic relation between the Ly$\alpha$ photons and FUV photons, and the strong correlation between the Ly$\alpha$ luminosity and FUV luminosity, the correlation between $\mathrm{r_{sc}^{iso}}$ and $\mathrm{L_{FUV}}$ is not surprising, at all. It should be noted that we see the aperture size (where $(S/N)_{Ly\alpha}$ drops below one) used for measuring the global properties of the galaxies (stellar mass, $\mathrm{L_{Ly\alpha}}$, $\mathrm{L_{FUV}}$, $\mathrm{f_{esc}}$, etc) strongly depends on $\mathrm{r_{sc}^{iso}}$. Consequently, at least part (or maybe all) of the observed correlation between $\mathrm{r_{sc}^{iso}}$, and the stellar mass, $\mathrm{L_{Ly\alpha}}$, and $\mathrm{L_{FUV}}$ can be because of this aperture effect. Galaxies with more extended Ly$\alpha$ emission in the halo will have a larger area with significant (S/N$>1$) emission and, therefore, a larger global aperture.
\par
Similar to $\mathrm{r_{sc}^{Ly\alpha}}$ (see Sec. \ref{rsc_discuss}), $\mathrm{r_{sc}^{iso}}$ correlates with the size of the SF regions. Finally, we see that $\mathrm{r_{sc}^{iso}}$ correlates with the centroid shift. This finding suggests Ly$\alpha$ morphology is more offset from the FUV morphology, and Ly$\alpha$ halos extend further away from the SF regions in galaxies with a more pronounced scattering of Ly$\alpha$ photons. The correlation with centroid shift is stronger for the isophotal scale length compared to $\mathrm{r_{sc}^{Ly\alpha}}$. The reason for this could be that i) the difference between stellar and Ly$\alpha$ morphology is better captured in the isophotal analysis, and ii) the isophotal analysis is done on a subset of the galaxies used for the circular analysis (see Sec.\ref{rsc_iso_res_sec}).
\subsection{Comparison with high$-z$ galaxies} \label{muse_discussion}
\begin{figure*}
 \centering
     \includegraphics[width=\textwidth]{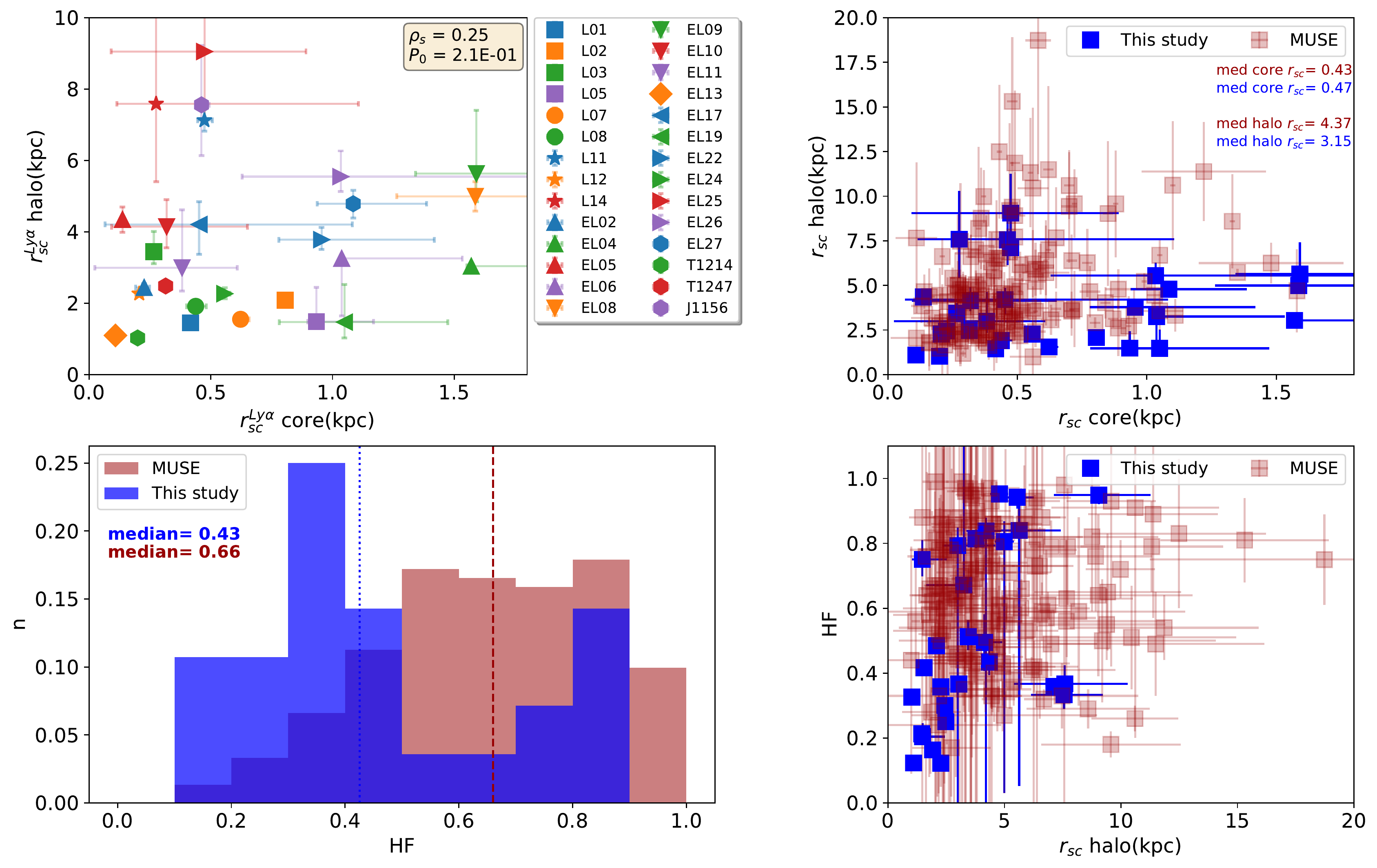}

  \caption{
          Our results and how they compare to the MUSE study. Top left: how the Ly$\alpha$ halo scale length is compared to the Ly$\alpha$ core extent. Top right: the Ly$\alpha$ halo extent versus the Ly$\alpha$ core extent for both MUSE and our sample, the core extent is somewhat similar. However, Ly$\alpha$ is more extended in the MUSE sample compared to ours. Bottom left: The distribution of the HF in MUSE and our study, the bimodal distribution in HF in our study is not present in the LAEs observed in the MUSE study. Moreover, the median of the HF in the MUSE study (0.66) is larger than the measured median in our study (0.43). Bottom right, HF versus the Ly$\alpha$ halo extent measured in our sample and MUSE. The space covered in the HF-$\mathrm{r_{sc}^{Ly\alpha}}$ space in our sample is within the space covered by MUSE.
          }
     \label{LARS_comp_MUSE_fig}
\end{figure*}
By comparing observations of low redshift galaxies with similar systems at high redshift, it may be possible to constrain changes to Ly$\alpha$ physics due to galaxy evolution. 
\par
The selection criteria used for the galaxies included in our sample is very similar to a standard LBG selection criterion, targeting UV bright galaxies with ongoing star formation. The Ly$\alpha$ imaging of LBG selected galaxies have mainly been done with narrow-band observations or stacking \citep[e.g.][]{Steidel2011ApJ, Momose2014MNRAS}. In fact, \cite{Guaita2015AA} compared the Ly$\alpha$ images from the original LARS sample (first 14 galaxies in our sample) to these two studies. However, the data used in these studies are deeper than our study by almost two orders of magnitudes. In other words, these studies reach fainter SB levels further out and consequently, the physical scales probed in these studies are larger than those in our study. For example, \cite{Steidel2011ApJ} probed a physical scale in the range of $ \geq 10$ kpc. Their data indicate that Ly$\alpha$ scale lengths in the inner halos (<20 kpc) are shorter ($\sim$ 10 kpc) than those for the outer halos ($\gtrapprox$ 20 kpc, rather probing the CGM), but still much longer than for the continuum ($\sim$ 3 kpc). Therefore, to compare galaxies on an individual level, we need to turn to Integral Field Spectroscopy (IFS) which can provide Ly$\alpha$ images of high$-z$ galaxies, albeit at significantly worse spatial resolution than for the nearby galaxies in this study. In the IFS Ly$\alpha$ surveys performed by VLT/MUSE \citep[e.g.][]{Bacon2015AA}, sources are detected based on their Ly$\alpha$ emission rather than UV brightness which means that the sample selection is somewhat different. Compared to a pure LBG selection, selection based on the Ly$\alpha$ emission will yield a sample with higher $\mathrm{EW_{Ly\alpha}}$ galaxies. This also applies for comparing the MUSE Ly$\alpha$ results to LARS and should be kept in mind for the following section, where we discuss some of our results and how they compare to the $z = $ 3-6 LAEs studied by \cite{Leclercq2017AA}. 
\par 
We begin by comparing the Ly$\alpha$ core and halo scale lengths (from fitting a double exponential function to the Ly$\alpha$ SB profiles, see Sec. \ref{pr_fi_sb}) measured from our sample and the results obtained for LAEs in \cite{Leclercq2017AA}. 
Figure \ref{LARS_comp_MUSE_fig} shows that similar to the MUSE findings, we see no relation between the measured Ly$\alpha$ halo scale lengths and the measured scale lengths in the core in our sample (top left panel). The parameter space covered by the LARS measurements is not similar to the MUSE findings (top right panel). We see that, unlike the core region where the fitted scale lengths coverage in MUSE and LARS are similar, and the measured medians are close, the galaxies in the MUSE study possess more extended Ly$\alpha$ halo, due to the larger FoV and possibly different selection criteria. 
The median of the Ly$\alpha$ core scale length in our sample is slightly higher than the MUSE sample, 0.47 kpc in our sample compared to the 0.43 kpc in the MUSE study. However, the median of the Ly$\alpha$ halo scale length in the MUSE sample is higher than our sample. The assessed median of the Ly$\alpha$ halo scale length in their sample is 4.37 kpc compared to 3.15 kpc in our sample.
\par
Furthermore, we see a difference both in the distribution and the median of the HF measured in our study compared to the MUSE study. The HF in our sample follows a bimodal distribution which is not observed in the LAEs observed in the MUSE studies (bottom left panel in Fig. \ref{LARS_comp_MUSE_fig}), and the measured median of the HF in our sample is 0.43, which is lower than the median of the HF of the LAEs in the MUSE study (0.66). This could be due to a selection effect; the galaxies studied in the MUSE study are generally brighter in Ly$\alpha$ compared to the galaxies in our sample (the median of the Ly$\alpha$ luminosity in the MUSE sample is almost one order of magnitude higher than our sample). 
\par
Finally, the bottom right panel shows a comparison of HF versus $\mathrm{r_{sc}^{Ly\alpha}}$ for our data to the same parameters for the MUSE survey. We find that the space coverage of our data in HF - $\mathrm{r_{sc}^{Ly\alpha}}$ space is within the space coverage of the high$-z$ LAEs studied in the MUSE study. The correlation between HF and $\mathrm{r_{sc}^{Ly\alpha}}$ in our sample is not seen in the MUSE results. We note that the MUSE observations follow the Ly$\alpha$ much further out than we can in most LARS galaxies. This may introduce unknown implications for our measurements. HST imaging of $z\sim$ 0.3--0.5 galaxies could help bridge this gap and constrain this relation further.
\par
It should also be noted that there are several differences between how we measure the Ly$\alpha$ extent of the core and halo compared to the MUSE study. We fit a double exponential function (see Eq. \ref{double_exp_eq}) where each component (core and halo) is described with a single exponential term. 
The MUSE study also uses a double exponential profile model. However, the core Ly$\alpha$ function is held constant and is set to an exponential fit derived from an HST FUV profile. The summed (core + halo) profile is also smoothed to match the ground-based spatial resolution.
\par
Moreover, many of the galaxies in our sample shows Ly$\alpha$ absorption in the centre, which is not seen in the high-$z$ LAEs. This could be because, in the MUSE study, regions were averaged over much larger distances (MUSE has a lower spatial resolution and the galaxies are further away), or a selection effect associated with their average higher Ly$\alpha$ equivalent widths, indicative of lower destruction of Ly$\alpha$ photons in the central regions.
\par
In our analysis, we tried to limit the effect of absorption, but it does affect the core characteristics. Moreover, the MUSE sample is selected on Ly$\alpha$ emission, whereas our sample selection is agnostic to the Ly$\alpha$ properties, and this difference likely has an impact on the comparison.
\begin{figure}
 \centering
    \includegraphics[width=\linewidth]{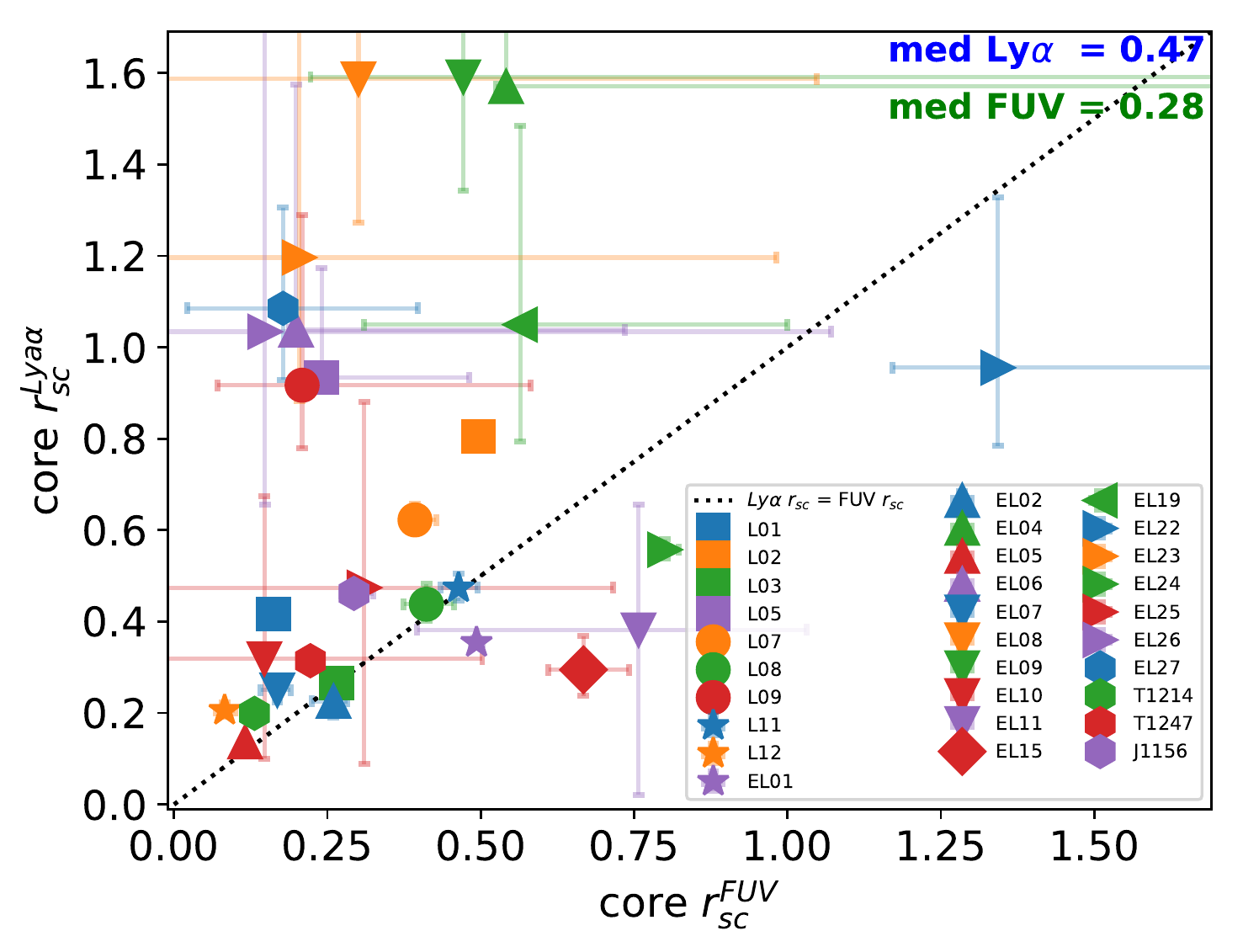}
  \caption{
          Ly$\alpha$ core $r_{sc}$ vs. FUV core $r_{sc}$. The data suggest that even in the core regions of the galaxies, Ly$\alpha$ is more extended than the FUV continuum. The measured median of the Ly$\alpha$ and FUV core scale lengths are displayed at the top left part of the figure with blue and green colours, respectively.
          }
     \label{lya_vs_fuv_core_fig}
\end{figure}
\par
The high-resolution HST data available for our sample also provides the opportunity to explore whether the assumption used in the high$-z$ studies that Ly$\alpha$ and FUV continuum scale lengths in the core are similar is valid or not.
In Fig. \ref{lya_vs_fuv_core_fig}, the Ly$\alpha$ core scale length is plotted against the FUV core scale length. Our data suggest that even in the core regions of the galaxies, Ly$\alpha$ is more extended than FUV. The measured median of the Ly$\alpha$ core scale length exceeds the FUV core scale length by more than 70$\%$. In other words, our data suggest that Ly$\alpha$ is more extended than the FUV not only in the halo, but also in the core regions. The MUSE study assumed them to be equal (since they cannot spatially resolve Ly$\alpha$ in the core) and have therefore likely underestimated the size of the Ly$\alpha$ core region. 
\par
Finally, to investigate how our galaxies would look like if they were at the same redshift as the MUSE sources (and observed with MUSE), we smoothed the Ly$\alpha$ maps of the LARS galaxies to match the MUSE seeing and physical scale at $z=$3.
In this step, the same method and the same fitting function used in the MUSE study was used for describing the convolved SB profiles. This comparison is not trivial because of the limited FoV of our HST observations. The spatial extent of the Ly$\alpha$ emission in the non-convolved maps is often so large that the smoothing moves most of the flux outside of the image, making it impossible to study the Ly$\alpha$ light distribution in the same way as in the MUSE analysis. The detailed information on Ly$\alpha$ in the core for the LARS galaxies is completely washed out by the smoothing. Also, smoothing galaxies with substantial absorption in the core result in profiles that are not comparable to the observed $z\sim3$ profiles. The failure of this test to give any useful information is due to the enormous disparity of spatial resolution and FoV size between LARS and the MUSE Ly$\alpha$ observations. At $z\sim3$ and with MUSE spatial resolution, most of the images of LARS galaxies only cover a few resolution elements. The situation is better for the galaxies at slightly higher redshift, and an intermediate redshift sample of HST observed Ly$\alpha$ emitters might be easier to use for comparison (Runnholm et al. in prep). 

\section{Summary} \label{conclusion_sec}
In this article, we presented a study of how Ly$\alpha$ emission is spatially distributed in a sample of 45 nearby star-forming galaxies observed with the HST. We started by examining different ways to describe the Ly$\alpha$ and FUV SB profiles of the galaxies and concluded that a single Sérsic profile could not describe the Ly$\alpha$ and FUV SB profiles of the galaxies in our sample. However, a double exponential function (Eq. \ref{double_exp_eq}) describes the majority of Ly$\alpha$, and FUV SB profiles well, in particular, the outermost part of the Ly$\alpha$ SB profiles, which is the main focus of this study. Following \cite{Bigiel2008AJ, Micheva2018AA} studies, FUV, and Ly$\alpha$ SB profiles were decomposed into core and halo parts by using a FUV-based SFRD threshold of 0.01 $\mathrm{M_\odot \, yr^{-1} \,  kpc^{-2}}$, and the Ly$\alpha$ HF was defined based on the fitted results in the Ly$\alpha$ halo SB profile and the observed luminosity in the core (see Eq. \ref{hf_eq}).
\par
We studied the light distribution of the galaxies in different wavebands such as FUV, Ly$\alpha$, and I band. The first and the second image moments were used to derive morphological parameters, such as the centroid shift between Ly$\alpha$ and FUV, and minor/major axis ratios (b/a) in Ly$\alpha$, FUV and I band and the difference between the measured position angles in Ly$\alpha$ and FUV ($\Delta$PAs) to study the Ly$\alpha$ morphology of the galaxies. Moreover, to characterise compactness and areal intensity of star formation, the size and average FUV SB of the regions with SFRD > 0.01 $\mathrm{M_\odot \, yr^{-1} \,  kpc^{-2}}$ were also measured.
\par
We also explored the galaxies in our sample to study the relations between the global Ly$\alpha$ observables: Ly$\alpha$ total luminosity, Ly$\alpha$ equivalent width and Ly$\alpha$ escape fraction; and the Ly$\alpha$ morphological properties: Ly$\alpha$ halo scale length, Ly$\alpha$ halo fraction, axis ratio, $\Delta$C, and $\Delta$PA. 
We observed a correlation between the Ly$\alpha$ luminosity and the axis ratio, suggesting that bright galaxies in Ly$\alpha$ appear rounder than the faint one. We also found that Ly$\alpha$ luminosity anti-correlates with the HF, meaning most Ly$\alpha$ luminosity is coming from the Ly$\alpha$ halos in galaxies that are faint in Ly$\alpha$. Our findings also suggest that faint LAEs have more emission coming out from their halos, and potentially put these objects among the Ly$\alpha$ low SB galaxies. Therefore, detecting and observing faint LAEs at high redshift is more challenging than what their global  LAE  fluxes would indicate.  Consequently, conclusions drawn from studying the bright LAEs only, might be biased. Our data suggest that there is no relation between the Ly$\alpha$ escape fraction and any of the Ly$\alpha$ morphological quantities. Finally, We saw a weak anti-correlation between $\mathrm{EW_{Ly\alpha}}$ and the HF, indicating either lower escape fractions or an intrinsically older stellar population.
\par
We also studied the interdependencies of the Ly$\alpha$ morphological quantities (Ly$\alpha$ halo scale length, Ly$\alpha$ halo fraction, axis ratio, $\Delta$C, and $\Delta$PA). Moreover, we investigated how the global parameters such as the stellar mass, the nebular reddening, and star-forming characteristics (size of the SF regions and the average intensity of the FUV SB) affect the Ly$\alpha$ morphology.
\par
We saw that $\mathrm{r_{sc}^{Ly\alpha}}$ correlates with the stellar mass, size of the SF regions, and the HF, and anti-correlates with the average FUV SB within the SF regions, and the axis ratio. These findings suggest that massive galaxies possess more extended Ly$\alpha$ halos, and the mass-size relation also holds for the Ly$\alpha$ emission. Moreover, galaxies with more extended star-forming regions also have larger Ly$\alpha$ halos, probably because the ionising photons cover larger space in the galaxies, and due to the scattering, Ly$\alpha$ halos get more extended.
The anti-correlation between the $\mathrm{r_{sc}^{Ly\alpha}}$ and $\mathrm{\overline{FUV_{SB}}}$ indicate that galaxies with high SFR density do not have very extended Ly$\alpha$ emission. Finally, the anti-correlation between $\mathrm{r_{sc}^{Ly\alpha}}$ and the axis ratio suggests that galaxies with more elongated Ly$\alpha$ morphology are also more extended in Ly$\alpha$. The strong correlation between $\mathrm{r_{sc}^{Ly\alpha}}$ and HF suggest that galaxies that have more extended Ly$\alpha$ halos, have higher HF, too. However, this could simply be due to the way that we define HF (see Eq. \ref{hf_eq}), and this correlation might reflect that these two quantities are implicitly related to each other. 
\par
Our data show that HF anti-correlates with the total FUV luminosity, the average FUV SB within the SF regions, and the axis ratio; and correlates with $\mathrm{r_{sc}^{Ly\alpha}}$. These findings imply that Ly$\alpha$ halos contribute more to the total Ly$\alpha$ luminosity in galaxies with lower star formation rates (lower FUV luminosity) and therefore also lower intrinsic Ly$\alpha$ luminosity. Moreover,  Ly$\alpha$ halos contribute more to the overall Ly$\alpha$ luminosity in SFGs galaxies with lower SFR density. Finally, we found that Ly$\alpha$ halos in galaxies with low HF are one average rounder.
\par
We found that the centroid shift correlates with the stellar mass and the size of the SF regions implying that the centroid shift correlates with two quantities which correlate with the $\mathrm{r_{sc}^{Ly\alpha}}$. This may suggest that the centroid shift grows in galaxies with larger $\mathrm{r_{sc}^{Ly\alpha}}$, since Ly$\alpha$ halo do not grow symmetrically, and larger Ly$\alpha$ halo means a larger difference between the Ly$\alpha$ and FUV measured centroid. However, we did not see a correlation between $\Delta$C and $\mathrm{r_{sc}^{Ly\alpha}}$ in our sample.
\par
We saw that $\Delta$PA anti-correlates with the stellar mass, $\mathrm{E(B-V)_n}$, and the size of the SF regions suggesting that Ly$\alpha$ and FUV misalignment is more significant in massive galaxies. 
The anti-correlation between $\Delta$PA and the nebular reddening is mainly because of a lack of dusty galaxies with high $\Delta$PA. Perhaps, this is because of the absorption of the Ly$\alpha$ photons by the dust particles resulting in less scattering of the Ly$\alpha$ photons. Consequently, Ly$\alpha$ photons follow the same path that the FUV photons take to escape the galaxies. The anti-correlation between $\Delta$PA and the size of the SF regions could be illustrated by more direct channels available for Ly$\alpha$ to escape in galaxies with larger SF regions. The Ly$\alpha$ photons are then escaping closer to where they were produced, and the overall emission is more similar to the FUV distribution.
\par
We saw that Ly$\alpha$ axis ratio correlates with $\mathrm{(b/a)_{FUV}}$, $\mathrm{\overline{FUV_{SB}}}$, and $\Delta$PA. The correlation between $\mathrm{(b/a)_{Ly\alpha}}$ and $\mathrm{(b/a)_{FUV}}$ suggests that Ly$\alpha$ distribution in the galaxies depends on the FUV distribution. We see that galaxies with high SFR density (high $\mathrm{\overline{FUV_{SB}}}$) appear rounder in their Ly$\alpha$ morphology (high $\mathrm{(b/a)_{Ly\alpha}}$), assuming that less scattering occurs in galaxies with high SFR density, these galaxies would also have less elongated Ly$\alpha$ morphology as well as less extended halos. Finally, the correlation between $\mathrm{(b/a)_{Ly\alpha}}$ and $\Delta$PA indicates that in galaxies where Ly$\alpha$ photons escape more in off-axis direction compared to the FUV morphology, the Ly$\alpha$ morphology is rounder and more symmetric.
\par
We also studied the extent of the Ly$\alpha$ halos through an isophotal approach. Since using circular apertures may give a skewed representation of the halos, because of the asymmetric nature of Ly$\alpha$ halos and the stellar morphology of the galaxies in our sample. We fit a simple exponential function to the Ly$\alpha$ isophotal SB profile of the halo to find the Ly$\alpha$ isophotal halo scale length ($\mathrm{r_{sc}^{iso}}$). Given that an isophotal profile is always narrower than a circular profile, the majority of the galaxies appear more extended using circular apertures.
Furthermore, we also looked at the relation between the measured isophotal Ly$\alpha$ halo scale length ($\mathrm{r_{sc}^{iso}}$) and Ly$\alpha$ observables, quantities used for studying the Ly$\alpha$ morphology, and also some of the global observables. 
We saw that $\mathrm{r_{sc}^{iso}}$ correlate with stellar mass, Ly$\alpha$ luminosity, FUV luminosity, the size of the star-forming regions, and the centroid shift. In other words, our data indicate that the size-mass relation in Ly$\alpha$ emission line holds using the isophotal Ly$\alpha$ scale lengths, too. Since we see the aperture size (where $(S/N)_{Ly\alpha}$ drops below one) used for measuring the global properties of the galaxies strongly depends on $\mathrm{r_{sc}^{iso}}$, at least part (or maybe all) of the observed correlation between $\mathrm{r_{sc}^{iso}}$, and the stellar mass, $\mathrm{L_{Ly\alpha}}$, and $\mathrm{L_{FUV}}$ can be because of this aperture effect. 
\par
Finally, we compared our results with the high$-z$ LAEs studied with MUSE \citep{Leclercq2017AA}. Whereas our sample selection is agnostic to the presence of Ly$\alpha$ emission, the MUSE sample is selected on Ly$\alpha$ emission, so some differences are to be expected. The parameter space covered by the core and halo Ly$\alpha$ scale lengths in our sample cover is similar to the LAEs in the MUSE study. Ly$\alpha$ core regions in our sample (median of the core scale length 0.47) extend similar to high$-z$ LAEs (median of the core scale length 0.43). However, Ly$\alpha$ halos in the LAEs are more extended (median of the halo scale length 4.37) compared to our sample (median of the halo scale length 3.15). Furthermore, Ly$\alpha$ halos in the LAEs in the MUSE study contribute more to the total Ly$\alpha$ luminosity (median of HF = 0.66) compared to the nearby SFGs in our sample (median of HF = 0.43).
Furthermore, the bimodal distribution observed in the HF in our sample is not observed in their sample. The comparison between the HF measured for the LAEs in our sample (median of 0.43) compared to the high$-z$ LAEs in the MUSE study ((median of 0.66)) indicate the Ly$\alpha$ halos in the nearby SFGs contribute less to the total Ly$\alpha$ luminosity compared to high$-z$ LAEs. 
\par
Utilising high-resolution data for the nearby galaxies in our sample, we tested whether the assumption that the FUV and Ly$\alpha$ scale length in cores are similar. Our results show that even in the core regions, Ly$\alpha$ is more extended, and the median of the measured core Ly$\alpha$ scale length is higher than $150\%$ the median of the measured FUV scale length. We provide all our measurements in electronic tables for general use and reference.
\begin{acknowledgements}
M.H. is a Fellow of the Knut and Alice Wallenberg Foundation. D.K. is supported by the Centre National d'Etudes Spatiales (CNES) / Centre National de la Recherche Scienti que (CNRS); convention no 131425. J.M.M.H. is funded by Spanish State Research Agency grants PID2019-107061GB-C61 and  MDM-2017-0737 (Unidad de Excelencia Maria de Maeztu CAB). A.V. acknowledges support from the European Research Council under grant agreement ERC-stg-757258 (TRIPLE) and the Swiss National Foundation under grant PP00P2 176808.
\end{acknowledgements}
\longtab{
\begin{landscape}
\begin{longtable}{lcccccccccccccl}
\caption{
    All the quantities measured in this study. 
    } \\
\hline \hline

ID \tablefootmark{ID} & FUV & Ly$\alpha$ & $r_{sc}^{FUV}$   & $r_{sc}^{Ly\alpha}$   & HF  & $Ly\alpha_{SB}$ & $\Delta$C   & $(b/a)_{Ly\alpha}$    & $(b/a)_{FUV}$   & $(b/a)_I$   & $\Delta$PA  & $r_{sc}^{iso}$  & $r_{SFRD > 0.01}$   & $\overline{FUV_{SB}}$ \\
& core &  core & (kpc) & (kpc) &  & $(10^{38}$& (kpc) &  &   & &    (deg) & (kpc) & (kpc) &  $(10^{40}$ \\
& $r_{sc}$ (kpc) & $r_{sc}$ (kpc)& & &  & $\mathrm{erg/s/kpc^2})$ & &   &   & & & & &  $\mathrm{erg/s/kpc^2/\si{\angstrom}})$ \\

\hline \\
\endfirsthead
\hline \hline
ID  & FUV & Ly$\alpha$ & $r_{sc}^{FUV}$   & $r_{sc}^{Ly\alpha}$   & HF  & $Ly\alpha_{SB}$ & $\Delta$C   & $(b/a)_{Ly\alpha}$    & $(b/a)_{FUV}$   & $(b/a)_I$   & $\Delta$PA  & $r_{sc}^{iso}$  & $r_{SFRD > 0.01}$   & $\overline{FUV_{SB}}$ \\
& core &  core & (kpc) & (kpc) &  & $(10^{38}$& (kpc) &  &   & &    (deg) & (kpc) & (kpc) &  $(10^{40}$ \\
& $r_{sc}$ (kpc) & $r_{sc}$ (kpc)& & &  & $\mathrm{erg/s/kpc^2})$ & &   &   & & & & &  $\mathrm{erg/s/kpc^2/\si{\angstrom}})$ \\
\hline \\
\endhead
\vspace*{2mm}
L01    &  $0.16_{-0.00}^{+0.03}$  &  $0.42_{-0.02}^{+0.01}$  &  $0.90_{-0.00}^{+0.28}$    &  $1.45_{-0.06}^{+0.04}$     &  $0.21_{-0.01}^{+0.01}$  &  $5.08 \pm 1.16$  &  1.79  &  0.68  &  0.41  &  0.42  &  14.66  &  $1.42 \pm 0.11$  &  2.64   &  0.08  \\
\vspace*{2mm}
L02    &  $0.50_{-0.00}^{+0.00}$  &  $0.81_{-0.02}^{+0.02}$  &  $1.96_{-0.06}^{+0.24}$    &  $2.08_{-0.08}^{+0.11}$     &  $0.48_{-0.01}^{+0.01}$  &  $2.58 \pm 0.91$  &  2.04  &  0.61  &  0.37  &  0.40  &  32.00  &  $1.00 \pm 0.11$  &  2.26   &  0.04  \\
\vspace*{2mm}
L03    &  $0.26_{-0.00}^{+0.00}$  &  $0.27_{-0.01}^{+0.01}$  &  $2.82_{-0.04}^{+0.04}$    &  $3.44_{-0.37}^{+0.49}$     &  $0.51_{-0.04}^{+0.05}$  &  $3.63 \pm 0.85$  &  0.10  &  0.70  &  0.68  &  0.54  &  5.99   &  $0.96 \pm 0.29$  &  3.85   &  0.01  \\
\vspace*{2mm}
L04    &  $0.21_{-0.00}^{+0.02}$  &  N/A                     &  $1.65_{-0.00}^{+0.02}$    &  N/A                        &  $0.00_{-0.00}^{+0.12}$  &  $1.96 \pm 1.06$  &  1.78  &  0.79  &  0.39  &  0.38  &  33.68  &  $0.60 \pm 0.13$  &  4.01   &  0.04  \\
\vspace*{2mm}
L05    &  $0.24_{-0.00}^{+0.00}$  &  $0.93_{-0.04}^{+0.23}$  &  $2.35_{-0.12}^{+0.13}$    &  $1.49_{-0.02}^{+1.00}$     &  $0.20_{-0.01}^{+0.04}$  &  $2.09 \pm 0.98$  &  0.34  &  0.89  &  0.68  &  0.43  &  17.78  &  $1.38 \pm 0.06$  &  2.25   &  0.15  \\
\vspace*{2mm}
L06    &  $0.21_{-0.00}^{+0.00}$  &  N/A                     &  $1.65_{-0.02}^{+0.02}$    &  N/A                        &  N/A                     &  $2.41 \pm 1.33$  &  1.04  &  0.40  &  0.57  &  0.49  &  0.45   &  N/A           &  1.87   &  0.02  \\
\vspace*{2mm}
L07    &  $0.39_{-0.00}^{+0.00}$  &  $0.62_{-0.02}^{+0.03}$  &  $2.01_{-0.22}^{+0.26}$    &  $1.55_{-0.05}^{+0.08}$     &  $0.42_{-0.01}^{+0.01}$  &  $4.76 \pm 1.08$  &  1.12  &  0.76  &  0.27  &  0.47  &  20.39  &  $1.27 \pm 0.07$  &  1.89   &  0.14  \\
\vspace*{2mm}
L08    &  $0.41_{-0.00}^{+0.00}$  &  $0.44_{-0.04}^{+0.04}$  &  $1.65_{-0.00}^{+0.01}$    &  $1.91_{-0.06}^{+0.07}$     &  $0.16_{-0.01}^{+0.01}$  &  $6.02 \pm 1.43$  &  2.17  &  0.56  &  0.66  &  0.71  &  5.49   &  $1.30 \pm 0.21$  &  5.07   &  0.03  \\
\vspace*{2mm}
L09    &  $0.21_{-0.00}^{+0.00}$  &  $0.92_{-0.14}^{+0.36}$  &  $6.94_{-0.04}^{+0.04}$    &  $60.33_{-28.14}^{+32.82}$  &  $1.00_{-1.00}^{+0.00}$  &  $7.37 \pm 1.19$  &  3.34  &  0.35  &  0.28  &  0.26  &  1.86   &  $2.80 \pm 0.13$  &  6.28   &  0.05  \\
\vspace*{2mm}
L10    &  $0.73_{-0.30}^{+0.02}$  &  $0.99_{-2.10}^{+2.43}$  &  $2.81_{-1.62}^{+0.81}$    &  $4.53_{-0.29}^{+38.38}$    &  $1.12_{-0.05}^{+0.07}$  &  $2.12 \pm 1.05$  &  2.04  &  0.54  &  0.29  &  0.49  &  15.45  &  $1.41 \pm 0.14$  &  3.03   &  0.03  \\
\vspace*{2mm}
L11    &  $0.46_{-0.01}^{+0.01}$  &  $0.47_{-0.03}^{+0.03}$  &  $6.01_{-0.05}^{+0.05}$    &  $7.12_{-0.30}^{+0.30}$     &  $0.36_{-0.02}^{+0.02}$  &  $1.76 \pm 1.08$  &  2.21  &  0.26  &  0.21  &  0.34  &  2.88   &  $4.05 \pm 0.29$  &  13.41  &  0.02  \\
\vspace*{2mm}
L12    &  $0.08_{-0.00}^{+0.00}$  &  $0.21_{-0.01}^{+0.02}$  &  $4.06_{-0.98}^{+1.04}$    &  $2.27_{-0.08}^{+0.07}$     &  $0.36_{-0.02}^{+0.02}$  &  $1.42 \pm 0.73$  &  0.89  &  0.78  &  0.81  &  0.69  &  0.89   &  $2.34 \pm 0.13$  &  4.28   &  0.13  \\
\vspace*{2mm}
L13    &  N/A                     &  N/A                     &  N/A                       &  N/A                        &  N/A                     &  $2.55 \pm 1.00$  &  0.74  &  0.80  &  0.71  &  0.50  &  0.48   &  $2.00 \pm 0.37$  &  11.72  &  0.02  \\
\vspace*{2mm}
L14    &  $0.35_{-0.04}^{+0.06}$  &  $0.28_{-0.17}^{+0.82}$  &  $6.94_{-4.94}^{+6.04}$    &  $7.59_{-2.16}^{+2.95}$     &  $0.37_{-0.04}^{+0.06}$  &  $1.97 \pm 0.85$  &  1.15  &  0.79  &  0.86  &  0.89  &  8.25   &  $3.97 \pm 0.19$  &  5.58   &  0.11  \\
\vspace*{2mm}
EL01   &  $0.49_{-0.00}^{+0.00}$  &  N/A                     &  $1.85_{-0.00}^{+0.00}$    &  N/A                        &  N/A                     &  $1.88 \pm 1.00$  &  0.21  &  0.74  &  0.66  &  0.88  &  6.67   &  $1.02 \pm 0.12$  &  4.57   &  0.04  \\
\vspace*{2mm}
EL02   &  $0.26_{-0.00}^{+0.00}$  &  $0.23_{-0.04}^{+0.02}$  &  $2.15_{-0.00}^{+0.00}$    &  $2.44_{-0.08}^{+0.11}$     &  $0.30_{-0.02}^{+0.02}$  &  $2.00 \pm 0.89$  &  1.15  &  0.81  &  0.85  &  0.78  &  82.24  &  $1.20 \pm 0.03$  &  4.28   &  0.04  \\
\vspace*{2mm}
EL03   &  $0.34_{-0.01}^{+0.01}$  &  N/A                     &  $2.55_{-0.01}^{+0.01}$    &  N/A                        &  $0.00_{-0.00}^{+0.17}$  &  $2.12 \pm 1.10$  &  0.34  &  0.45  &  0.47  &  0.49  &  1.32   &  $3.43 \pm 0.23$  &  8.84   &  0.02  \\
\vspace*{2mm}
EL04   &  $0.54_{-0.05}^{+0.04}$  &  $1.57_{-0.02}^{+0.03}$  &  $1.25_{-0.00}^{+0.00}$    &  $3.04_{-0.04}^{+0.05}$     &  $0.37_{-0.01}^{+0.02}$  &  $2.43 \pm 1.21$  &  0.72  &  0.58  &  0.60  &  0.45  &  5.54   &  $1.36 \pm 0.14$  &  3.62   &  0.05  \\
\vspace*{2mm}
EL05   &  $0.12_{-0.00}^{+0.00}$  &  $0.14_{-0.01}^{+0.00}$  &  $3.32_{-0.02}^{+0.02}$    &  $4.35_{-0.40}^{+0.34}$     &  $0.43_{-0.04}^{+0.04}$  &  $2.01 \pm 1.17$  &  2.17  &  0.61  &  0.66  &  0.66  &  1.05   &  $1.68 \pm 0.46$  &  5.19   &  0.02  \\
\vspace*{2mm}
EL06   &  $0.20_{-0.00}^{+0.00}$  &  $1.04_{-0.00}^{+0.57}$  &  $1.96_{-0.01}^{+0.01}$    &  $3.26_{-1.63}^{+0.08}$     &  $0.67_{-0.00}^{+0.70}$  &  $2.69 \pm 1.09$  &  0.89  &  0.80  &  0.50  &  0.67  &  3.31   &  N/A           &  3.44   &  0.02  \\
\vspace*{2mm}
EL07   &  $0.17_{-0.00}^{+0.00}$  &  N/A                     &  $1.51_{-0.01}^{+0.01}$    &  N/A                        &  N/A                     &  $1.67 \pm 0.94$  &  1.51  &  0.41  &  0.40  &  0.42  &  3.19   &  $0.83 \pm 0.25$  &  2.83   &  0.03  \\
\vspace*{2mm}
EL08   &  $0.30_{-0.00}^{+0.01}$  &  $1.59_{-0.32}^{+0.74}$  &  $2.03_{-0.01}^{+0.01}$    &  $5.00_{-0.40}^{+0.44}$     &  $0.81_{-0.77}^{+0.06}$  &  $2.06 \pm 1.08$  &  1.76  &  0.45  &  0.75  &  0.74  &  12.80  &  $1.10 \pm 0.16$  &  4.04   &  0.01  \\
\vspace*{2mm}
EL09   &  $0.47_{-0.00}^{+0.00}$  &  $1.59_{-0.25}^{+1.78}$  &  $1.05_{-0.02}^{+0.02}$    &  $5.63_{-0.79}^{+1.78}$     &  $0.84_{-0.78}^{+0.08}$  &  $1.97 \pm 1.04$  &  0.56  &  0.64  &  0.82  &  0.67  &  15.92  &  N/A           &  2.82   &  0.03  \\
\vspace*{2mm}
EL10   &  $0.15_{-0.00}^{+0.00}$  &  $0.32_{-0.25}^{+0.38}$  &  $3.09_{-0.03}^{+0.03}$    &  $4.14_{-0.61}^{+0.75}$     &  $0.49_{-0.09}^{+0.09}$  &  $2.78 \pm 1.31$  &  1.02  &  0.23  &  0.19  &  0.24  &  3.92   &  N/A           &  4.47   &  0.01  \\
\vspace*{2mm}
EL11   &  $0.76_{-0.00}^{+0.00}$  &  $0.38_{-0.36}^{+0.27}$  &  $1.59_{-0.04}^{+0.04}$    &  $2.99_{-0.59}^{+1.69}$     &  $0.79_{-0.80}^{+0.05}$  &  $1.91 \pm 0.99$  &  0.61  &  0.63  &  0.62  &  0.72  &  13.21  &  $0.90 \pm 0.12$  &  2.48   &  0.03  \\

EL12   &  $0.31_{-0.01}^{+0.01}$  &  N/A                     &  $2.12_{-0.01}^{+0.01}$    &  N/A                        &  N/A                     &  $2.19 \pm 1.07$  &  0.06  &  0.39  &  0.42  &  0.38  &  0.02   &  N/A           &  4.12   &  0.01  \\
\vspace*{2mm}
EL13   &  $0.10_{-0.00}^{+0.00}$  &  $0.11_{-0.00}^{+0.00}$  &  $5.30_{-1.18}^{+2.36}$    &  $1.10_{-0.00}^{+0.00}$     &  $0.00_{-0.00}^{+0.17}$  &  $2.23 \pm 1.13$  &  0.08  &  0.89  &  0.90  &  0.81  &  8.78   &  $0.62 \pm 0.07$  &  1.14   &  0.15  \\
\vspace*{2mm}
EL14   &  $0.74_{-0.01}^{+0.03}$  &  N/A                     &  $32.89_{-0.33}^{+19.42}$  &  N/A                        &  N/A                     &  $1.80 \pm 0.99$  &  0.04  &  0.87  &  0.90  &  0.93  &  22.73  &  N/A           &  2.43   &  0.03  \\
\vspace*{2mm}
EL15   &  $0.67_{-0.01}^{+0.01}$  &  N/A                     &  $8.59_{-0.55}^{+0.61}$    &  N/A                        &  N/A                     &  $2.03 \pm 0.96$  &  0.97  &  0.51  &  0.62  &  0.60  &  26.57  &  $0.73 \pm 0.08$  &  9.06   &  0.00  \\
\vspace*{2mm}
EL16   &  $0.46_{-0.01}^{+0.01}$  &  N/A                     &  $2.55_{-0.03}^{+0.03}$    &  N/A                        &  N/A                     &  $2.14 \pm 0.96$  &  0.63  &  0.27  &  0.35  &  0.48  &  5.89   &  N/A           &  6.47   &  0.00  \\
\vspace*{2mm}
EL17   &  $0.38_{-0.01}^{+0.01}$  &  $0.45_{-0.40}^{+0.57}$  &  $1.88_{-0.02}^{+0.02}$    &  $4.21_{-0.82}^{+0.63}$     &  $0.84_{-0.85}^{+0.04}$  &  $1.96 \pm 1.18$  &  0.61  &  0.32  &  0.34  &  0.39  &  11.48  &  N/A           &  3.27   &  0.01  \\
\vspace*{2mm}
EL18   &  $0.14_{-0.00}^{+0.00}$  &  N/A                     &  $1.72_{-0.03}^{+0.03}$    &  N/A                        &  N/A                     &  $3.03 \pm 1.10$  &  0.27  &  0.13  &  0.25  &  0.26  &  2.37   &  N/A           &  2.11   &  0.01  \\
\vspace*{2mm}
EL19   &  $0.56_{-0.00}^{+0.00}$  &  $1.05_{-0.28}^{+0.41}$  &  $0.80_{-0.01}^{+0.02}$    &  $1.47_{-0.43}^{+1.02}$     &  $0.75_{-0.06}^{+0.06}$  &  $2.05 \pm 0.95$  &  0.72  &  0.64  &  0.48  &  0.47  &  62.53  &  N/A           &  1.67   &  0.03  \\
\vspace*{2mm}
EL20   &  $0.66_{-0.24}^{+0.00}$  &  N/A                     &  $14.68_{-7.75}^{+54.97}$  &  N/A                        &  $0.00_{-0.00}^{+0.17}$  &  $1.29 \pm 0.86$  &  0.17  &  0.66  &  0.56  &  0.91  &  2.92   &  $1.03 \pm 0.10$  &  2.18   &  0.02  \\
\vspace*{2mm}
EL21   &  $0.36_{-0.01}^{+0.01}$  &  N/A                     &  $1.82_{-0.04}^{+0.04}$    &  N/A                        &  N/A                     &  $2.29 \pm 1.12$  &  0.61  &  0.35  &  0.21  &  0.28  &  19.49  &  N/A           &  1.33   &  0.01  \\
\vspace*{2mm}
EL22   &  $1.34_{-0.02}^{+0.03}$  &  $0.96_{-0.17}^{+0.48}$  &  $13.02_{-1.77}^{+3.56}$   &  $3.78_{-0.27}^{+0.32}$     &  $0.81_{-0.03}^{+0.03}$  &  $1.86 \pm 0.95$  &  1.15  &  0.56  &  0.48  &  0.45  &  7.36   &  $1.50 \pm 0.13$  &  5.19   &  0.03  \\
\vspace*{2mm}
EL23   &  $0.20_{-0.00}^{+0.00}$  &  N/A                     &  $5.01_{-0.02}^{+0.02}$    &  N/A                        &  N/A                     &  $2.29 \pm 1.11$  &  0.12  &  0.74  &  0.83  &  0.81  &  1.62   &  N/A           &  9.75   &  0.01  \\
\vspace*{2mm}
EL24   &  $0.80_{-0.00}^{+0.00}$  &  $0.56_{-0.02}^{+0.02}$  &  $4.07_{-0.57}^{+0.84}$    &  $2.27_{-0.15}^{+0.16}$     &  $0.12_{-0.02}^{+0.02}$  &  $1.52 \pm 0.81$  &  0.71  &  0.78  &  0.85  &  0.88  &  15.48  &  $1.47 \pm 0.06$  &  3.38   &  0.05  \\
\vspace*{2mm}
EL25   &  $0.31_{-0.00}^{+0.00}$  &  $0.47_{-0.35}^{+0.41}$  &  $4.81_{-0.03}^{+0.03}$    &  $9.05_{-1.82}^{+2.40}$     &  $0.95_{-0.03}^{+0.02}$  &  $1.89 \pm 1.04$  &  2.25  &  0.63  &  0.76  &  0.71  &  17.17  &  $2.31 \pm 0.80$  &  5.42   &  0.01  \\
\vspace*{2mm}
EL26   &  $0.15_{-0.00}^{+0.00}$  &  $1.03_{-0.43}^{+1.02}$  &  $3.37_{-0.02}^{+0.03}$    &  $5.55_{-0.42}^{+0.64}$     &  $0.94_{-0.03}^{+0.01}$  &  $1.82 \pm 0.91$  &  1.01  &  0.51  &  0.43  &  0.43  &  11.43  &  N/A           &  4.67   &  0.01  \\
\vspace*{2mm}
EL27   &  $0.18_{-0.10}^{+0.02}$  &  $1.09_{-0.14}^{+0.22}$  &  $4.28_{-0.05}^{+0.04}$    &  $4.79_{-0.41}^{+0.41}$     &  $0.95_{-0.01}^{+0.01}$  &  $3.74 \pm 1.22$  &  1.40  &  0.72  &  0.68  &  0.73  &  2.80   &  $2.19 \pm 0.60$  &  3.48   &  0.02  \\
\vspace*{2mm}
EL28   &  $0.39_{-0.00}^{+0.00}$  &  N/A                     &  $2.44_{-0.01}^{+0.01}$    &  N/A                        &  N/A                     &  $2.22 \pm 1.12$  &  1.24  &  0.32  &  0.82  &  0.81  &  28.47  &  N/A           &  6.75   &  0.01  \\
\vspace*{2mm}
T1214  &  $0.13_{-0.01}^{+0.01}$  &  $0.20_{-0.02}^{+0.01}$  &  $0.98_{-0.03}^{+0.03}$    &  $1.03_{-0.07}^{+0.07}$     &  $0.33_{-0.01}^{+0.01}$  &  $2.22 \pm 0.84$  &  1.20  &  0.80  &  0.49  &  0.40  &  27.91  &  $0.74 \pm 0.06$  &  1.78   &  0.03  \\
\vspace*{2mm}
T1247  &  $0.22_{-0.00}^{+0.00}$  &  $0.31_{-0.00}^{+0.00}$  &  $0.93_{-0.00}^{+0.00}$    &  $2.48_{-0.03}^{+0.03}$     &  $0.25_{-0.01}^{+0.00}$  &  $7.85 \pm 1.06$  &  1.18  &  0.95  &  0.85  &  0.54  &  28.00  &  $2.27 \pm 0.13$  &  4.17   &  0.18  \\
\vspace*{2mm}
J1156  &  $0.29_{-0.07}^{+0.14}$  &  $0.46_{-0.02}^{+0.03}$  &  $2.89_{-0.11}^{+0.12}$    &  $7.56_{-1.53}^{+1.79}$     &  $0.33_{-0.04}^{+0.05}$  &  $3.88 \pm 1.37$  &  1.84  &  0.87  &  0.57  &  0.42  &  36.75  &  $4.25 \pm 0.23$  &  7.88   &  0.04  \\
\hline
\label{measured_paparm_this_study}
\end{longtable}
\tablefoot{
           ID: The the LARS ID, FUV core $r_{sc}$ (kpc): measured scale length for the core in the FUV SB profiles, Ly$\alpha$ core $r_{sc}$ (kpc): measured scale length for the core in the Ly$\alpha$ SB profiles, $\mathrm{r_{sc}^{FUV}}$ (kpc): measured scale length for the halo in the FUV profiles, $\mathrm{r_{sc}^{Ly\alpha}}$ scale length for the halo in the Ly$\alpha$ profiles, HF: measured Ly$\alpha$ halo fraction (upper error provided for L04, EL03, EL13, and EL20 are based on the observed HF), $Ly\alpha_{SB}$: Ly$\alpha$ SB of the faintest (furthest) annulus bin (or upper limit if S/N < 2) measured in the Voronoi tessellated maps,$\Delta$C: measured centroid shift between Ly$\alpha$ and FUV, $\mathrm{(b/a)_{Ly\alpha}}$: Ly$\alpha$ axis ratio, $\mathrm{(b/a)_{FUV}}$: FUV axis ratio, $\mathrm{(b/a)_I}$: I band axis ratio, $\Delta$PA (deg): difference between measured position angle for Ly$\alpha$ and FUV, $\mathrm{r_{sc}^{iso}}$: measured Ly$\alpha$ scale length through isophotal approach, $\mathrm{r_{SFRD > 0.01}}$ the area where the  SFRD > 0.01 $\mathrm{M_\odot \, yr^{-1} \,  kpc^{-2}}$ presented by the equivalent radius (the radius of a circle with the same area), $\mathrm{\overline{FUV_{SB}}}$: the average FUV SB measured within these region with SFRD > 0.01.
           }
\end{landscape}
}

\longtab{
\begin{longtable}{rcccccc}
\caption{
    Global quantities characterising the Ly$\alpha$ physics of each galaxy, used from Melinder et al. (in prep).
    } \\
\hline \hline

ID  & Mass  & $L_{Ly\alpha}$  & $L_{FUV}$   & ${f_{esc}}$   & $\mathrm{E(B - V)_n}$   & $\mathrm{EW_{Ly\alpha}}$ \\
  & ($10^8$ $M_\odot$)    & ($10^{39} \,  \mathrm{erg/s}$) &  ($10^{39} \, \mathrm{erg/s}$) &  &          &                   \\

\hline \\
\endfirsthead
\hline \hline
ID  & Mass  & $L_{Ly\alpha}$  & $L_{FUV}$   & ${f_{esc}}$   & $\mathrm{E(B - V)_n}$   & $\mathrm{EW_{Ly\alpha}}$ \\
  & ($10^8$ $M_\odot$)    & ($10^{39} $erg/s) & ($10^{39} $erg/s) &  &          &                   \\

\hline \\
\endhead

\vspace*{2mm}
L01    &  $162.28_{-3.96}^{+3.91}$       &  $890.89_{-8.92}^{+8.87}$        &  $19.47_{-0.04}^{+0.04}$   &  $0.14 \pm 0.00$  &  $0.16_{-0.00}^{+0.00}$  &  $45.76_{-0.47}^{+0.53}$   \\
\vspace*{2mm}
L02    &  $57.52_{-2.71}^{+3.09}$        &  $427.41_{-8.36}^{+9.33}$        &  $7.26_{-0.04}^{+0.04}$    &  $0.30 \pm 0.01$  &  $0.01_{-0.01}^{+0.01}$  &  $58.84_{-1.19}^{+1.38}$   \\
\vspace*{2mm}
L03    &  $227.66_{-13.89}^{+75.22}$     &  $172.52_{-6.71}^{+7.39}$        &  $5.71_{-0.03}^{+0.04}$    &  $0.00 \pm 0.00$  &  $0.89_{-0.02}^{+0.01}$  &  $30.24_{-1.34}^{+1.52}$   \\
\vspace*{2mm}
L04    &  $110.68_{-1.41}^{+1.00}$       &  $51.80_{-4.30}^{+4.82}$         &  $13.50_{-0.02}^{+0.02}$   &  $0.01 \pm 0.00$  &  $0.17_{-0.01}^{+0.01}$  &  $3.84_{-0.31}^{+0.35}$    \\
\vspace*{2mm}
L05    &  $86.32_{-3.83}^{+2.80}$        &  $713.90_{-11.58}^{+11.49}$      &  $25.77_{-0.06}^{+0.05}$   &  $0.14 \pm 0.00$  &  $0.14_{-0.00}^{+0.00}$  &  $27.70_{-0.44}^{+0.51}$   \\
\vspace*{2mm}
L06    &  $1.86_{-0.04}^{+0.05}$         &  $1.51_{-1.33}^{+1.34}$          &  $1.02_{-0.01}^{+0.01}$    &  $0.01 \pm 0.01$  &  $0.07_{-0.02}^{+0.01}$  &  $1.48_{-1.27}^{+1.27}$    \\
\vspace*{2mm}
L07    &  $90.16_{-3.46}^{+2.76}$        &  $689.98_{-11.49}^{+14.35}$      &  $17.02_{-0.06}^{+0.07}$   &  $0.11 \pm 0.00$  &  $0.25_{-0.01}^{+0.01}$  &  $40.54_{-0.84}^{+0.92}$   \\
\vspace*{2mm}
L08    &  $1034.79_{-13.30}^{+17.69}$    &  $393.60_{-12.66}^{+11.25}$      &  $23.12_{-0.07}^{+0.07}$   &  $0.00 \pm 0.00$  &  $0.95_{-0.02}^{+0.02}$  &  $17.02_{-0.54}^{+0.48}$   \\
\vspace*{2mm}
L09    &  $584.38_{-7.43}^{+5.71}$       &  $650.95_{-27.23}^{+28.36}$      &  $58.62_{-0.11}^{+0.15}$   &  $0.02 \pm 0.00$  &  $0.35_{-0.01}^{+0.01}$  &  $11.10_{-0.47}^{+0.48}$   \\
\vspace*{2mm}
L10    &  $181.20_{-3.44}^{+4.88}$       &  $11.00_{-10.40}^{+12.43}$       &  $11.34_{-0.09}^{+0.13}$   &  $0.00 \pm 0.00$  &  $0.70_{-0.05}^{+0.04}$  &  $0.97_{-0.59}^{+0.72}$    \\
\vspace*{2mm}
L11    &  $1505.59_{-25.33}^{+34.20}$    &  $1816.66_{-62.44}^{+69.72}$     &  $87.96_{-0.51}^{+0.55}$   &  $0.07 \pm 0.01$  &  $0.48_{-0.05}^{+0.05}$  &  $20.65_{-0.45}^{+0.49}$   \\
\vspace*{2mm}
L12    &  $428.14_{-47.39}^{+34.40}$     &  $1700.37_{-52.71}^{+60.54}$     &  $83.88_{-0.37}^{+0.28}$   &  $0.03 \pm 0.00$  &  $0.64_{-0.01}^{+0.01}$  &  $20.27_{-0.66}^{+0.76}$   \\
\vspace*{2mm}
L13    &  $372.59_{-73.06}^{+65.76}$     &  $-1973.12_{-463.78}^{+331.67}$  &  $121.53_{-3.13}^{+2.72}$  &  $0.00 \pm 0.01$  &  $0.53_{-0.03}^{+0.03}$  &  $-16.24_{-3.41}^{+2.30}$  \\
\vspace*{2mm}
L14    &  $382.88_{-99.69}^{+85.73}$     &  $5666.14_{-288.93}^{+237.99}$   &  $112.50_{-1.05}^{+1.19}$  &  $0.27 \pm 0.02$  &  $0.10_{-0.01}^{+0.01}$  &  $50.37_{-2.49}^{+1.76}$   \\
\vspace*{2mm}
EL01   &  $775.48_{-4.09}^{+4.15}$       &  $549.51_{-9.26}^{+7.22}$        &  $24.21_{-0.03}^{+0.03}$   &  $0.01 \pm 0.00$  &  $0.52_{-0.01}^{+0.00}$  &  $22.70_{-0.40}^{+0.32}$   \\
\vspace*{2mm}
EL02   &  $298.01_{-1.42}^{+2.28}$       &  $356.20_{-11.01}^{+6.03}$       &  $26.48_{-0.05}^{+0.05}$   &  $0.07 \pm 0.00$  &  $0.00_{-0.02}^{+0.01}$  &  $13.45_{-0.41}^{+0.23}$   \\
\vspace*{2mm}
EL03   &  $811.76_{-11.51}^{+9.90}$      &  $125.32_{-15.10}^{+18.33}$      &  $27.45_{-0.09}^{+0.07}$   &  $0.01 \pm 0.00$  &  $0.42_{-0.01}^{+0.01}$  &  $4.56_{-0.52}^{+0.64}$    \\
\vspace*{2mm}
EL04   &  $332.54_{-3.34}^{+2.98}$       &  $470.76_{-11.80}^{+14.87}$      &  $25.66_{-0.05}^{+0.06}$   &  $0.07 \pm 0.00$  &  $0.26_{-0.01}^{+0.01}$  &  $18.34_{-0.47}^{+0.59}$   \\
\vspace*{2mm}
EL05   &  $690.61_{-6.66}^{+9.05}$       &  $581.87_{-16.72}^{+20.98}$      &  $22.18_{-0.08}^{+0.08}$   &  $0.13 \pm 0.01$  &  $0.15_{-0.02}^{+0.03}$  &  $26.24_{-0.79}^{+0.99}$   \\
\vspace*{2mm}
EL06   &  $147.25_{-4.44}^{+3.16}$       &  $138.17_{-13.58}^{+9.84}$       &  $10.22_{-0.05}^{+0.07}$   &  $0.06 \pm 0.01$  &  $0.29_{-0.04}^{+0.04}$  &  $13.52_{-1.26}^{+0.96}$   \\
\vspace*{2mm}
EL07   &  $169.03_{-11.54}^{+9.08}$      &  $111.78_{-11.60}^{+17.71}$      &  $11.31_{-0.08}^{+0.08}$   &  $0.04 \pm 0.00$  &  $0.07_{-0.02}^{+0.02}$  &  $9.88_{-0.98}^{+1.50}$    \\
\vspace*{2mm}
EL08   &  $383.49_{-7.17}^{+7.49}$       &  $126.02_{-8.20}^{+10.37}$       &  $8.21_{-0.05}^{+0.04}$    &  $0.01 \pm 0.00$  &  $0.72_{-0.05}^{+0.06}$  &  $15.35_{-0.96}^{+1.24}$   \\
\vspace*{2mm}
EL09   &  $46.77_{-2.55}^{+3.25}$        &  $44.69_{-6.46}^{+6.76}$         &  $8.14_{-0.03}^{+0.03}$    &  $0.04 \pm 0.01$  &  $0.17_{-0.02}^{+0.02}$  &  $5.49_{-0.78}^{+0.82}$    \\
\vspace*{2mm}
EL10   &  $187.77_{-6.65}^{+6.24}$       &  $98.47_{-14.34}^{+8.12}$        &  $5.47_{-0.06}^{+0.05}$    &  $0.02 \pm 0.00$  &  $0.70_{-0.05}^{+0.05}$  &  $18.01_{-2.62}^{+1.37}$   \\
\vspace*{2mm}
EL11   &  $254.25_{-10.50}^{+12.87}$     &  $72.80_{-10.54}^{+8.08}$        &  $7.31_{-0.04}^{+0.04}$    &  $0.06 \pm 0.01$  &  $0.20_{-0.04}^{+0.04}$  &  $9.96_{-1.38}^{+1.07}$    \\
\vspace*{2mm}
EL12   &  $269.60_{-9.77}^{+5.50}$       &  $-27.69_{-9.37}^{+7.53}$        &  $6.17_{-0.03}^{+0.05}$    &  $0.00 \pm 0.00$  &  $0.75_{-0.05}^{+0.05}$  &  $-4.49_{-1.34}^{+1.12}$   \\
\vspace*{2mm}
EL13   &  $67.81_{-1.13}^{+1.07}$        &  $262.00_{-8.41}^{+7.86}$        &  $6.66_{-0.03}^{+0.04}$    &  $0.20 \pm 0.01$  &  $0.22_{-0.02}^{+0.02}$  &  $39.33_{-1.48}^{+1.34}$   \\
\vspace*{2mm}
EL14   &  $59.23_{-3.97}^{+3.25}$        &  $5.21_{-4.94}^{+7.11}$          &  $5.13_{-0.03}^{+0.02}$    &  $0.00 \pm 0.00$  &  $0.14_{-0.02}^{+0.02}$  &  $1.02_{-0.87}^{+1.29}$    \\
\vspace*{2mm}
EL15   &  $108.95_{-4.73}^{+3.61}$       &  $73.78_{-11.90}^{+9.91}$        &  $4.08_{-0.05}^{+0.05}$    &  $0.11 \pm 0.02$  &  $0.09_{-0.05}^{+0.06}$  &  $18.07_{-2.78}^{+2.33}$   \\
\vspace*{2mm}
EL16   &  $62.02_{-3.08}^{+2.24}$        &  $2.99_{-4.89}^{+6.50}$          &  $2.52_{-0.03}^{+0.03}$    &  $0.01 \pm 0.01$  &  $0.27_{-0.05}^{+0.04}$  &  $1.19_{-1.82}^{+2.53}$    \\
\vspace*{2mm}
EL17   &  $129.19_{-3.80}^{+3.80}$       &  $77.06_{-10.21}^{+8.48}$        &  $3.98_{-0.04}^{+0.05}$    &  $0.15 \pm 0.05$  &  $0.02_{-0.11}^{+0.10}$  &  $19.34_{-2.48}^{+2.01}$   \\
\vspace*{2mm}
EL18   &  $54.31_{-18.26}^{+21.32}$      &  $20.21_{-6.28}^{+7.88}$         &  $2.24_{-0.03}^{+0.03}$    &  $0.03 \pm 0.01$  &  $0.32_{-0.06}^{+0.06}$  &  $9.02_{-2.47}^{+3.27}$    \\

EL19   &  $32.75_{-2.17}^{+2.11}$        &  $31.90_{-4.16}^{+4.39}$         &  $2.67_{-0.02}^{+0.02}$    &  $0.07 \pm 0.01$  &  $0.06_{-0.04}^{+0.05}$  &  $11.94_{-1.52}^{+1.67}$   \\
\vspace*{2mm}
EL20   &  $403.01_{-15.74}^{+12.28}$     &  $22.97_{-4.66}^{+3.65}$         &  $2.70_{-0.02}^{+0.02}$    &  $0.03 \pm 0.01$  &  $0.21_{-0.04}^{+0.04}$  &  $8.50_{-1.68}^{+1.33}$    \\
\vspace*{2mm}
EL21   &  $20.91_{-1.56}^{+1.39}$        &  $12.23_{-4.04}^{+4.46}$         &  $1.24_{-0.02}^{+0.02}$    &  $0.05 \pm 0.02$  &  $0.36_{-0.11}^{+0.11}$  &  $9.83_{-3.14}^{+3.58}$    \\
\vspace*{2mm}
EL22   &  $96.93_{-13.38}^{+15.80}$      &  $220.01_{-15.53}^{+18.82}$      &  $5.25_{-0.10}^{+0.07}$    &  $0.06 \pm 0.01$  &  $0.05_{-0.04}^{+0.03}$  &  $41.94_{-0.57}^{+0.68}$   \\
\vspace*{2mm}
EL23   &  $501.11_{-11.75}^{+8.32}$      &  $120.74_{-26.20}^{+24.26}$      &  $22.27_{-0.14}^{+0.12}$   &  $0.03 \pm 0.01$  &  $0.00_{-0.02}^{+0.04}$  &  $5.42_{-1.12}^{+1.05}$    \\
\vspace*{2mm}
EL24   &  $852.02_{-12.94}^{+14.01}$     &  $481.31_{-13.37}^{+16.66}$      &  $19.71_{-0.07}^{+0.09}$   &  $0.00 \pm 0.00$  &  $1.33_{-0.03}^{+0.04}$  &  $24.42_{-0.71}^{+0.81}$   \\
\vspace*{2mm}
EL25   &  $345.56_{-23.49}^{+20.01}$     &  $187.18_{-20.56}^{+24.99}$      &  $17.01_{-0.11}^{+0.13}$   &  $0.10 \pm 0.01$  &  $0.04_{-0.02}^{+0.03}$  &  $11.00_{-1.17}^{+1.44}$   \\
\vspace*{2mm}
EL26   &  $447.84_{-7.45}^{+8.86}$       &  $213.54_{-12.17}^{+18.90}$      &  $10.31_{-0.09}^{+0.08}$   &  $0.07 \pm 0.01$  &  $0.02_{-0.02}^{+0.03}$  &  $20.72_{-1.11}^{+1.72}$   \\
\vspace*{2mm}
EL27   &  $193.44_{-3.22}^{+2.49}$       &  $206.38_{-10.89}^{+17.09}$      &  $10.85_{-0.08}^{+0.08}$   &  $0.12 \pm 0.01$  &  $0.02_{-0.03}^{+0.03}$  &  $19.02_{-1.05}^{+1.56}$   \\
\vspace*{2mm}
EL28   &  $122.67_{-4.65}^{+4.55}$       &  $42.91_{-13.24}^{+9.91}$        &  $11.76_{-0.08}^{+0.06}$   &  $0.01 \pm 0.00$  &  $0.20_{-0.04}^{+0.04}$  &  $3.65_{-1.07}^{+0.82}$    \\
\vspace*{2mm}
T1214  &  $7.24_{-4.47}^{+6.02}$         &  $242.38_{-11.07}^{+7.86}$       &  $2.91_{-0.04}^{+0.05}$    &  $0.25 \pm 0.01$  &  $0.00_{-0.01}^{+0.01}$  &  $83.26_{-12.87}^{+8.48}$  \\
\vspace*{2mm}
T1247  &  $171.60_{-10.00}^{+13.61}$     &  $2637.56_{-28.77}^{+40.56}$     &  $102.80_{-0.17}^{+0.19}$  &  $0.06 \pm 0.00$  &  $0.27_{-0.00}^{+0.00}$  &  $25.66_{-0.31}^{+0.40}$   \\
\vspace*{2mm}
J1156  &  $3514.19_{-721.95}^{+376.47}$  &  $10929.57_{-684.05}^{+640.22}$  &  $188.65_{-5.72}^{+5.03}$  &  $0.24 \pm 0.03$  &  $0.19_{-0.03}^{+0.03}$  &  $57.94_{-4.47}^{+4.43}$   \\

\hline
\label{global_measurements}
\end{longtable}
\tablefoot{ID: ID of the galaxies in our sample, Mass: stellar mass of the galaxies, $\mathrm{L_{Ly\alpha}}$: total Ly$\alpha$ luminosity, $\mathrm{L_{FUV}}$L total FUV luminosity, $\mathrm{f_{esc}}$: Ly$\alpha$ escape fraction, $\mathrm{E(B -V)_n}$: nebular reddening, $\mathrm{EW_{Ly\alpha}}$: Ly$\alpha$ equivalent width. }
}


\bibliographystyle{aa.bst} 
\bibliography{aa.bib} 

\begin{appendix} 
\section{Ly$\alpha$ and FUV profile and fits in the circular aperture} \label{radia_analysis_appdix}
In this section, the Ly$\alpha$ and FUV profiles and the fitted double exponential, next to the FUV, and Ly$\alpha$ maps of all the galaxies in our sample are presented. In each figure, the left panels show the Ly$\alpha$ and FUV profiles and the fitted model to them. In these panels, blue and green represent Ly$\alpha$, and FUV, respectively. The dark and light colours distinguish the core from the halo, respectively. For those bins with S/N < 1, the 1$\sigma$ upper limit are displayed, and the bins with 1 $\leq$ S/N < 2 (data points not used in the fits) are displayed with the empty square symbol. In the bottom left of the left panels, the fitted Ly$\alpha$ and FUV halo scale lengths and their corresponding error bars are printed with sky blue and light green colours, respectively. The middle and right panels show the FUV and Ly$\alpha$ maps of each galaxy, respectively. The blue rings represent the largest radii where a circle centred at the FUV brightest pixel fits within each map, used as the last bin where the photometry was performed on, and the green ring shows the radius where the SFRD drops below 0.01 $\mathrm{M_\odot \, yr^{-1} \,  kpc^{-2}}$.
\begin{figure*}[t!]
 \centering 
 	\includegraphics[width=\textwidth] {LARS01_profile_fuv_lya_img.jpeg}
 	\includegraphics[width=\textwidth] {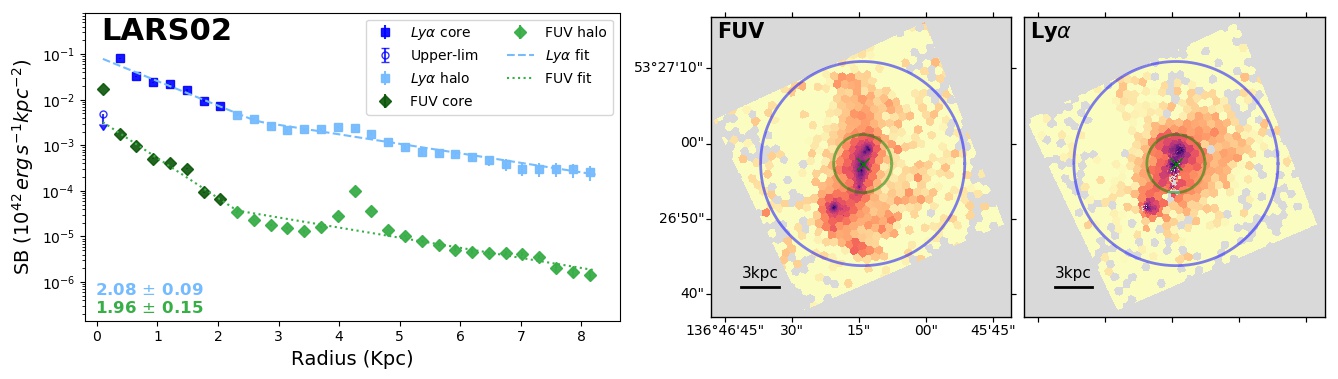}
 	\includegraphics[width=\textwidth] {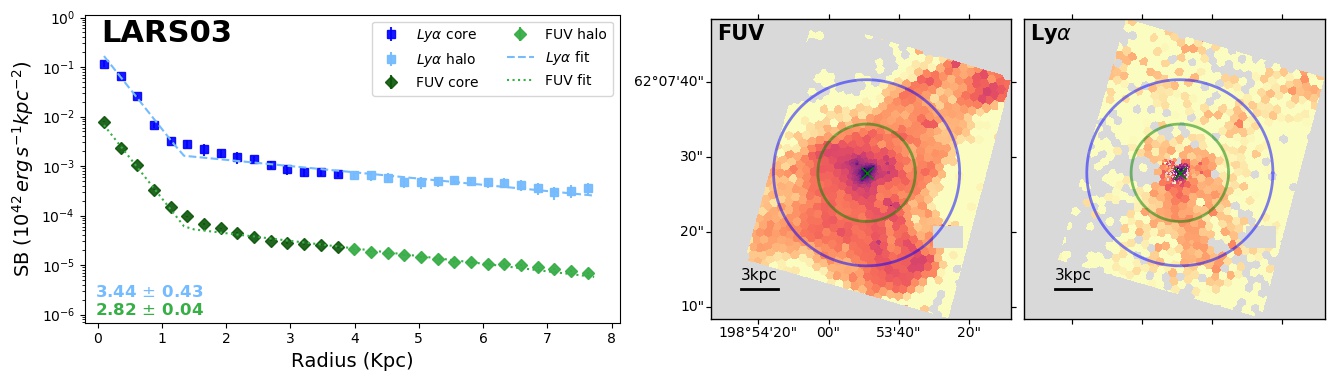}
 	\includegraphics[width=\textwidth] {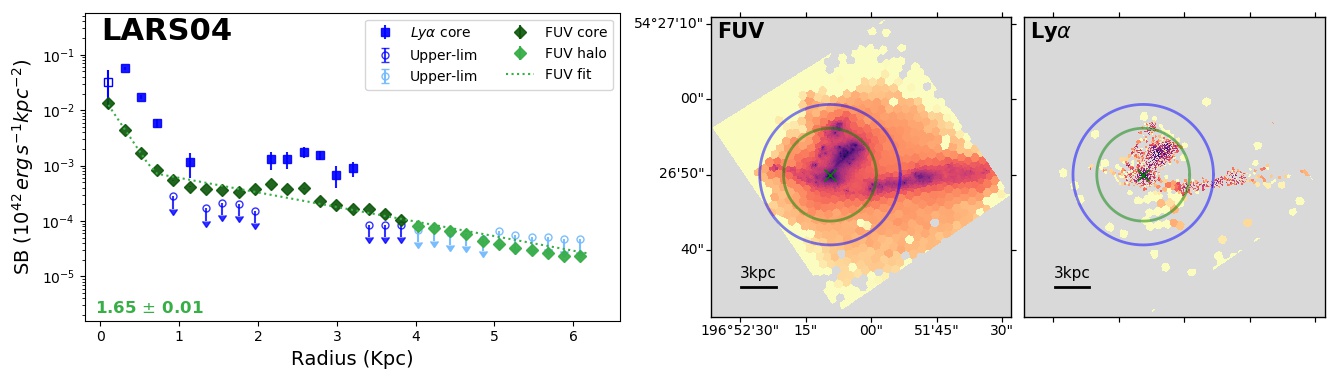}
	
\caption{
        Same as Fig. \ref{paper_example_lya_fuv_radial}, but for LARS01 - LARS04.
        }
\end{figure*} 
\begin{figure*}[t!]

 \centering 
 	\includegraphics[width=\textwidth] {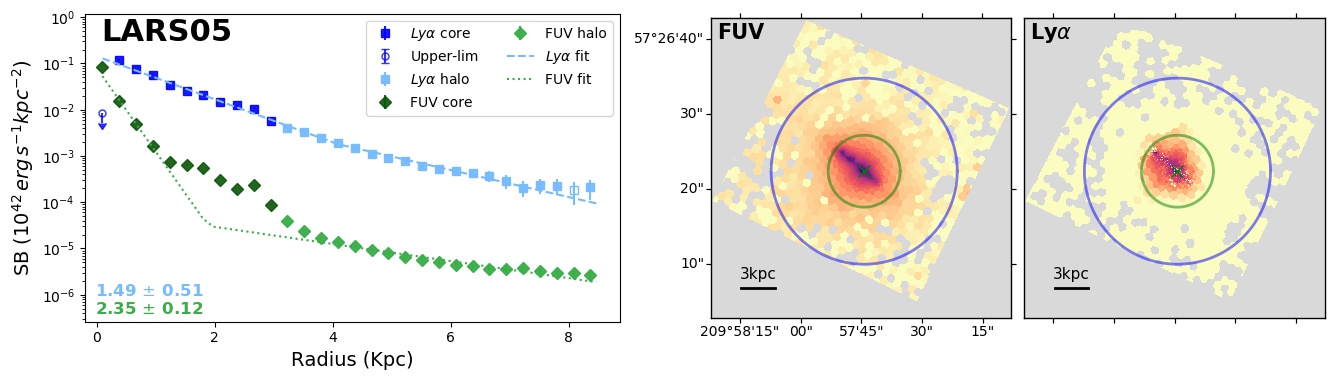}
 	\includegraphics[width=\textwidth] {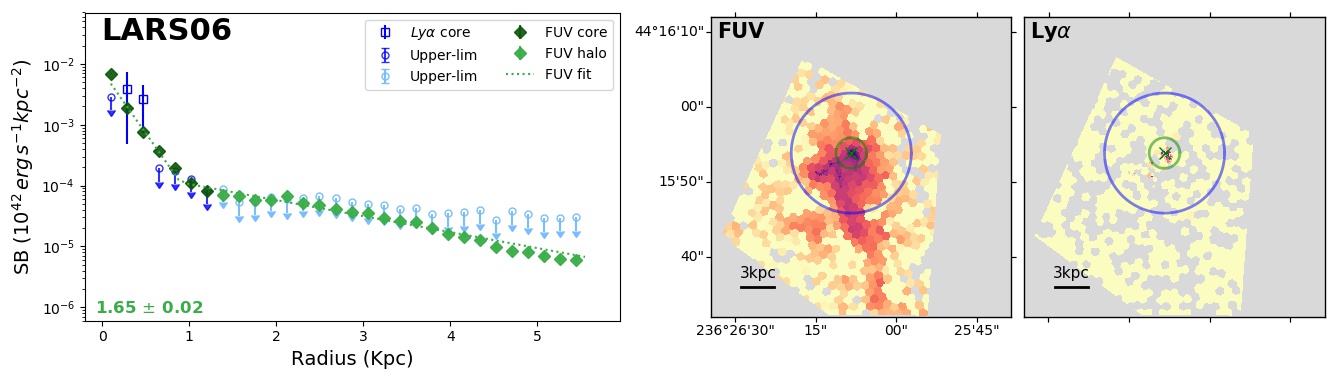}
 	\includegraphics[width=\textwidth] {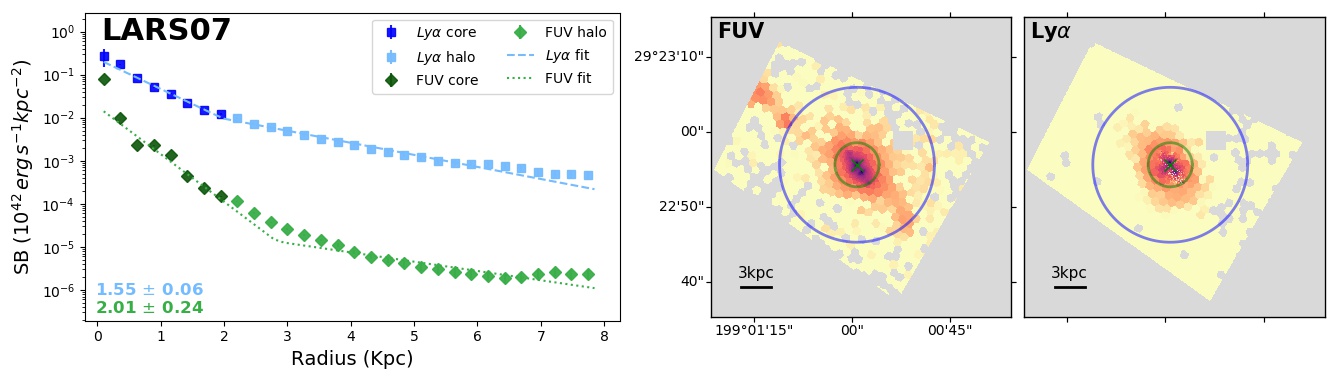}
 	\includegraphics[width=\textwidth] {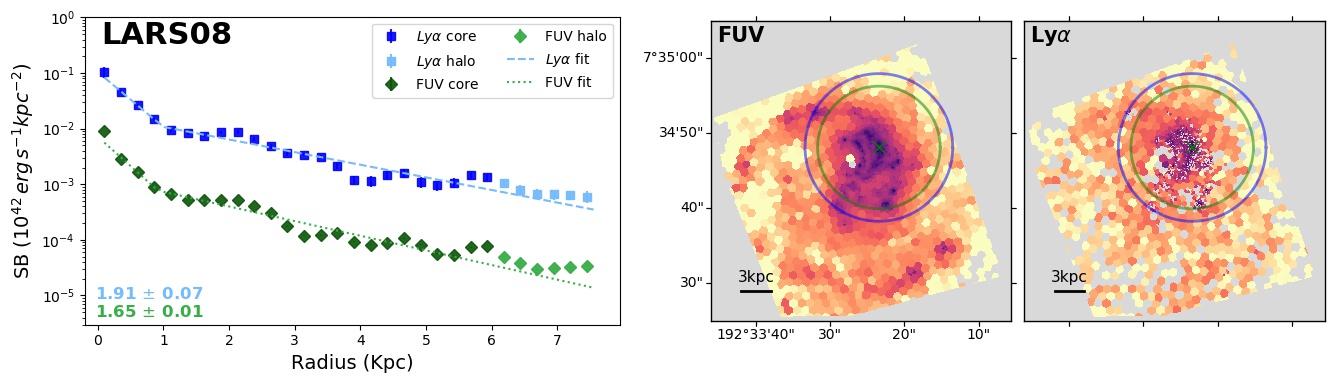}
	
\caption{
        Same as Fig. \ref{paper_example_lya_fuv_radial}, but for LARS05 - LARS08.
        }
\end{figure*} 
\begin{figure*}[t!]

 \centering 
 	\includegraphics[width=\textwidth] {LARS09_profile_fuv_lya_img.jpeg}
 	\includegraphics[width=\textwidth] {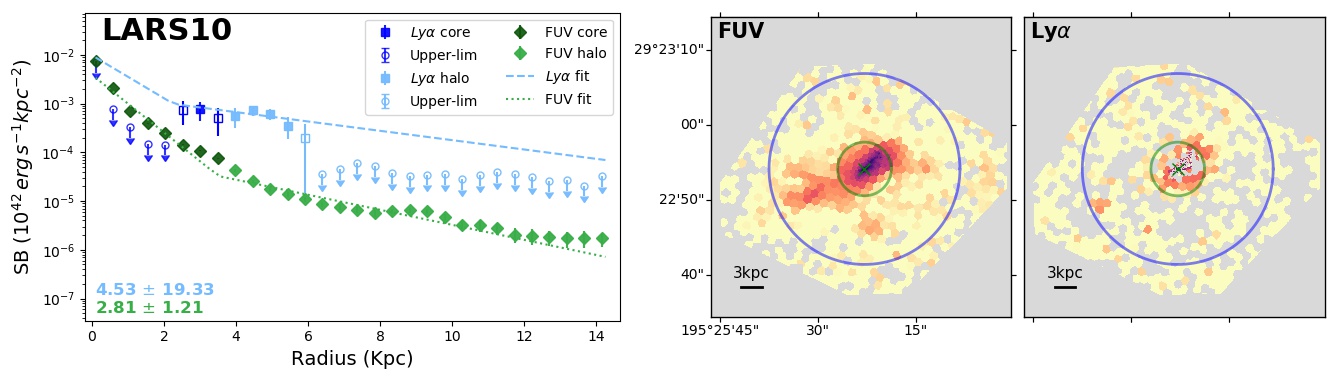}
 	\includegraphics[width=\textwidth] {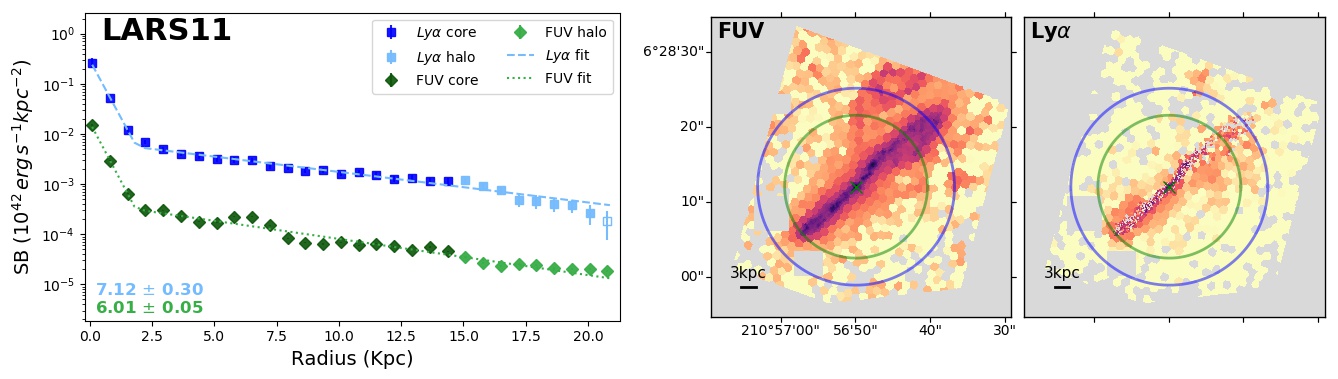}
 	\includegraphics[width=\textwidth] {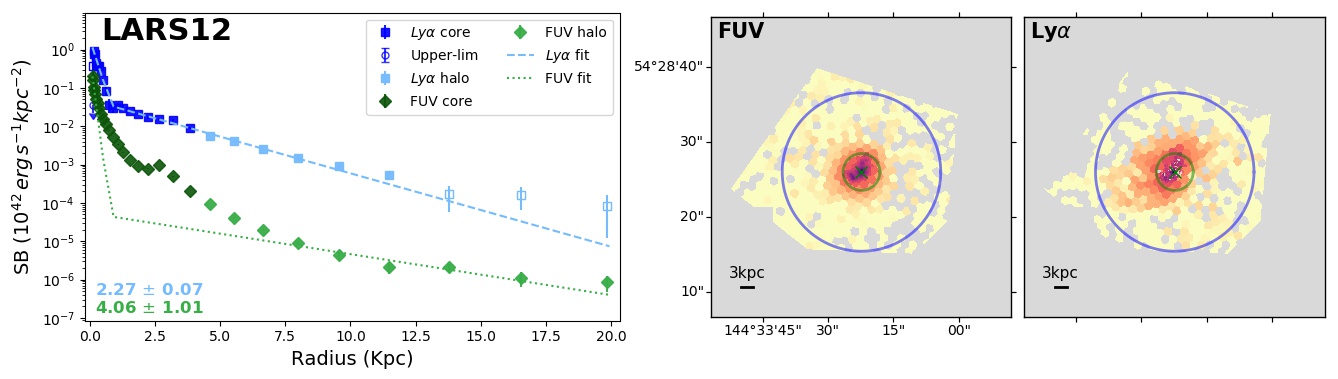}

\caption{
        Same as Fig. \ref{paper_example_lya_fuv_radial}, but for LARS09 - LARS12.
        }
\end{figure*} 
\begin{figure*}[t!]

 \centering 
 	\includegraphics[width=\textwidth] {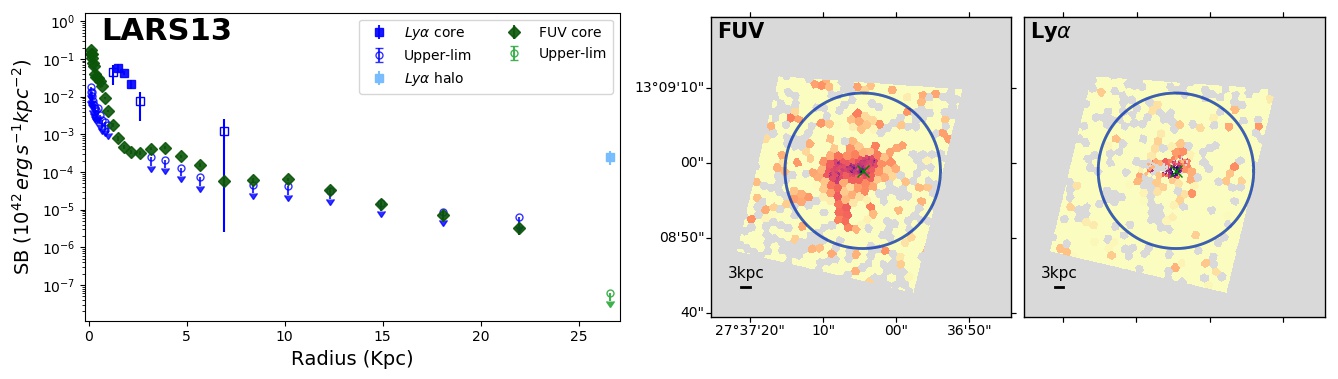}
 	\includegraphics[width=\textwidth] {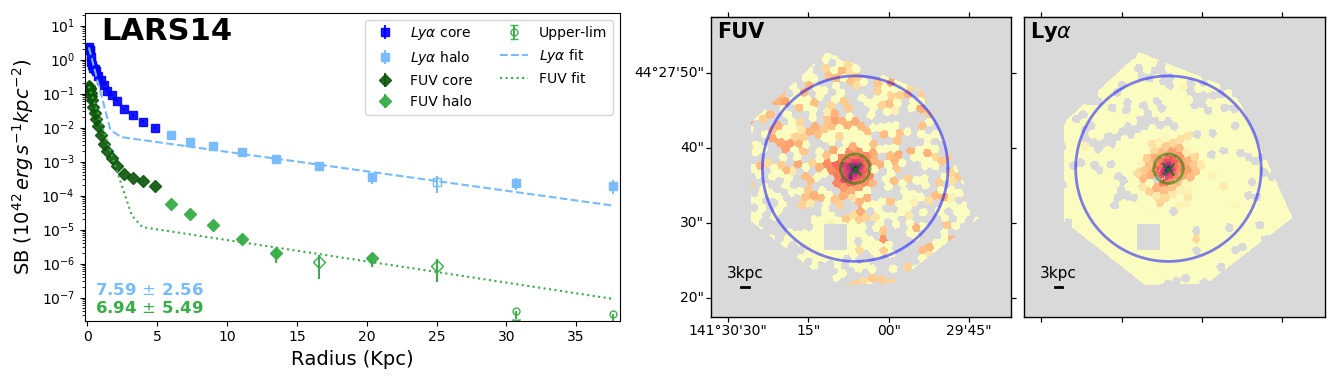}
 	\includegraphics[width=\textwidth] {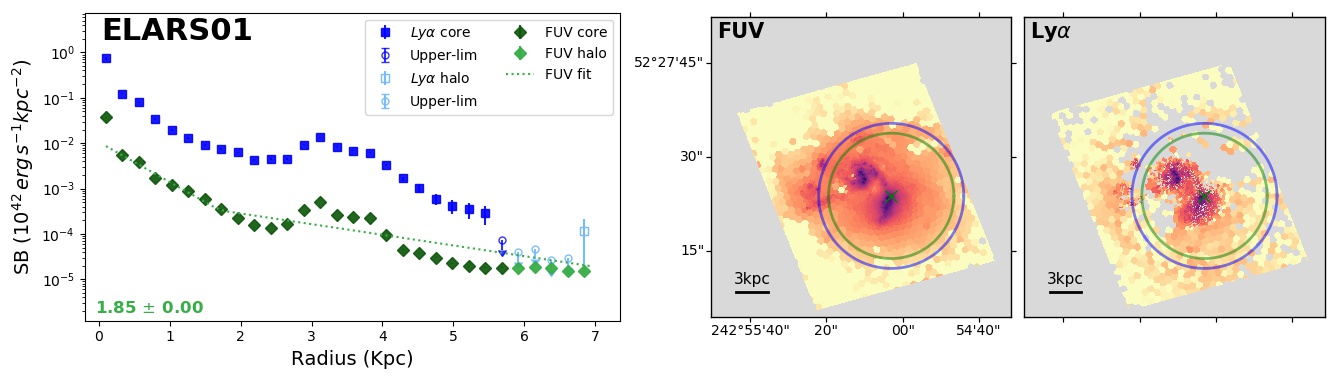}
 	\includegraphics[width=\textwidth] {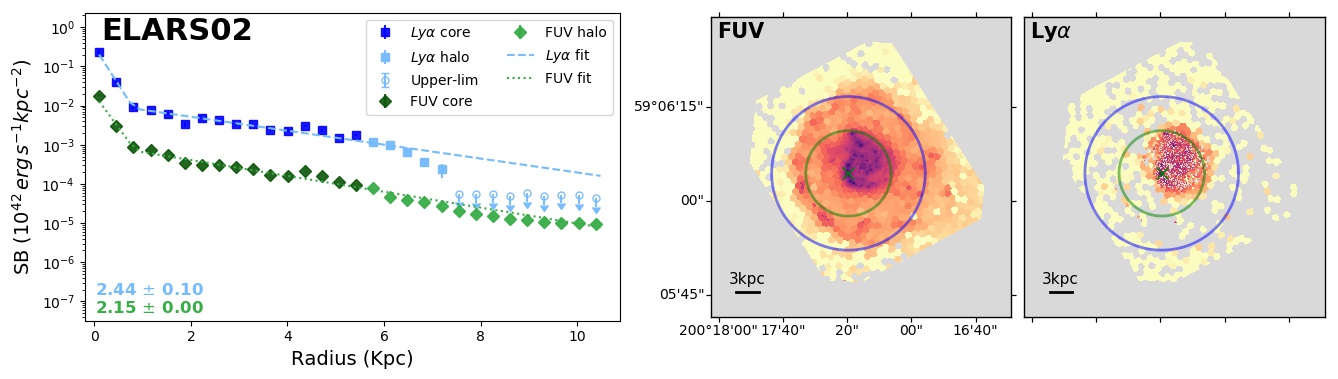}

\caption{
        Same as Fig. \ref{paper_example_lya_fuv_radial}, but for LARS13, LARS14, ELARS01, and ELARS02.
        }
\end{figure*} 
\begin{figure*}[t!]

 \centering 
 	\includegraphics[width=\textwidth] {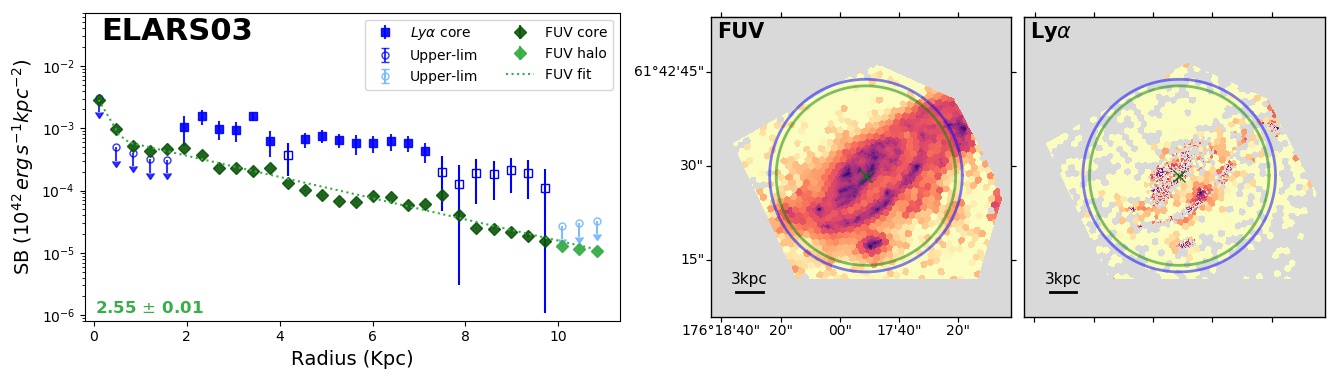}
 	\includegraphics[width=\textwidth] {ELARS04_profile_fuv_lya_img.jpeg}
 	\includegraphics[width=\textwidth] {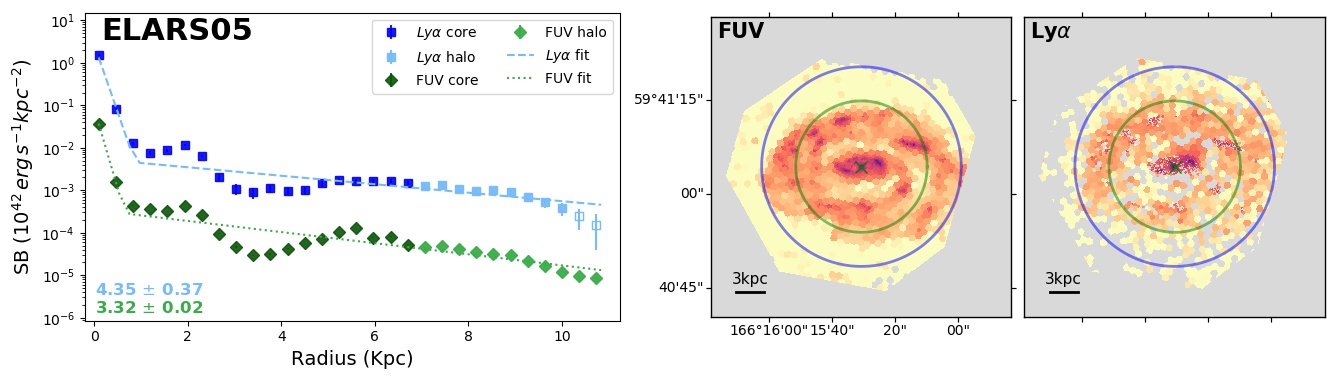}
 	\includegraphics[width=\textwidth] {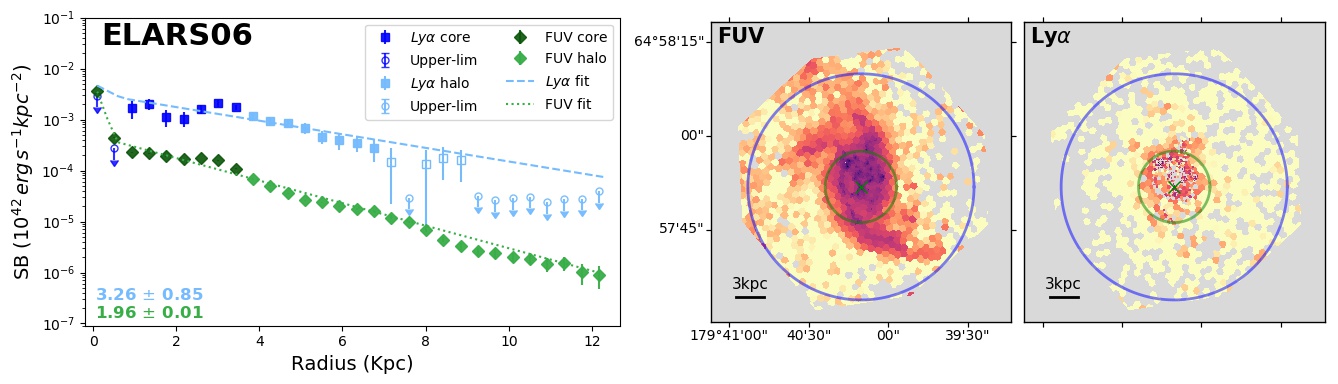}

\caption{
        Same as Fig. \ref{paper_example_lya_fuv_radial}, but for ELARS03 - LARS06.
        }
\end{figure*}
\begin{figure*}[t!]

 \centering 
 	\includegraphics[width=\textwidth] {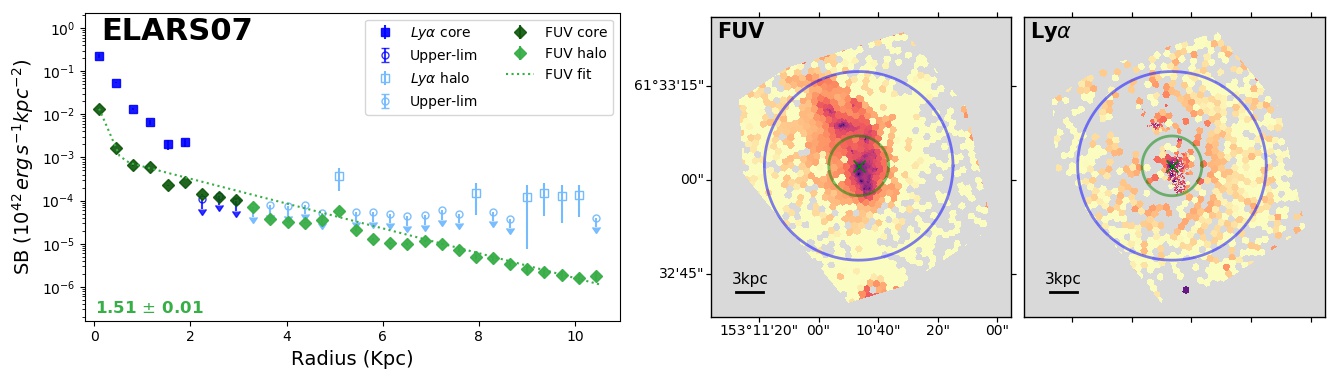}
 	\includegraphics[width=\textwidth] {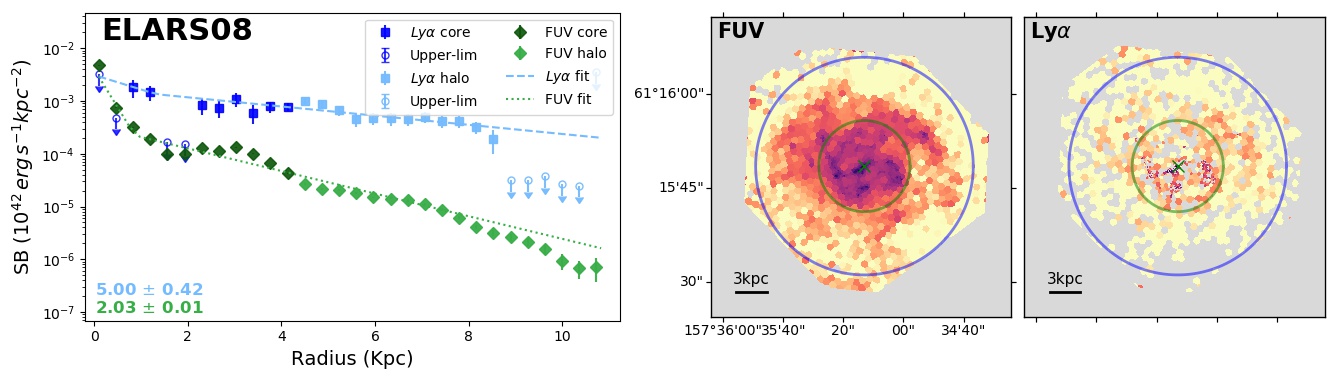}
 	\includegraphics[width=\textwidth] {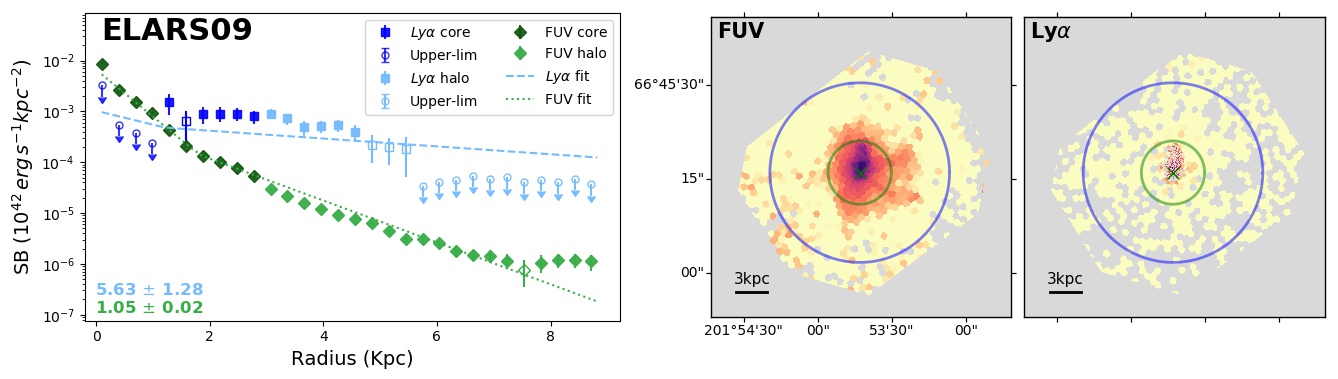}
 	\includegraphics[width=\textwidth] {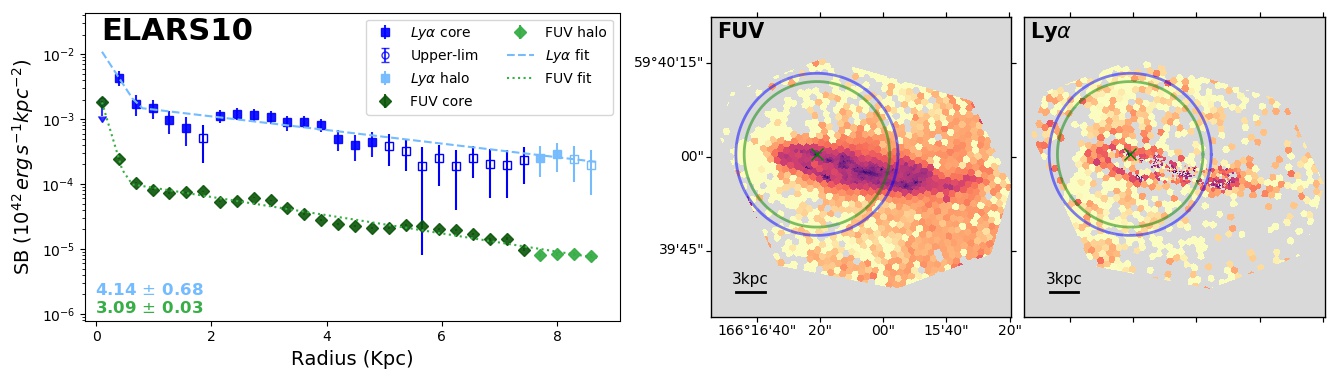}

\caption{
        Same as Fig. \ref{paper_example_lya_fuv_radial}, but for ELARS07 - ELARS10.
        }
\end{figure*} 
\begin{figure*}[t!]

 \centering 
 	\includegraphics[width=\textwidth] {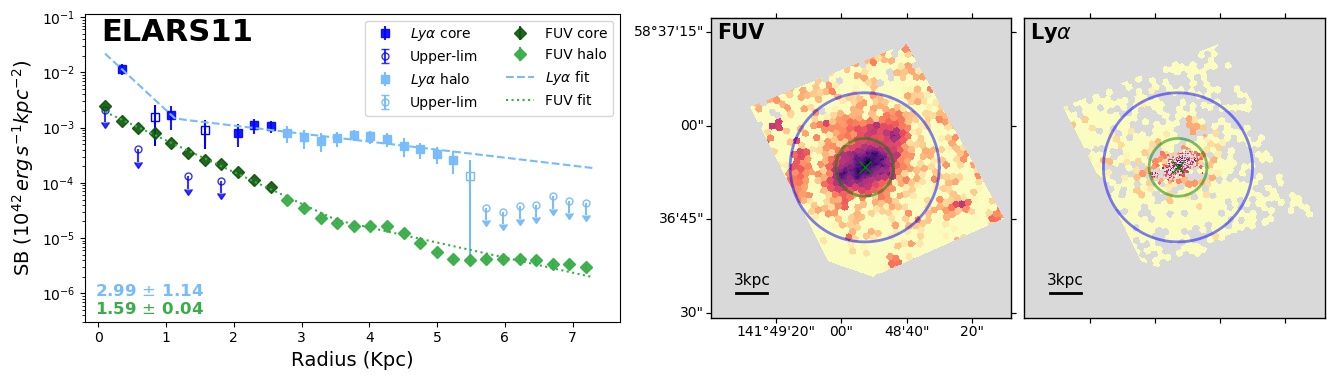}
 	\includegraphics[width=\textwidth] {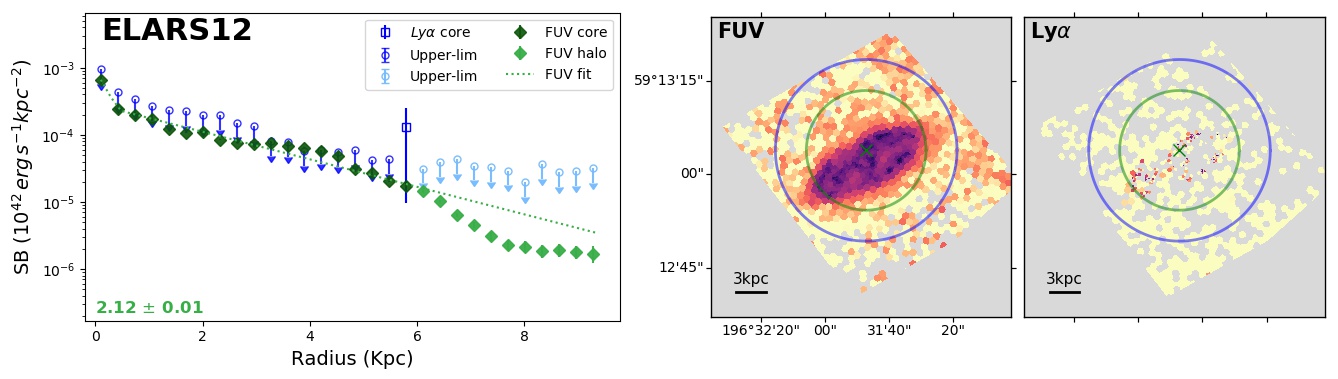}
 	\includegraphics[width=\textwidth] {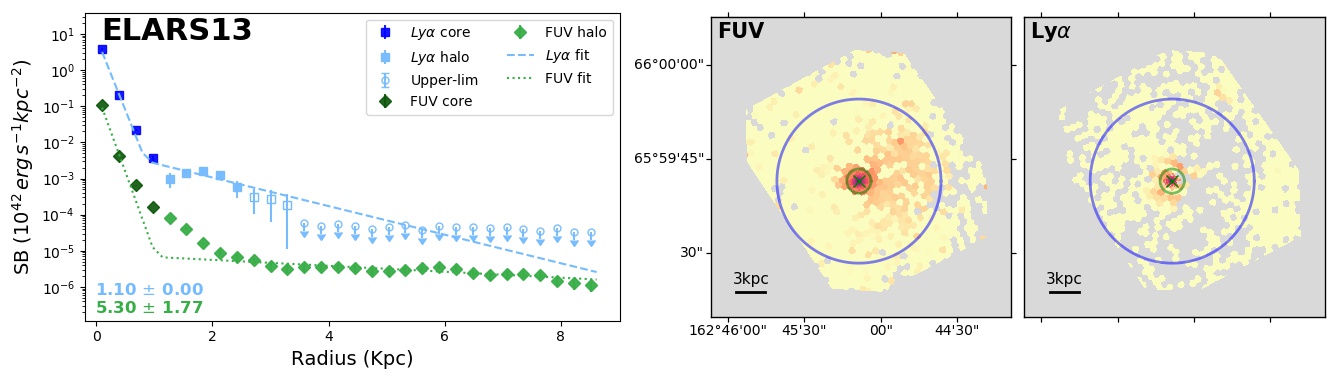}
 	\includegraphics[width=\textwidth] {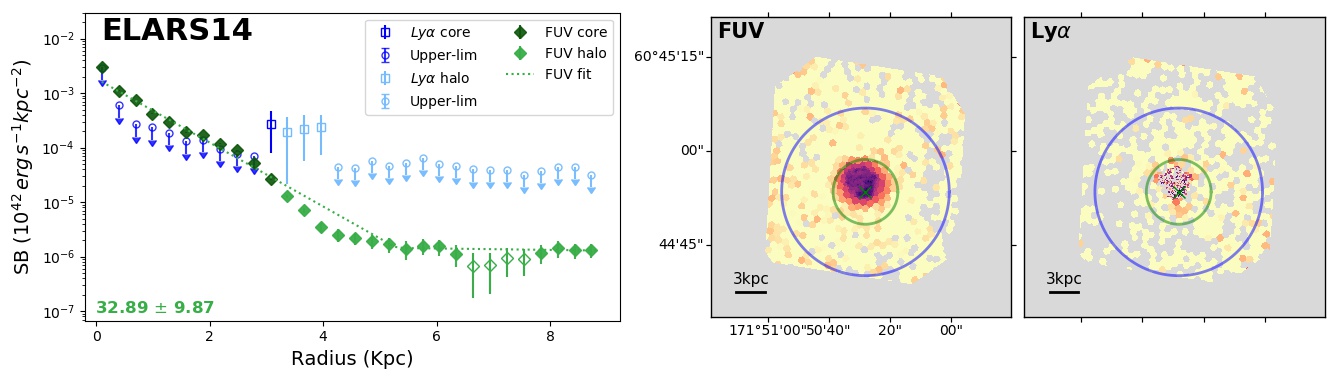}

\caption{
        Same as Fig. \ref{paper_example_lya_fuv_radial}, but for ELARS11 - ELARS14.
        }
\end{figure*}
\begin{figure*}[t!]

 \centering 
  	\includegraphics[width=\textwidth] {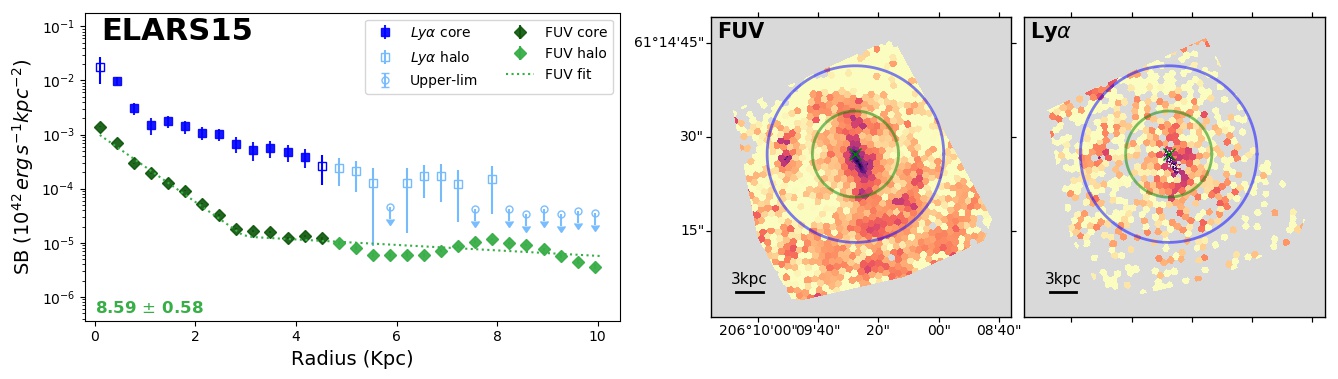}
 	\includegraphics[width=\textwidth] {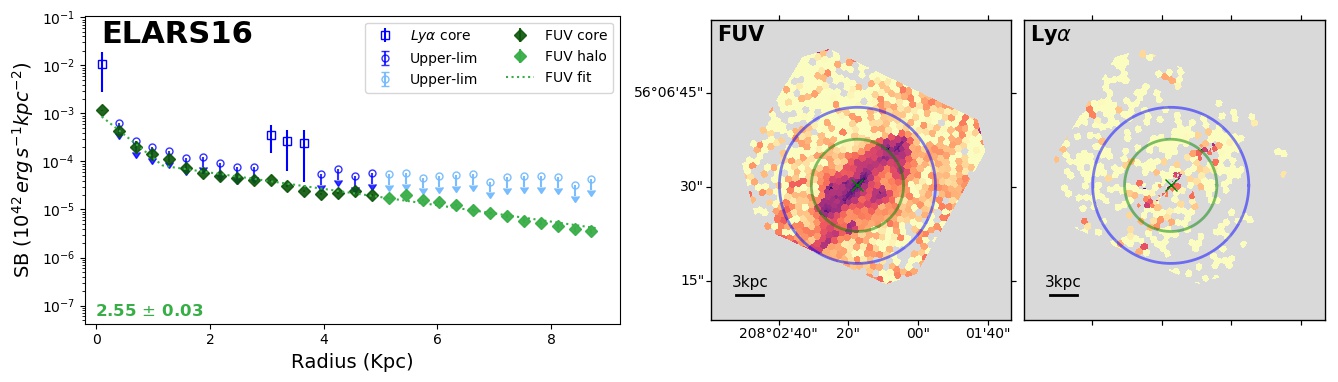}
  	\includegraphics[width=\textwidth] {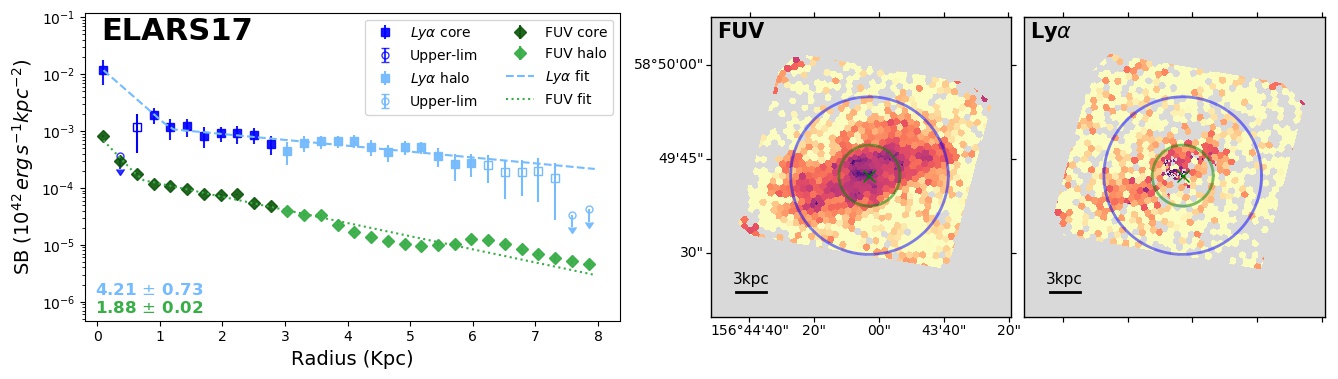}
  	\includegraphics[width=\textwidth] {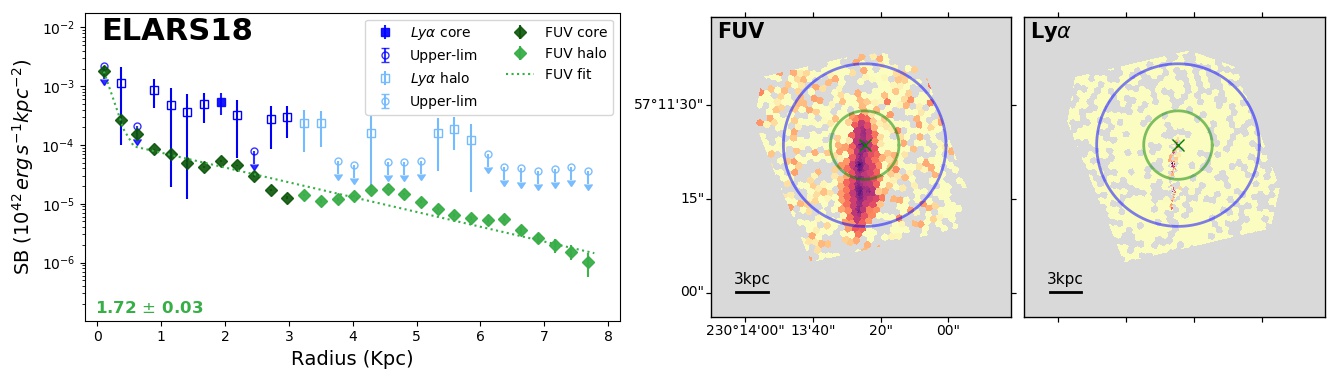}

\caption{
        Same as Fig. \ref{paper_example_lya_fuv_radial}, but for ELARS15 - ELARS18.
        }
\end{figure*} 
\begin{figure*}[t!]

 \centering 
 	\includegraphics[width=\textwidth] {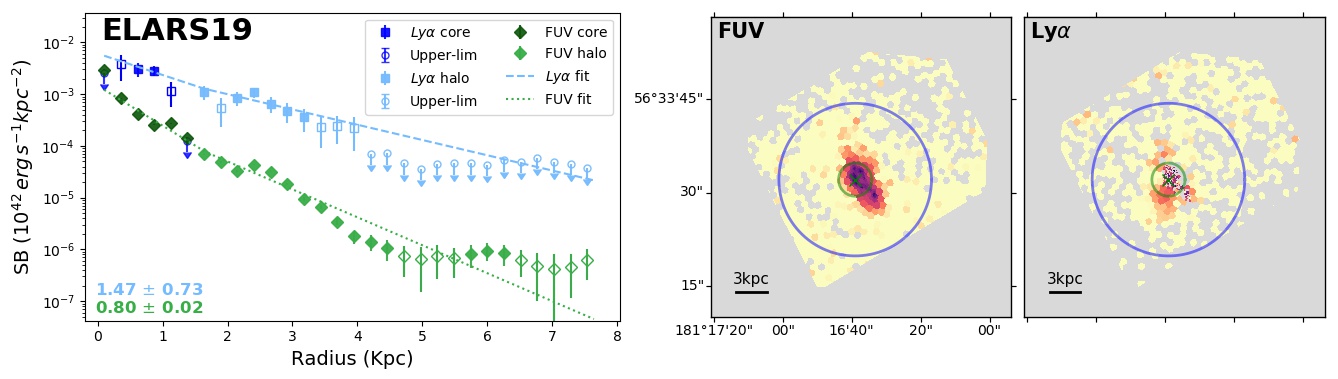}
 	\includegraphics[width=\textwidth] {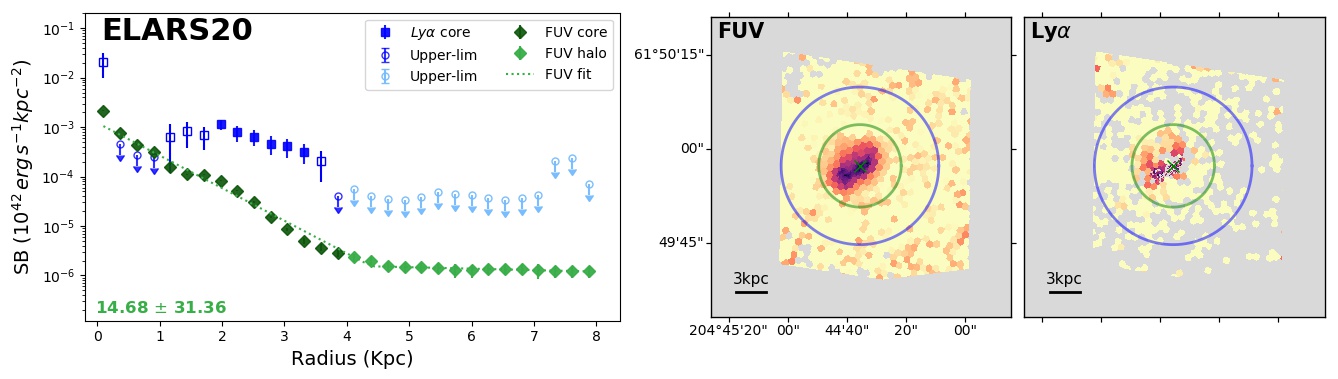}
 	\includegraphics[width=\textwidth] {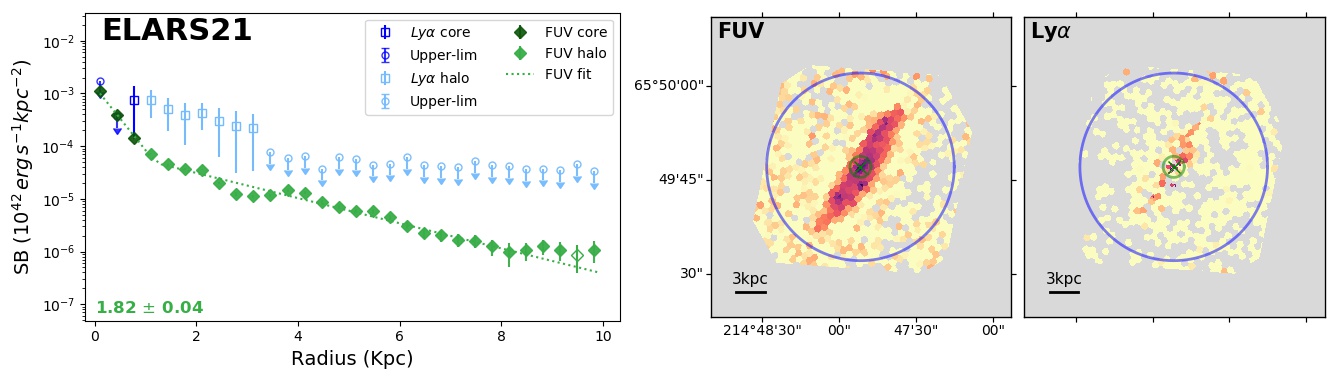}
 	\includegraphics[width=\textwidth] {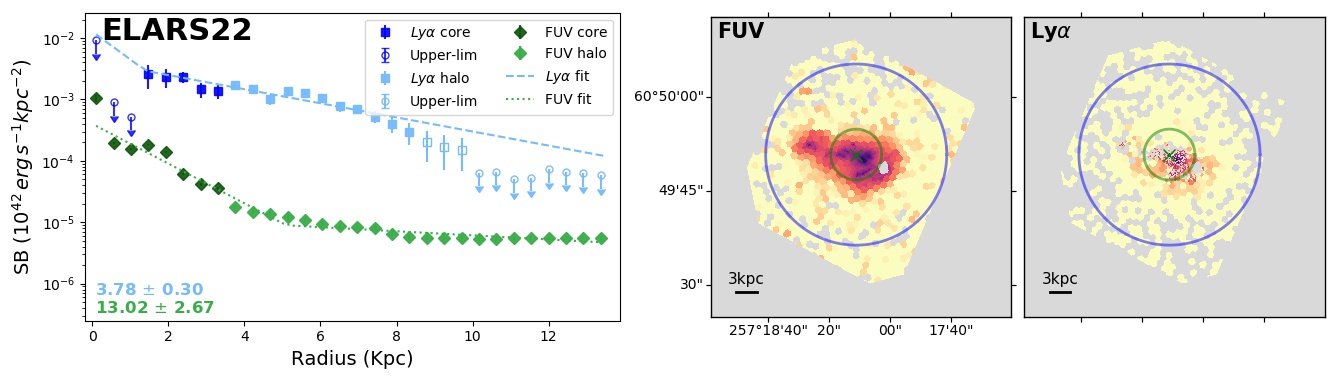}
   
\caption{
        Same as Fig. \ref{paper_example_lya_fuv_radial}, but for ELARS19 - ELARS22.
        }
\end{figure*} 
\begin{figure*}[t!]

 \centering 
 	\includegraphics[width=\textwidth] {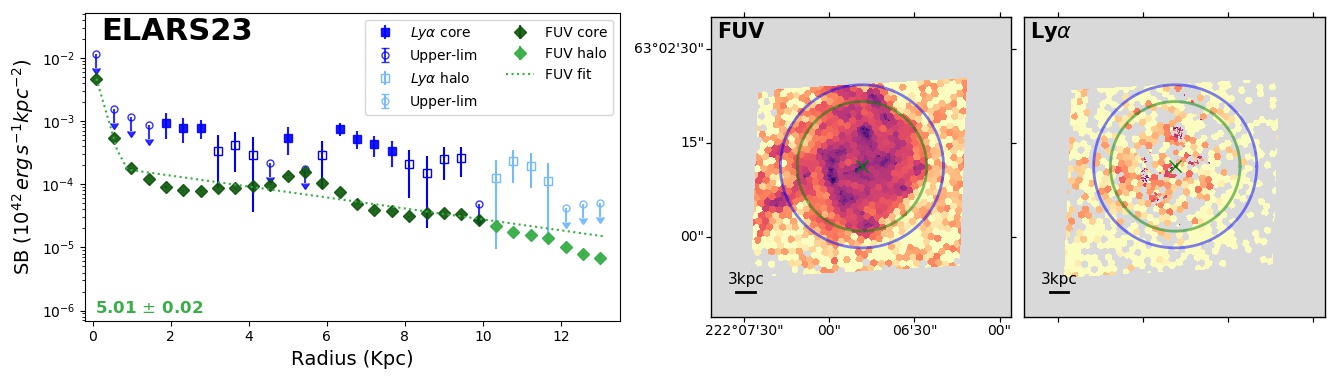}
 	\includegraphics[width=\textwidth] {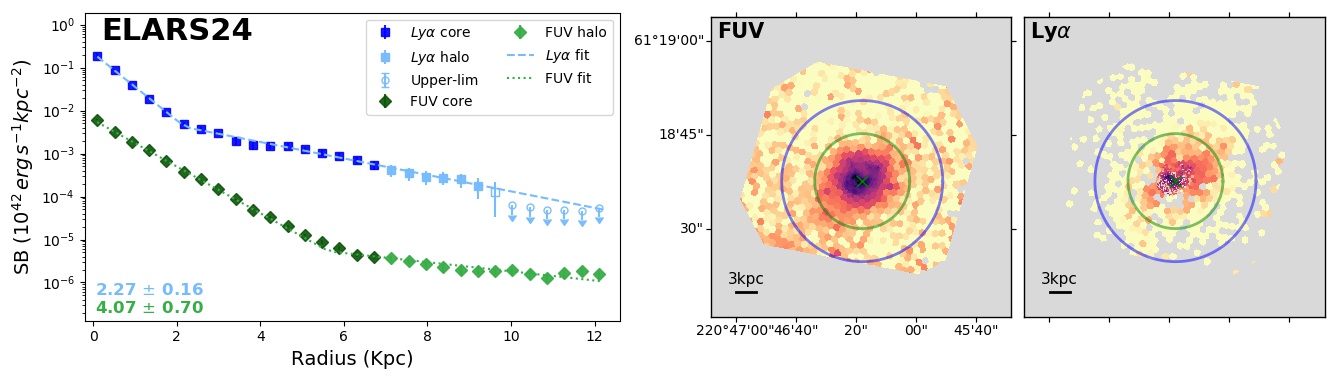}
 	\includegraphics[width=\textwidth] {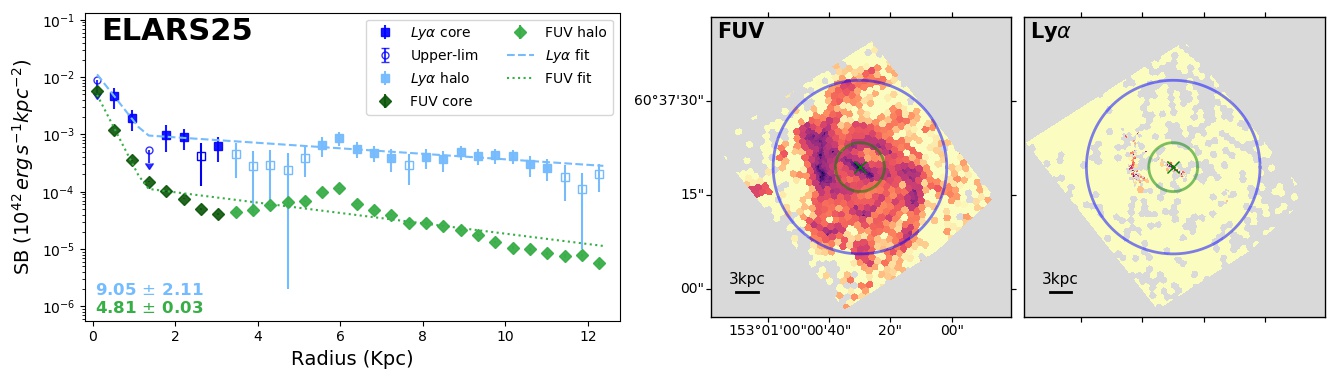}
 	\includegraphics[width=\textwidth] {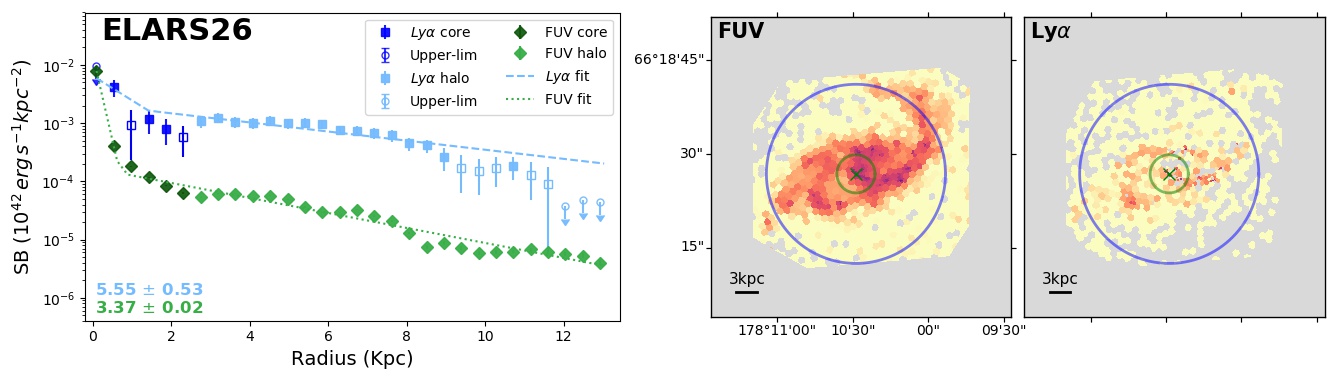}

\caption{
        Same as Fig. \ref{paper_example_lya_fuv_radial}, but for ELARS23 - ELARS26.
        }
\end{figure*} 
\begin{figure*}[t!]

 \centering 
 	\includegraphics[width=\textwidth] {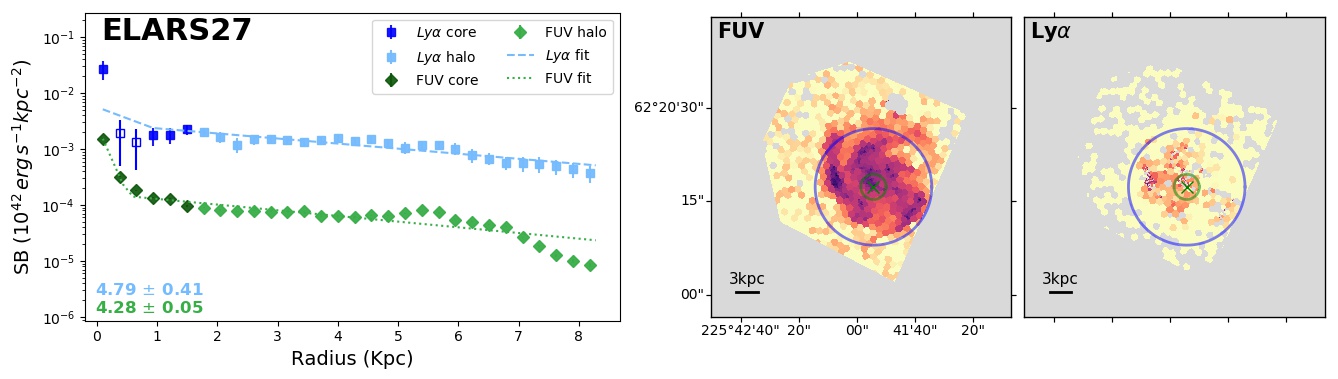}
 	\includegraphics[width=\textwidth] {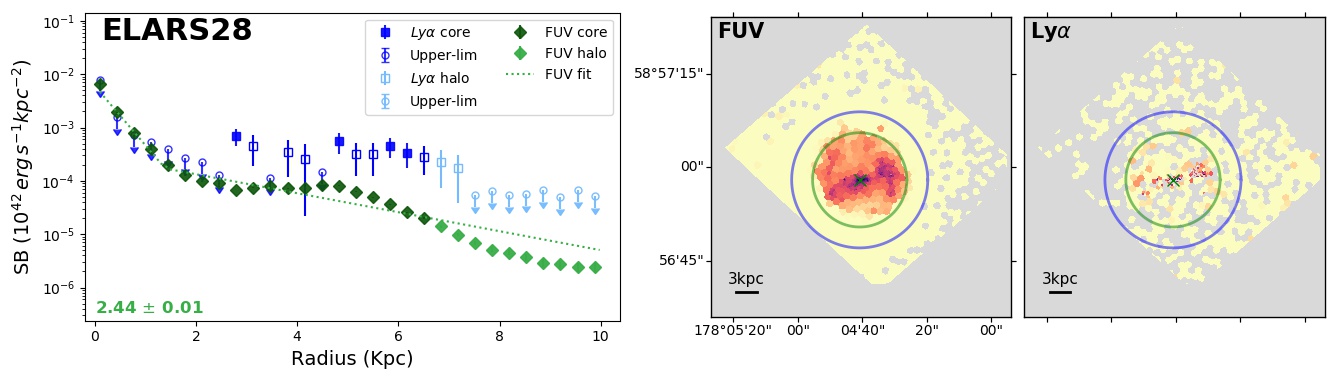}
 	\includegraphics[width=\textwidth] {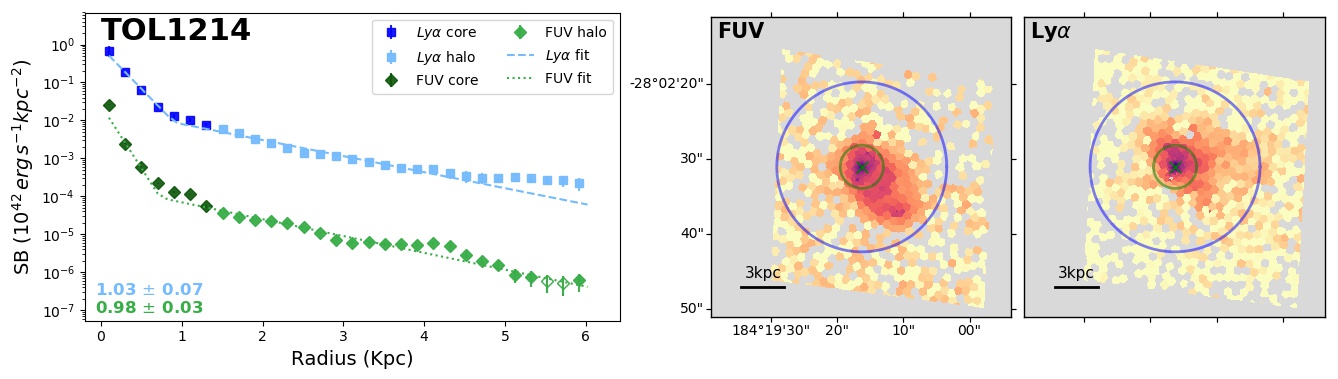}
 	\includegraphics[width=\textwidth] {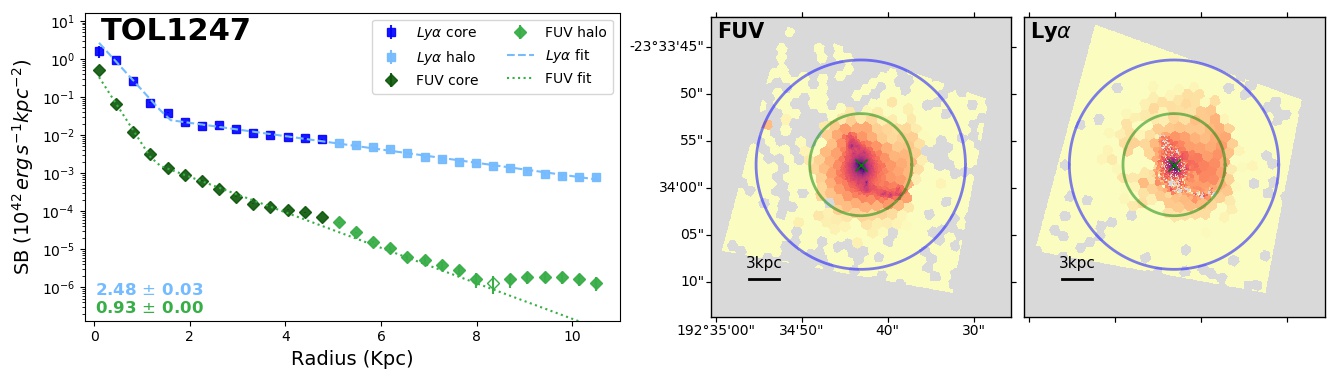}

\caption{
        Same as Fig. \ref{paper_example_lya_fuv_radial}, but for ELARS27, ELARS28, Tol1214, and Tol1247.
        }
\end{figure*} 
\begin{figure*}[t!]

 \centering 
 	\includegraphics[width=\textwidth] {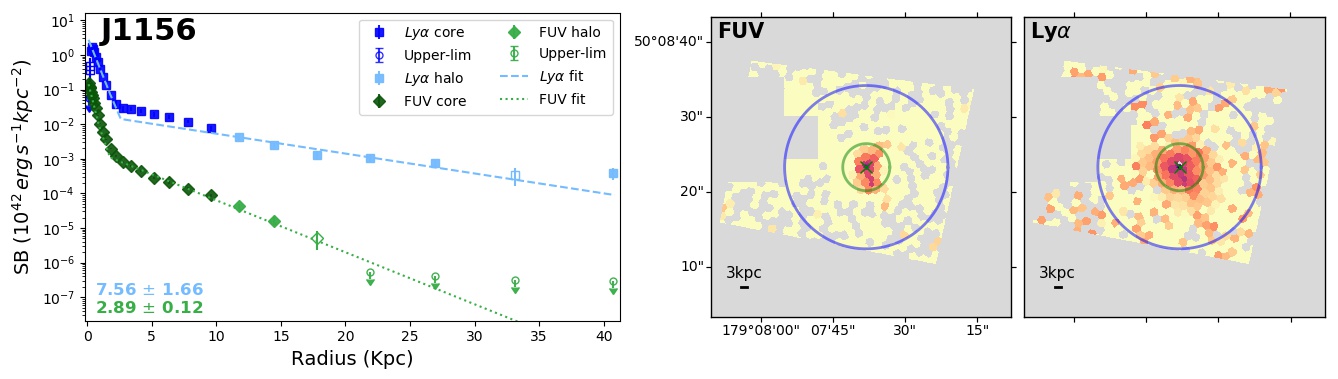}

\caption{Same as Fig. \ref{paper_example_lya_fuv_radial}, but for J1156}
\end{figure*} 
\clearpage
\section{Systematic uncertainties on the halo exponential fits} \label{bgest_appdix}
\begin{figure*}[hb]
 \centering 
    \includegraphics[width=\textwidth] {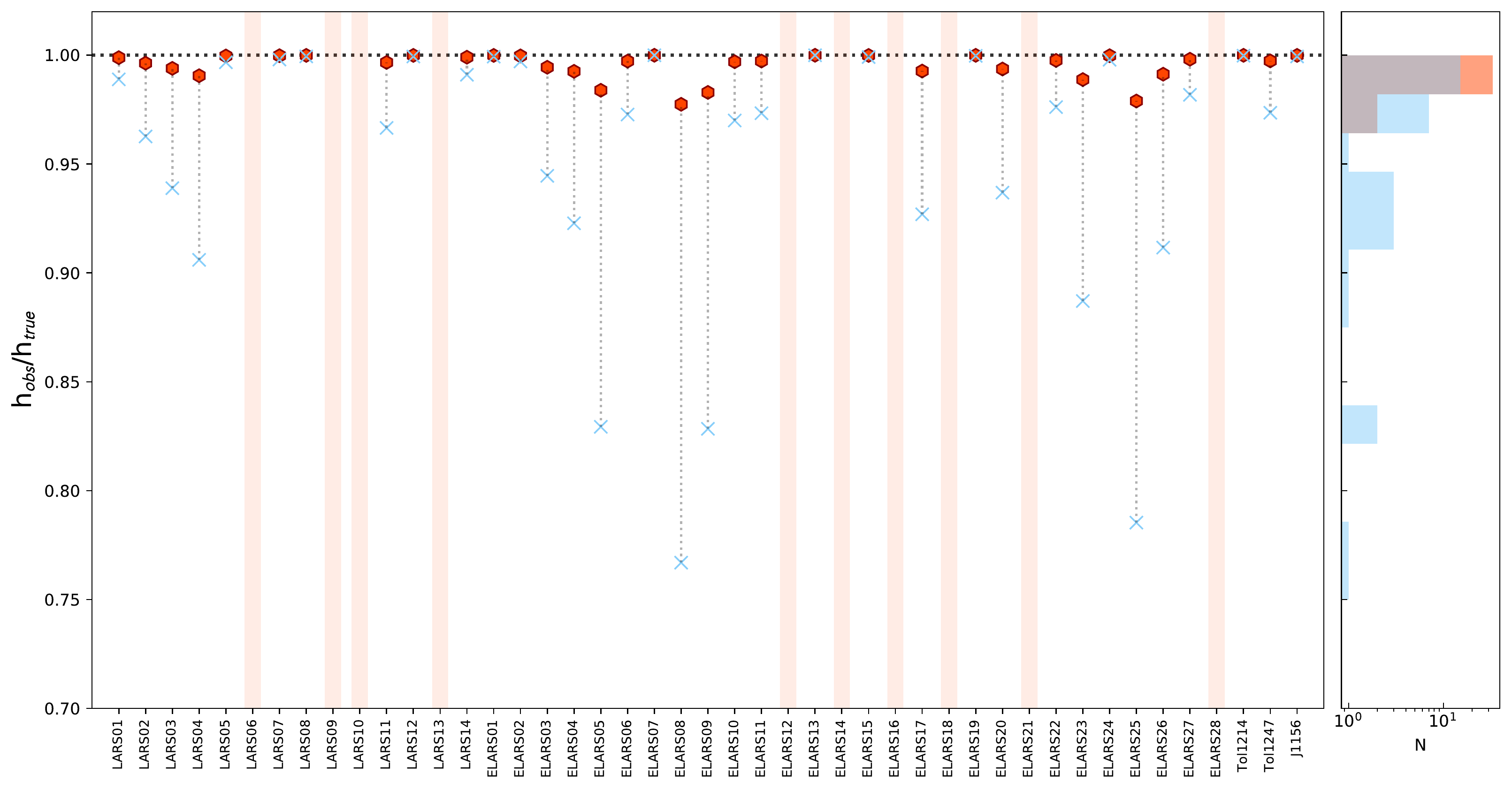}
\caption{
        The effect of over-subtracting Ly$\alpha$ on the fitted halo scale lengths. Galaxies marked with a vertical stripe are excluded from the sample due to low signal-to-noise in the halo (see Sec. \ref{rsc_result_sec}). Crosses show the worst case assumption of all subtracted background flux being Ly$\alpha$, and hexagons show the ratios when assuming 10\% of the background being Ly$\alpha$.
        }
    \label{bgest_fig}
\end{figure*}
In this section, we investigate the systematic uncertainties on the halo exponential fitting resulting from a possible over-subtraction of Ly$\alpha$ in the maps. Given the uncertainties in the core regions because of the Ly$\alpha$ continuum absorption, background over-subtraction is likely a negligible effect for the core profile fitting. As noted in the text, we make two assumptions to perform this estimate: (i) the halo profile can be fitted with an exponentially declining function, and (ii) the over-subtracted background is equal to the background flux subtracted during data reduction. In practice, the first assumption means that we cannot estimate this effect for the galaxies excluded from the sample in Sec. \ref{rsc_result_sec}. As mentioned in the main text, the second assumption is the worst possible case and thus provides an upper bound on the uncertainty. 
\par
For the halo exponential profile, adding back in the over-subtracted flux can be written as:
\begin{equation}
    f(r) = f_{obs}(r) + f_B,
\end{equation}
where $f(r)$ is the total surface brightness profile, $f_{obs}(r)$ is the exponential profile modelled on the subtracted data, and $f_B$ is the background level. Assumption (i) above means that $f_B$ cannot be much larger than $f_{obs}(r)$. Under assumption (ii) above $f_B$ is purely Ly$\alpha$, and should also thus follow the exponential function, but evaluated at the radius $r=r_{max}$, which is always larger than the maximum radius used to obtain the fit. Hence,
\begin{equation}
    f(r)/f(r_1) = e^{-r/h_{obs}} + e^{-r_{max}/h_{true}},   
\end{equation}
where we have normalised the function by the surface brightness at the innermost point of the halo, $f(r_1)$. The scale length is denoted by $h_{true}$. With knowledge of the true scale length we can calculate the resulting scale length ($h_{obs}$) for the profile with the background flux removed.
\begin{equation}
    f_{obs}(r) = f(r_1) e^{-r/h_{obs}} = f(r_1)\left( e^{-r/h_{true}} - e^{-r_{max}/h_{true}} \right)
\end{equation}
The scale lengths are found by fitting an exponential to the halo data points between the radii $r_1$ and $r_2$ (where the signal-to-noise of the Ly$\alpha$ surface brightness drops below 2). For this estimate we 
calculate the scale length from a simple line "fit" (in logarithmic space) to $f_{obs}$ between $r_1$ and $r_2$:
\begin{equation}
\label{bgest_eq}
    h_{obs} = \frac{r_2 - r_1}{\ln{\frac{e^{-(r_1-r_{max})/h_{true}}-1}{e^{-(r_2-r_{max})/h_{true}}-1}}}
\end{equation}
To get a first order estimate of the effect we use the best-fit halo scale length, $r_{sc,\mathrm{Ly}\alpha}$ in place of $h_{true}$ and calculate the $h_{obs}/r_{sc,\mathrm{Ly}\alpha}$ ratio. A ratio lower than one indicates that the fitted scale length has been underestimated due to over-subtraction of the background. The result for all of the galaxies is shown in Fig. \ref{bgest_fig}, where we show the ratios assuming that all of the background is Ly$\alpha$ (crosses), and with 10\% of the background being Ly$\alpha$ (hexagons). We note that for all but 4 galaxies (ELARS05, ELARS08, ELARS09, and ELARS25), the systematic uncertainties on the scale lengths are less than 10\%, even in the worst case. The worst case assumption puts an absolute upper bound on the uncertainty and likely over-estimates the systematic effects substantially. LARS09 is here marked as an excluded galaxy, even though it has a bright enough halo to get a profile fit (see Section~\ref{rsc_result_sec}). Because it has an extremely large fitted scale length, the over-subtraction uncertainty becomes very large ($\sim$70\%).
\par
In Table~\ref{bgest_tab} we show the subtracted background levels in luminosity units (also shown in Figures~\ref{paper_example_lya_fuv_radial} and Appendix~\ref{radia_analysis_appdix}) and the associated statistical uncertainty of this level. The third column lists the relative systematic (positive) error on the total Ly$\alpha$ luminosities ( $\delta L_{\mathrm{Ly}\alpha,\mathrm{sys}} = \left[\mathrm{L_{Ly\alpha}}+\mathrm{BG}_{\mathrm{Ly}\alpha}\right]/\mathrm{L_{Ly\alpha}}$) assuming that 10/100 \% of the background is intrinsic $\mathrm{Ly}\alpha$ emission. The table also lists the scale length ratios ($h_{obs}/r_{sc,\mathrm{Ly}\alpha}$) calculated above.

\begin{table*}

\caption{
        Background fluxes and systematic uncertainty estimates on Ly$\alpha$ total luminosity and profile scale length.
        }
\label{bgest_tab}        
\centering
\begin{tabular}{l l c c c c}
\hline
\hline
ID    & 	SB$_{\mathrm{bg}}$	& $\delta L_{\mathrm{Ly}\alpha,\mathrm{sys}}$, 100\% & $\delta L_{\mathrm{Ly}\alpha,\mathrm{sys}}$, 10\%	& $h_{obs}/r_{sc,\mathrm{Ly}\alpha}$, 100\% & $h_{obs}/r_{sc,\mathrm{Ly}\alpha}$, 10\%\\
 	  & 	10$^{38}$ erg/s/kpc$^2$ &  	& &		\\ 
\hline
\hline
LARS01  &  5.47$\pm$ 0.097  & 1.09  & 1.01&   0.989  & 0.999\\
LARS02  &  2.19$\pm$ 0.068  & 1.11  & 1.01&   0.963  & 0.996\\
LARS03  &  8.07$\pm$ 0.12   & 1.85  & 1.09&   0.939  & 0.994\\
LARS04  &  2.10$\pm$ 0.45   & 2.43  & 1.14&   0.906  & 0.991\\
LARS05  &  2.49$\pm$ 0.058  & 1.08  & 1.01&   0.997  & 1.00 \\
LARS06  &  2.32$\pm$ 0.058  & 1.13  & 1.01&   NA     & NA   \\
LARS07  &  1.81$\pm$ 0.050  & 1.05  & 1.00&   0.998  & 1.00 \\
LARS08  &  5.55$\pm$ 0.079  & 1.25  & 1.02&   0.999  & 1.00 \\
LARS09  &  7.59$\pm$ 0.10   & 1.61  & 1.06&   NA     & NA   \\
LARS10  &  3.45$\pm$ 0.062  & 4.57  & 1.36&   NA     & NA   \\
LARS11  &  4.73$\pm$ 0.082  & 1.35  & 1.04&   0.967  & 0.997\\
LARS12  &  3.99$\pm$ 0.075  & 1.29  & 1.03&   0.999  & 1.00 \\
LARS13  &  4.16$\pm$ 0.080  & NA    & NA  &   NA     & NA   \\
LARS14  &  4.05$\pm$ 0.084  & 1.32  & 1.03&   0.991  & 0.999\\
ELARS01 &  5.78$\pm$ 0.10   & 1.16  & 1.02&   0.999  & 1.00 \\
ELARS02 &  3.42$\pm$ 0.069  & 1.17  & 1.02&   0.997  & 1.00 \\
ELARS03 &  13.6$\pm$ 0.27   & 4.23  & 1.32&   0.945  & 0.994\\
ELARS04 &  14.0$\pm$ 0.23   & 1.86  & 1.09&   0.923  & 0.993\\
ELARS05 &  1.85$\pm$ 0.075  & 1.12  & 1.01&   0.829  & 0.984\\
ELARS06 &  8.55$\pm$ 0.17   & 2.61  & 1.16&   0.973  & 0.997\\
ELARS07 &  16.7$\pm$ 0.23   & 5.98  & 1.50&   1.00   & 1.00 \\
ELARS08 &  15.3$\pm$ 0.23   & 3.94  & 1.29&   0.766  & 0.977\\
ELARS09 &  42.1$\pm$ 0.52   & 10.2  & 1.92&   0.828  & 0.983\\
ELARS10 &  5.61$\pm$ 0.11   & 2.32  & 1.13&   0.970  & 0.997\\
ELARS11 &  39.9$\pm$ 0.27   & 7.52  & 1.65&   0.973  & 0.997\\
ELARS12 &  37.4$\pm$ 0.23   & NA    & NA  &   NA     & NA   \\
ELARS13 &  7.61$\pm$ 0.11   & 1.10  & 1.01&   1.00   & 1.00 \\
ELARS14 &  17.2$\pm$ 0.16   & 19.4  & 2.84&   NA     & NA   \\
ELARS15 &  6.21$\pm$ 0.096  & 2.74  & 1.17&   0.999  & 1.00 \\
ELARS16 &  2.41$\pm$ 0.048  & 4.76  & 1.38&   NA     & NA   \\
ELARS17 &  7.84$\pm$ 0.11   & 2.80  & 1.18&   0.927  & 0.994\\
ELARS18 &  7.41$\pm$ 0.094  & 5.27  & 1.43&   NA     & NA   \\
ELARS19 &  13.6$\pm$ 0.12   & 3.36  & 1.24&   1.00   & 1.00 \\
ELARS20 &  5.49$\pm$ 0.080  & 2.07  & 1.11&   0.937  & 0.994\\
ELARS21 &  5.34$\pm$ 0.077  & 2.47  & 1.15&   NA     & NA   \\
ELARS22 &  42.4$\pm$ 0.31   & 7.13  & 1.61&   0.976  & 0.998\\
ELARS23 &  4.58$\pm$ 0.073  & 2.64  & 1.16&   0.887  & 0.988\\
ELARS24 &  8.95$\pm$ 0.10   & 1.59  & 1.06&   0.998  & 1.00 \\
ELARS25 &  21.7$\pm$ 0.16   & 6.50  & 1.55&   0.785  & 0.979\\
ELARS26 &  11.0$\pm$ 0.11   & 3.21  & 1.22&   0.912  & 0.991\\
ELARS27 &  24.1$\pm$ 0.19   & 3.45  & 1.25&   0.982  & 0.998\\
ELARS28 &  7.15$\pm$ 0.095  & 3.95  & 1.29&   NA     & NA   \\
T1214   &  4.97$\pm$ 0.083  & 1.22  & 1.02&   1.00   & 1.00 \\
T1247   &  4.08$\pm$ 0.10   & 1.05  & 1.01&   0.974  & 0.997\\
J1156   &  1.11$\pm$ 0.022  & 1.05  & 1.01&   0.999  & 1.00 \\ 
\hline
\hline
\end{tabular}
\end{table*}
\clearpage
\section{Ly$\alpha$ morphology}\label{lya_morphology_appdix}
In this section, for each galaxy we represent the regions used to determine the morphological parameters used in this study ($\Delta$C, $\mathrm{(b/a)_{Ly\alpha}}$, $\mathrm{(b/a)_{FUV}}$, $\mathrm{(b/a)_I}$, and $\Delta$PA). These regions are determined on the non-binned images smoothed with the kernel size corresponding to a certain physical scale specified in Table. \ref{kernel_size_morph}. The blue contour represent the regions that are either brighter than $\mathrm{SB_{Ly\alpha}} = 1.5 \times 10^{39} \, \mathrm{erg/s/kpc^2}$ (except J1156, $\mathrm{SB_{Ly\alpha}} = 5.0 \times 10^{39} \, \mathrm{erg/s/kpc^2}$), or the SFRD is higher than 0.1 $\mathrm{M_\odot \, yr^{-1} \,  kpc^{-2}}$. The green, and red contours represent the regions that are brighter than $\mathrm{SB_{FUV}} = 2.5 \times 10^{37} \, \mathrm{erg/s/kpc^2/\si{\angstrom}}$, $\mathrm{SB_{I\, band}} = 1.5 \times 10^{37} \, \mathrm{erg/s/kpc^2/\si{\angstrom}}$ (The I band SB threshold used for ELARS01 is different and is equal to $\mathrm{SB_{i\, band}} = 1. \times 10^{38} \, \mathrm{erg/s/kpc^2/\si{\angstrom}}$) all displayed on the Ly$\alpha$ images. The blue, green, and red crosses represent the measured centroids (non-weighted first image moment) in the specified regions for Ly$\alpha$, FUV, and I band, respectively. The PA determined in each region is also displayed with blue dashed, green dotted, and red dash-dotted lines for Ly$\alpha$, FUV, and I band, respectively. The measured axis ratios determined in each region are given in the lower left part of the panels with blue, green, and red for Ly$\alpha$, FUV, and I band, respectively. Finally, the centroid shift($\Delta$C) between Ly$\alpha$ and FUV, and difference between the measured PA of Ly$\alpha$, and FUV ($\Delta$PA) are given on the lower right side of the each panel.
\begin{figure*}[t!]

 \centering 
    \includegraphics[width=\textwidth] {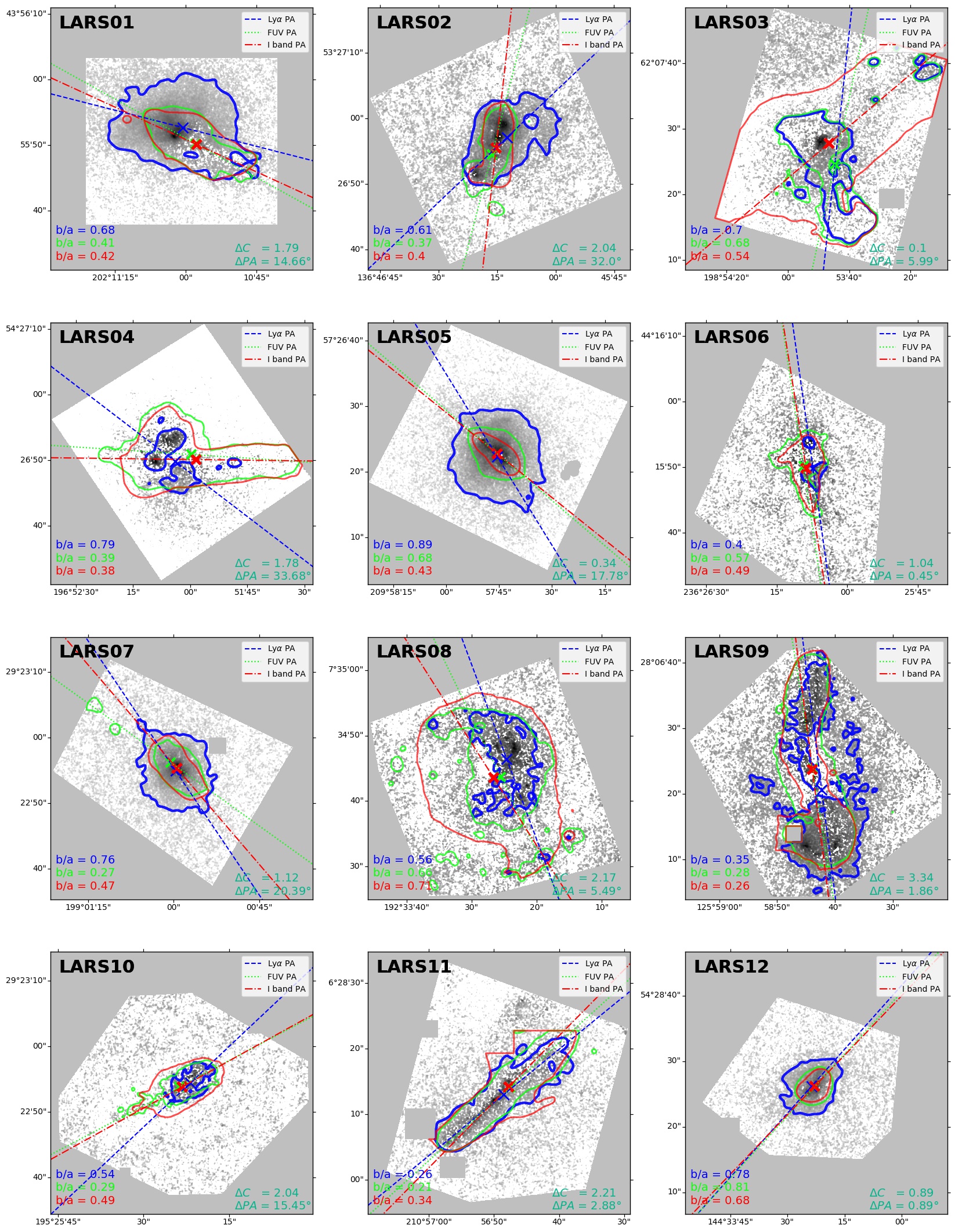}
\caption{
        Same as Fig. \ref{morph_paper_example}, but for LARS01 - LARS12.
        }
\end{figure*}
\begin{figure*}[t!]

 \centering 
    \includegraphics[width=\textwidth] {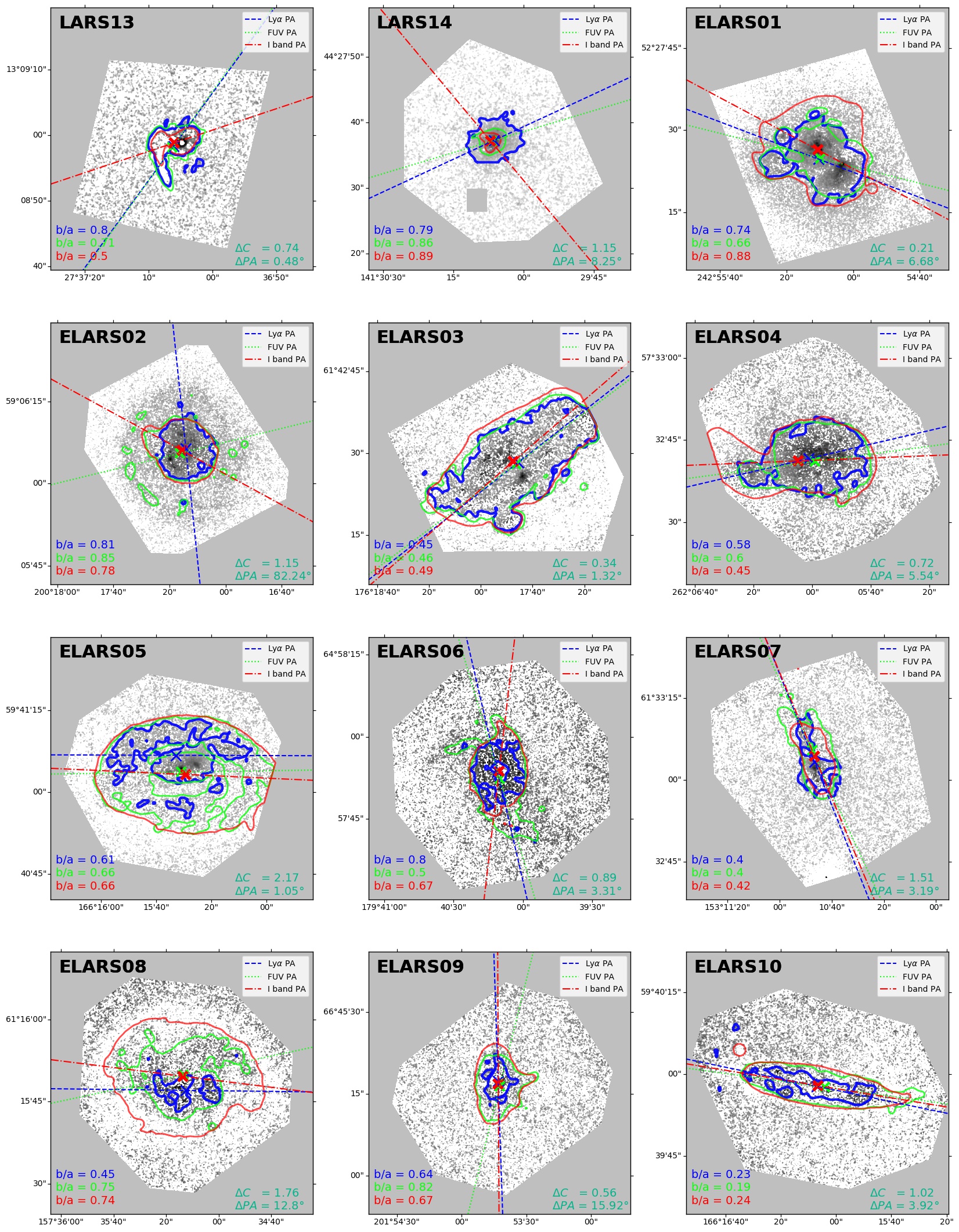}

\caption{
        Same as Fig. \ref{morph_paper_example}, but for LARS13, LARS14, ELARS01 - ELARS10.
        }

\end{figure*} 
\begin{figure*}[t!]

 \centering 
    \includegraphics[width=\textwidth] {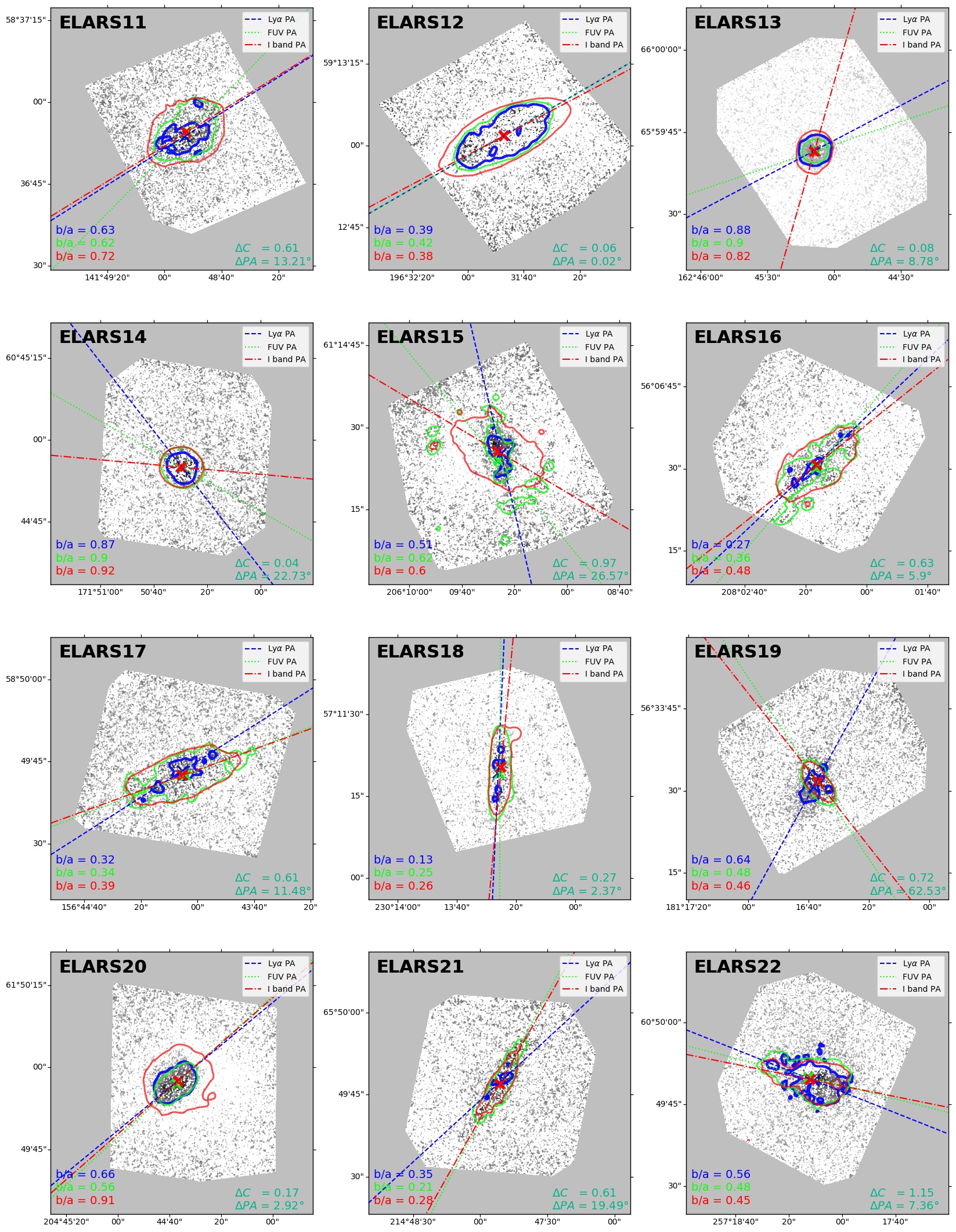}
    
\caption{
        Same as Fig. \ref{morph_paper_example}, but for ELARS11, ELARS22.
        }

\end{figure*} 
\begin{figure*}[t!]

 \centering 
    \includegraphics[width=\textwidth] {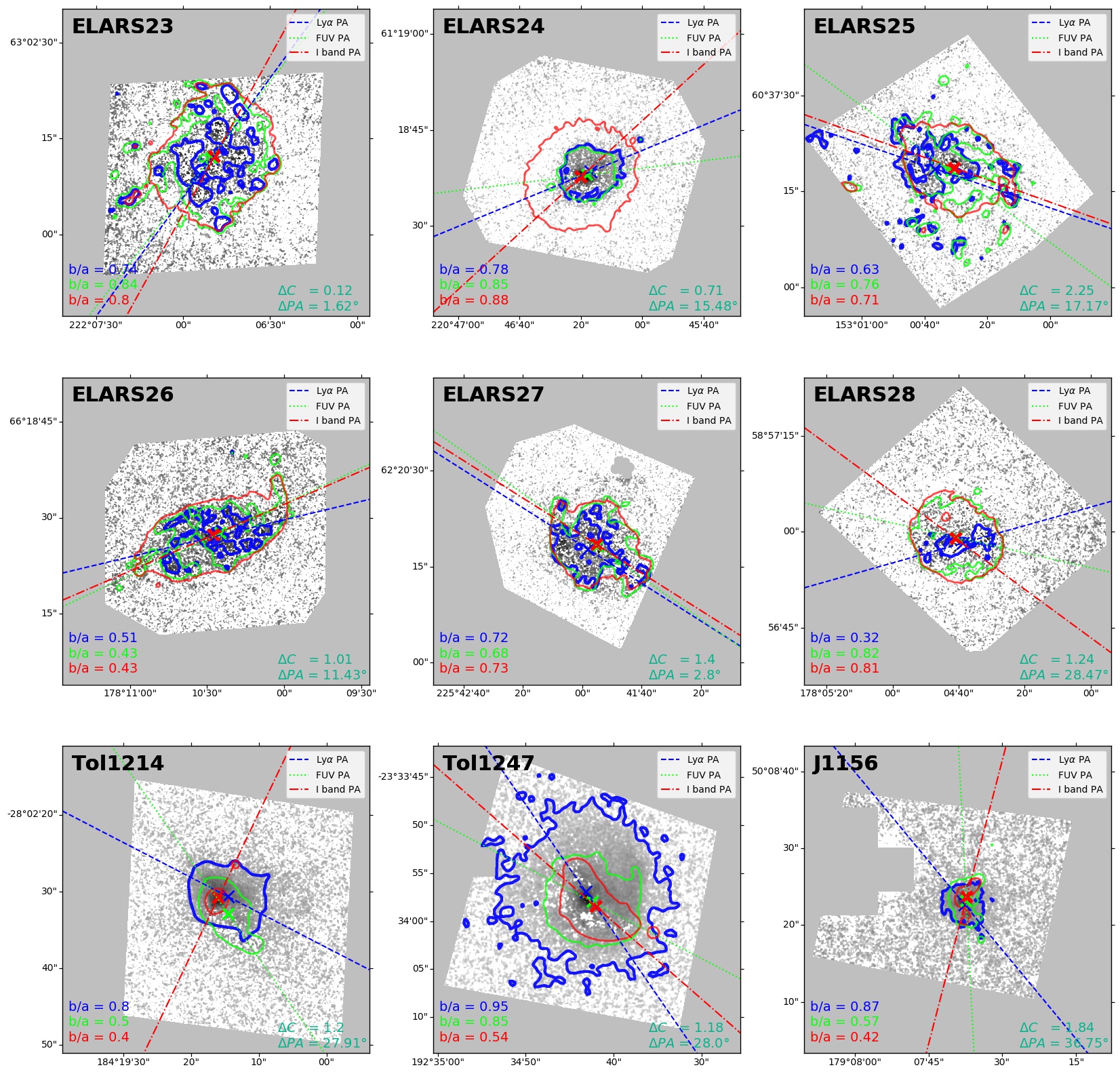}

\caption{
        Same as Fig. \ref{morph_paper_example}, but for ELARS23 - ELARS28, Tol1214, Tol1247, and J1156. The Ly$\alpha$ SB limits used for J1156 is different from the rest of the sample and is equal to $5.0 \times 10^{39} \, \mathrm{erg/s/kpc^2}$.
        }

\end{figure*} 
\clearpage
\section{Extent of the halo through isophotal approach }\label{r_iso_appx}
In this section, we present the faint Ly$\alpha$ isophotes used to study the extent of the Ly$\alpha$ halo through the isophotal approach for each galaxy. In each figure, in the top panels the region within the isophotal levels of even multipliers (1, 2, 3, 4) of the faintest limit ($1.5 \times 10^{39} \, \mathrm{erg/s/kpc^2}$, except J1156 $5.0 \times 10^{39} \, \mathrm{erg/s/kpc^2}$).
are denoted by red, orange, green and blue contours, all displayed on the Ly$\alpha$ maps of the galaxies. The bottom panels show the data points corresponding to these isophotes displayed with the same colour, and a single exponential fit (Eq. \ref{single_exp_iso}) to the points. The fitted scale lengths (and the measured error bar determined from the fit) are given in the lower left corner. The blue diamond represents the measured Ly$\alpha$ SB in the binned images at the inner most region (usually within r = 0.1 kpc from the brightest FUV point).
\begin{figure*}[t!]

 \centering 
    \includegraphics[width=\textwidth] {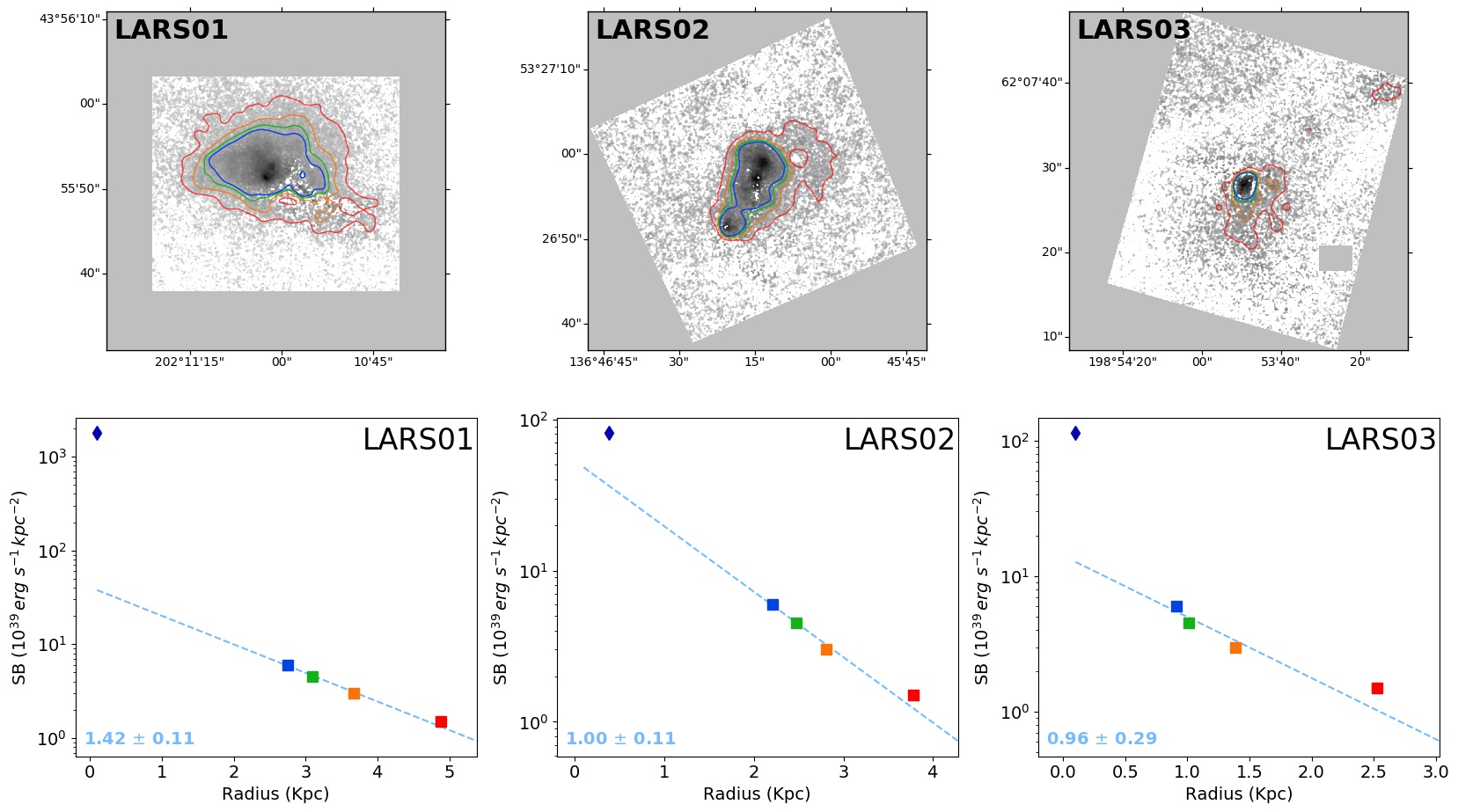}
    \includegraphics[width=\textwidth] {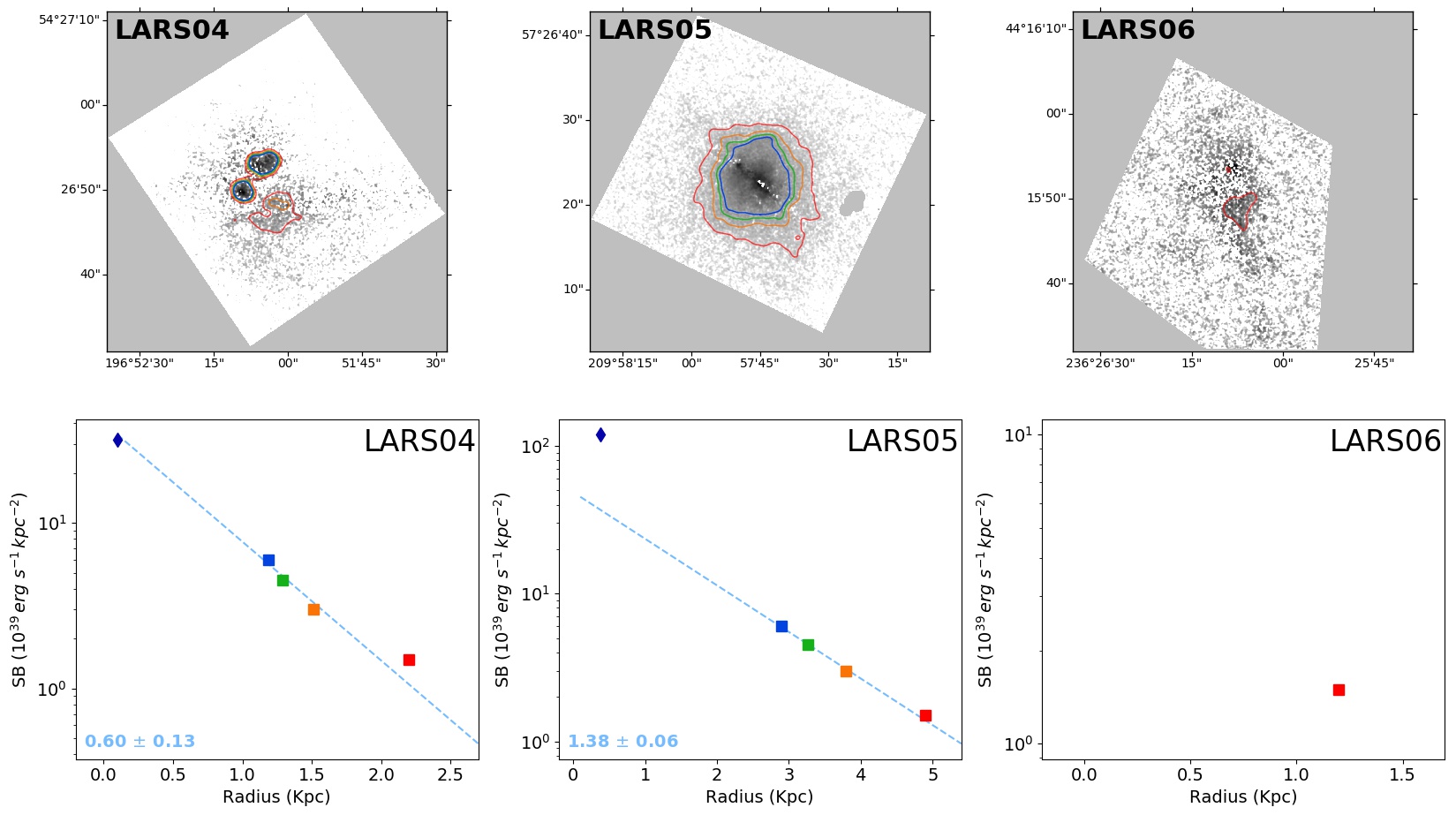}

\caption{
        Same as Fig. \ref{r_iso_morph_fig_main_txt}, but for LARS01 - LARS06.
        }
\end{figure*}
\begin{figure*}[t!]

 \centering 
    \includegraphics[width=\textwidth] {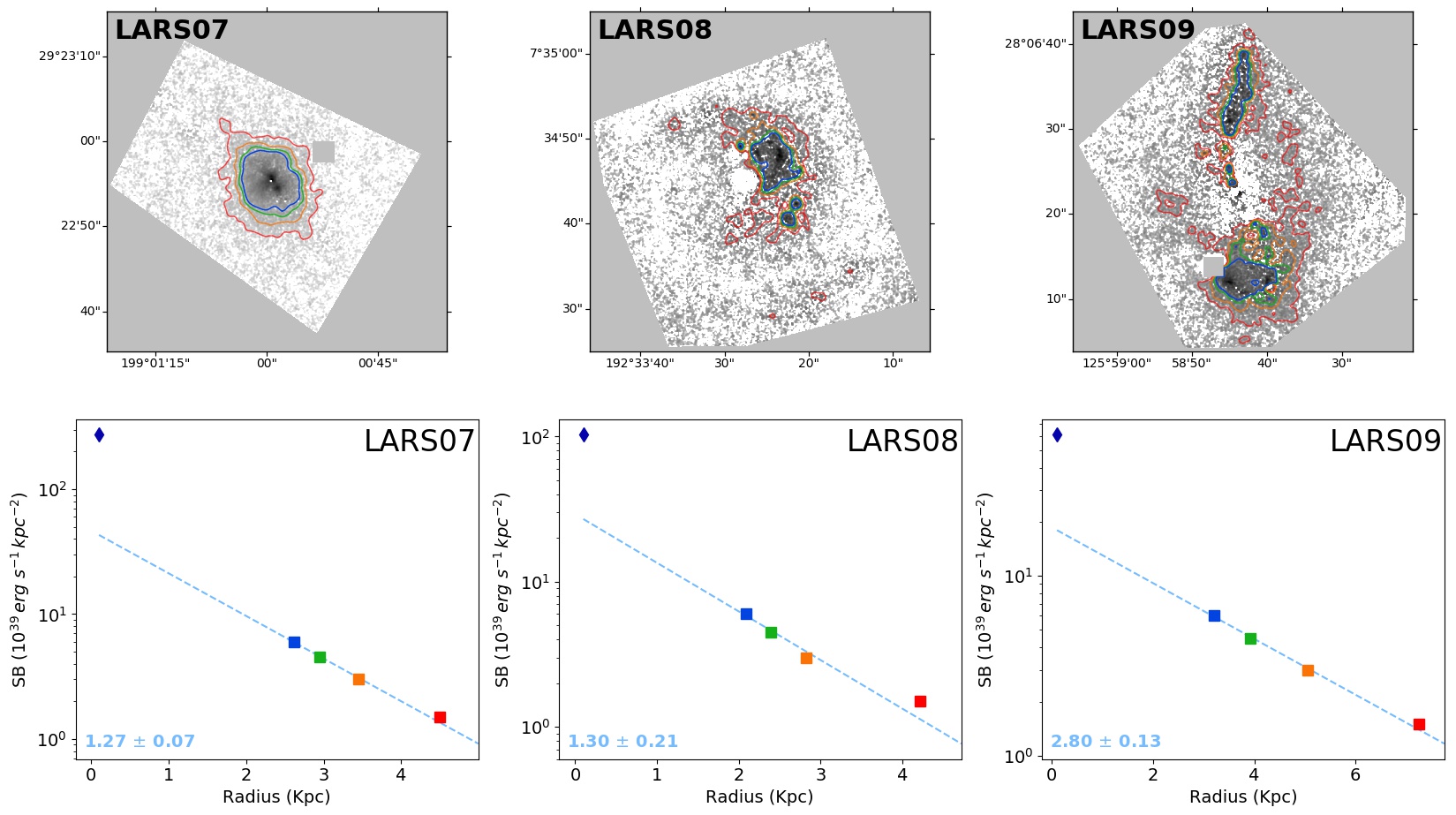}
    \includegraphics[width=\textwidth] {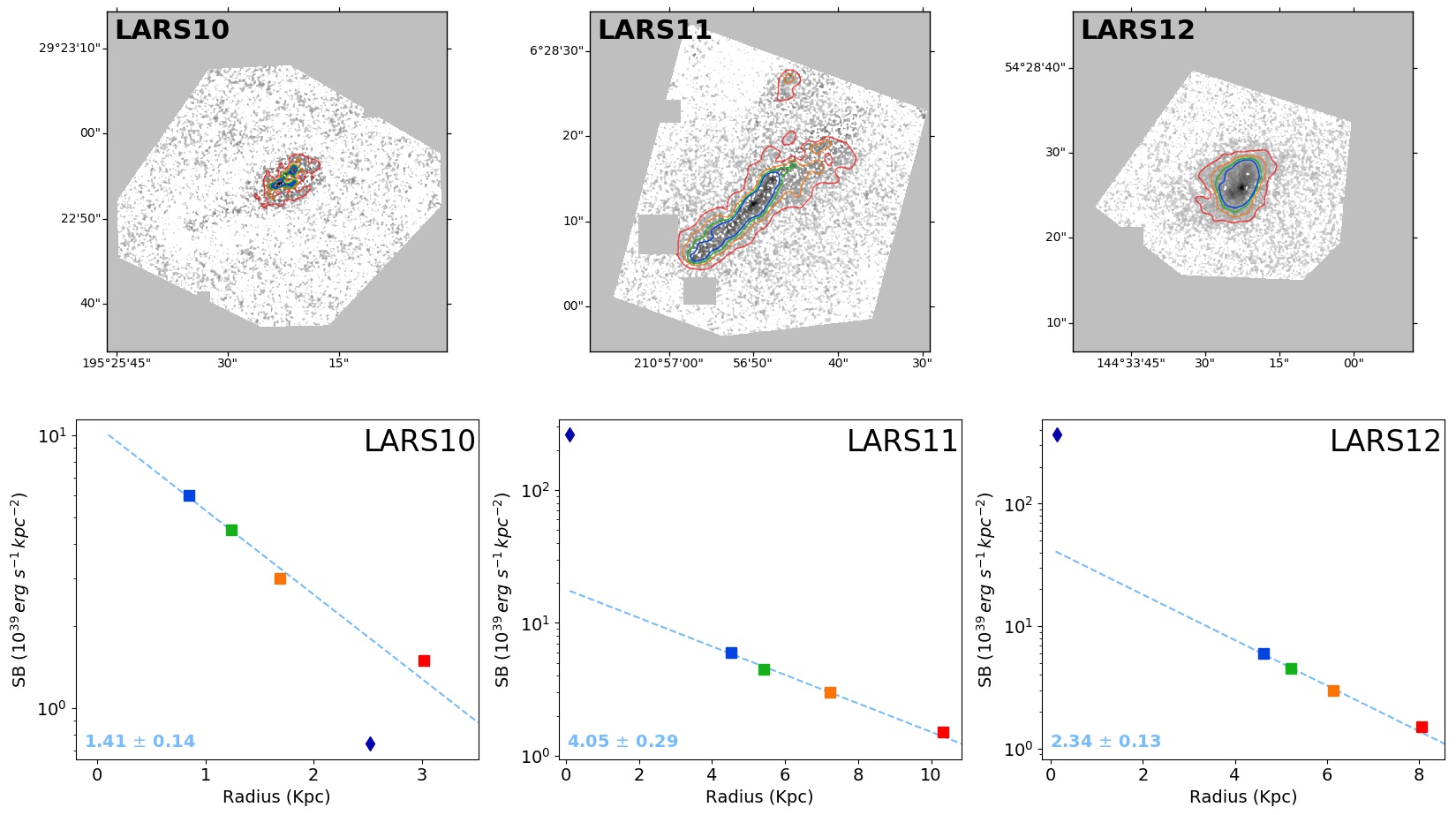}

\caption{
        Same as Fig. \ref{r_iso_morph_fig_main_txt}, but for LARS07 - LARS12.
        }

\end{figure*} 
\begin{figure*}[t!]

 \centering 
    \includegraphics[width=\textwidth] {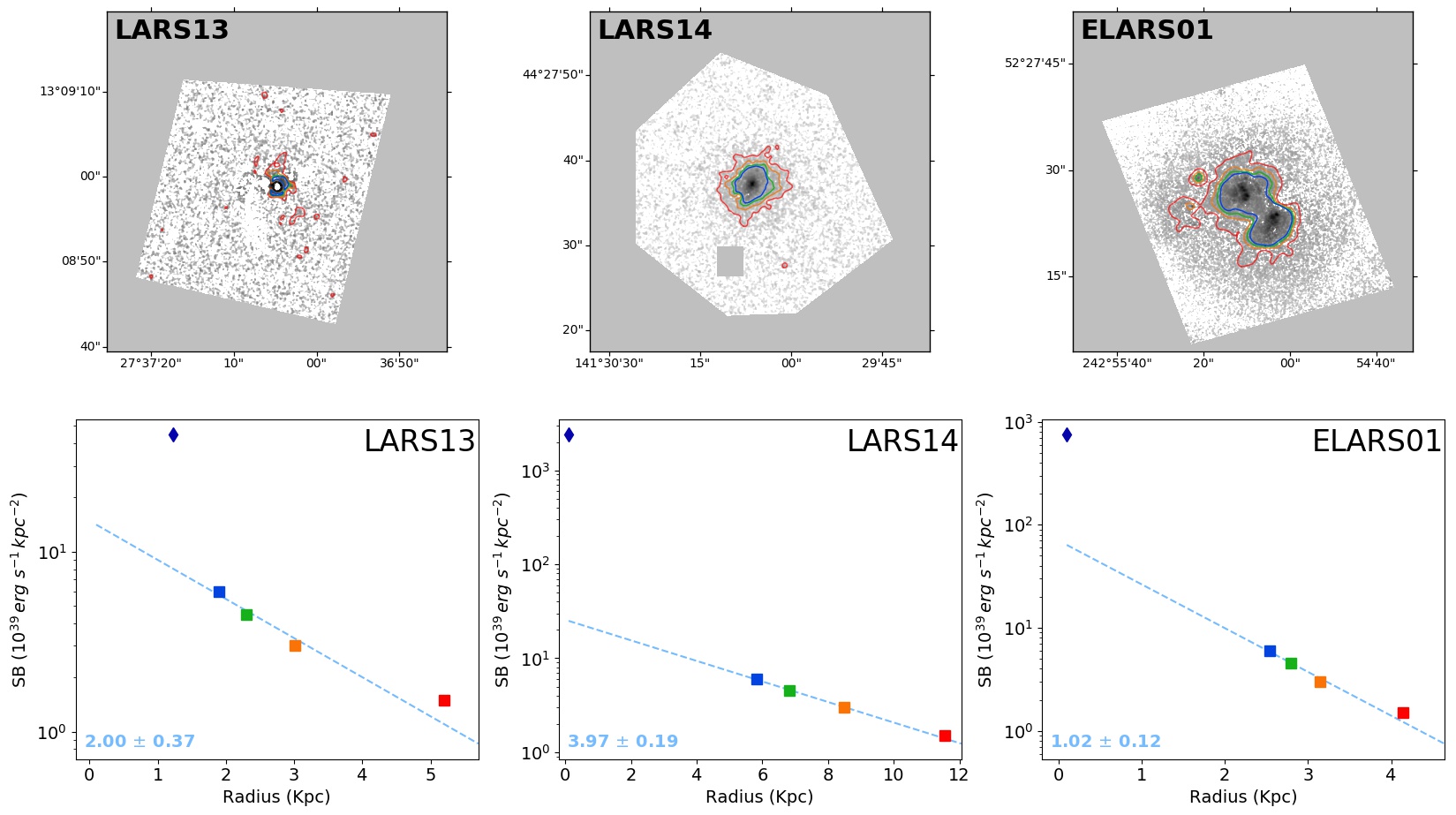}
    \includegraphics[width=\textwidth] {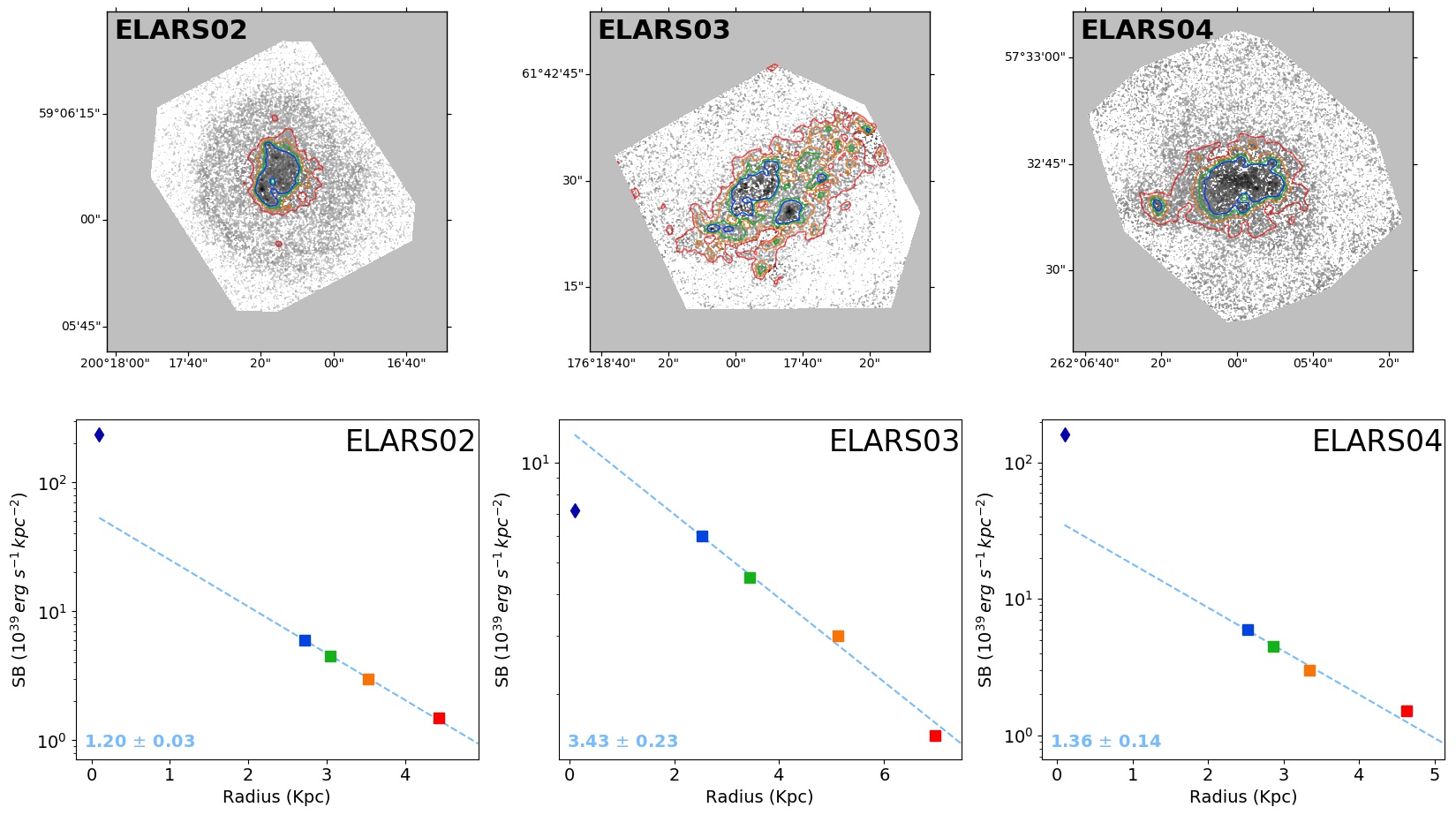}

\caption{
        Same as Fig. \ref{r_iso_morph_fig_main_txt}, but for LARS13, LARS14 and ELARS01- ELARS04.
        }

\end{figure*} 
\begin{figure*}[t!]

 \centering 
    \includegraphics[width=\textwidth] {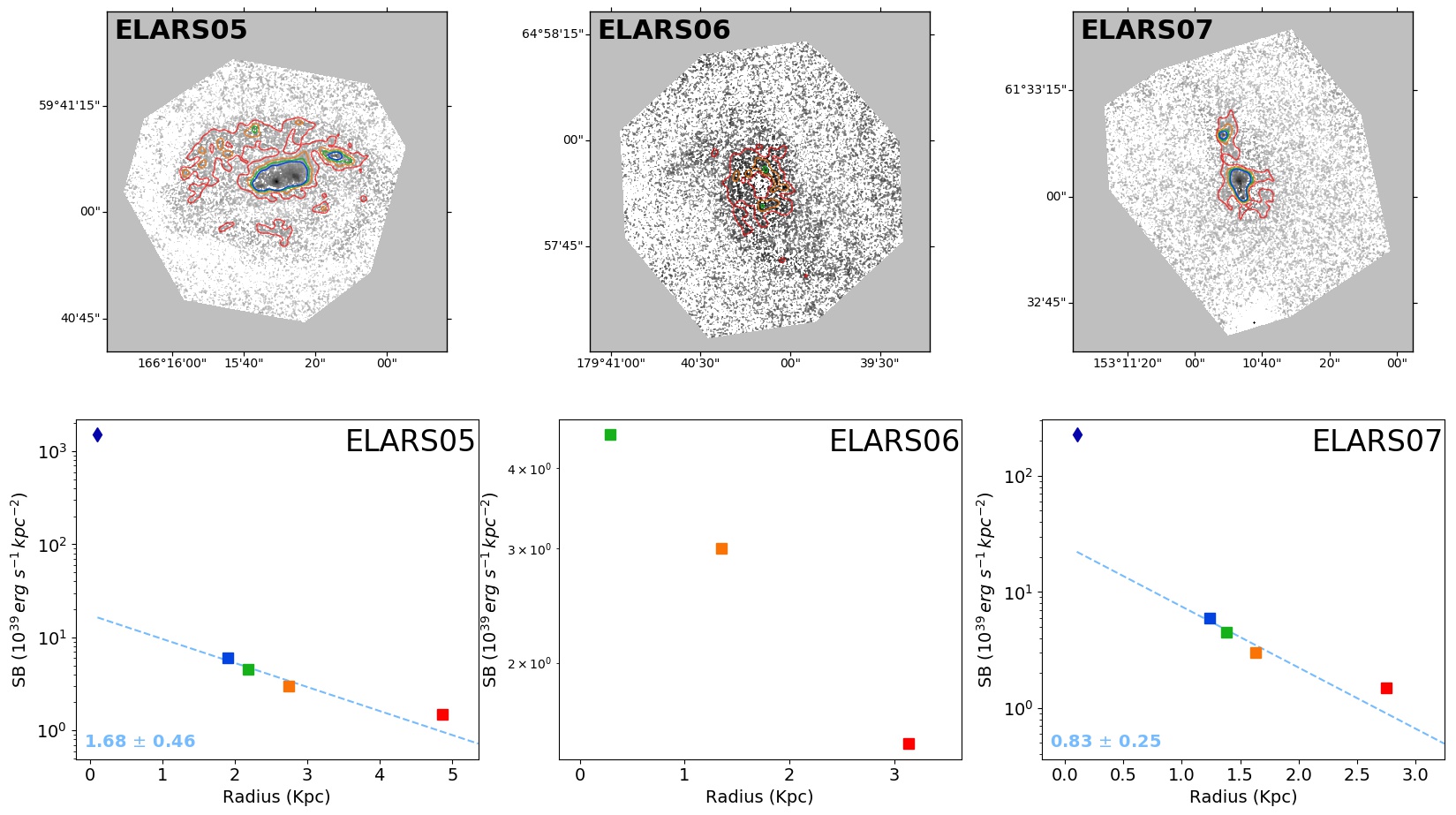}
    \includegraphics[width=\textwidth] {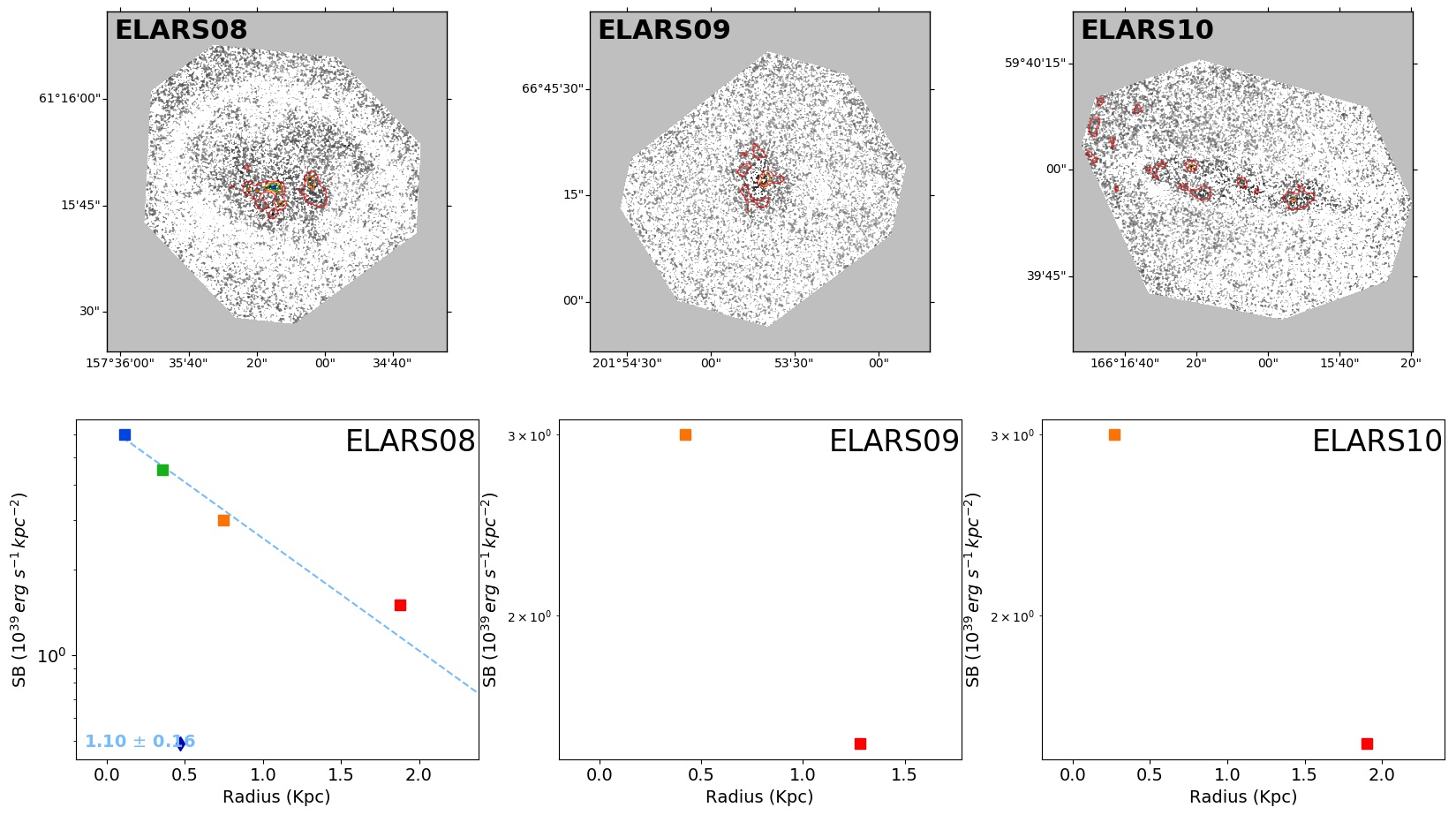}

\caption{
        Same as Fig. \ref{r_iso_morph_fig_main_txt}, but for ELARS05 - ELARS10.
        }

\end{figure*} 
\begin{figure*}[t!]

 \centering 
    \includegraphics[width=\textwidth] {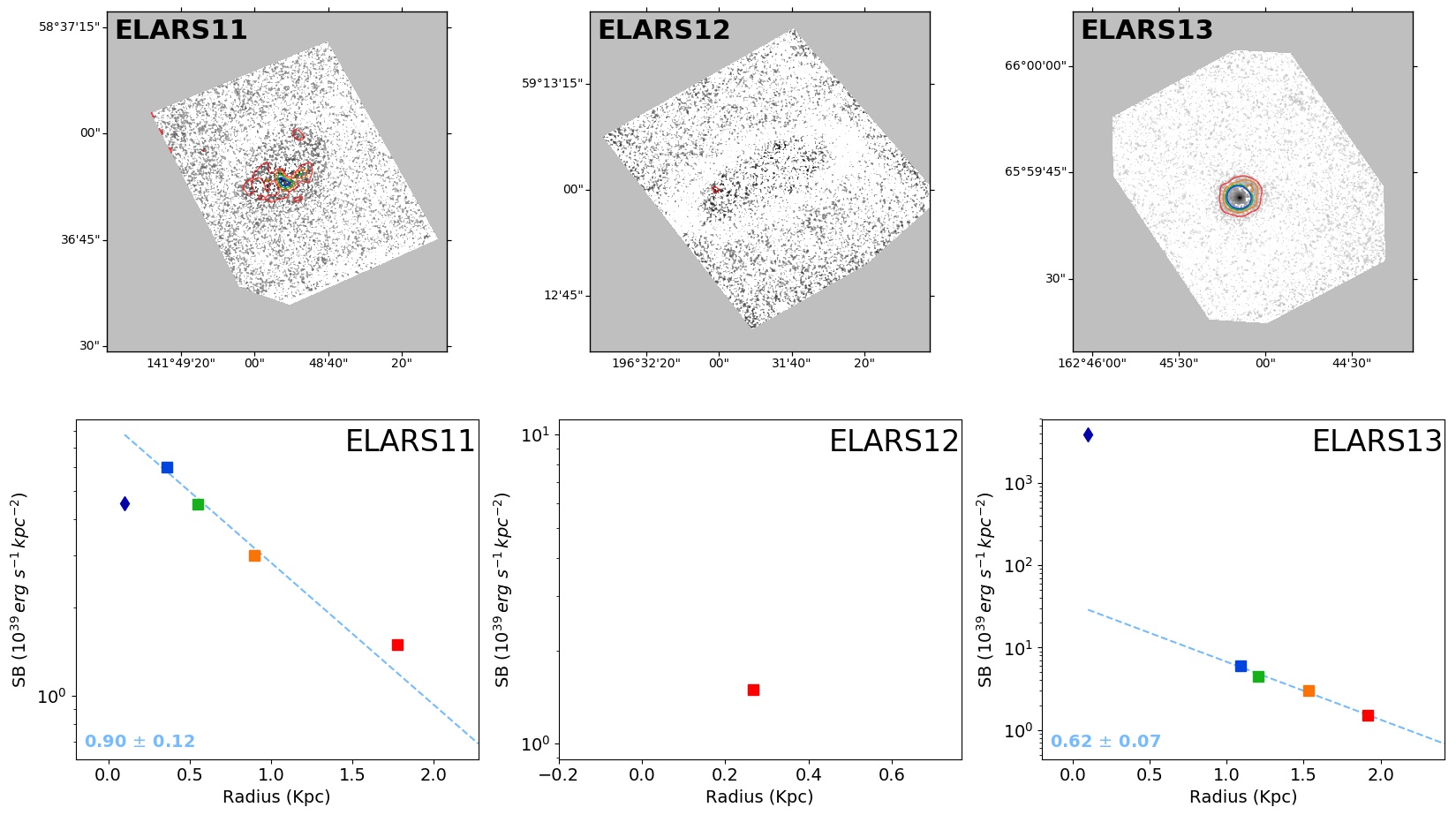}
    \includegraphics[width=\textwidth] {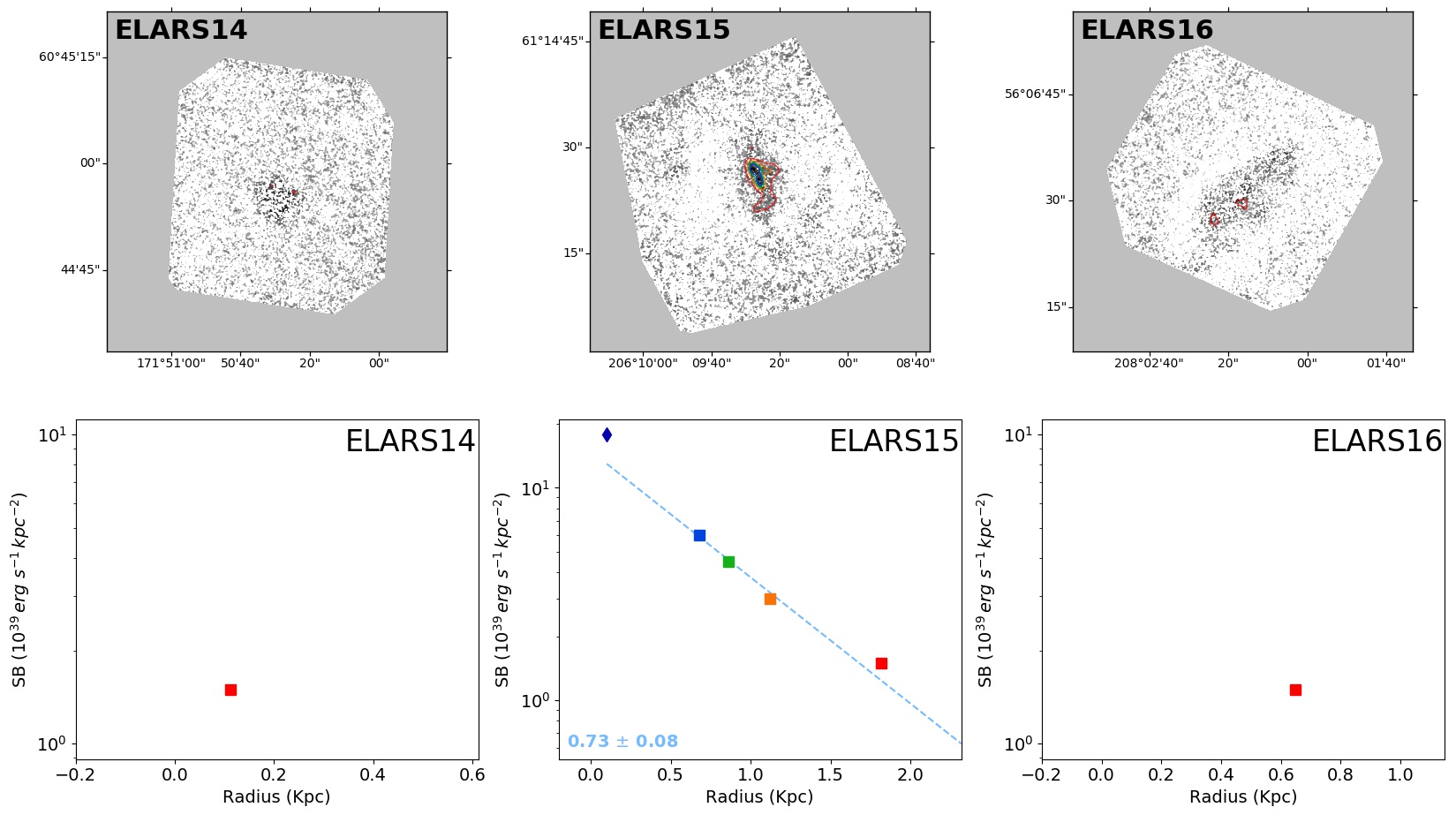}

\caption{
        Same as Fig. \ref{r_iso_morph_fig_main_txt}, but for ELARS11 - ELARS16.
        }

\end{figure*} 
\begin{figure*}[t!]

 \centering 
    \includegraphics[width=\textwidth] {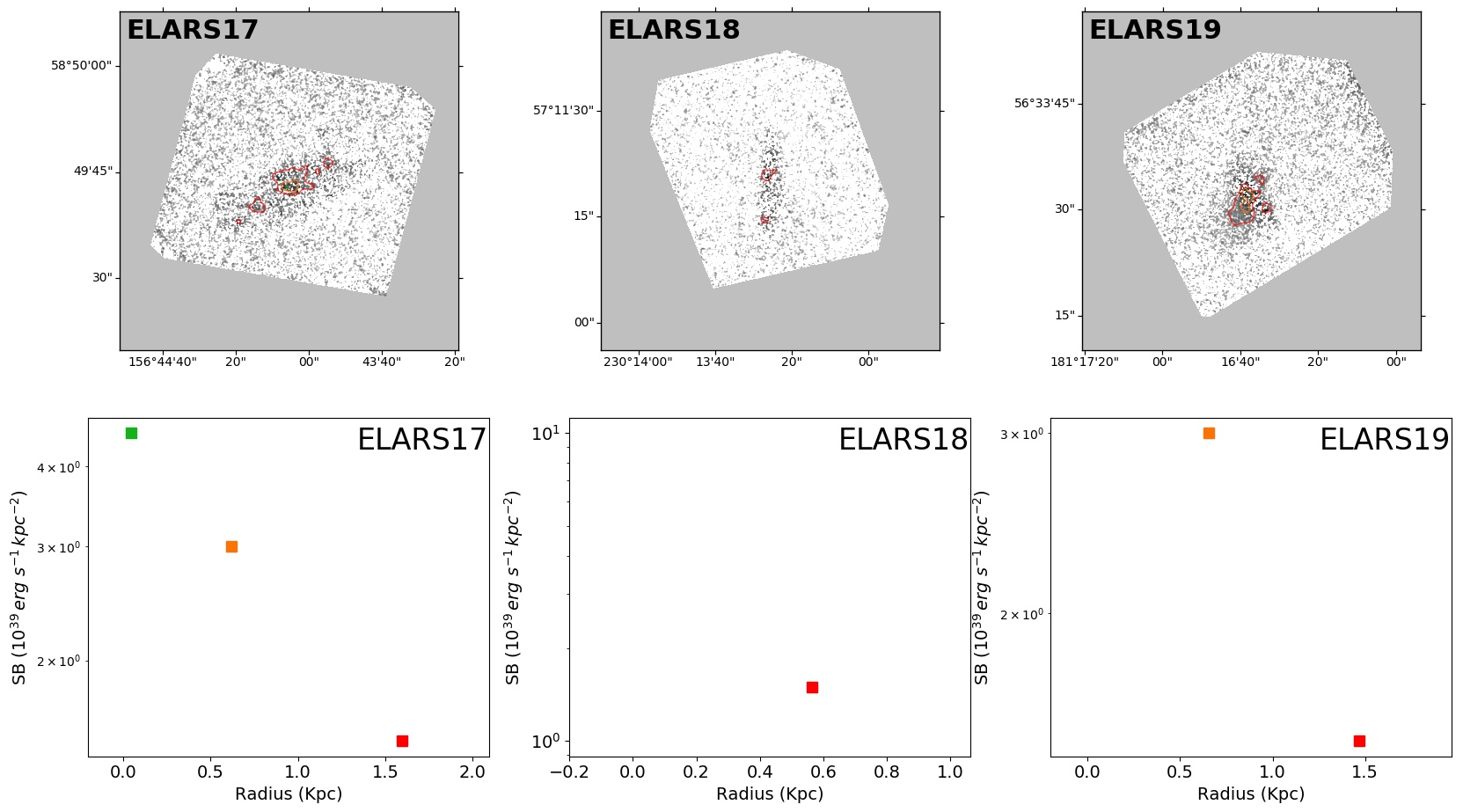}
    \includegraphics[width=\textwidth] {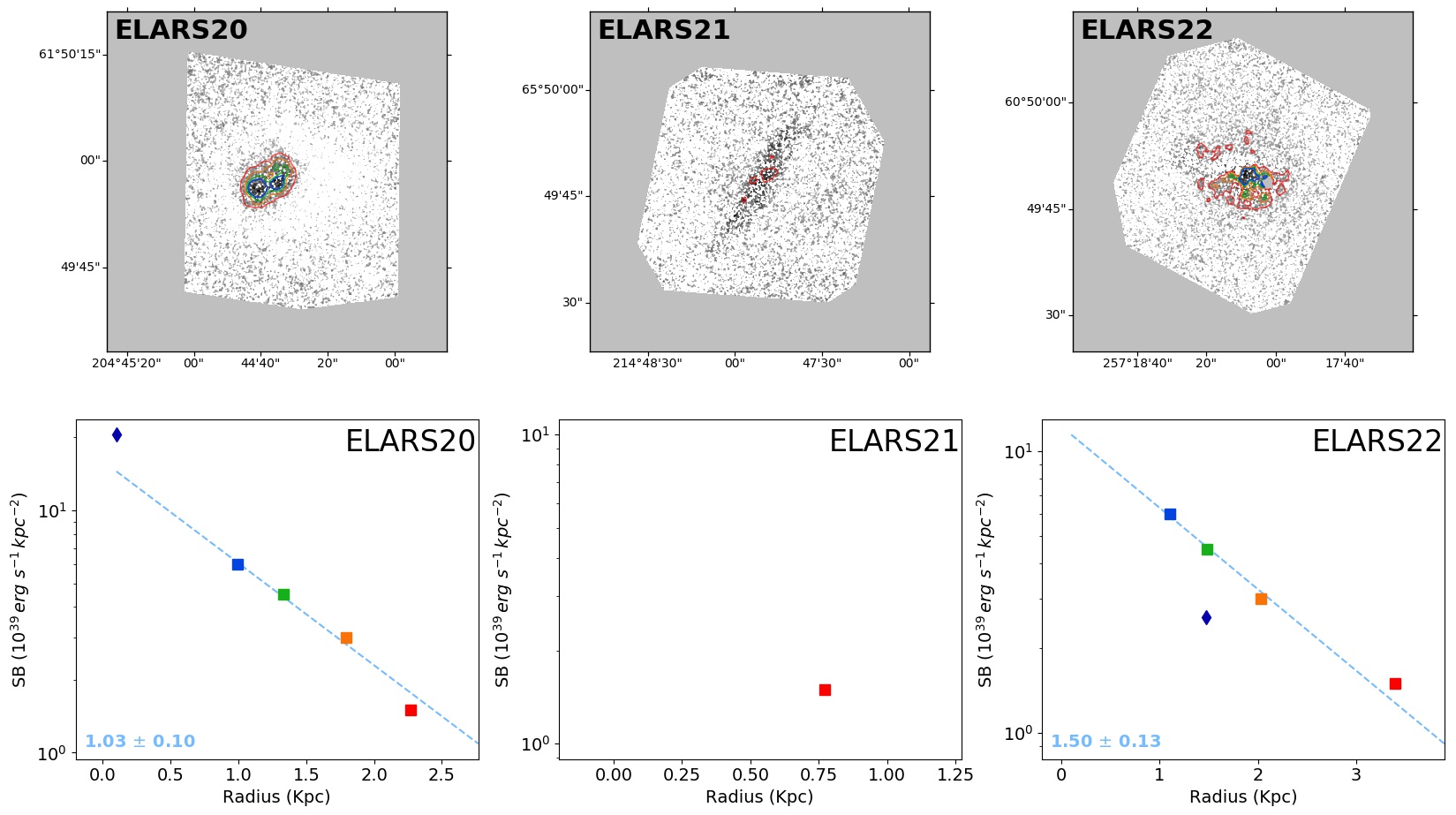}

\caption{
        Same as Fig. \ref{r_iso_morph_fig_main_txt}, but for ELARS17 - ELARS22.
        }

\end{figure*} 
\begin{figure*}[t!]

 \centering 
    \includegraphics[width=\textwidth] {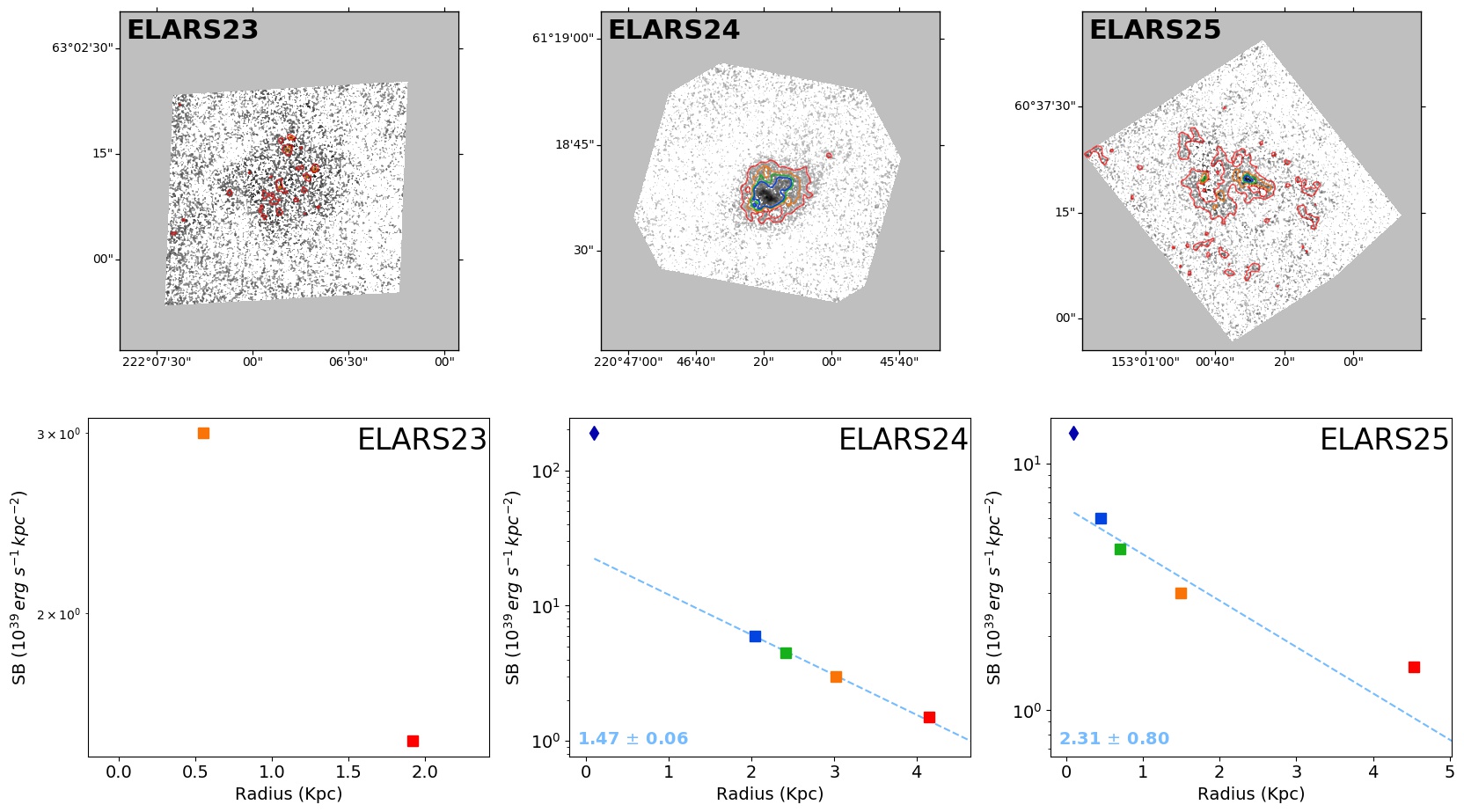}
    \includegraphics[width=\textwidth] {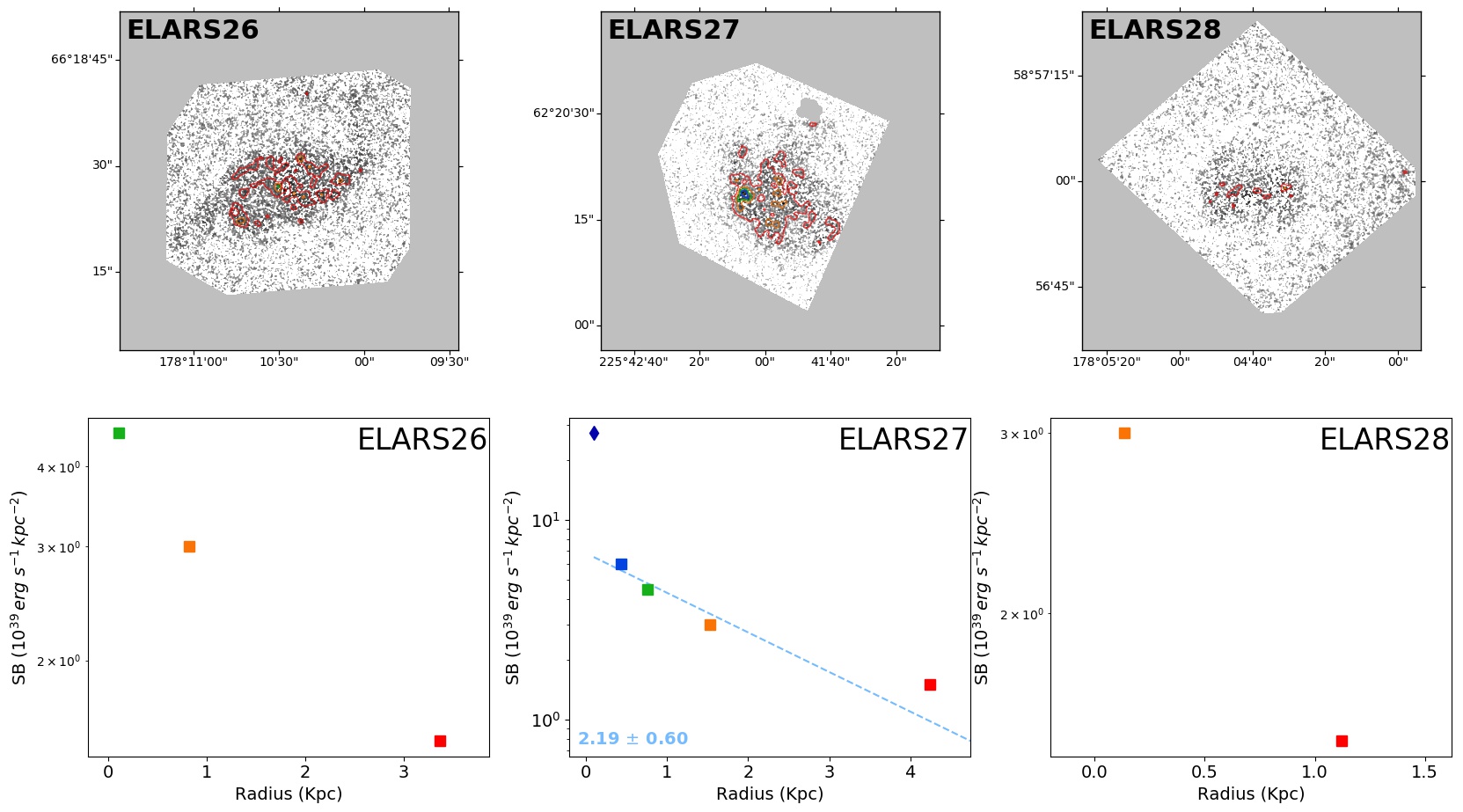}

\caption{
        Same as Fig. \ref{r_iso_morph_fig_main_txt}, but for ELARS23 - ELARS28.
        }

\end{figure*} 
\begin{figure*}[t!]

 \centering 
    \includegraphics[width=\textwidth] {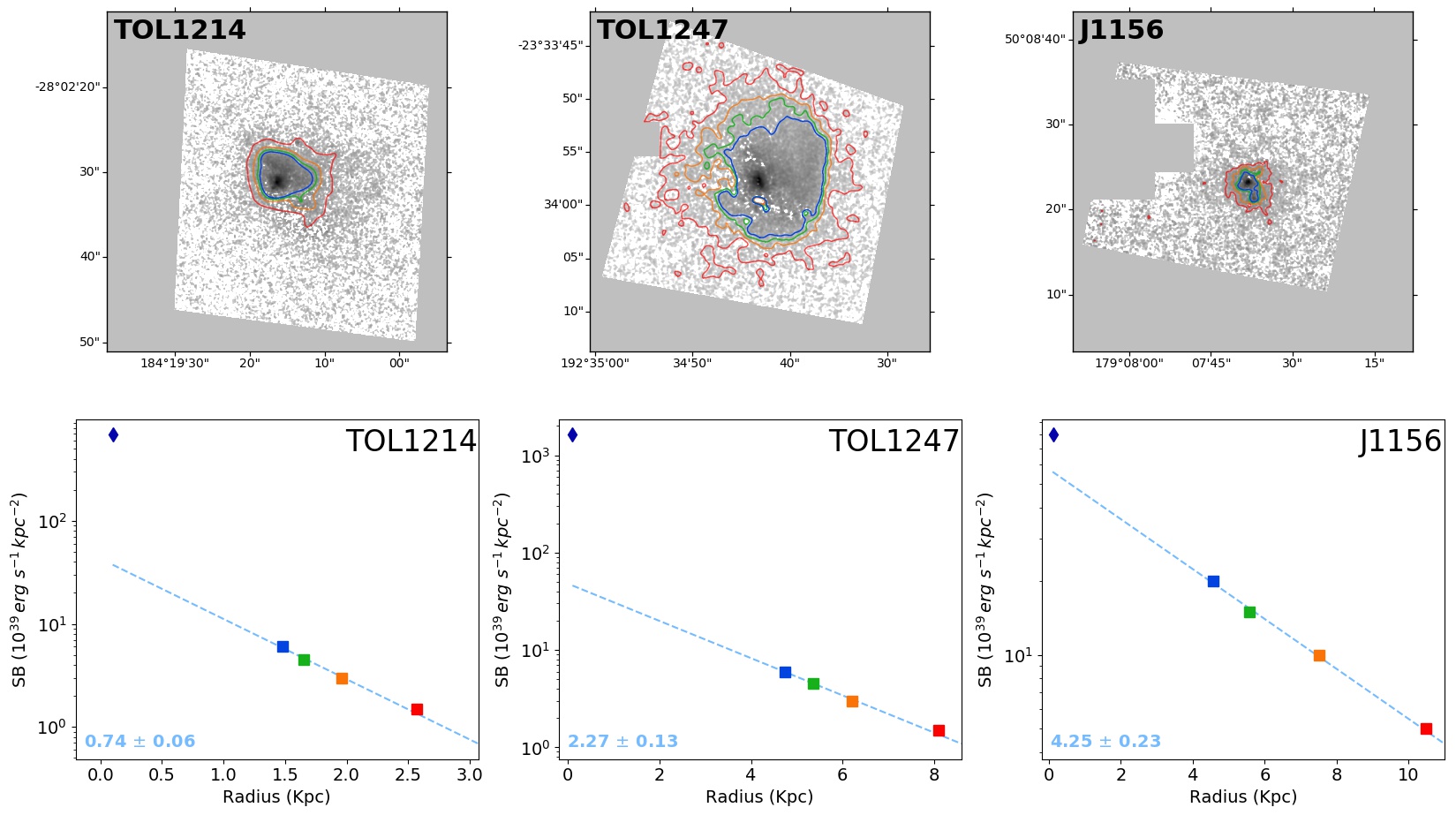}

\caption{
        Same as Fig. \ref{r_iso_morph_fig_main_txt}, but for TOL1214, TOL1247, and J1156. The Ly$\alpha$ FEI used for J1156 is different from the rest of the sample and is equal to $5.0 \times 10^{39} \, \mathrm{erg/s/kpc^2}$). 
        }

\end{figure*} 
\clearpage
\section{Star forming properties}\label{fuv_morphology_appdix}
In this section, we represent the regions where the SFRD > 0.01 $\mathrm{M_\odot \, yr^{-1} \,  kpc^{-2}}$ used for studying the SF properties of the host galaxies in our sample. We used two parameters for studying the SF properties of the galaxies, the area where the  SFRD > 0.01 $\mathrm{M_\odot \, yr^{-1} \,  kpc^{-2}}$ presented by the equivalent radius $\mathrm{r_{SFRD>0.01}}$ (the radius of a circle with the same area), and the average FUV SB ($\mathrm{\overline{FUV_{SB}}}$) by measuring the FUV SB within these regions and divided by the area. The measured $\mathrm{\overline{FUV_{SB}}}$ are given on the lower left side of each panel.
\begin{figure*}[t!]

 \centering 
    \includegraphics[width=\textwidth] {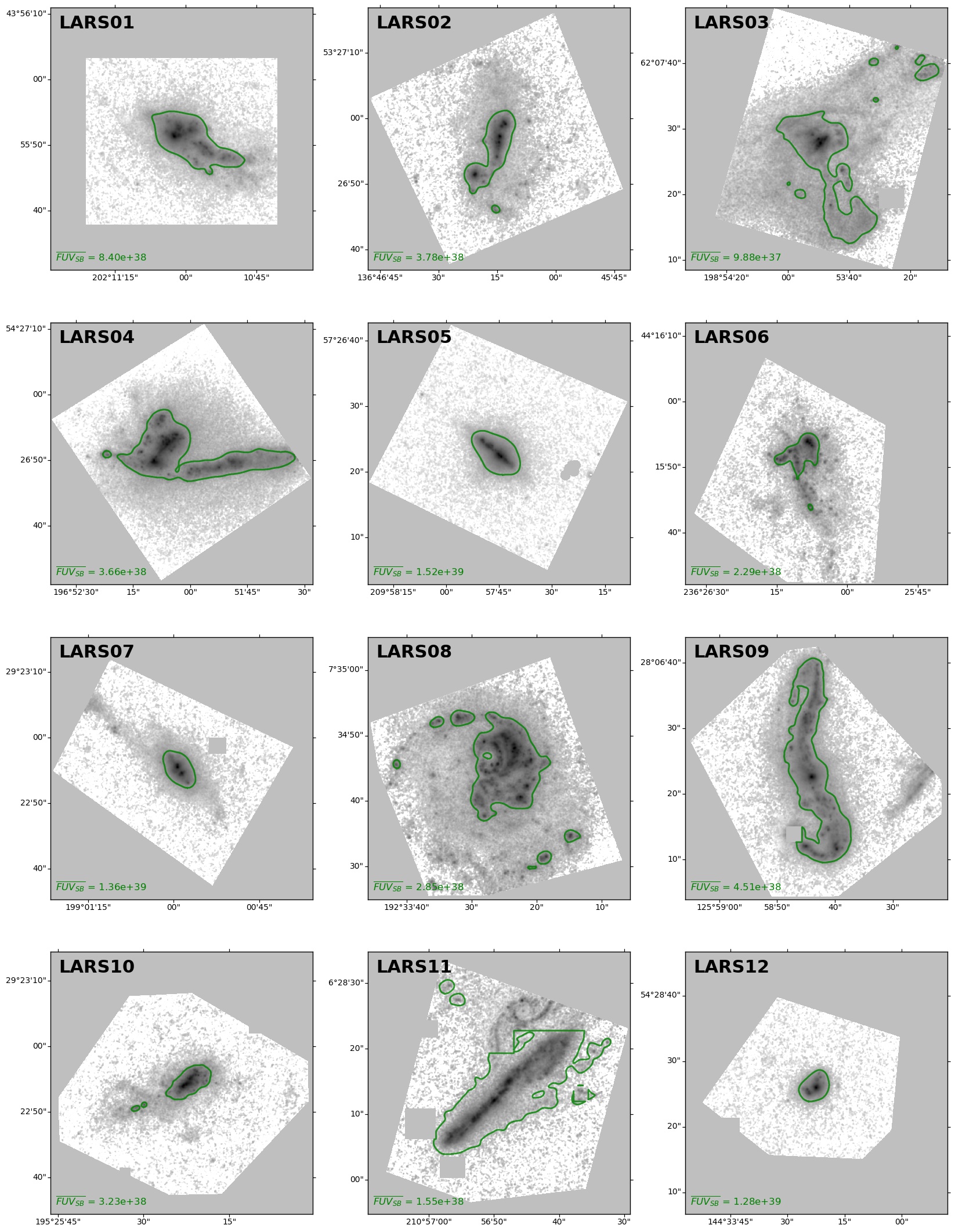}
\caption{
        Same as Fig. \ref{fuv_morph}, but for LARS01 - LARS12.
        }
\end{figure*}
\begin{figure*}[t!]

 \centering 
    \includegraphics[width=\textwidth] {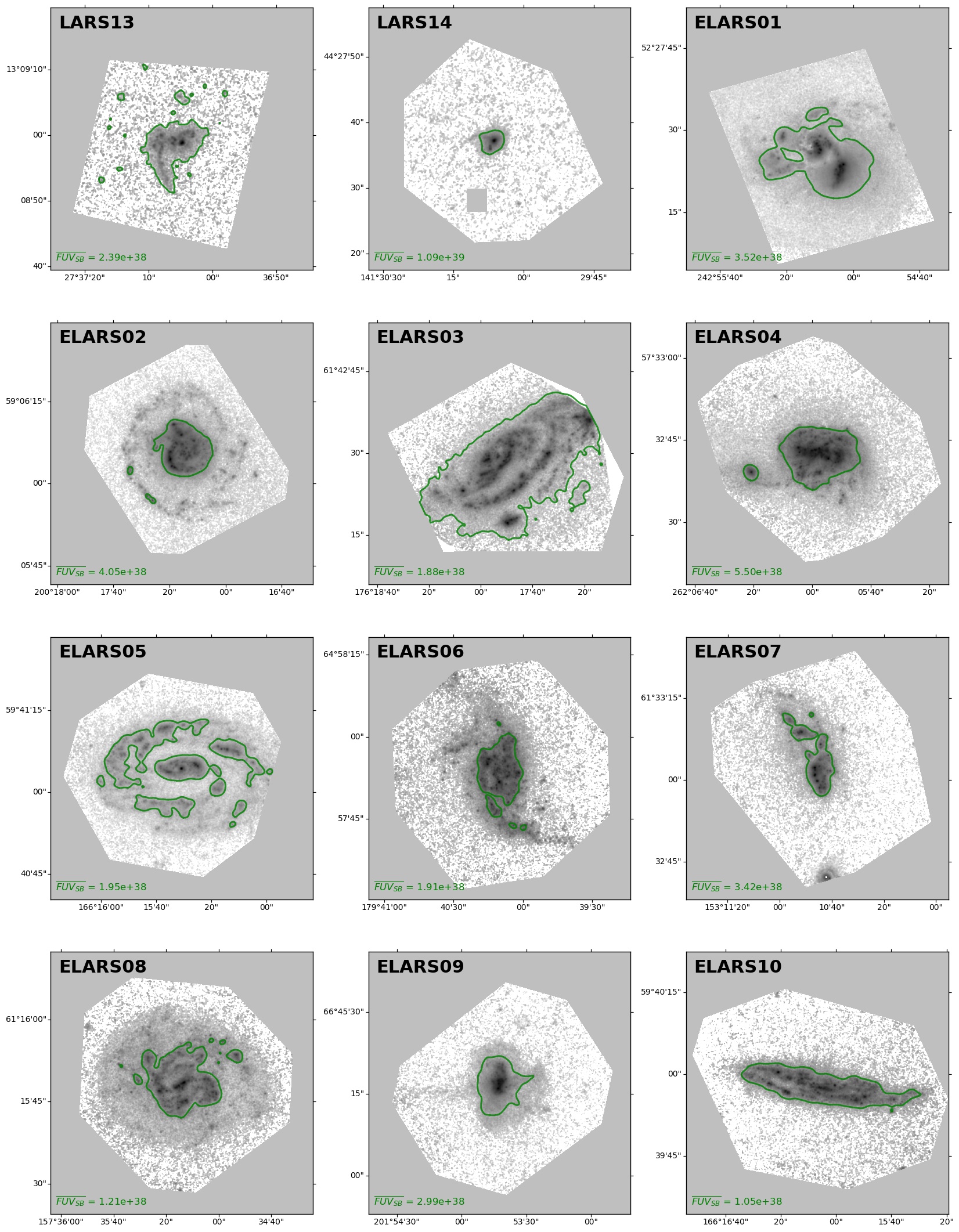}

\caption{
        Same as Fig. \ref{fuv_morph}, but for LARS13, LARS14, ELARS01 - ELARS10.
        }

\end{figure*} 
\begin{figure*}[t!]

 \centering 
    \includegraphics[width=\textwidth] {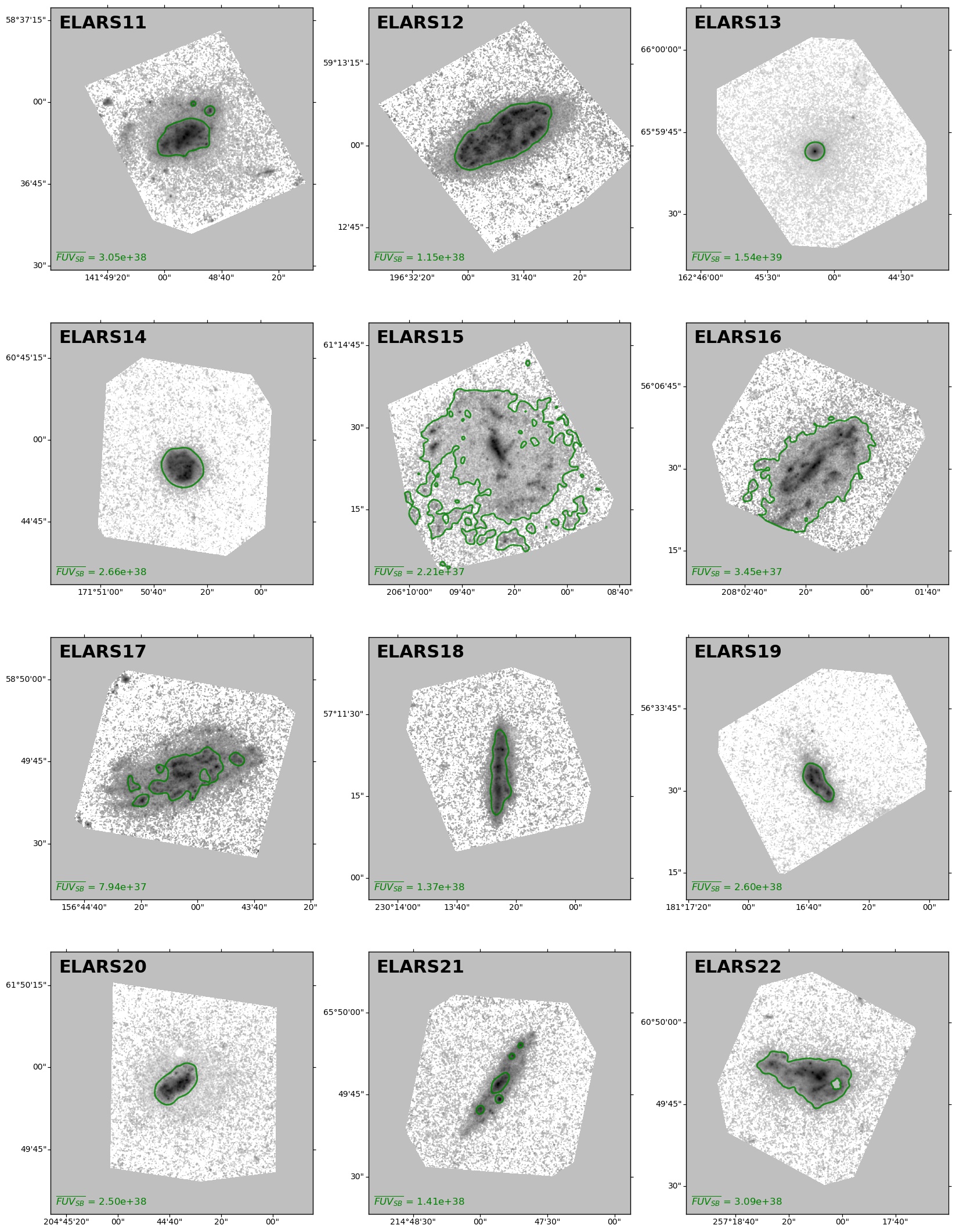}
    
\caption{
        Same as Fig. \ref{fuv_morph}, but for ELARS11, ELARS22.
        }

\end{figure*} 
\begin{figure*}[t!]

 \centering 
    \includegraphics[width=\textwidth] {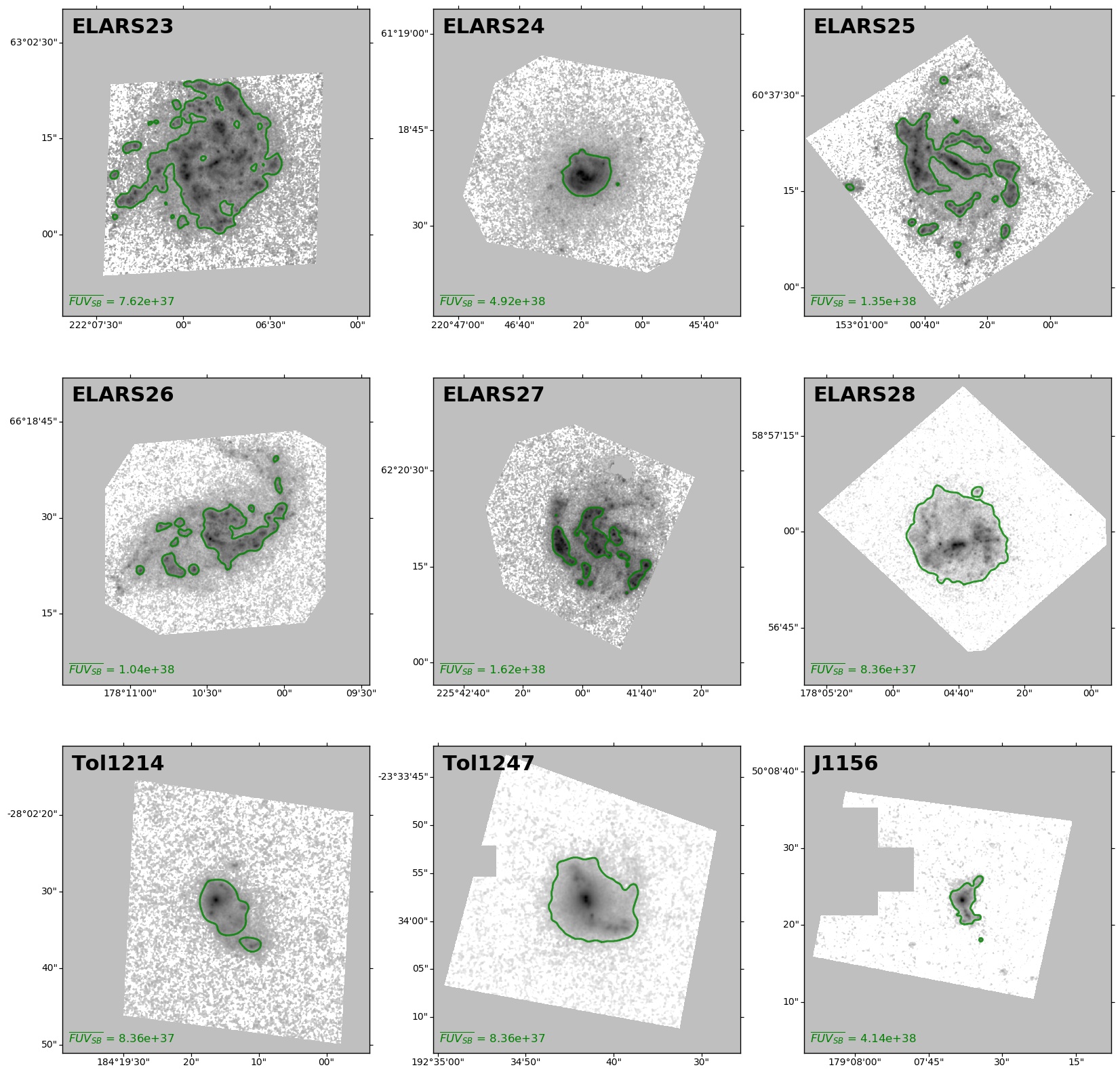}

\caption{
        Same as Fig. \ref{fuv_morph}, but for ELARS23 - ELARS28, Tol1214, Tol1247, and J1156.
        }

\end{figure*} 
\clearpage
\section{Ly$\alpha$ observables versus Ly$\alpha$ morphology}\label{lya_obs_appx}
In this section, we present how the Ly$\alpha$ observables: Ly$\alpha$ luminosity ($\mathrm{L_{Ly\alpha}}$), Ly$\alpha$ escape fraction ($\mathrm{f_{esc}}$), and Ly$\alpha$ equivalent width ($\mathrm{EW_{Ly\alpha}}$), vary with the quantities used to study the Ly$\alpha$ halo properties or Ly$\alpha$ morphology of the galaxies: Ly$\alpha$ halo scale length $\mathrm{r_{sc}^{Ly\alpha}}$, Ly$\alpha$ halo fraction (HF), morphological parameters, such as: axis ratio ($\mathrm{(b/a)_{Ly\alpha}}$), $\Delta$C, $\Delta$PA.
\begin{figure*}[t!]
 \centering 
    \includegraphics[width=\textwidth]{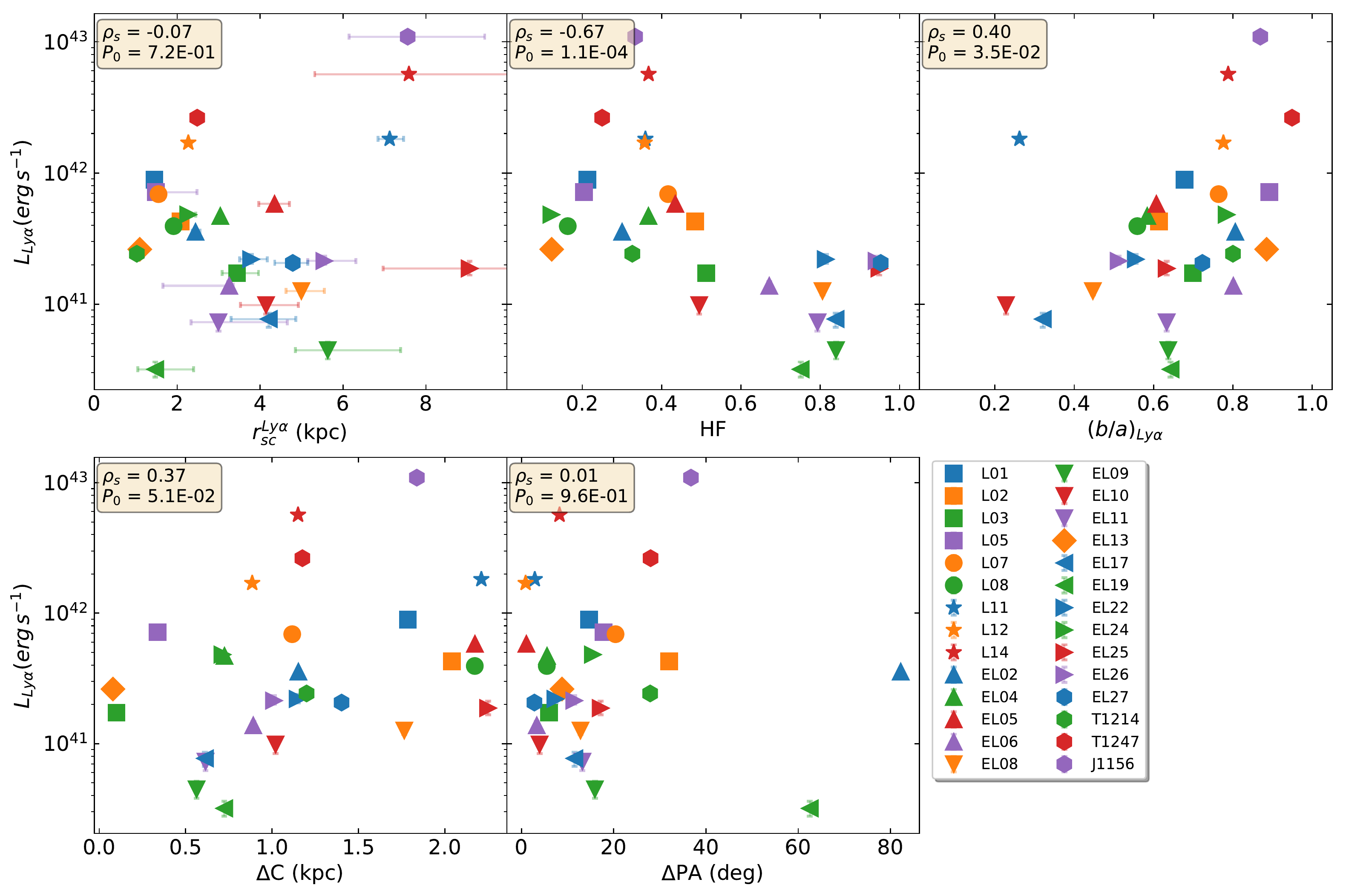}

\caption{
        How the total observed Ly$\alpha$ luminosity ($\mathrm{L_{Ly\alpha}}$) varies with the quantities used to study the Ly$\alpha$ morphology of the galaxies.
        }

\label{L_lya_vs_morph_appx_fig}
\end{figure*} 

\begin{figure*}[t!]
 \centering 
    \includegraphics[width=\textwidth] {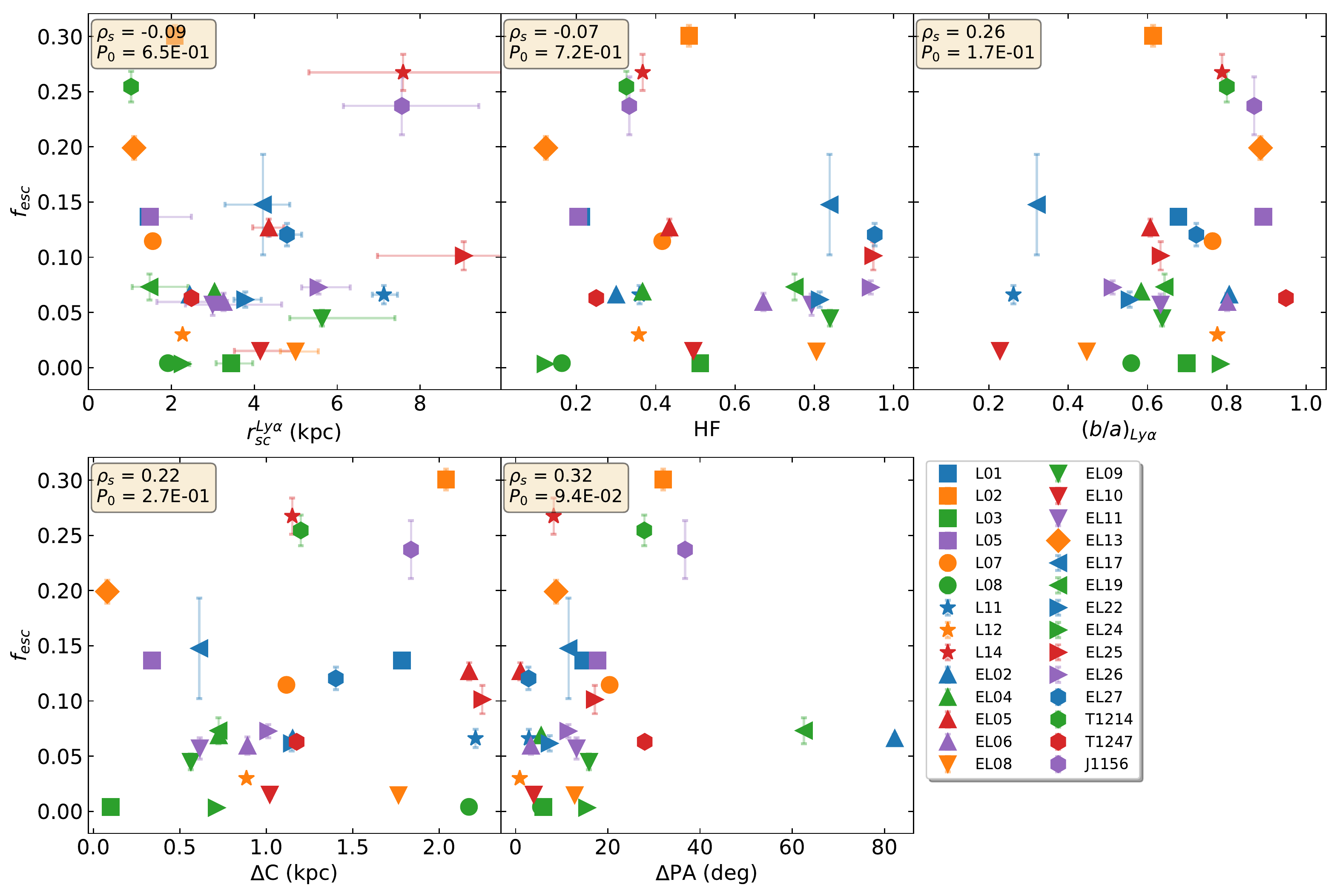}

\caption{
        How Ly$\alpha$ escape fraction ($\mathrm{f_{esc}}$) varies with the quantities used to study the Ly$\alpha$ morphology of the galaxies.
        }
\label{fesc_vs_morph_appx_fig}
\end{figure*} 

\begin{figure*}[t!]
 \centering 
    \includegraphics[width=\textwidth]{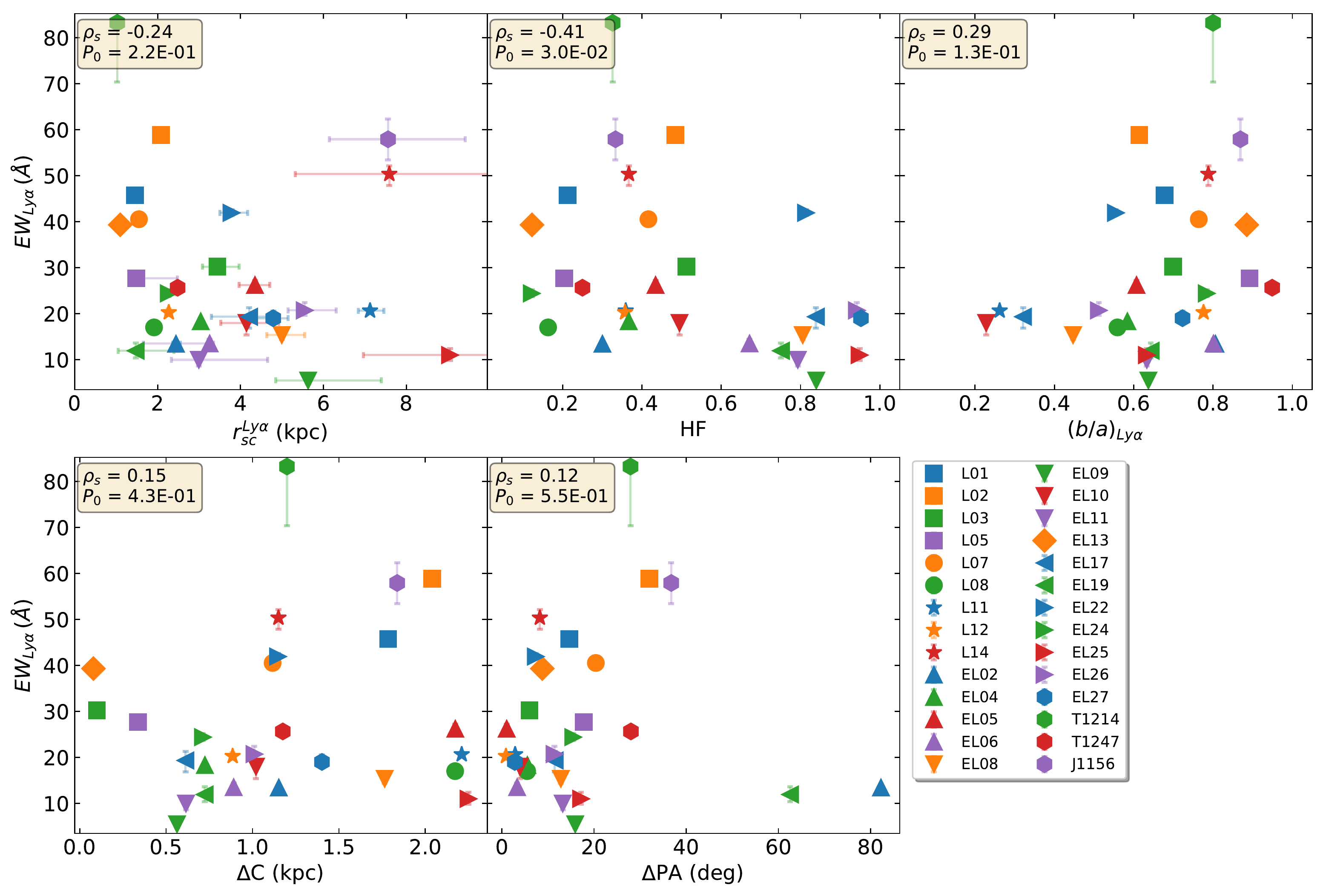}

\caption{
        How Ly$\alpha$ equivalent width ($\mathrm{EW_{Ly\alpha}}$) varies with the quantities used to study the Ly$\alpha$ morphology of the galaxies.
        }
\label{EW_vs_morph_appx_fig}
\end{figure*} 
\clearpage
\section{Ly$\alpha$ morphology versus the stellar properties}\label{lya_morph_appx}
In this section, we present how the quantities used to study the Ly$\alpha$ morphology and Ly$\alpha$ halo properties: Ly$\alpha$ halo scale length ($\mathrm{r_{sc}^{Ly\alpha}}$), Ly$\alpha$ halo fraction (HF), axis ratio ($\mathrm{(b/a)_{Ly\alpha}}$), the centroid shift ($\Delta$C), and the difference in position angles of Ly$\alpha$ and FUV ($\Delta$PA) vary with i) other Ly$\alpha$ morphology quantities, in addition to ii) some of the global quantities: stellar mass, and the nebular reddening,  iii) star-forming characteristics: the size of the SF regions, and the average intensity of the FUV SB.
\begin{figure*}[t!]
 \centering 
    \includegraphics[width=\textwidth] {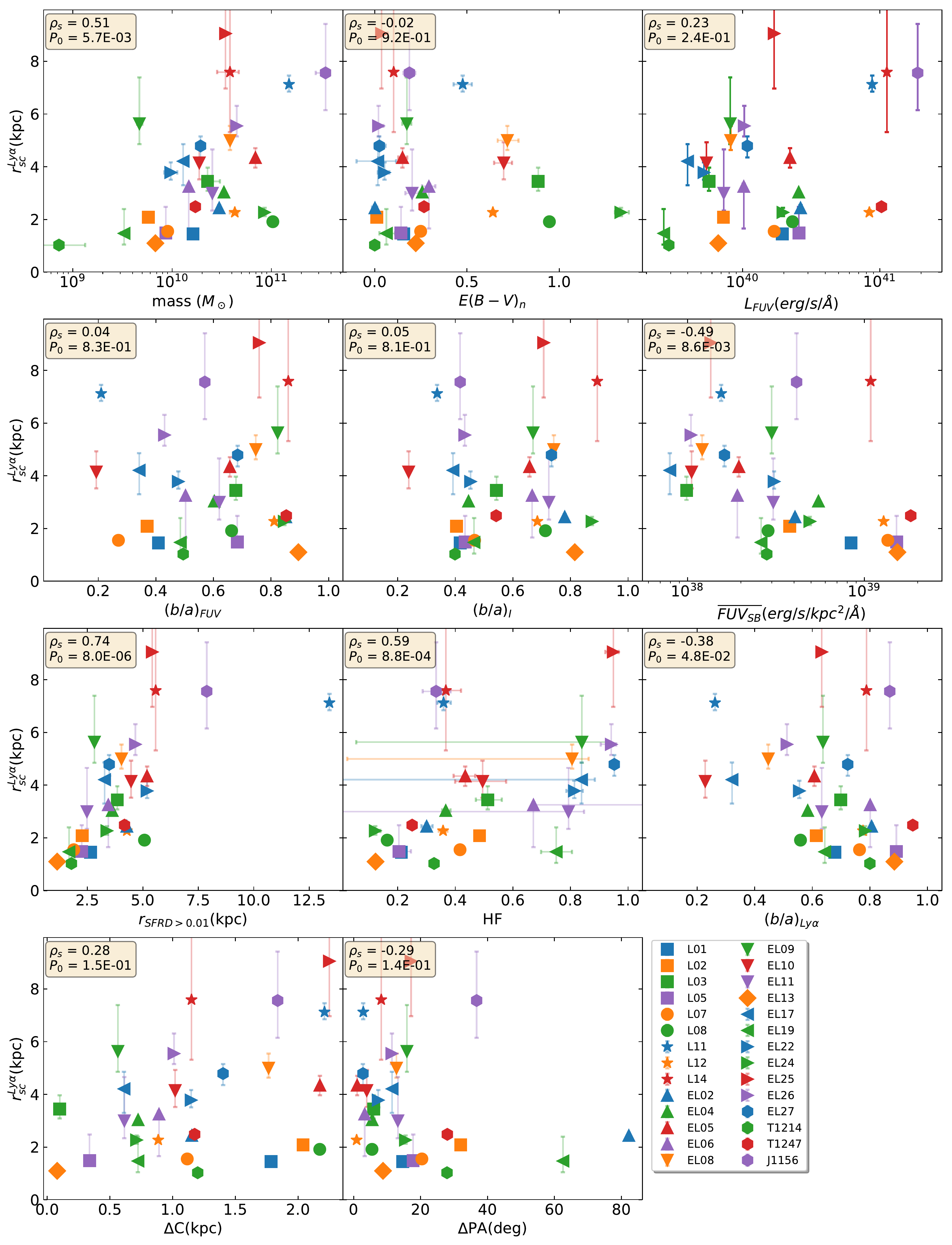}

\caption{
        How Ly$\alpha$ halo scale length ($\mathrm{r_{sc}^{Ly\alpha}}$) varies with other Ly$\alpha$ morphology quantities, and the stellar properties of the galaxies.
        }

    \label{rsc_vs_stellar_prop_appx_fig}
\end{figure*} 

\begin{figure*}[t!]
 \centering 
    \includegraphics[width=\textwidth] {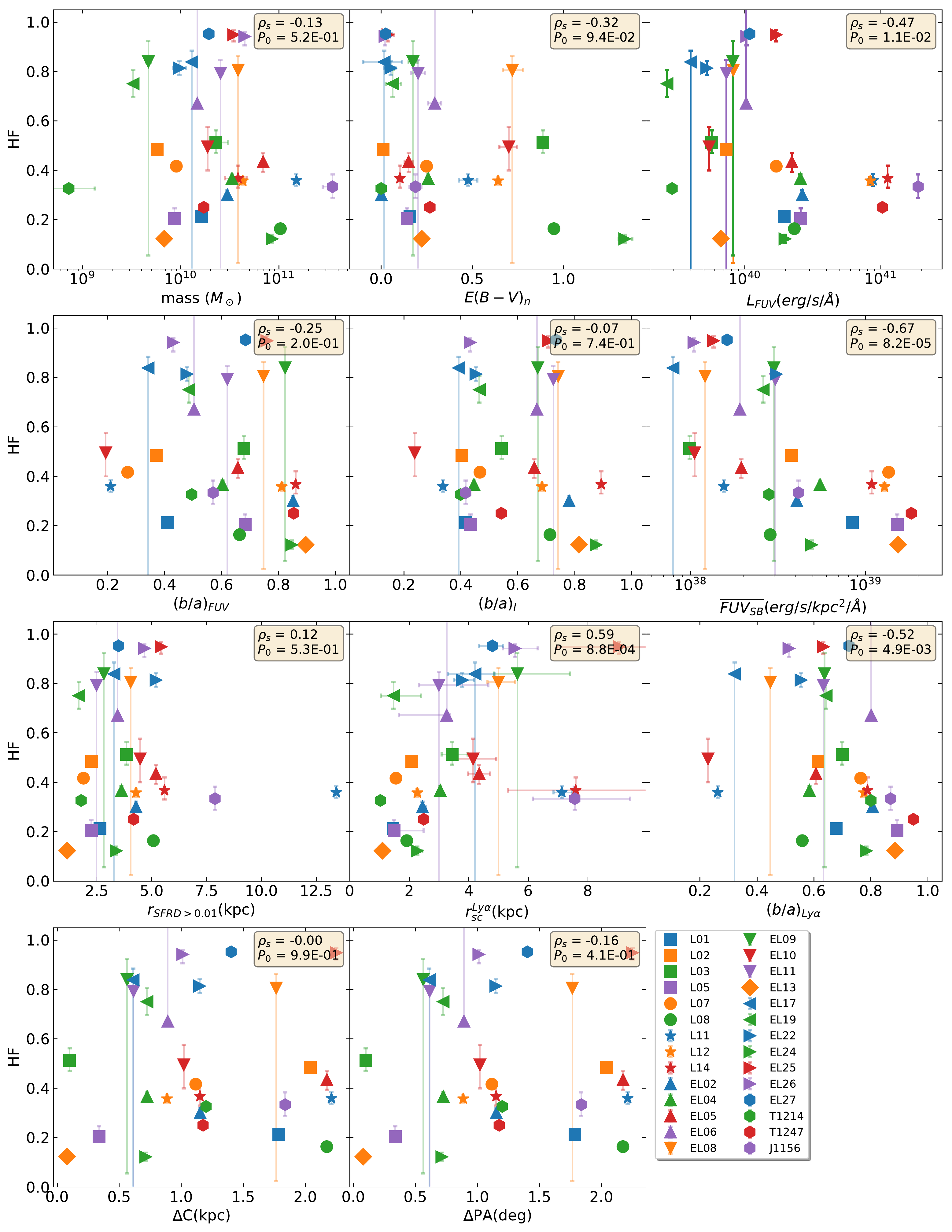}

\caption{
        How the Ly$\alpha$ halo fraction (HF) varies with other Ly$\alpha$ morphology quantities, and the stellar properties of the galaxies.
        }
        
\label{hf_vs_stellar_prop_appx_fig}
\end{figure*} 

\begin{figure*}[t!]
 \centering 
    \includegraphics[width=\textwidth] {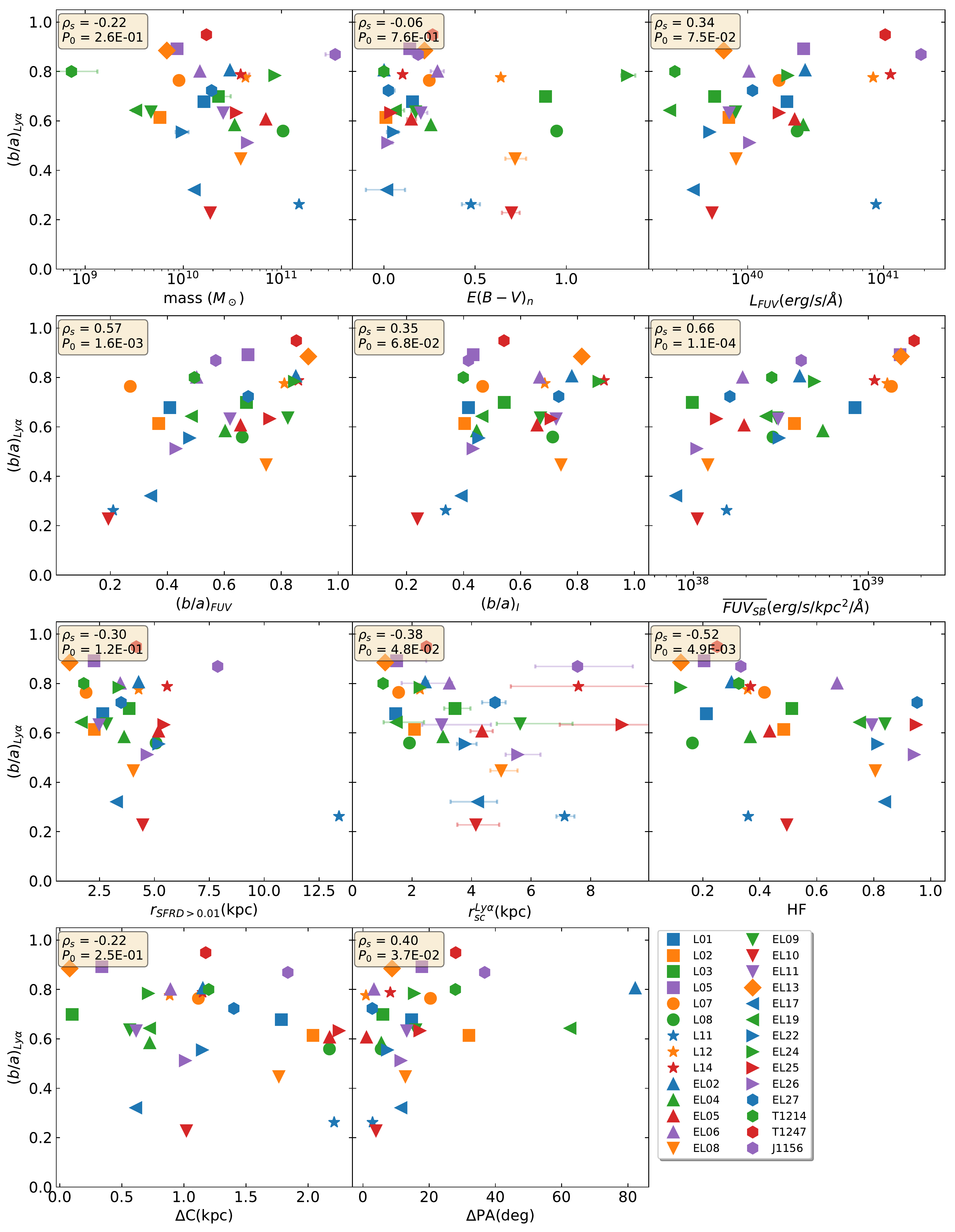}

\caption{
        How Ly$\alpha$ axis ratio ($\mathrm{(b/a)_{Ly\alpha}}$) varies with other Ly$\alpha$ morphology quantities, and the stellar properties of the galaxies.
        }
 
\label{b_a_vs_stellar_prop_appx_fig}
\end{figure*} 

\begin{figure*}[t!]
 \centering 
    \includegraphics[width=\textwidth] {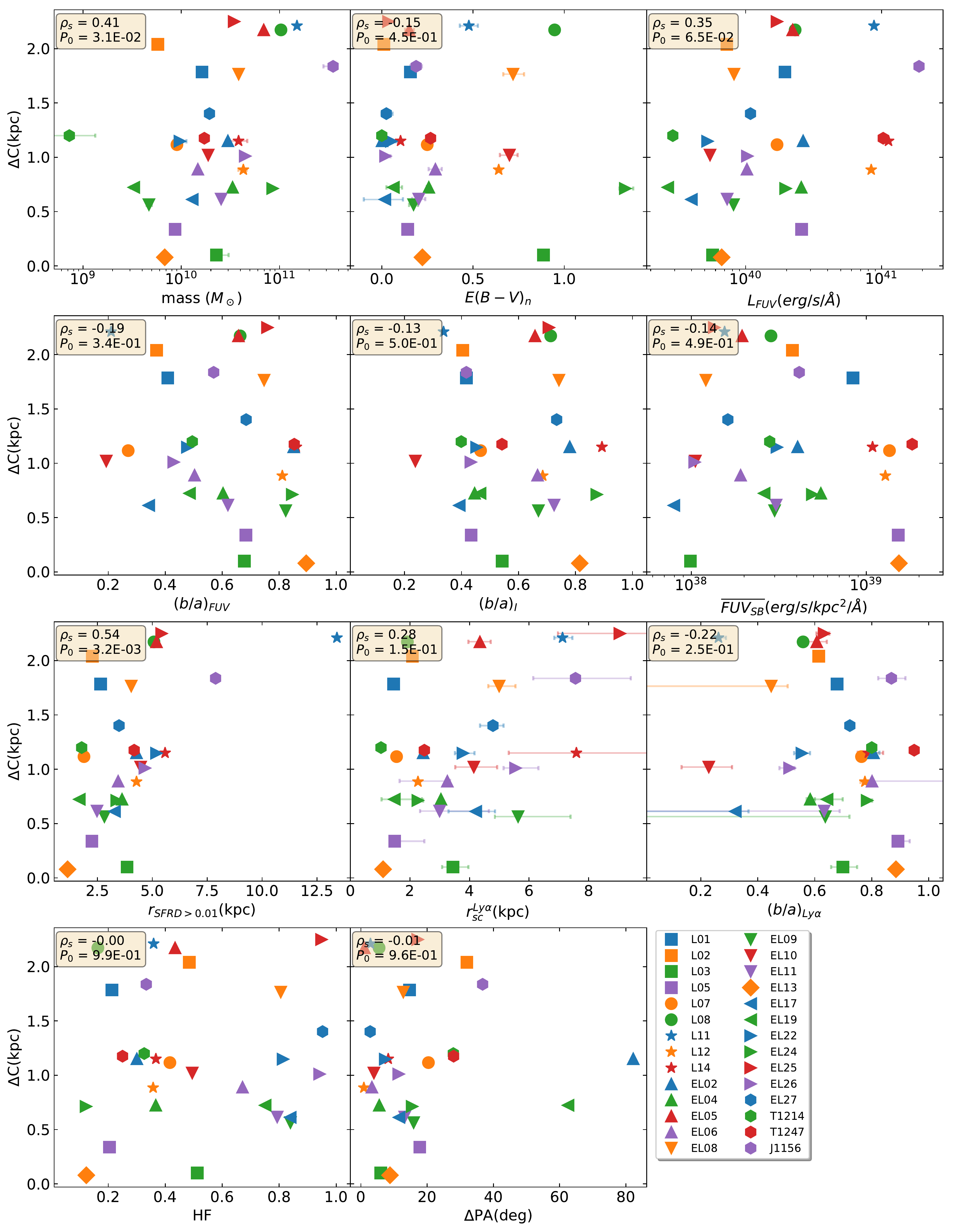}

\caption{
        How the centroid shift ($\Delta$C) varies with other Ly$\alpha$ morphology quantities, and the stellar properties of the galaxies.
        }
\label{cf_vs_stellar_prop_appx_fig}
\end{figure*} 

\begin{figure*}[t!]
 \centering 
    \includegraphics[width=\textwidth] {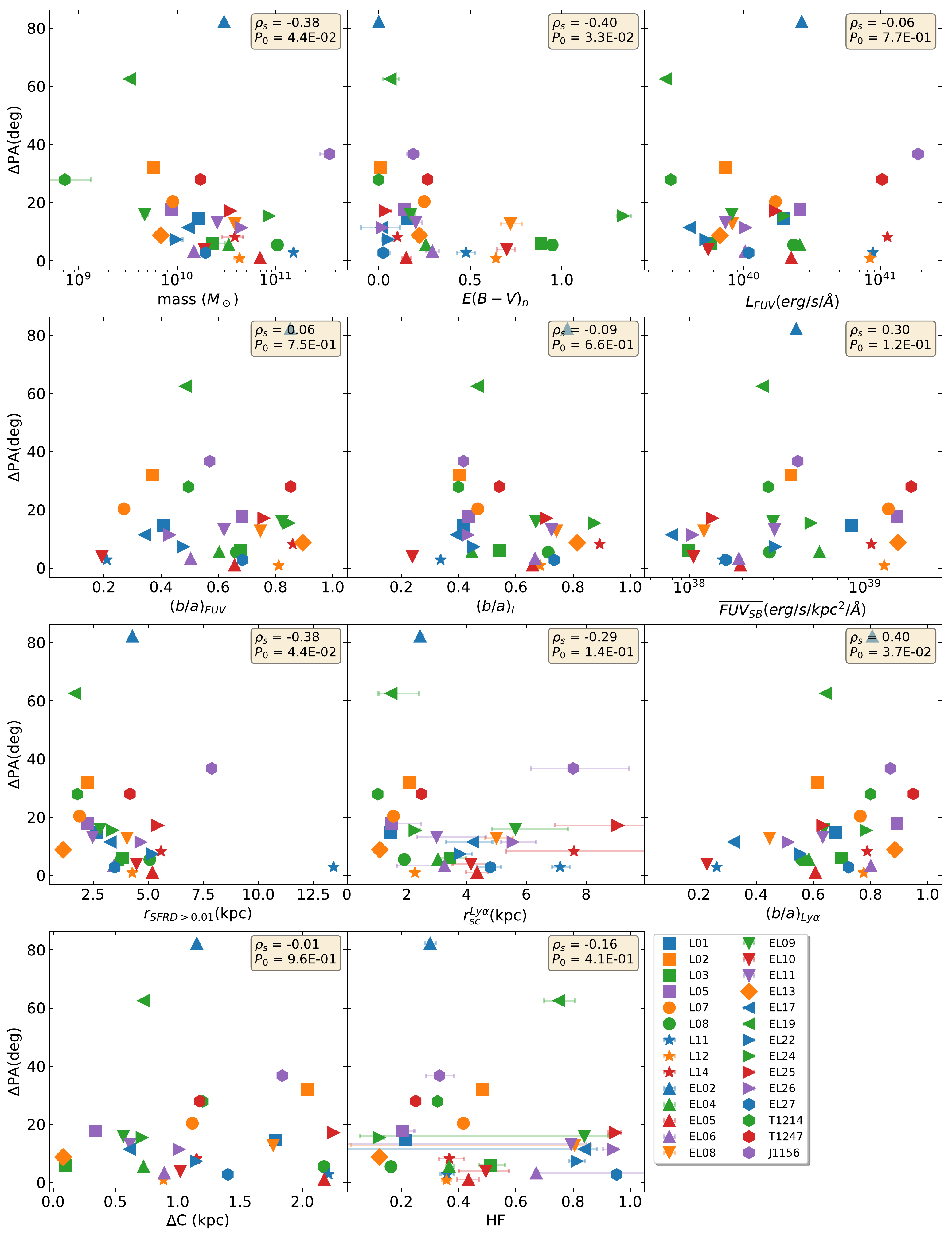}

\caption{
        How the difference in position angles of Ly$\alpha$ and FUV ($\Delta$PA) varies with other Ly$\alpha$ morphology quantities, and the stellar properties of the galaxies.
        }
        
\label{delta_pa_vs_stellar_prop_appx_fig}
\end{figure*} 
\clearpage
\section{Ly$\alpha$ isophotal halo scale length}\label{r_iso_appx_vs_all}
In this section, we present how the Ly$\alpha$ halo scale length measured from the isophotal approach ($\mathrm{r_{sc}^{iso}}$) varies with i) the Ly$\alpha$ morphology and Ly$\alpha$ halo properties: Ly$\alpha$ halo fraction (HF), axis ratio ($\mathrm{(b/a)_{Ly\alpha}}$), the centroid shift ($\Delta$C), and the difference in position angles of Ly$\alpha$ and FUV ($\Delta$PA) vary with ii) some of the global quantities: stellar mass, and the nebular reddening,  iii) star-forming characteristics: the size of the SF regions, and the average intensity of the FUV SB.
\begin{figure*}[t!]
 \centering 
    \includegraphics[width=\textwidth] {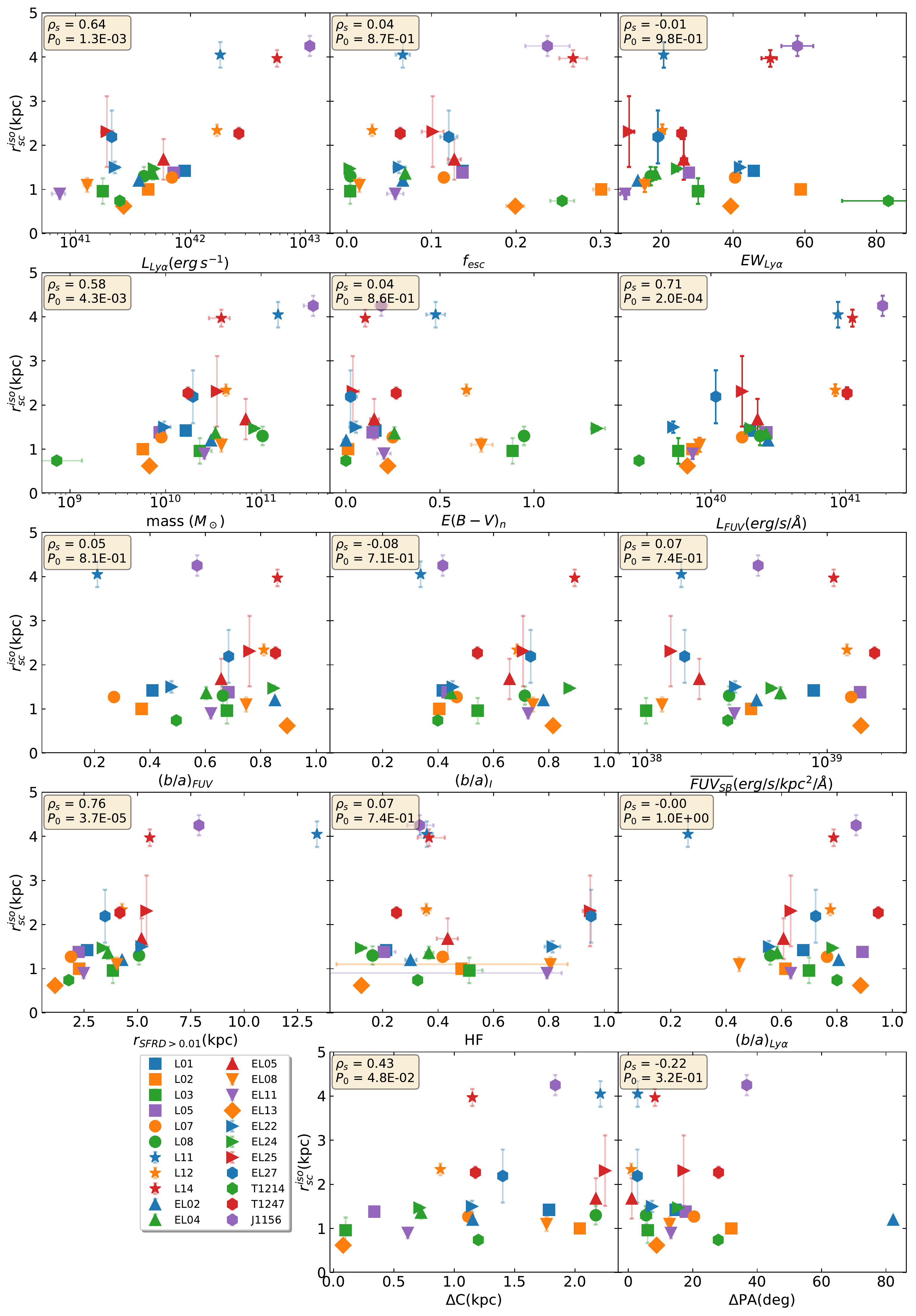}

\caption{
        How the Ly$\alpha$ halo scale length measured from the isophotal approach ($\mathrm{r_{sc}^{iso}}$) varies with the Ly$\alpha$ morphology quantities, and the stellar properties of the galaxies.
        }
        
\label{r_iso_vs_all_appx}
\end{figure*} 

\end{appendix}

\end{document}